\documentclass{article}
\usepackage[utf8]{inputenc}
\usepackage[margin=1in]{geometry}

\usepackage[bottom]{footmisc}
\usepackage{microtype}
\usepackage{graphicx}
\usepackage{subfigure}
\usepackage{booktabs}
\usepackage{natbib}
\setcitestyle{open={(},close={)}}
\usepackage{array}
\newcolumntype{?}{!{\vrule width 1pt}}

\usepackage[colorlinks=true,citecolor=blue]{hyperref}

\usepackage{amsmath,amsthm,amsfonts,amssymb,mathdots,mathtools,array,mathrsfs,bm,bbm,stmaryrd,graphicx,subfigure,xcolor}
\usepackage{algpseudocode,algorithm,algorithmicx}
\usepackage{breakcites}

\usepackage[T1]{fontenc}
\usepackage{enumerate}
\usepackage{inputenc}

\usepackage{graphicx} 
\usepackage{subfigure}

\usepackage{booktabs,balance}
\usepackage{rotating}
\usepackage{boldline}
\usepackage{makecell}
\usepackage{multirow}
\usepackage{balance,comment}

\usepackage{tikz}

\newtheorem{theorem}{Theorem}[section]

\newtheorem{definition}{Definition}[section]

\newcommand{\bsmat}{\begin{bmatrix} }
\newcommand{\esmat}{\end{bmatrix} }

\DeclareMathOperator*{\argmin}{argmin}

\usepackage{environ}
\NewEnviron{smallequation}{%
    \begin{equation}
    \scalebox{0.97}{$\BODY$}
    \end{equation}
    }
    \NewEnviron{smallalign}{%
    \begin{equation}
    \scalebox{0.97}{$\BODY$}
    \end{equation}
    }

\begin{document}

\title{\bf C-MinHash:\\ Practically Reducing Two Permutations to Just One }

\author{\textbf{Xiaoyun Li and Ping Li} \\\\
Cognitive Computing Lab\\
Baidu Research\\
10900 NE 8th St. Bellevue, WA 98004, USA\\
  \texttt{\{xiaoyunli,\ liping11\}@baidu.com}
}

\date{\vspace{0.5in}}
\maketitle

\begin{abstract}\vspace{0.1in}

\noindent Traditional minwise hashing (MinHash) requires applying $K$ independent permutations to estimate the Jaccard similarity in massive binary (0/1) data, where $K$ can be (e.g.,) 1024 or even larger, depending on applications. The recent work on C-MinHash~\citep{CMH2Perm2021} has shown, with rigorous proofs, that only {\bf two} permutations are needed. An initial permutation is applied to break whatever structures which might exist in the data, and a second permutation is re-used $K$ times to produce $K$ hashes,  via a circulant shifting fashion.  \cite{CMH2Perm2021} has proved that, perhaps surprisingly,  even though the $K$ hashes are correlated, the estimation variance is strictly smaller than the variance of the traditional MinHash.

\vspace{0.1in}

\noindent It has been demonstrated in~\cite{CMH2Perm2021} that the initial permutation in C-MinHash is indeed necessary. For the ease of theoretical analysis, they have used two independent permutations. In this paper, we show that one can actually simply use one permutation. That is, one single permutation is used for both the initial pre-processing step to break the structures in the data and the circulant hashing step to generate $K$ hashes. Although the theoretical analysis becomes very complicated, we are able to explicitly write down the expression for the expectation of the estimator. The new estimator  is no longer unbiased but the bias is extremely small and has essentially no impact on the estimation accuracy (mean square errors). An extensive set of experiments are provided to verify our claim for using just~one~permutation.
\end{abstract}

\newpage

\section{Background}

Given two binary data vectors $\bm v, \bm w \in \{0,1\}^D$ in $D$ dimensions, the Jaccard (resemblance) similarity $J(\bm v, \bm w)$ is defined as
\begin{align}  \label{def:a,f}
    J(\bm v,\bm w)\triangleq \frac{a}{f}, \hspace{0.2in} \text{ where }\     a\triangleq\sum_{i=1}^D \mathbbm 1\{v_i=w_i=1\}, \hspace{0.2in} f\triangleq\sum_{i=1}^D \mathbbm 1\{v_i=1\ \text{or}\ w_i=1\}.
\end{align}
Of course, practical applications may need to deal with billions or  thousands of billions of data vectors, not just two vectors. How to effectively store/transmit/retrieve the data and how to efficiently compute or estimate similarities among data vectors has always been a long-lasting research and engineering challenge.

\vspace{0.1in}

Minwise hashing (MinHash)~\citep{Proc:Broder,Proc:Broder_WWW97,Proc:Broder_STOC98,Proc:Li_Church_EMNLP05,Article:Li_Konig_CACM11} is a standard technique for efficiently estimating the Jaccard similarity in massive binary data. Classical MinHash requires applying $K$ (independent) random permutations on each data vector to produce $K$ hash values. The recent work by~\cite{CMH2Perm2021} proposed using just two random permutations: an initial permutation breaks whatever structures in the data vector and a second permutation is re-used $K$ times to generate $K$ hashes. They proved their surprising (and rather involved) theoretical finding that the estimation variance in their scheme is actually strictly smaller than the variance of the original MinHash.

\vspace{0.1in}

\cite{CMH2Perm2021} named their method as C-MinHash-$(\sigma,\pi)$, where $\sigma$ stands for the initial permutation and $\pi$ for the second permutation. A natural and immediate question to ask is why we really need two permutations. Indeed, \cite{CMH2Perm2021} also analyzed and experimented with C-MinHash-$(0,\pi)$, where ``0'' means that the initial permutation was not used. They reported that the performance of C-MinHash-$(0,\pi)$ was not satisfactory, because natural datasets typically do exhibit various structures.

\vspace{0.1in}

In this paper, we propose C-MinHash-$(\pi,\pi)$. That is, we just use one permutation $\pi$ for both the ``initial'' permutation to break the existing structures in the data and the ``second'' permutation to generate $K$ hashes. In~\cite{CMH2Perm2021}, they used two independent permutation mainly for simplifying the theoretical analysis, otherwise the complicated dependency would make the analysis challenging.  In this work, we are able to write down explicitly the sophisticated expression for the  expectation (mean) of the estimator for C-MinHash-$(\pi,\pi)$. Although the estimator is no longer strictly unbiased, the bias is so small that it can be safely neglected. We verify this claim via an extensive set of experiments.

\section{Review of MinHash and C-MinHash}

\begin{algorithm}[h]{
	\textbf{Input:} Binary data vector $\bm v\in\{0,1\}^D$, \hspace{0.2in}$K$ independent permutations $\pi_1,...,\pi_K$: $[D]\rightarrow[D]$.

\vspace{0.1in}	
	\textbf{Output:} $K$ hash values $h_1(\bm v),...,h_K(\bm v)$.
	
	\vspace{0.05in}
	
	For $k=1$ to $K$
	
	\vspace{0.05in}
	
	\hspace{0.2in}$h_k(\bm v) \leftarrow \min_{i:v_i\neq 0} \pi_k(i)$

	\vspace{0.05in}	
	End For
	\vspace{0.05in}
	}\caption{Minwise-hashing (MinHash) }
	\label{alg:MinHash}
\end{algorithm}

Algorithm~\ref{alg:MinHash} describes the procedure for the classical MinHash, using the example of one vector $\bm v\in\{0,1\}^D$. Note that the same set of permutations would be needed for all data vectors. After having generated $K$ hashes for both $\bm v, \bm w\in\{0,1\}^D$, i.e., $h_k(\bm v)$, $h_k(\bm w)$, $k=1, 2, ..., K$, the estimator of $J(\bm v, \bm w)$, i.e., the Jaccard similarity between $\bm v$ and $\bm w$, is simply
\begin{align}\label{MH-estimator}
    &\hat J_{MH}(\bm v,\bm w)=\frac{1}{K}\sum_{k=1}^K \mathbbm 1\{h_k(\bm v)=h_k(\bm w)\},\\
    \label{MH-var}
    &\mathbb E[\hat J_{MH}]=J,\hspace{0.4in} Var[\hat J_{MH}]=\frac{J(1-J)}{K}.
\end{align}

\begin{algorithm}[h]
{
	\textbf{Input:} Binary data vector $\bm v\in\{0,1\}^D$, \hspace{0.15in} Permutation vectors $\pi$ and $\sigma$: $[D]\rightarrow[D]$
	
	\vspace{0.05in}
	
	\textbf{Output:} Hash values $h_1(\bm v),...,h_K(\bm v)$
	
	\vspace{0.1in}

    Initial permutation: $\bm v'$ = $\sigma(\bm v)$
	
	\vspace{0.05in}
	
	For $k=1$ to $K$
	
	\vspace{0.05in}
	
	\hspace{0.2in}Shift $\pi$ circulantly rightwards by $k$ units: $\pi_k=\pi_{\rightarrow k}$
	
	\vspace{0.05in}
	
	\hspace{0.2in}$h_k(\bm v)\leftarrow \min_{i:v_i'\neq 0} \pi_{\rightarrow k}(i)$

	\vspace{0.05in}
	
	End For
	\vspace{0.05in}

	}\caption{C-MinHash-$(\sigma,\pi)$}
	\label{alg:C-MinHash initial}
\end{algorithm}

\begin{figure}[h]
\centering
\includegraphics[width=5.2in]{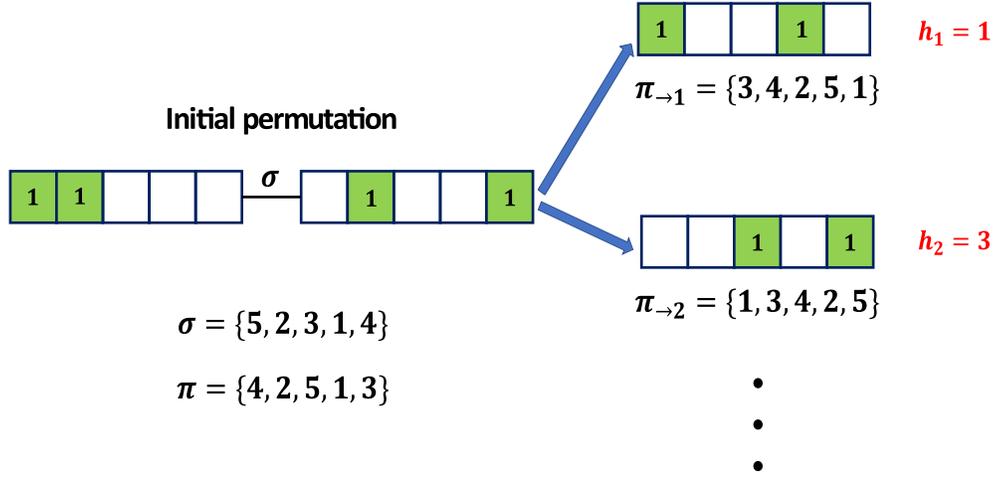}

\caption{An illustrative example of C-MinHash-$(\sigma,\pi)$ using two independent permutations: $\sigma$ for initial permutation and $\pi$ for generating $K$ hashes. The original data vector has two non-zeros, $v_1=v_2=1$. In this example, $h_1(\bm v)=1$, $h_2(\bm v)=3$.}
\label{fig:CMH}
\end{figure}

Algorithm~\ref{alg:C-MinHash initial} presents the procedure for C-MinHash-$(\sigma,\pi)$ developed in~\cite{CMH2Perm2021}, where the initial permutation $\sigma$ is first applied to break the  structures in the data and the second permutation $\pi$ is re-used $K$ times to produce $K$ hashes in a circulant shifting fashion.  Figure~\ref{fig:CMH} illustrates the procedure using a concrete example. The unbiased estimator of C-MinHash-$(\sigma,\pi)$ is then
\begin{align}  \label{est:2-perm}
    \hat J_{\sigma,\pi}(\bm v,\bm w)=\frac{1}{K}\sum_{k=1}^K \mathbbm 1\{h_k(\bm v)=h_k(\bm w)\},
\end{align}
where $h_k$'s are the hash values generated by Algorithm~\ref{alg:C-MinHash initial}. An interesting theoretical result was proved in~\cite{CMH2Perm2021} on the \textit{uniform superiority} of C-MinHash-$(\sigma,\pi)$ over the original MinHash in terms of the Jaccard estimation variance.

\begin{theorem}\citep{CMH2Perm2021} \label{theo:2 perm}
 It holds that $Var[\hat J_{\sigma,\pi}(\bm v,\bm w)]<Var[\hat J_{MH}(\bm v,\bm w)]$, $\forall \bm v,\bm w\in\{0,1\}^D$,  $K\leq D$, and $J\in (0,1)$.
\end{theorem}

The above result is surprising because it gives an example of less work leads to better performance. It was also shown in~\cite{CMH2Perm2021}  that the initial permutation $\sigma$ is necessary otherwise the estimation accuracy typically would drop due to the existing structures in the original data. This theoretical result can be beneficial in the design of hashing methods. For example, to ensure the estimation accuracy strictly follows the theory,  a naive implementation of the original MinHash  would be simply to store $K$ permutations: $[D]\rightarrow[D]$. When $D=2^{30}$ (which might be sufficient for many applications), it is unrealistic to store $K$ such permutations if $K=1024$. On the other hand, it is probably trivial to store just two such permutations.

\section{C-MinHash-$(\pi,\pi)$:  Reducing Two Permutations to Just One}
In this paper, we propose C-MinHash-$(\pi,\pi)$, by using just one permutation $\pi$ for both initially shuffling  the data and generating $K$ hashes via circulant shifting. Algorithm~\ref{alg:C-MinHash 1perm} is almost the same as Algorithm~\ref{alg:C-MinHash initial}.

\begin{algorithm}[h]
{
	\textbf{Input:} Binary data vector $\bm v\in\{0,1\}^D$, \hspace{0.15in} A permutation vector $\pi$: $[D]\rightarrow[D]$
	
	\vspace{0.05in}
	
	\textbf{Output:} Hash values $h_1(\bm v),...,h_K(\bm v)$
	
	\vspace{0.1in}

    Initial permutation: $\bm v'$ = $\pi(\bm v)$
	
	\vspace{0.05in}
	
	For $k=1$ to $K$
	
	\vspace{0.05in}
	
	\hspace{0.2in}Shift $\pi$ circulantly rightwards by $k$ units: $\pi_k=\pi_{\rightarrow k}$
	
	\vspace{0.05in}
	
	\hspace{0.2in}$h_k(\bm v)\leftarrow \min_{i:v_i'\neq 0} \pi_{\rightarrow k}(i)$

	\vspace{0.05in}
	
	End For
	\vspace{0.05in}

	}\caption{C-MinHash-$(\pi,\pi)$}
	\label{alg:C-MinHash 1perm}
\end{algorithm}

Analogously, we have the Jaccard estimator for C-MinHash-$(\pi,\pi)$:
\begin{align}  \label{est:1-perm}
    \hat J_{\pi,\pi}=\frac{1}{K}\sum_{k=1}^K \mathbbm 1\{h_k(\bm v)=h_k(\bm w)\},
\end{align}
with $h_k$'s are the hash values output by  Algorithm~\ref{alg:C-MinHash 1perm}. Strictly speaking,  $\hat J_{\pi,\pi}$ is no longer unbiased, but we will show that the bias has essentially no impact on the estimation accuracy in terms of the mean square error: MSE = bias$^2$ + variance.  While the theoretical analysis for C-MinHash-$(\sigma,\pi)$ in~\cite{CMH2Perm2021} was already rather involved, analyzing C-MinHash-$(\pi,\pi)$ becomes much more difficult. Nevertheless, we are able to at least derive the explicit expression for the expectation (mean) of the estimator:
\begin{align}
\mathbb{E}[ \hat J_{\pi,\pi}] = \frac{1}{K}\sum_{k=1}^K\mathbb E[\mathbbm 1\{h_k(\bm v)=h_k(\bm w)\}].
\end{align}

\begin{definition}\label{def-1}
For two binary data vectors $\bm v,\bm w\in \{0,1\}^D$, define the \textbf{location vector} as $\bm x\in\{O,\times,-\}^D$, with $\bm x_i$ being ``$O$'',``$\times$'',``$-$'' when $\bm v_i=\bm w_i=1$, $\bm v_i+\bm w_i=1$ and $\bm v_i=\bm w_i=0$, respectively.
\end{definition}

The collision of hash samples can be concretely described by the location vector $\bm x$ after permutation. If the first ``$O$'' appears before the first ``$\times$'' (counting from small to large), then the hash sample collides.   We will use hyper($N,m,n_1,...,n_p$) to denote the $p$-dimensional hyper-geometric distribution, where $N$ is the total number of instances, $m$ is number of draws and $n_i$, $i=1,...,p$ is the size of each class.

\begin{theorem} {\label{theo:mean-M2}}
Assume $K\leq D$ and let $a,f$ be defined as (\ref{def:a,f}). The location vector $\bm x$ is defined in Definition~\ref{def-1}. Denote $\mathcal B_1=\{i:\bm x_i=O\}$, $\mathcal B_2=\{i:\bm x_i=\times\}$ and $\mathcal B_3=\{i:\bm x_i=-\}$. For $a\leq j\leq D$ and $1\leq k\leq K$, define
\begin{align*}
    &\mathcal A_-(j)=\{\bm x_i:(i+k-1\  mod\ D)+1\leq j\},\quad \mathcal A_+(j)=\{\bm x_i:(i+k-1\  mod\ D)+1> j\}.
\end{align*}
Let $n_{-,1}(j)=|\{\bm x_i=O:i\in\mathcal A_-(j)\}|$ be the number of ``$O$'' points in $\mathcal A_-(j)$. Analogously let $n_{-,2}(j),n_{-,3}(j)$ be the number of ``$\times$'' and ``$-$'' points in $\mathcal A_-(j)$, and $n_{+,1}(j),n_{+,2}(j),n_{+,3}(j)$ be the number of ``$O$'', ``$\times$'' and ``$-$'' points in $\mathcal A_+(j)$. Then, for the $k$-th C-MinHash-$(\pi,\pi)$ hash collision indicator,
\begin{align*}
    &\mathbb E[\mathbbm 1\{h_k(\bm v)=h_k(\bm w)\}]=\sum_{j=1}^D\sum_{Z\in \Theta_j}P_j(Z)\cdot \Big\{ \sum_{i=1}^3\Psi_i(j)+ \sum_{i\in\mathcal B_1}(1-\frac{z_{-,1}}{n_{-,1}(i^*)})\tilde P_1\Big\},
\end{align*}
where $P_j(Z)$ is the density function of $Z=(z_{-,k}|_1^3,z_{+,k}|_1^3)$ which follows hyper($D,D-f,n_{-,k}(j)|_1^3,n_{+,k}(j)|_1^3$), with domain $\Theta_j$. For $q=1,2,3$, denote $\mathbbm 1^\#_q=\mathbbm 1\{j^\#\in \mathcal B_q\}$, and

\begin{align*}
    &\Psi_q(j)=\sum_{i\in\mathcal B_q,j<i^*}\sum_{p=1}^3 \mathbbm 1^\#_p\big(1-\frac{z_{-,p}}{n_{-,p}(j)}+\mathbbm 1\{q=3\}(2\frac{z_{-,p}}{n_{-,p}(j)}-1) \big) (1-\frac{z_{+,q}}{n_{+,q}(j)})\tilde P_q \bar J_q \\
    &\hspace{0.3in} +\sum_{i\in\mathcal B_q,j>i^*} \Big[ \mathbbm 1^\#_q (1-\frac{z_{-,q}}{n_{-,q}(j)})(1-\frac{z_{-,q}}{n_{-,q}(j)-1})+\sum_{p\neq q}^3 \mathbbm 1^\#_p (1-\frac{z_{-,q}}{n_{-,q}(j)})(1-\frac{z_{-,p}}{n_{-,p}(j)}) \Big]\tilde P_q J^*,
\end{align*}
where $i^*=(i+k-1\ mod\ D)+1$,  $i^\#=(i-k-1\ mod\ D)+1$, $\forall i$. Define $J^*=\frac{a-r_1}{f-r_1-r_2}$, and
\begin{align*}
    &\tilde P_1=\tilde P_2=\frac{1}{r_1+r_2}\frac{\binom{b_0}{r_3}\binom{D-j-b_0}{r_1+r_2-1}}{\binom{D-f}{r_3}\binom{f}{r_1+r_2}},\hspace{0.1in} \tilde P_3=\frac{1}{r_3}\frac{\binom{b_0}{r_3-1}\binom{D-j-b_0}{r_1+r_2}}{\binom{D-f}{r_3}\binom{f}{r_1+r_2}},\\
    &\bar J_q=\frac{r_1-\mathbbm 1\{q=1\}}{D-j-b_0}+(1-\frac{r_1+r_2-\mathbbm 1\{q\neq 3\}}{D-j-b_0})J^*,\  q=1,2,3,
\end{align*}
where $b_0=\sum_{k=1}^3z_{+,k}$, $r_1=a-z_{-,1}-z_{+,1}$, $r_2=f-a-z_{-,2}-z_{+,2}$ and $r_3=D-f-z_{-,3}-z_{+,3}$.

\end{theorem}

\vspace{0.1in}

Theorem~\ref{theo:mean-M2} says that   $\mathbb E[\mathbbm 1\{h_k(\bm v)=h_k(\bm w)\}]$ would be different for different $k$. For each $k$, the absolute value of the bias is typically very small, and furthermore, this bias would be averaged out with $K$ hash samples (as illustrated in Figure~\ref{fig:bias} and Figure~\ref{fig:mean}). Also, from the proof (see Appendix~\ref{sec:append proof}) we can know that when $a$ and $f$ are fixed, as $D$ increases, $\mathbb E [\hat J_{\pi,\pi}]$ would converge to $J$.


\begin{figure}[b!]

\vspace{-0.2in}

  \begin{center}
   \mbox{
    \includegraphics[width=2.1in]{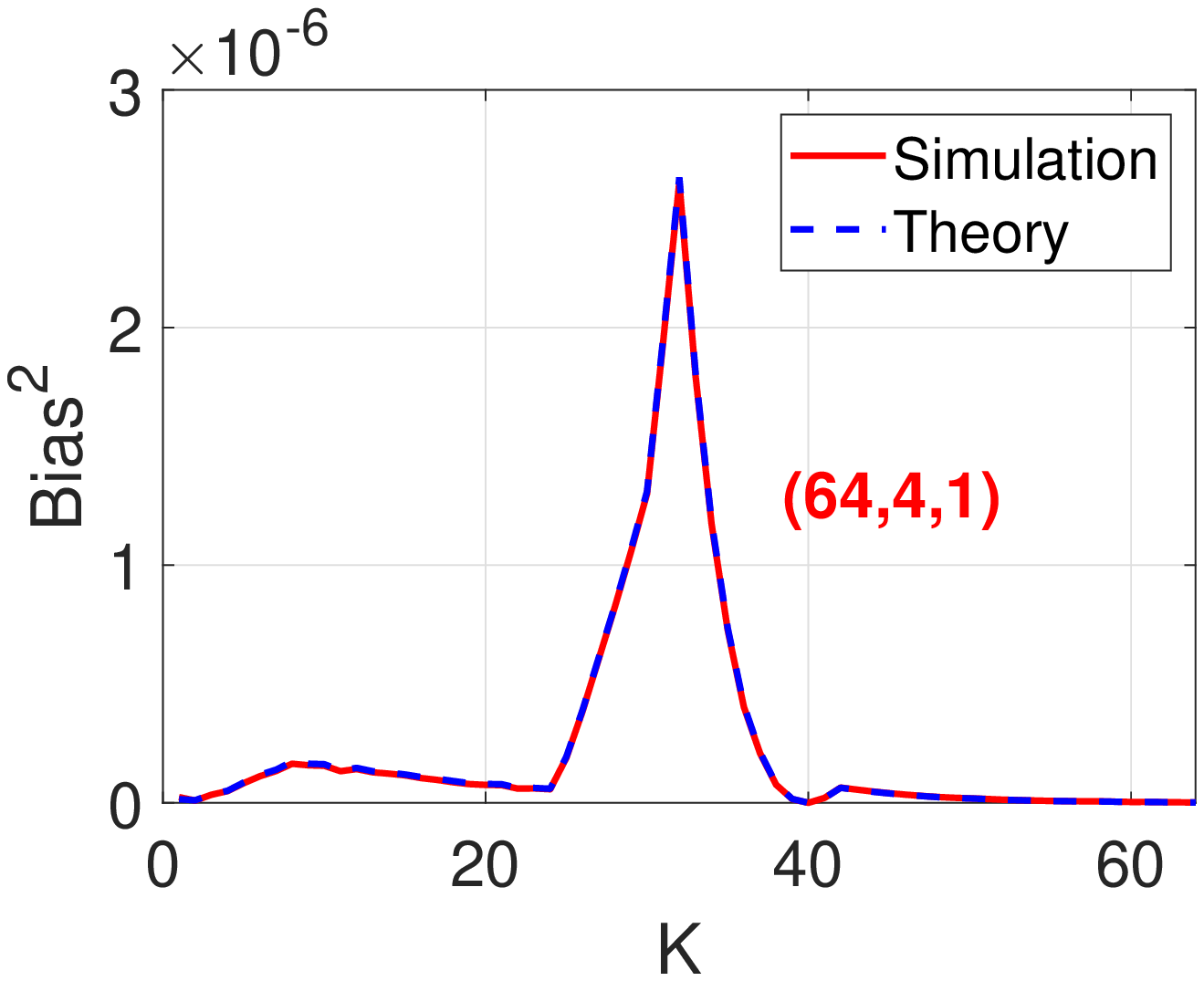}
    \includegraphics[width=2.1in]{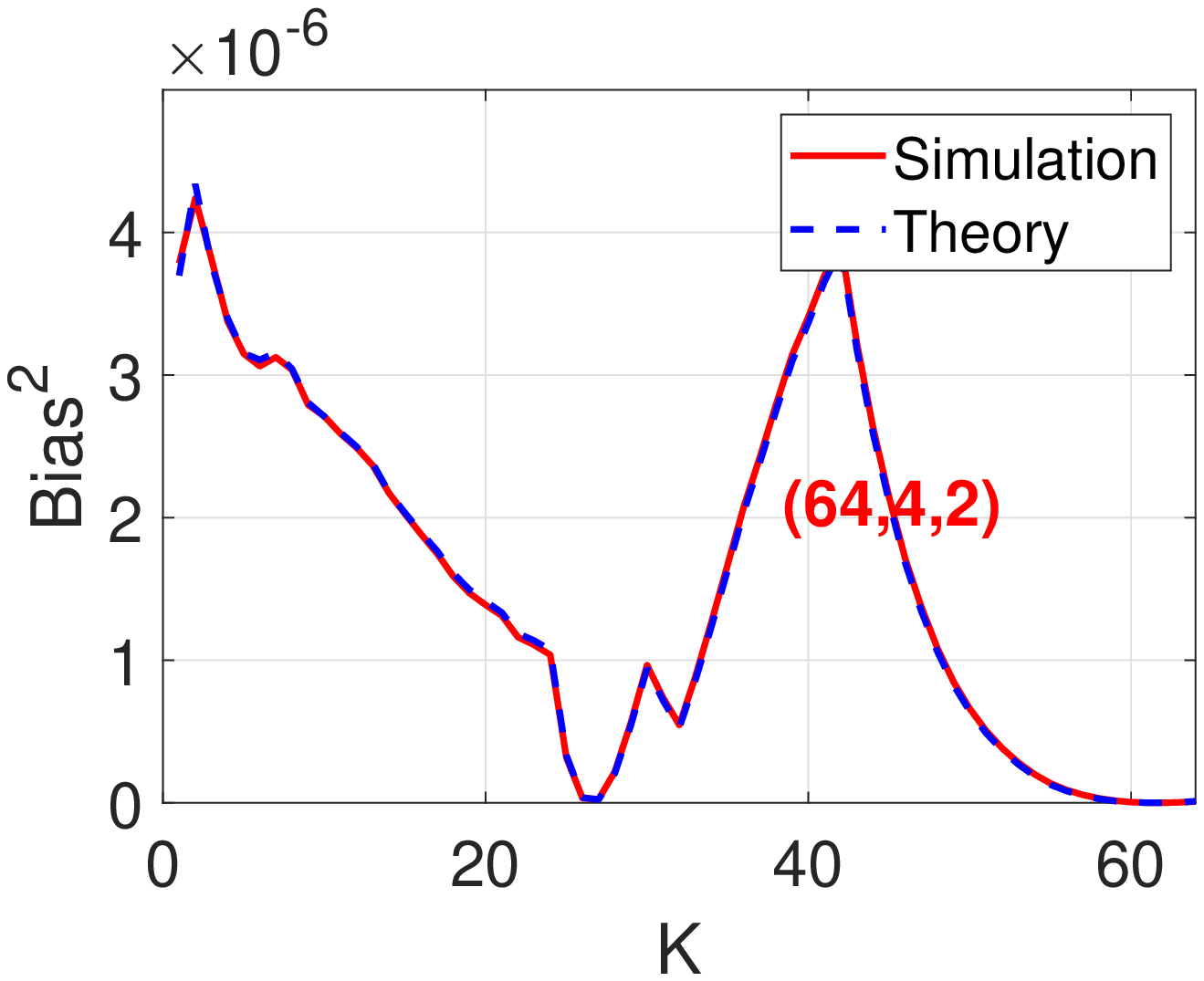}
    \includegraphics[width=2.1in]{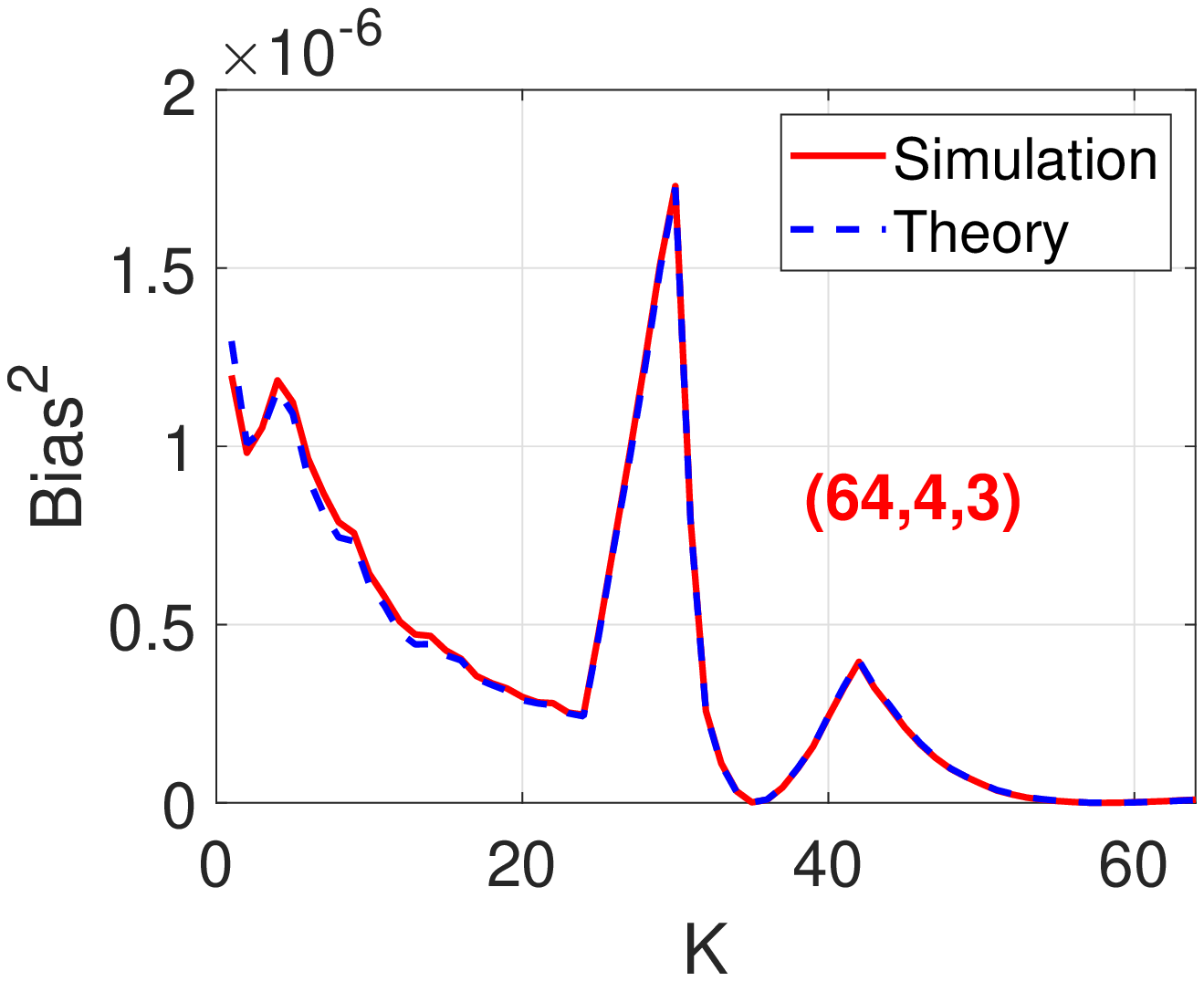}
    }
    \mbox{
    \includegraphics[width=2.1in]{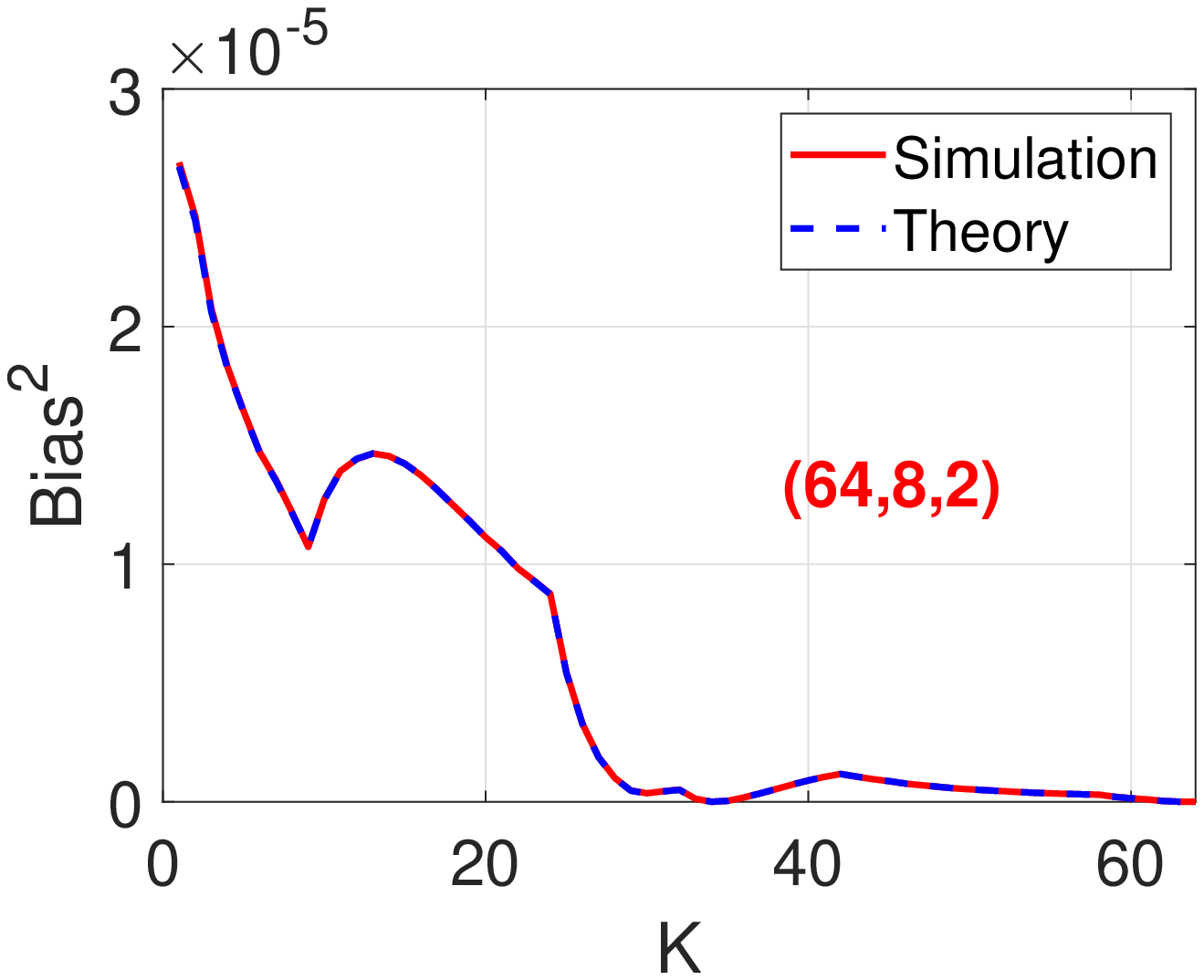}
    \includegraphics[width=2.1in]{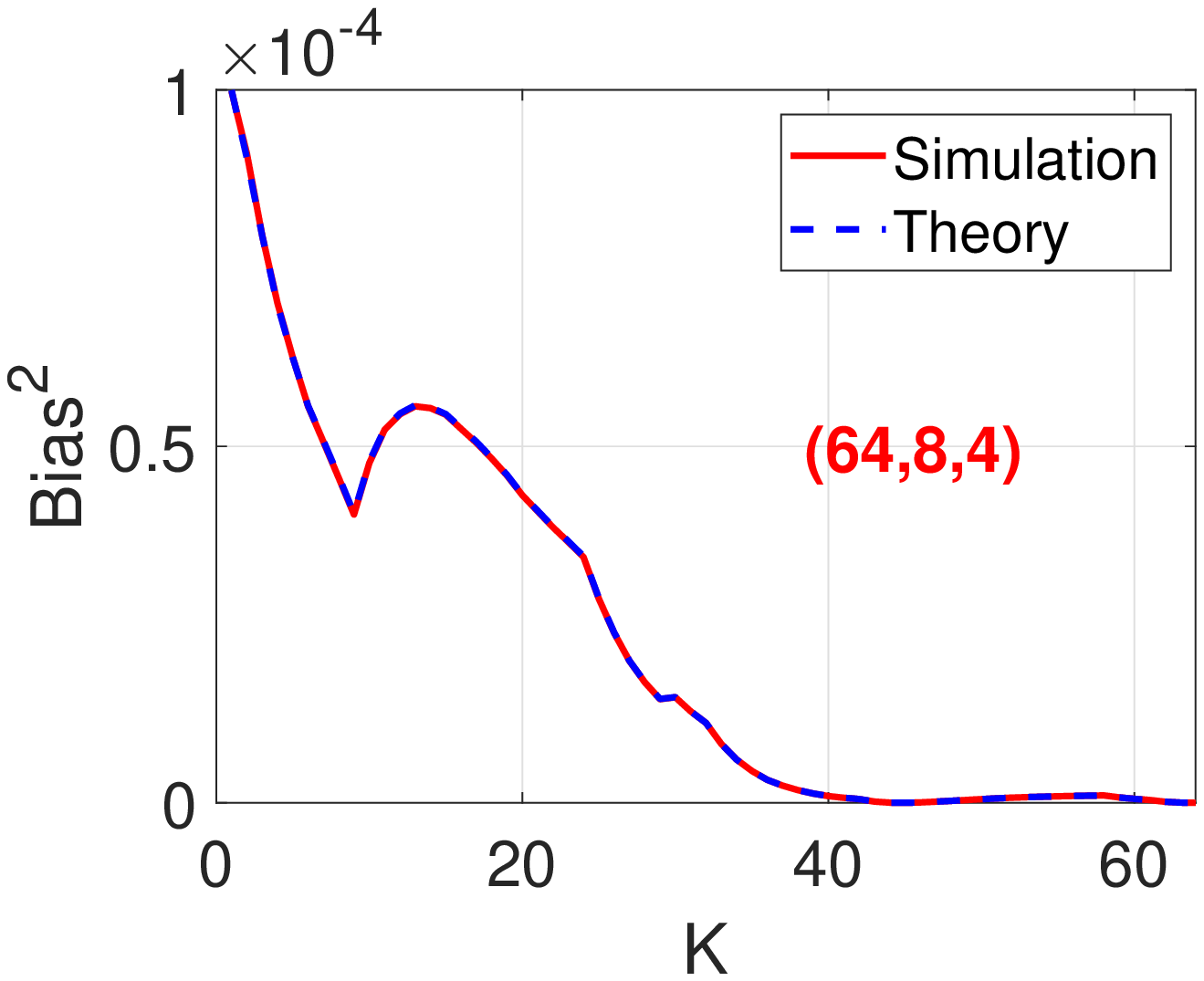}
    \includegraphics[width=2.1in]{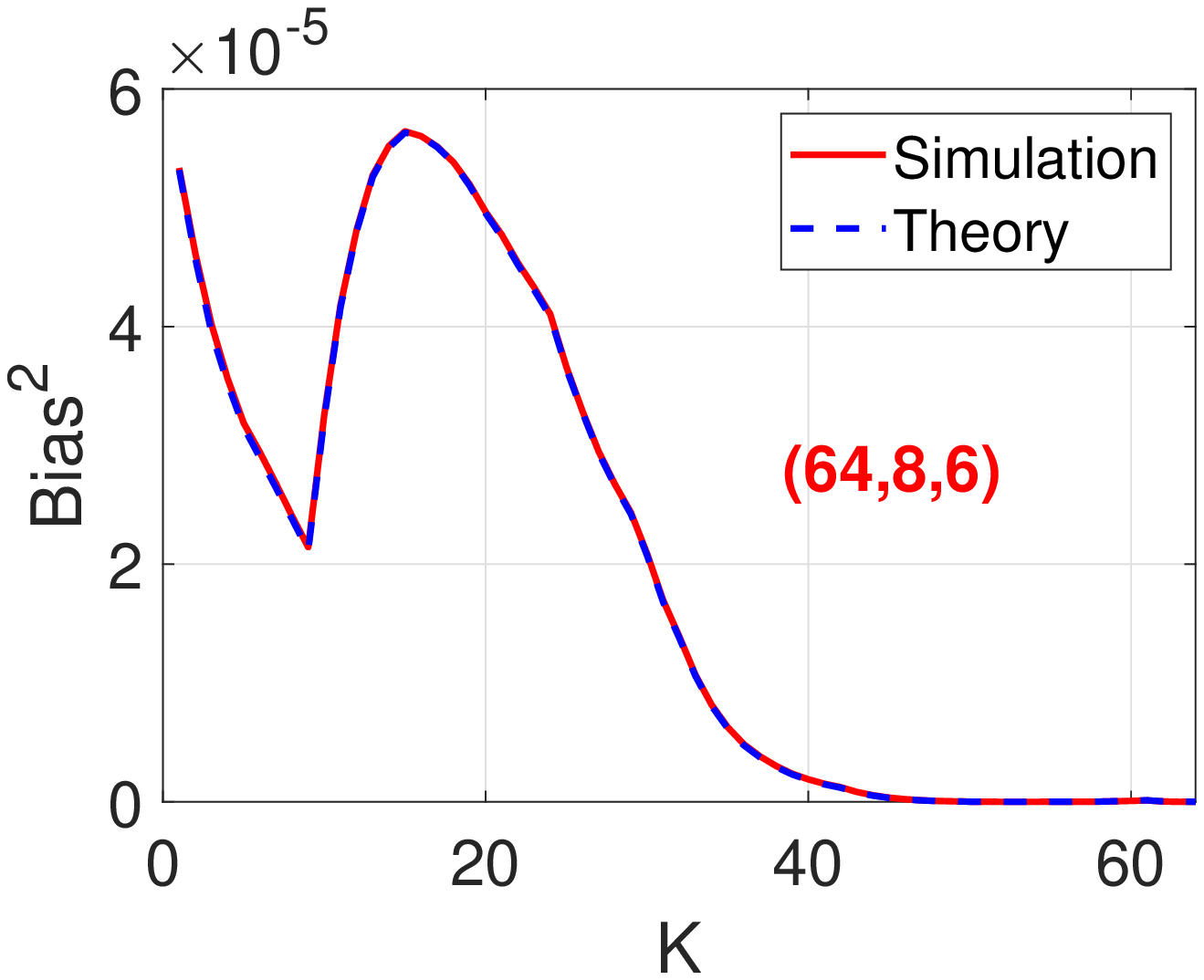}
    }
    \mbox{
    \includegraphics[width=2.1in]{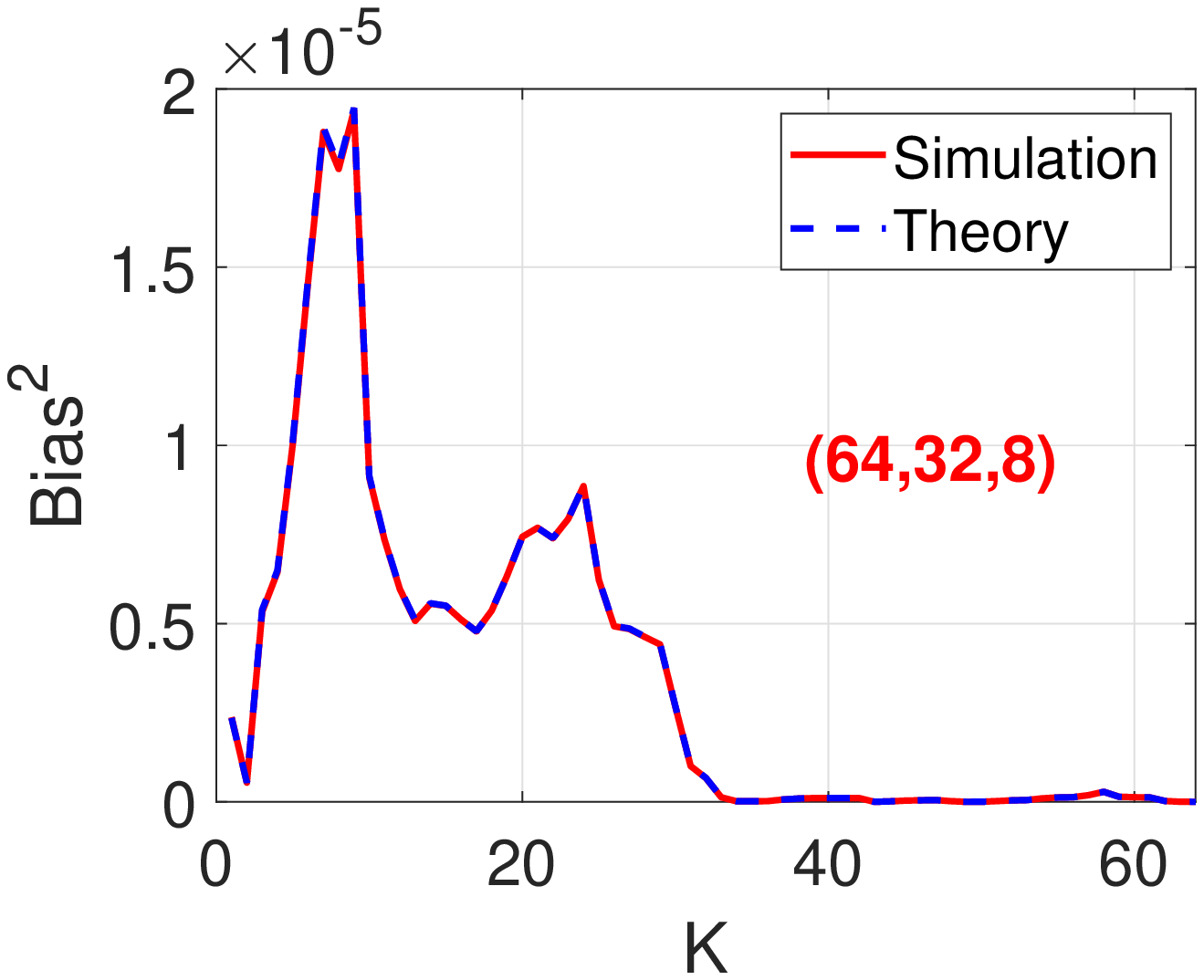}
    \includegraphics[width=2.1in]{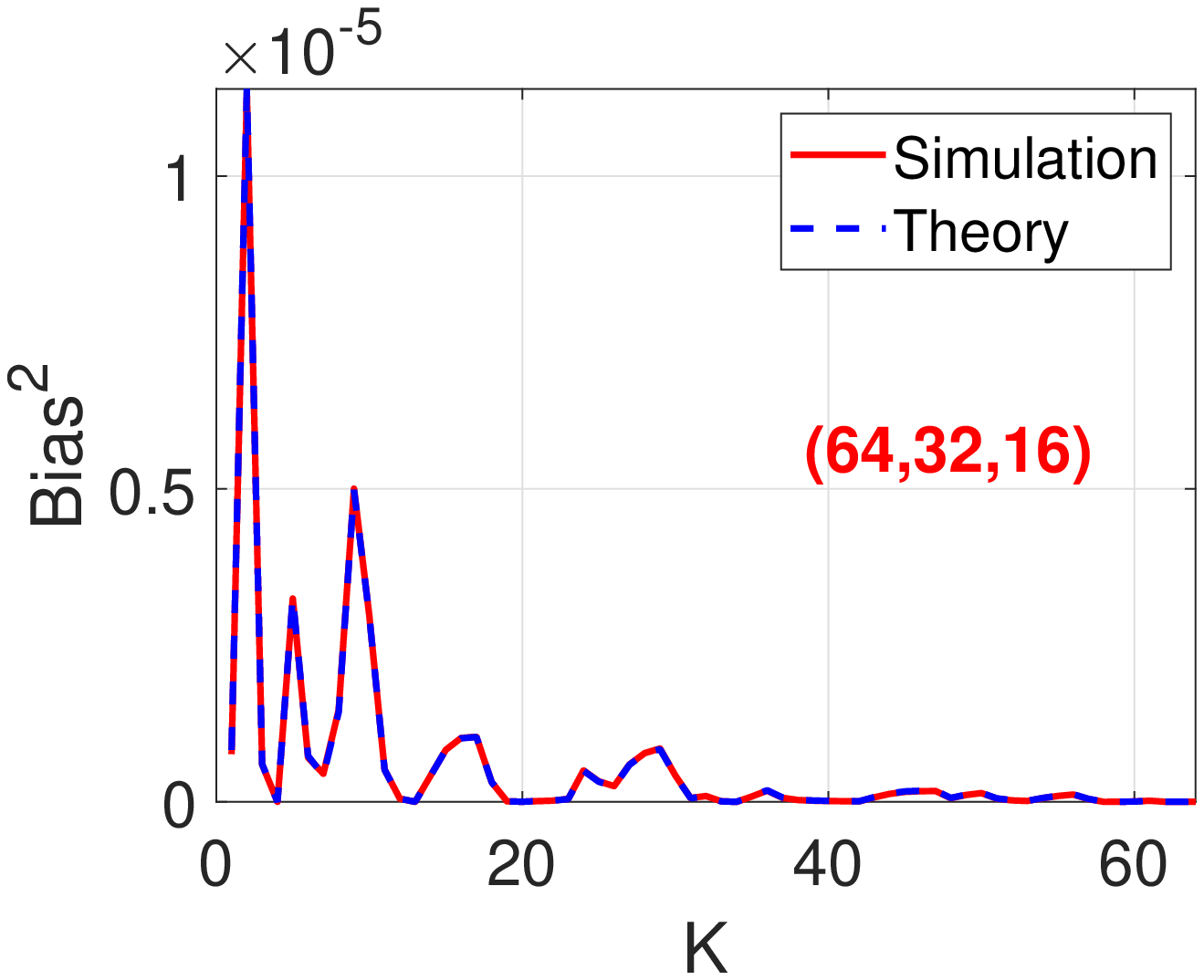}
    \includegraphics[width=2.1in]{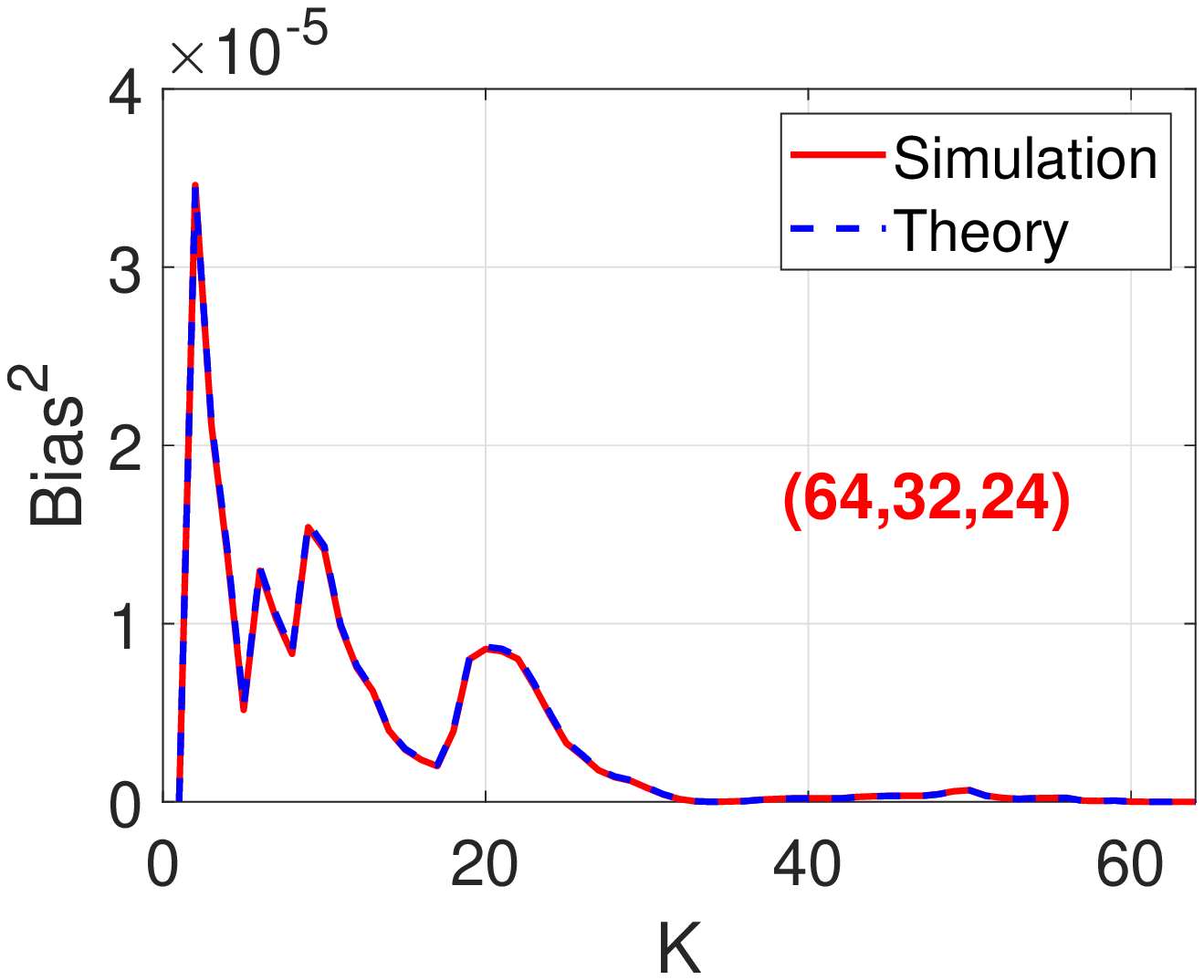}
    }
  \end{center}
  \vspace{-0.26in}
  \caption{bias$^2 = (\mathbb E[\hat J_{\pi,\pi}]-J)^2$ vs. number of hashes $K$ on simulated data pairs with various $(D,f,a)$ where $D=64$. Red solid curve is the empirical bias$^2$, and the blue dashed curve is based on Theorem~\ref{theo:mean-M2}. }
  \label{fig:bias}\vspace{-0.34in}
\end{figure}

\vspace{0.1in}
Figure~\ref{fig:bias} presents numerical examples to validate the theory and demonstrate the magnitude of bias$^2$ (recall MSE = bias$^2$ + variance). We simulate data pairs with a series of $(D,f,a)$ values, where the dimension is fixed as $D=64$ and we vary $f$ and $a$ (recall $a\triangleq\sum_{i=1}^D \mathbbm 1\{v_i=w_i=1\},\  f\triangleq\sum_{i=1}^D \mathbbm 1\{v_i=1\ \text{or}\ w_i=1\}$.) The non-zero entries are randomly assigned. The simulations match perfectly Theorem~\ref{theo:mean-M2}. As we can see, bias$^2$ is very small ($10^{-5}$ or even smaller) and approaches 0 as $K$ increases (i.e., the averaging effect).

\vspace{0.1in}

Furthermore, Figure~\ref{fig:mean} plots $\mathbb E[\mathbbm 1\{h_k(\bm v)=h_k(\bm w)\}]$ for every $k$,  $1\leq k\leq K$. Again, the simulation results perfectly match the theory in Theorem~\ref{theo:mean-M2}. As $\mathbb E[\mathbbm 1\{h_k(\bm v)=h_k(\bm w)\}]$ can be positive or negative, the overall bias $\mathbb{E}[ \hat J_{\pi,\pi}] = \frac{1}{K}\sum_{k=1}^K\mathbb E[\mathbbm 1\{h_k(\bm v)=h_k(\bm w)\}]$ would approach 0 as $K$ increases, as verified in Figure~\ref{fig:bias}.

\vspace{0.1in}

\begin{figure}[H]
	\begin{center}
		\mbox{
		\includegraphics[width=2.1in]{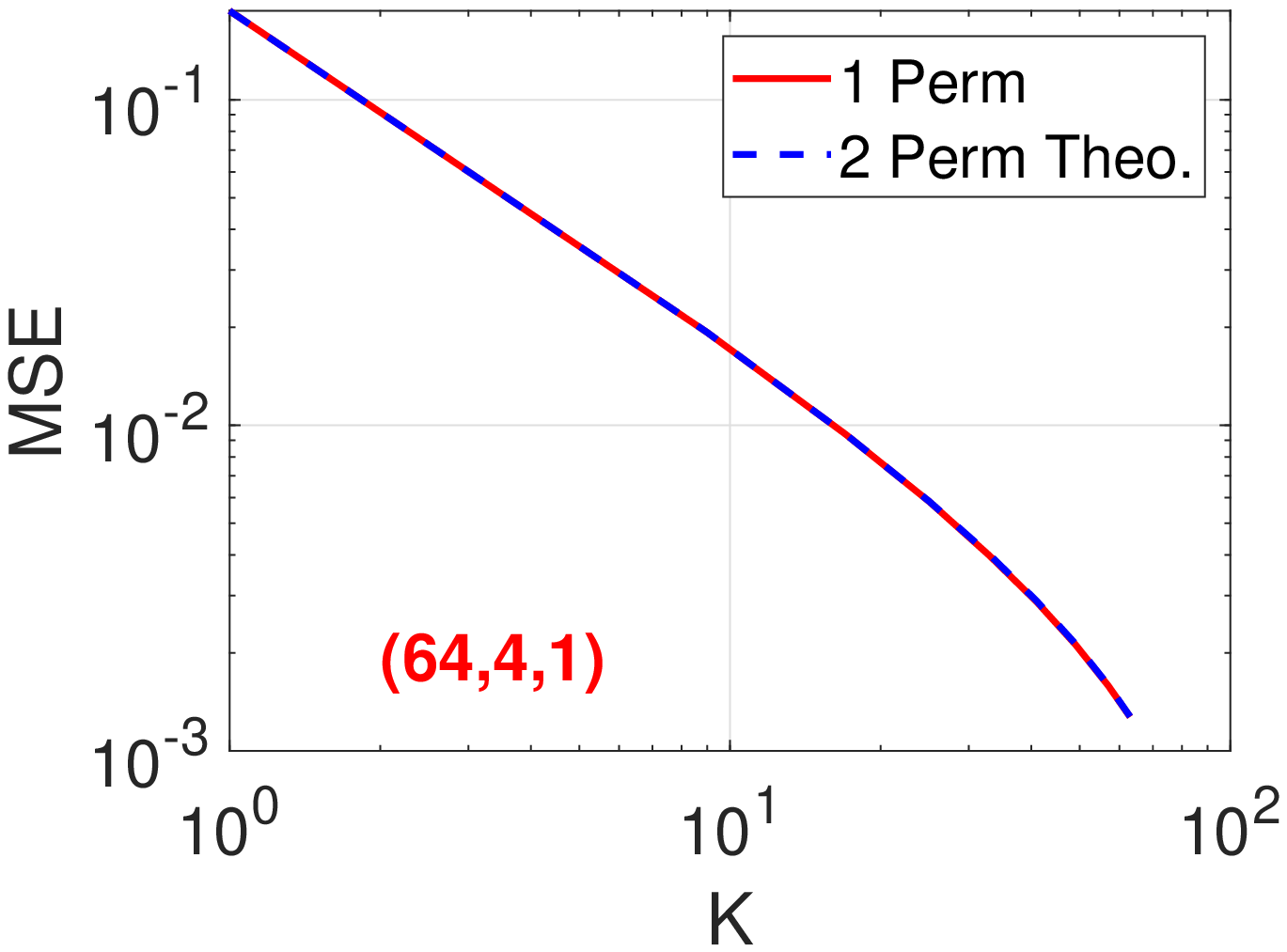}
		\includegraphics[width=2.1in]{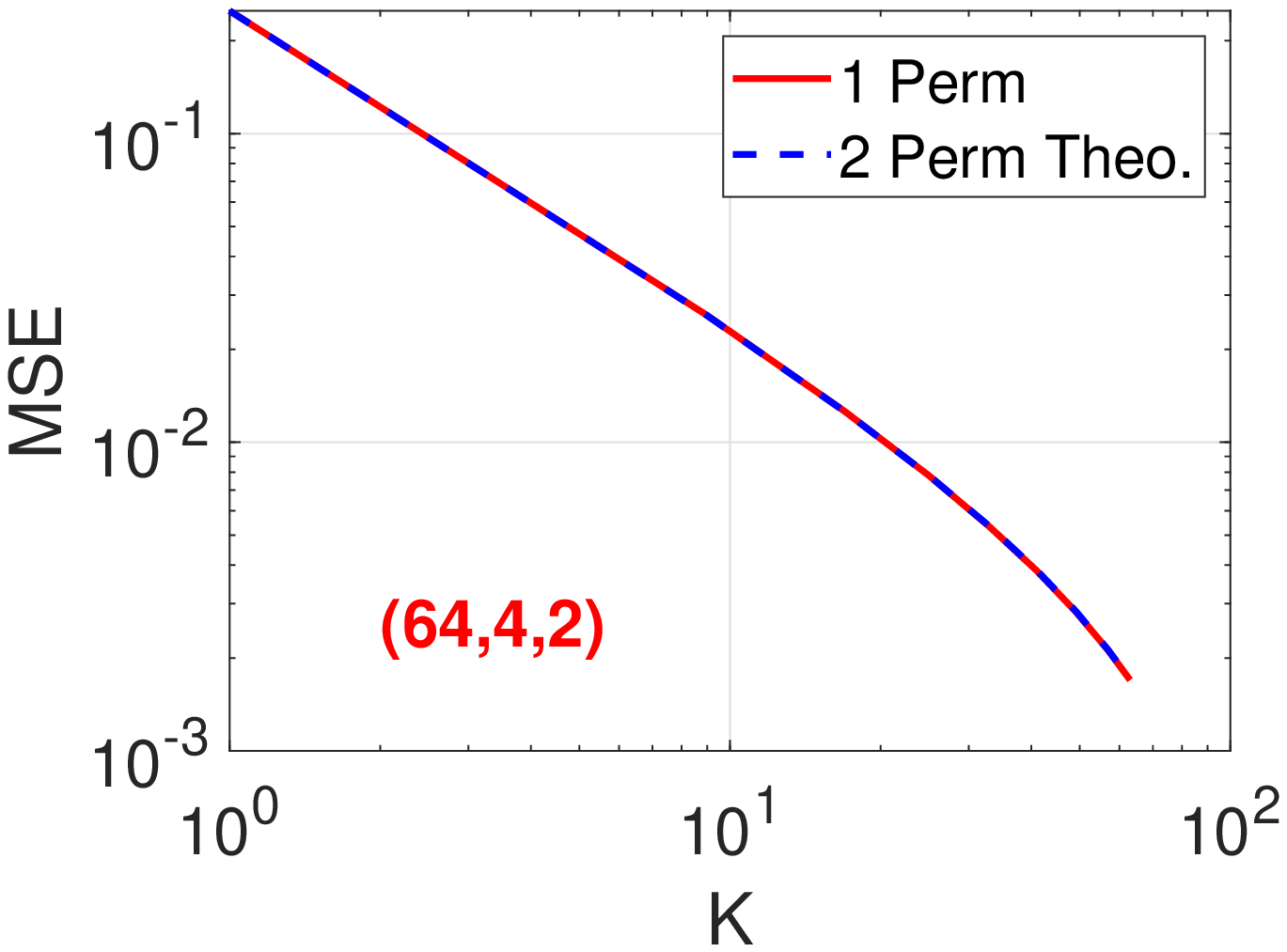}
		\includegraphics[width=2.1in]{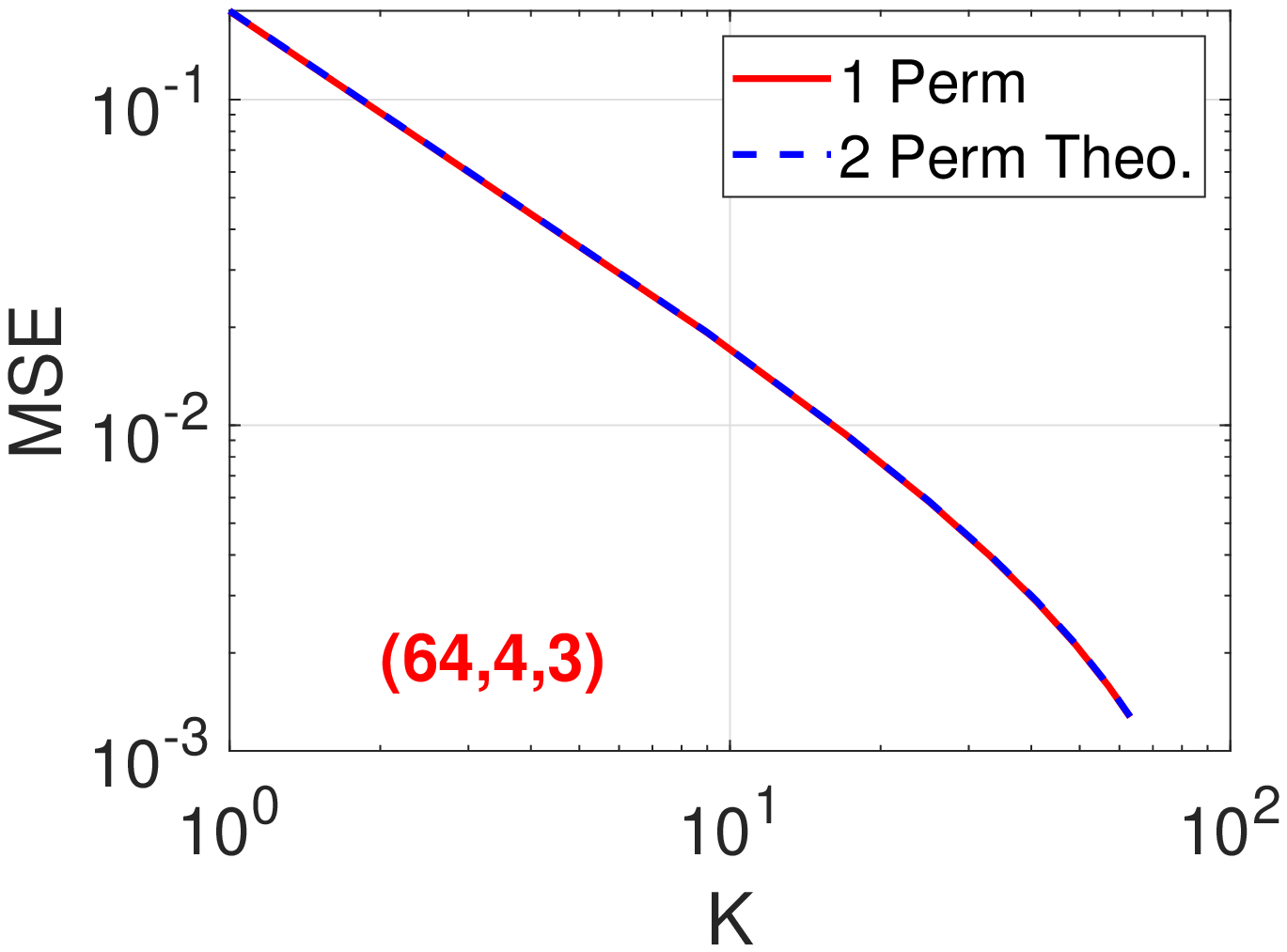}
		}
		\mbox{
		\includegraphics[width=2.1in]{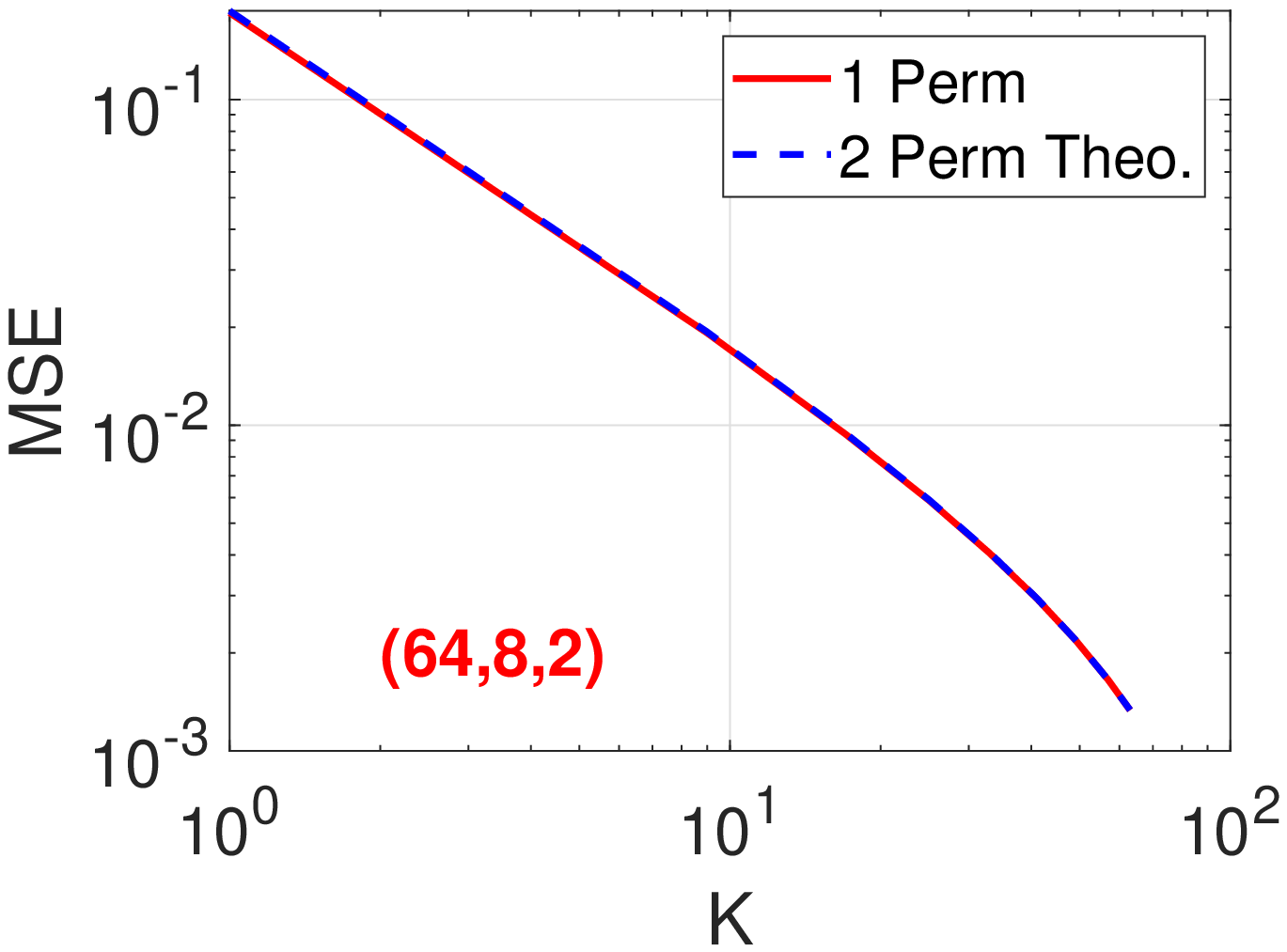}
		\includegraphics[width=2.1in]{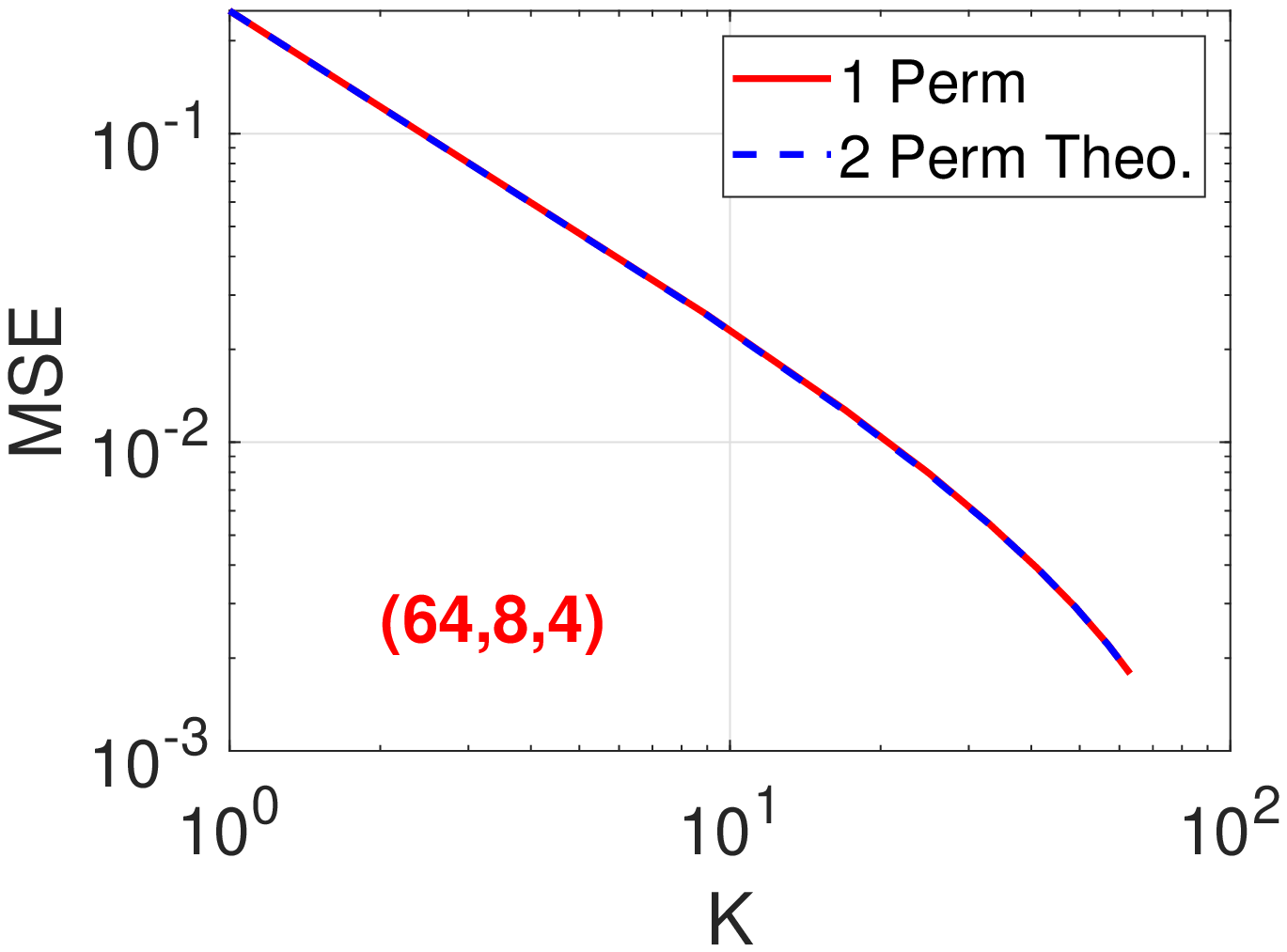}
		\includegraphics[width=2.1in]{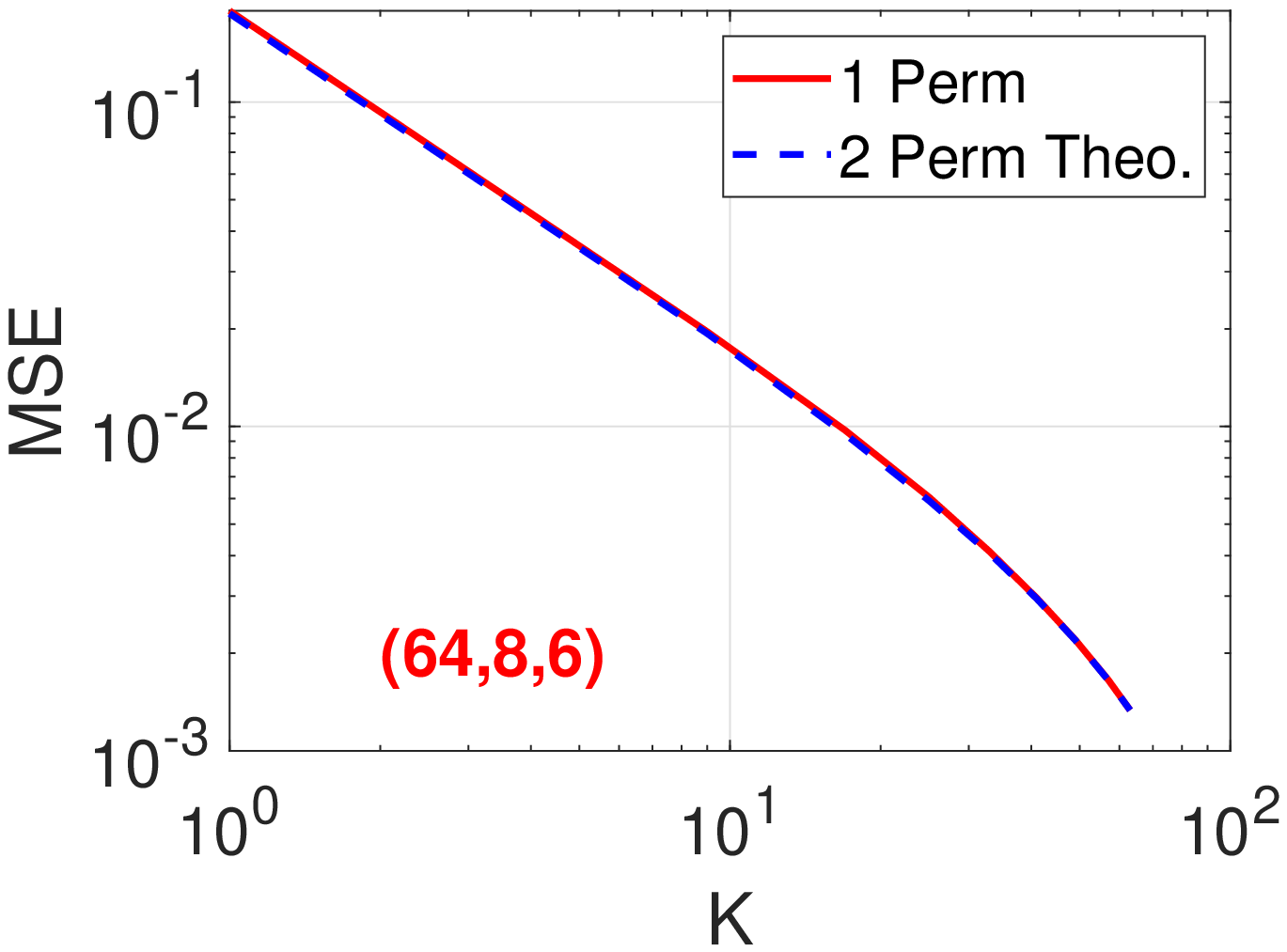}
		}
		\mbox{
		\includegraphics[width=2.1in]{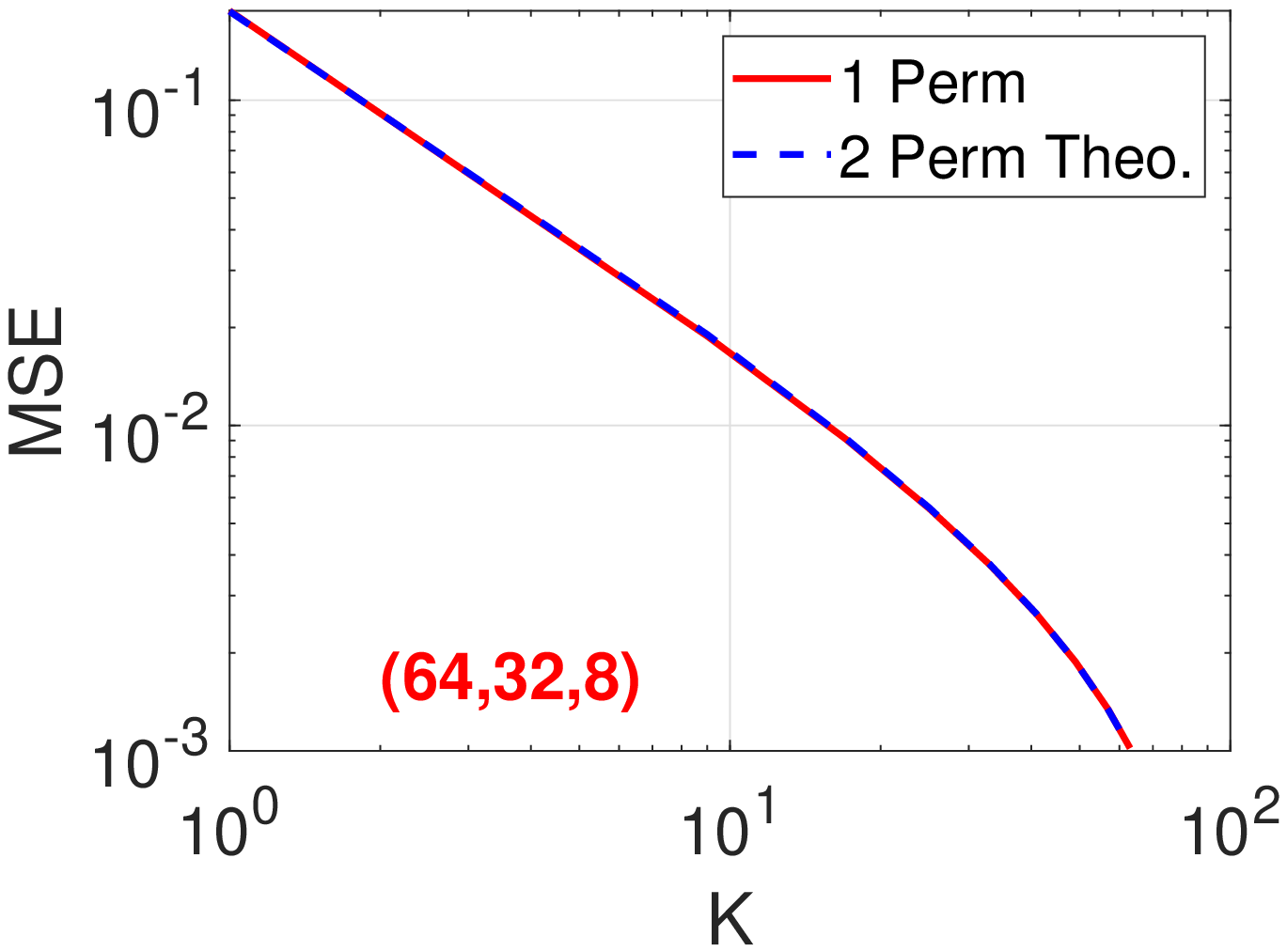}
		\includegraphics[width=2.1in]{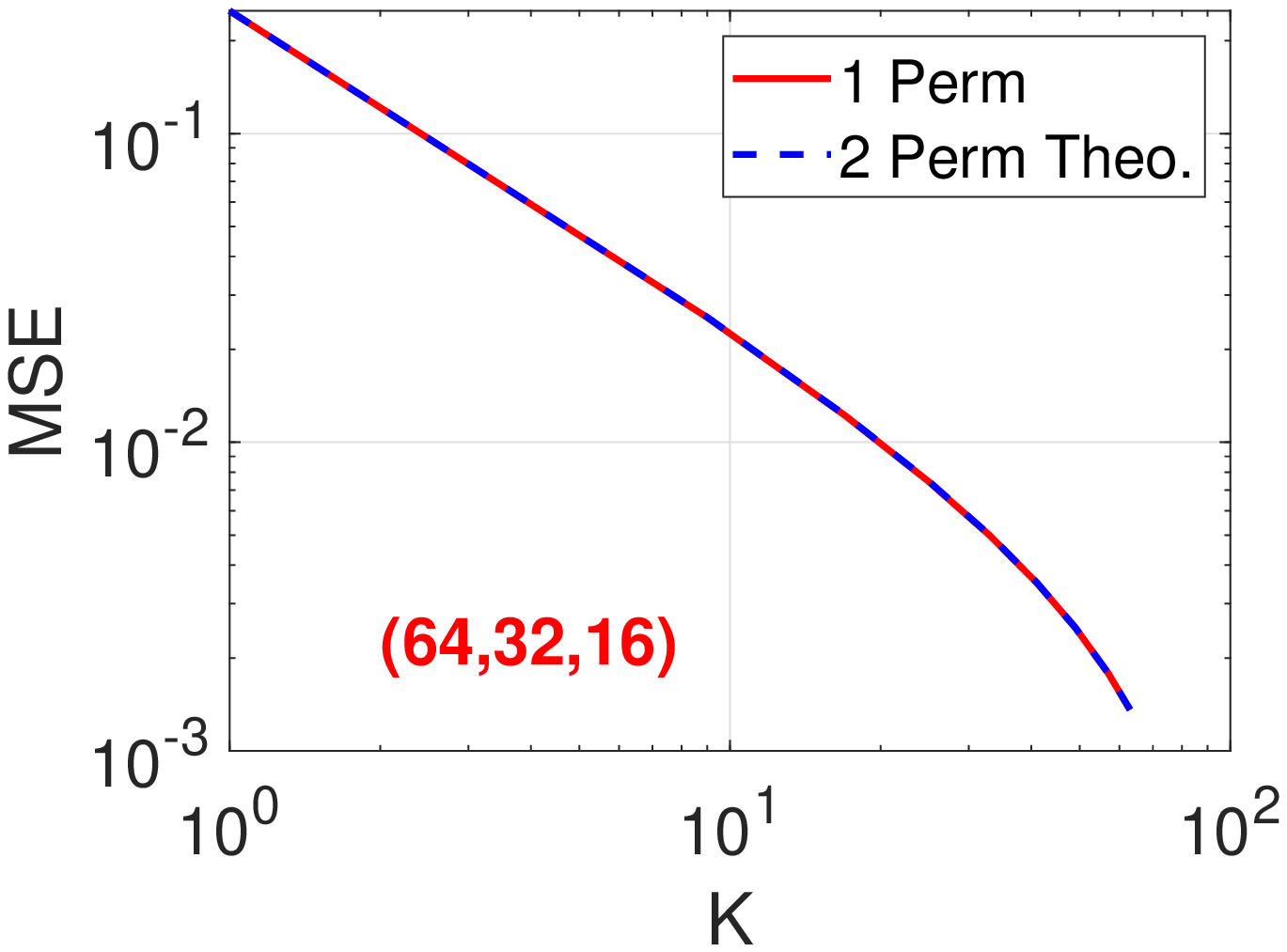}
		\includegraphics[width=2.1in]{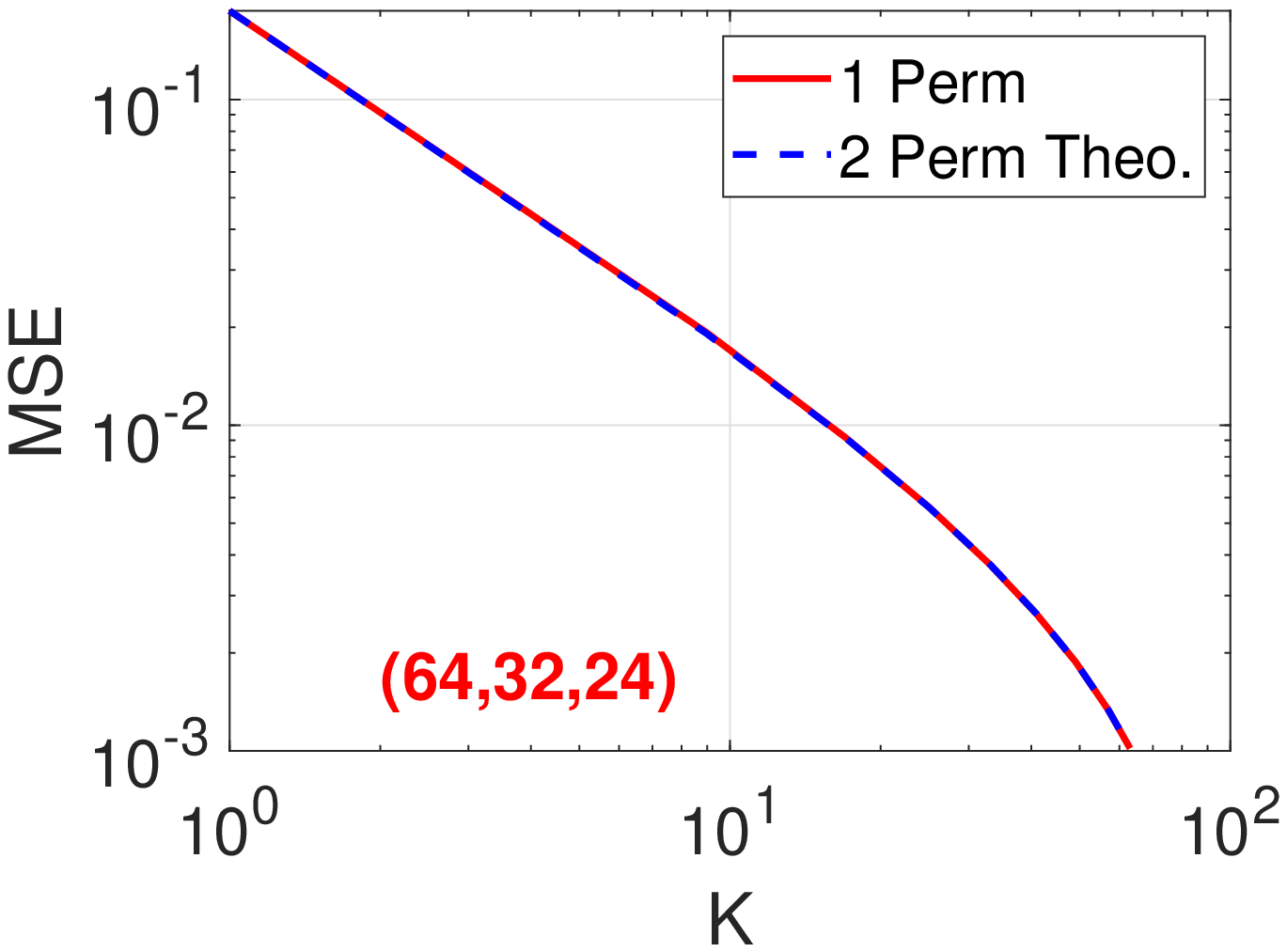}
		}
		\mbox{\hspace{-0.12in}
		\includegraphics[width=2.05in]{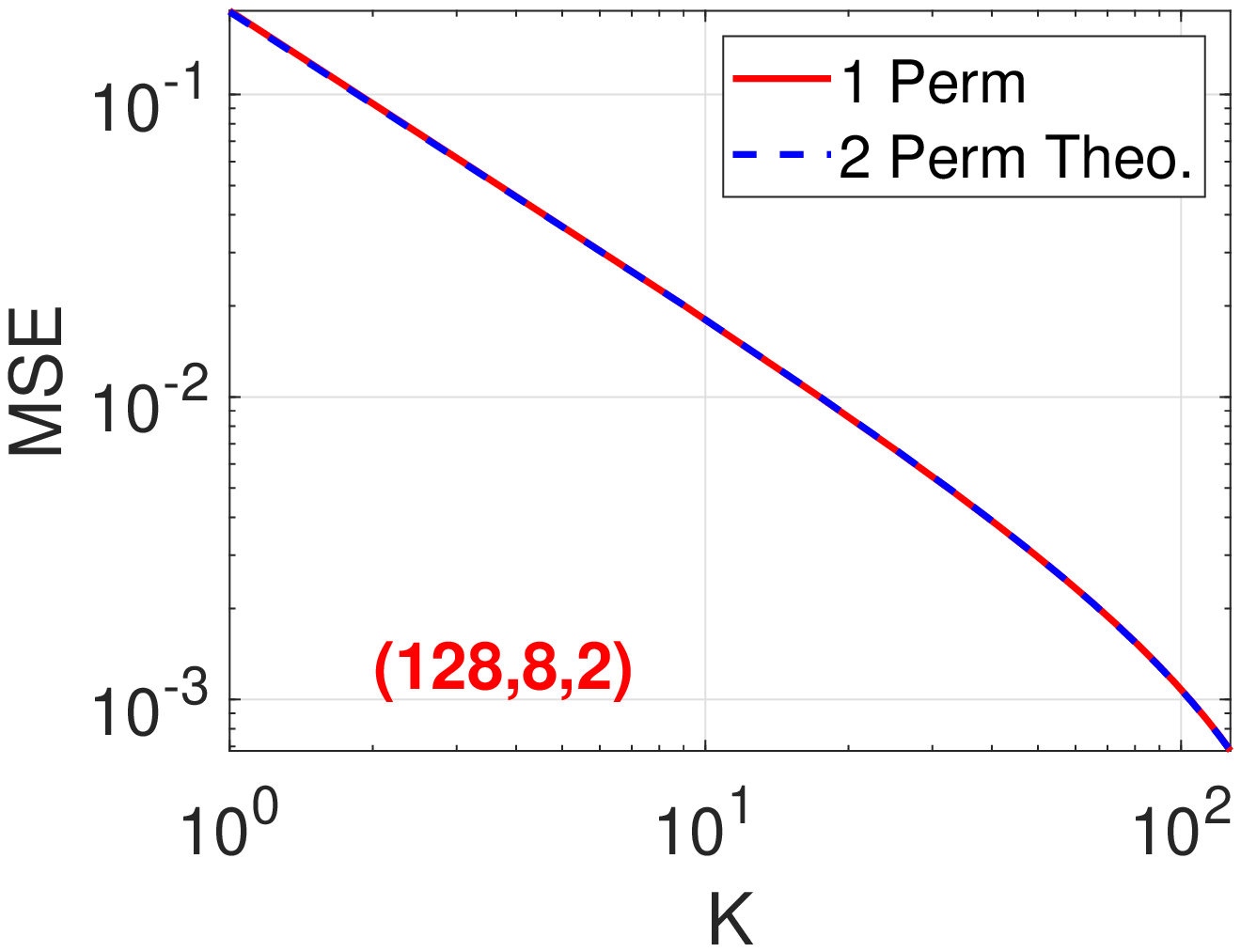}\hspace{0.04in}
		\includegraphics[width=2.05in]{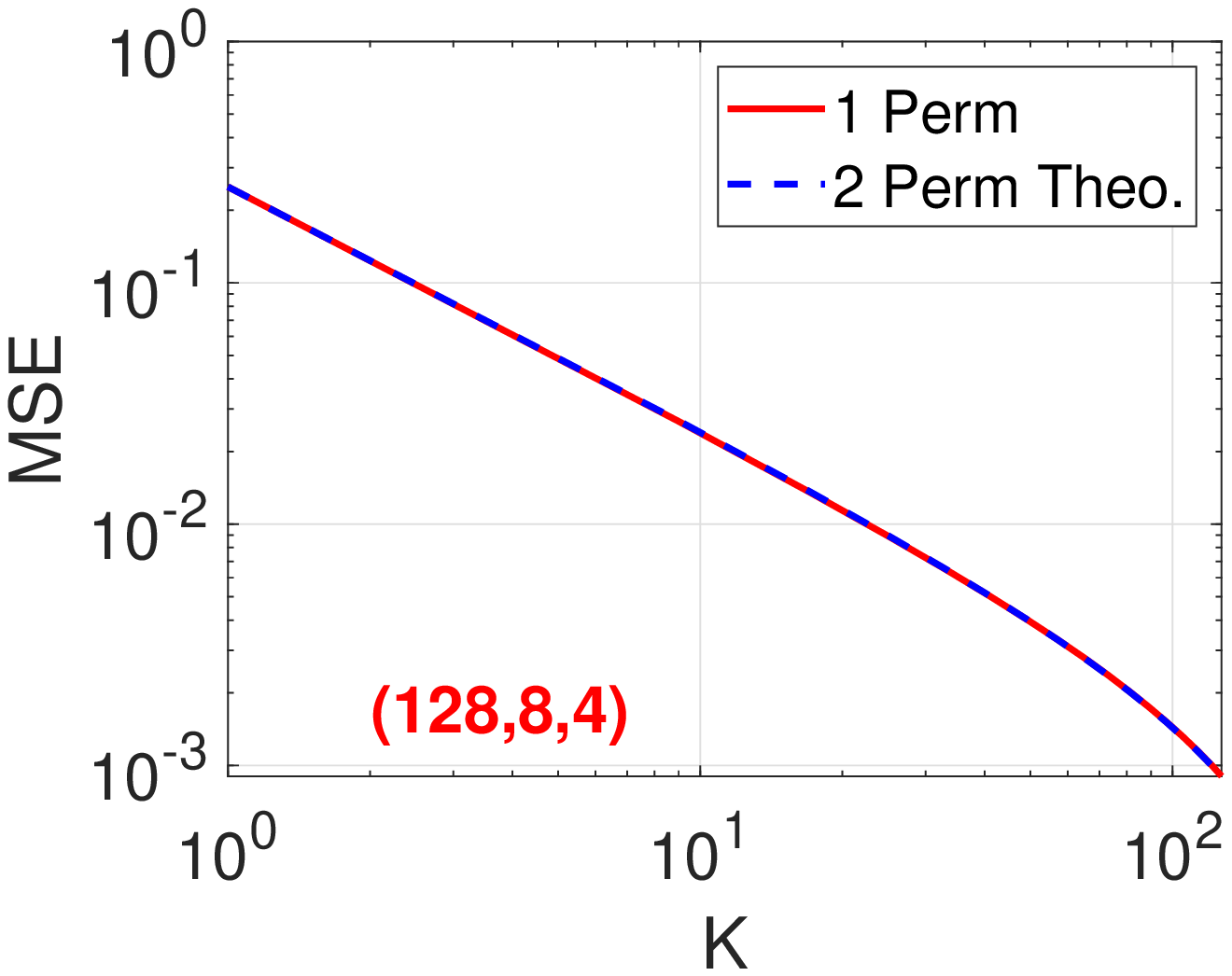}\hspace{0.04in}
		\includegraphics[width=2.05in]{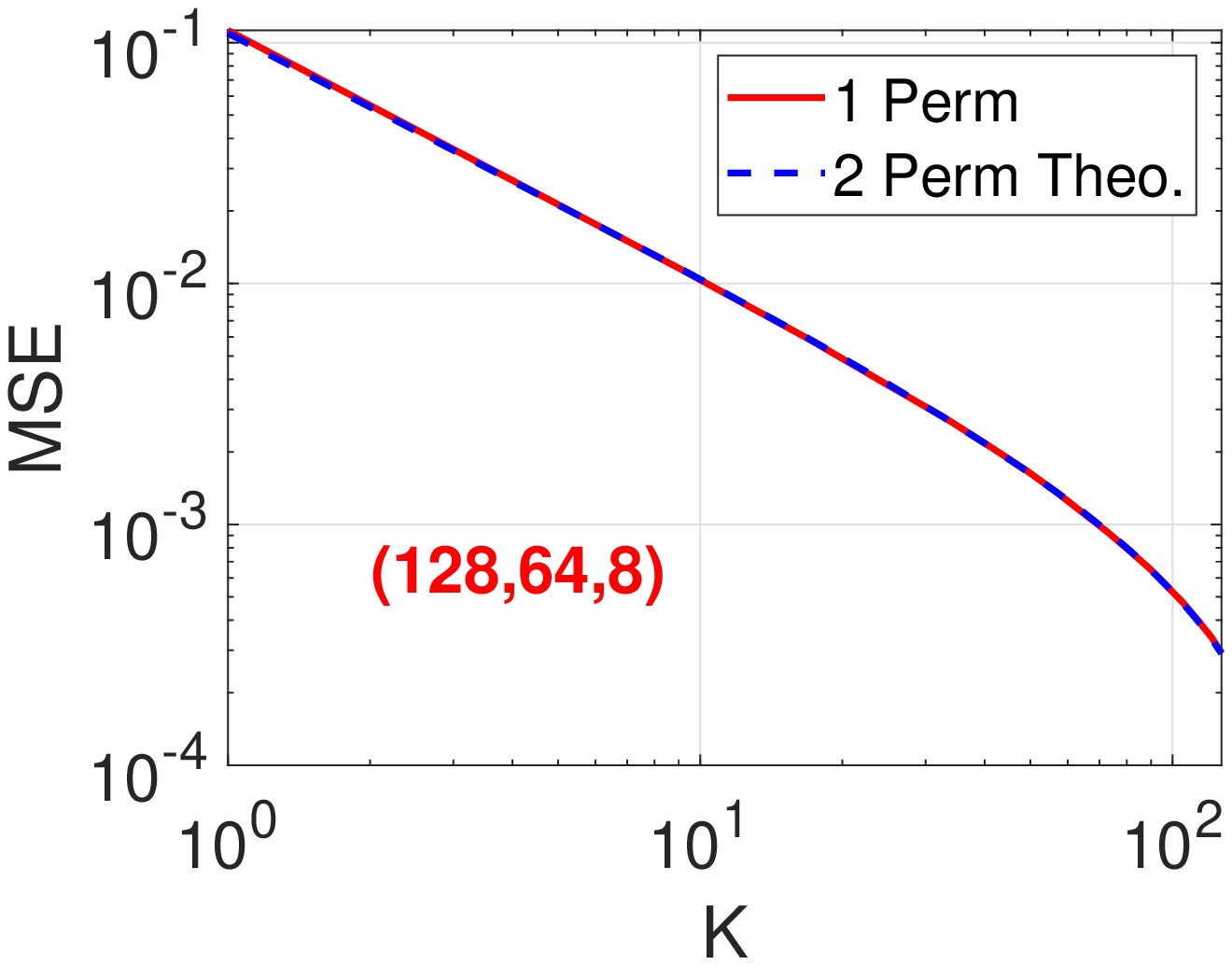}
		}
		\mbox{\hspace{-0.12in}
		\includegraphics[width=2.05in]{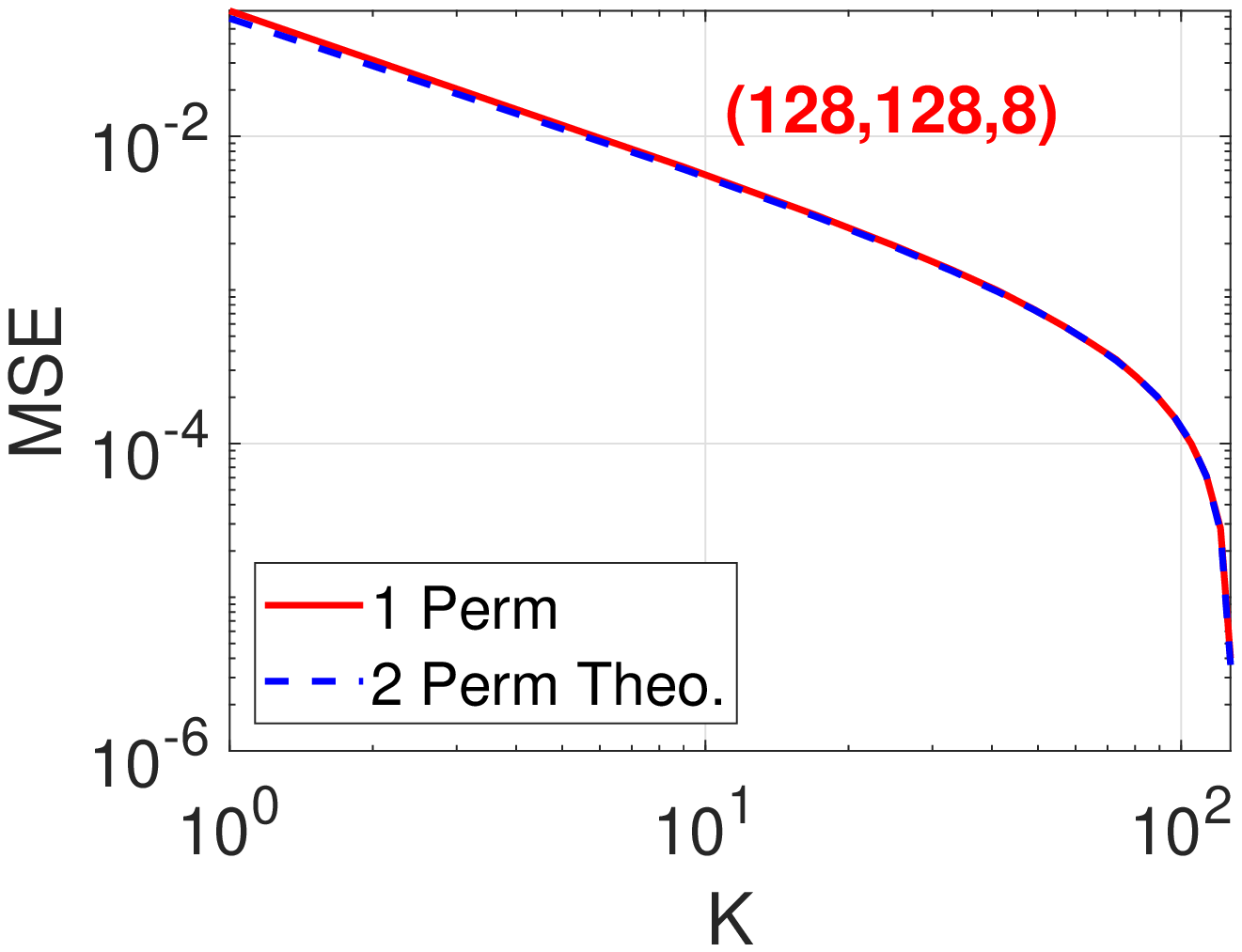}\hspace{0.04in}
		\includegraphics[width=2.05in]{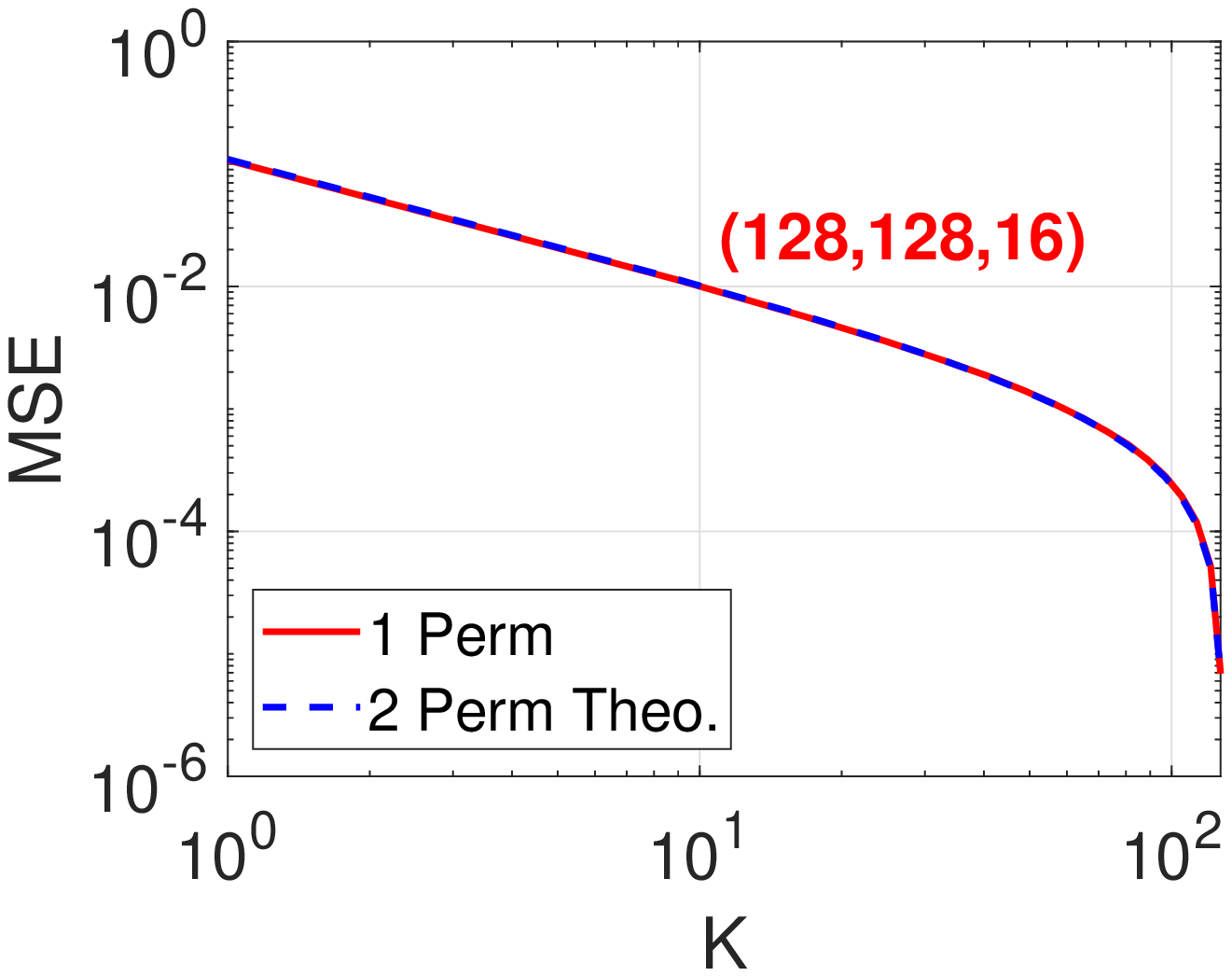}\hspace{0.04in}
		\includegraphics[width=2.05in]{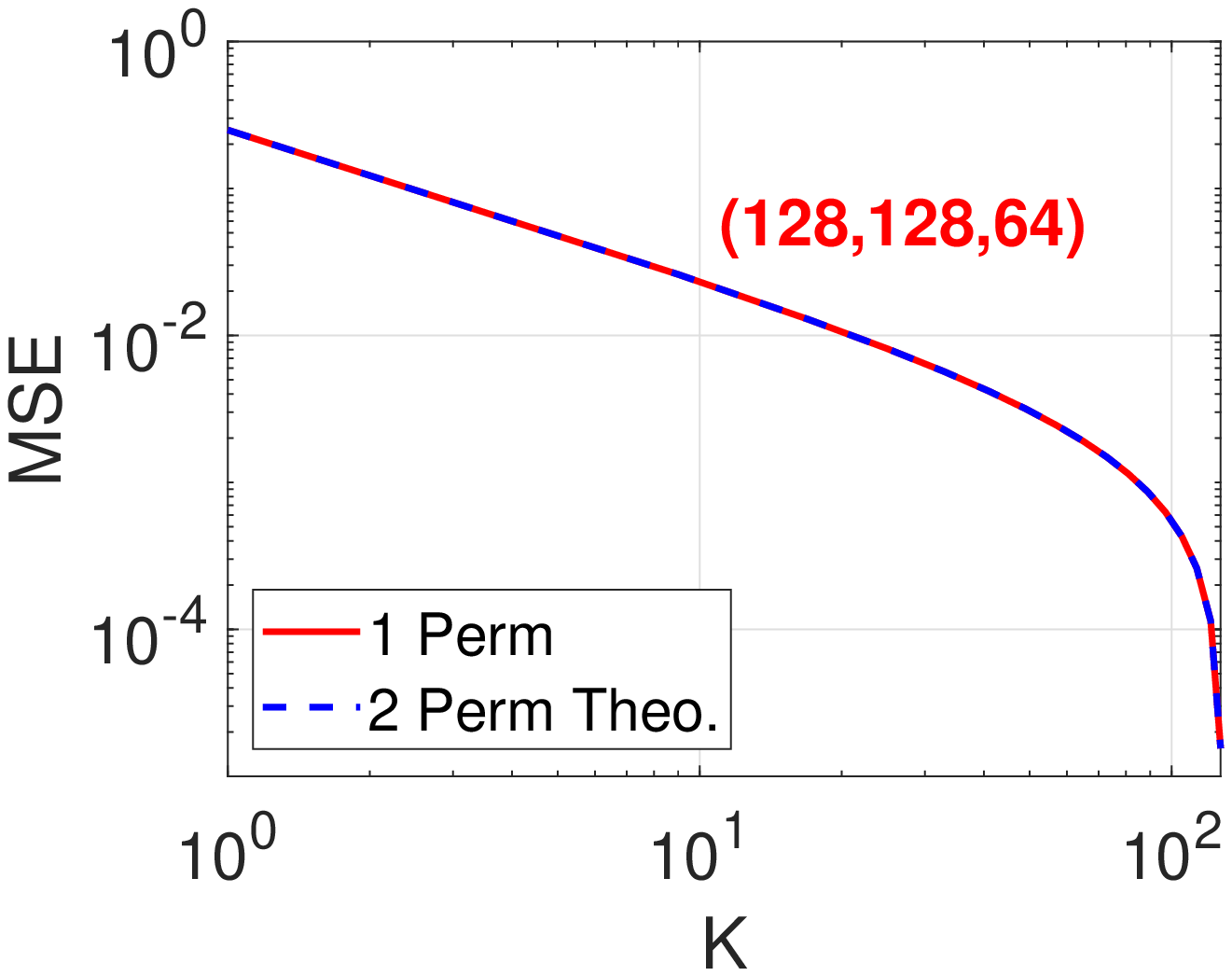}
		}		
	\end{center}
	\vspace{-0.2in}
	\caption{MSE = bias$^2$ + variance  vs.  $K$ on simulated data pairs. The first 3 rows correspond to the cases in Figure~\ref{fig:bias}. We plot the empirical MSEs for C-MinHash-$(\pi,\pi)$ (red, solid, ``1 Perm'') together with the theoretical variances of C-MinHash-$(\sigma,\pi)$ (blue, dashed, ``2 Perm''). The curves overlap in all cases. Bottom two rows have a special locational structure: the location vector $\bm x$ is such that $a$ ``$O$'''s are followed by $(f-a)$ ``$\times$'''s and then followed by $(D-f)$ ``$-$'''s sequentially.}
	\label{fig:MSE-method 2}\vspace{-0.1in}
\end{figure}

Figure~\ref{fig:MSE-method 2} plots the MSE = bias$^2$+variance, for comparing the empirical MSEs of C-MinHash-$(\pi,\pi)$ with the theoretical variances of C-MinHash-$(\sigma,\pi)$  in~\cite{CMH2Perm2021}. Besides the cases in Figure~\ref{fig:bias}, the bottom two rows of Figure~\ref{fig:MSE-method 2}   provide more examples where the data vector pairs have a special locational structure. In all figures, the overlapping MSE curves essentially verify our claim that we just need one permutation~$\pi$.

\begin{figure}[t]
  \begin{center}
   \mbox{
    \includegraphics[width=2.1in]{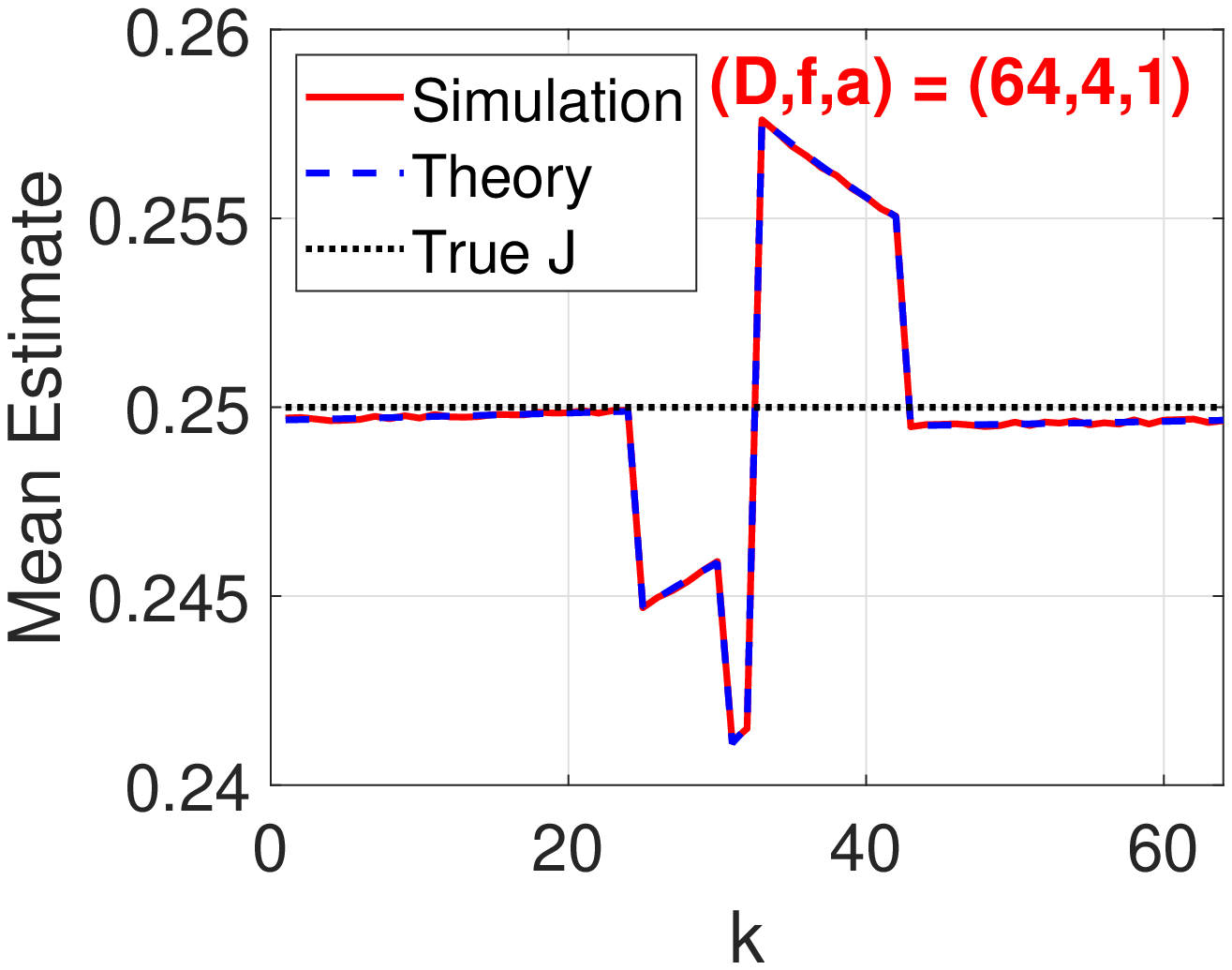}
    \includegraphics[width=2.1in]{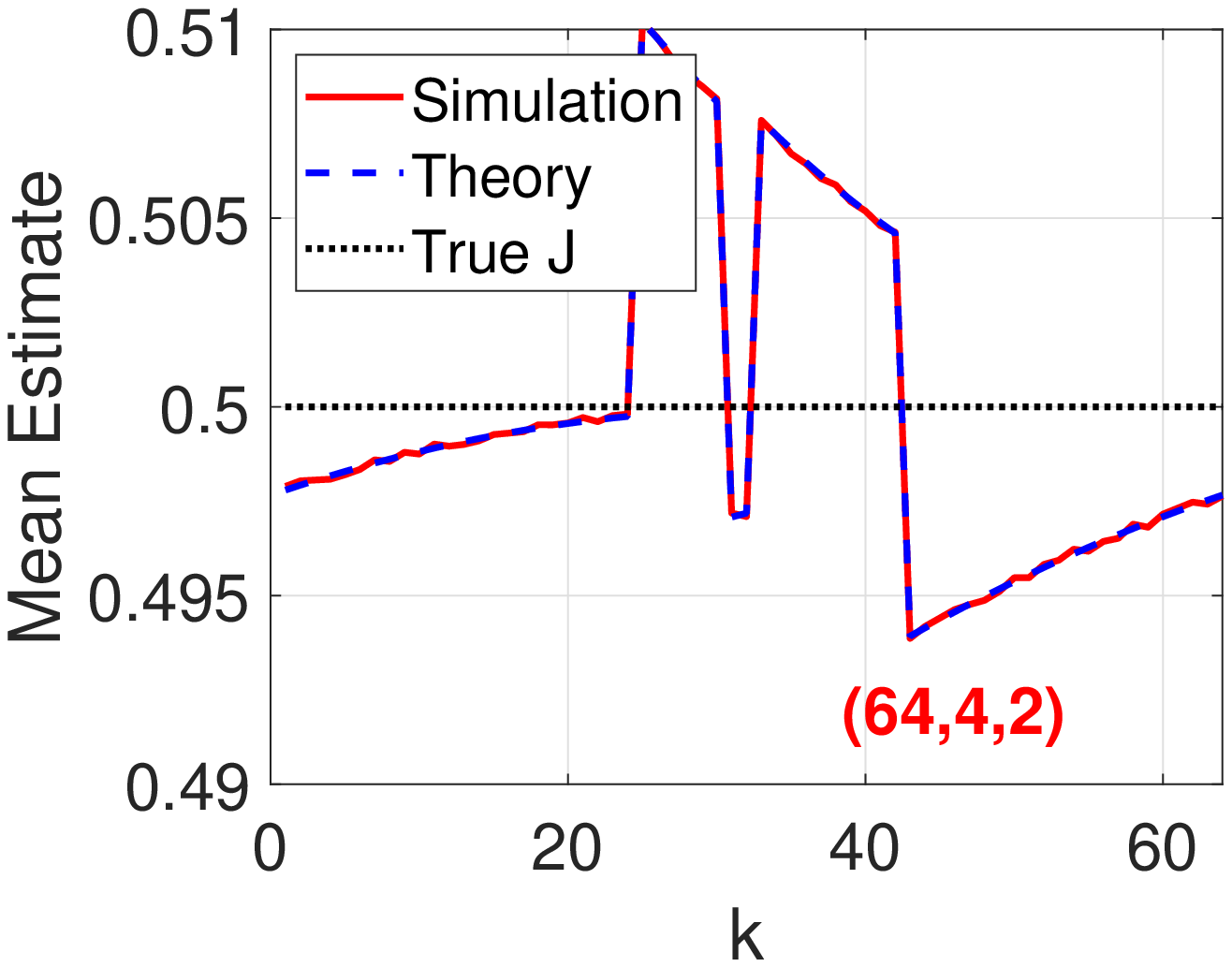}
    \includegraphics[width=2.1in]{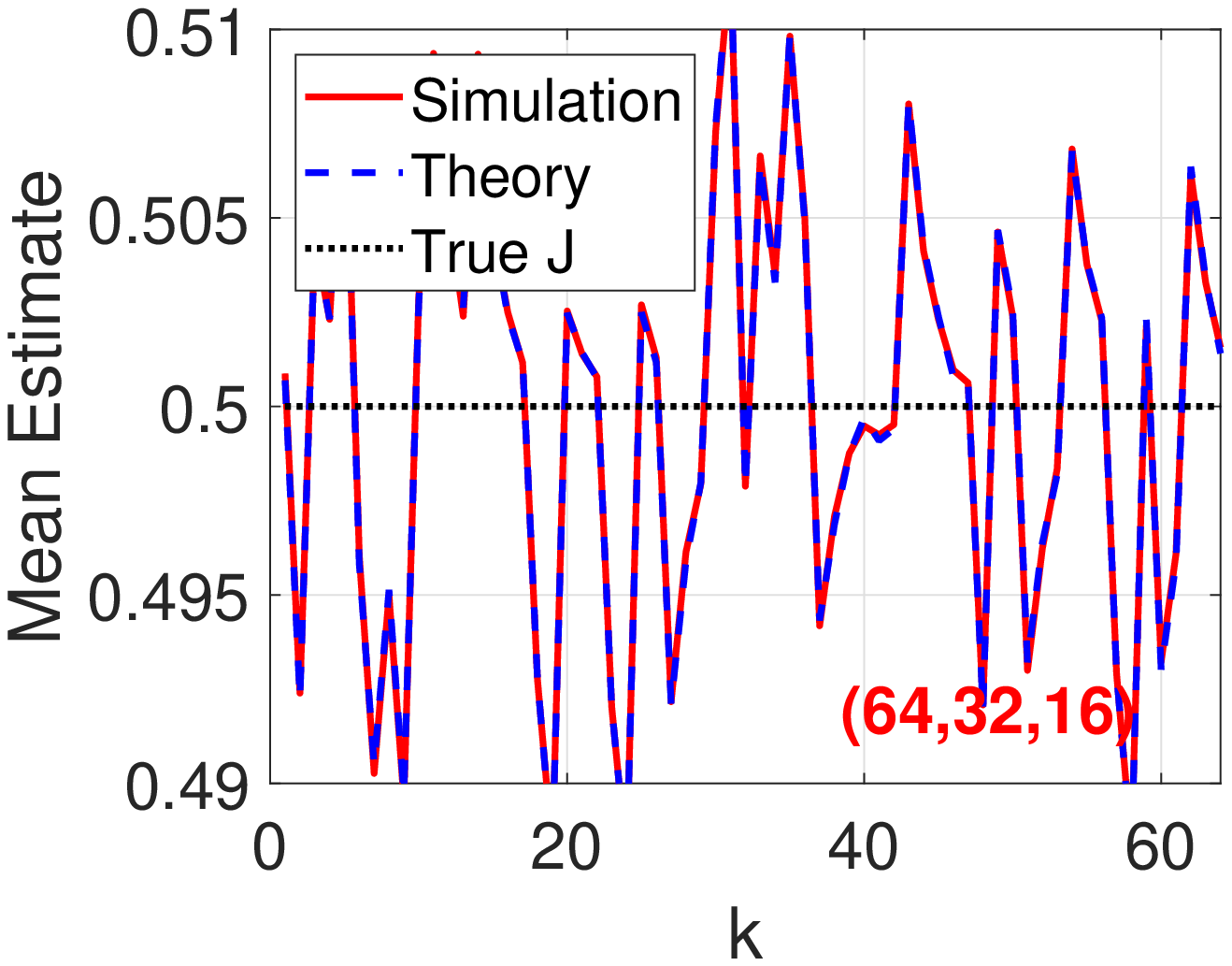}
    }
  \end{center}
  \vspace{-0.25in}
  \caption{$\mathbb E[\mathbbm 1\{h_k(\bm v)=h_k(\bm w)\}]$ for each $1\leq k\leq K$ on simulated data pairs with various $(D,f,a)$ where $D=64$. Clearly, simulations (red, solid) match perfectly the theory (blue, dashed) by  Theorem~\ref{theo:mean-M2}. }
  \label{fig:mean}\vspace{-0.1in}
\end{figure}

\section{Further Empirical Verification}  \label{sec:experiments}

In this section, we present extensive experiments on real data, to further validate that  C-MinHash-$(\pi,\pi)$ and  C-MinHash-$(\sigma,\pi)$ perform equivalently in Jaccard estimation. We first use 120 pairs of word vectors from the ``Words'' dataset~\citep{Proc:Li_Church_EMNLP05} to once again validate that the MSE of  C-MinHash-$(\pi,\pi)$ basically matches the theoretical variance of C-MinHash-$(\sigma,\pi)$ (which is strictly unbiased).

\subsection{MSE Comparisons on Words Dataset}  \label{sec:experiment MSE}

The ``Words'' dataset~\citep{Proc:Li_Church_EMNLP05} (which is publicly available) contains a large number of word vectors, with the $i$-th entry indicating whether this word appears in the $i$-th document, for a total of $D=2^{16}$ documents. The key statistics of the 120 selected word pairs are presented in Table~\ref{tab:word pairs}. Those 120 pairs of words are more or less randomly selected except that we make sure they cover a wide spectrum of data distributions. Denote $d$ as the number of non-zero entries in the vector. Table~\ref{tab:word pairs} reports the density $\tilde d=d/D$ for each word vector, ranging from 0.0006 to 0.6. The Jaccard similarity $J$ ranges from 0.002 to 0.95.

\begin{figure}[H]
  \begin{center}
   \mbox{
    \includegraphics[width=2.1in]{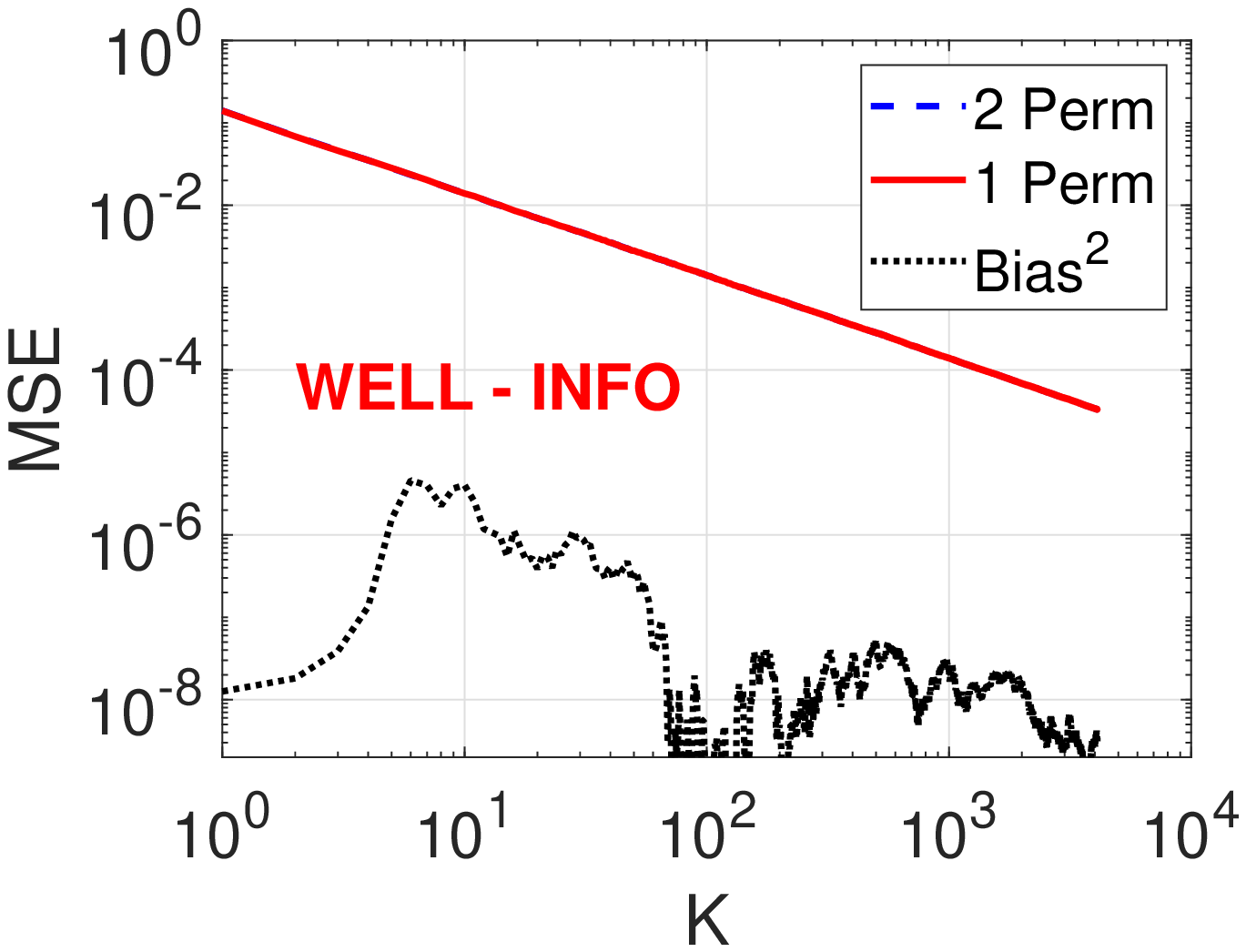}
    \includegraphics[width=2.1in]{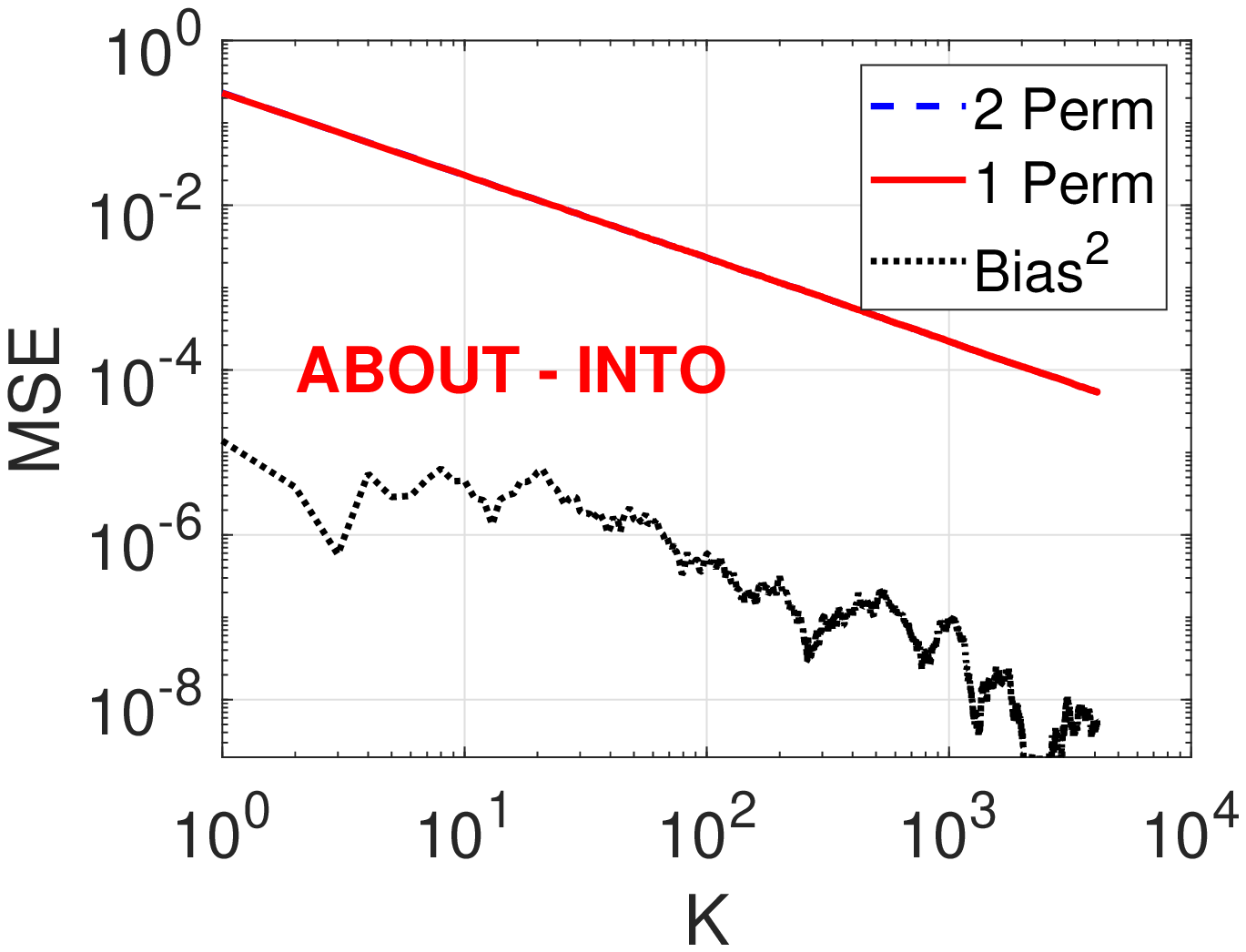}
    \includegraphics[width=2.1in]{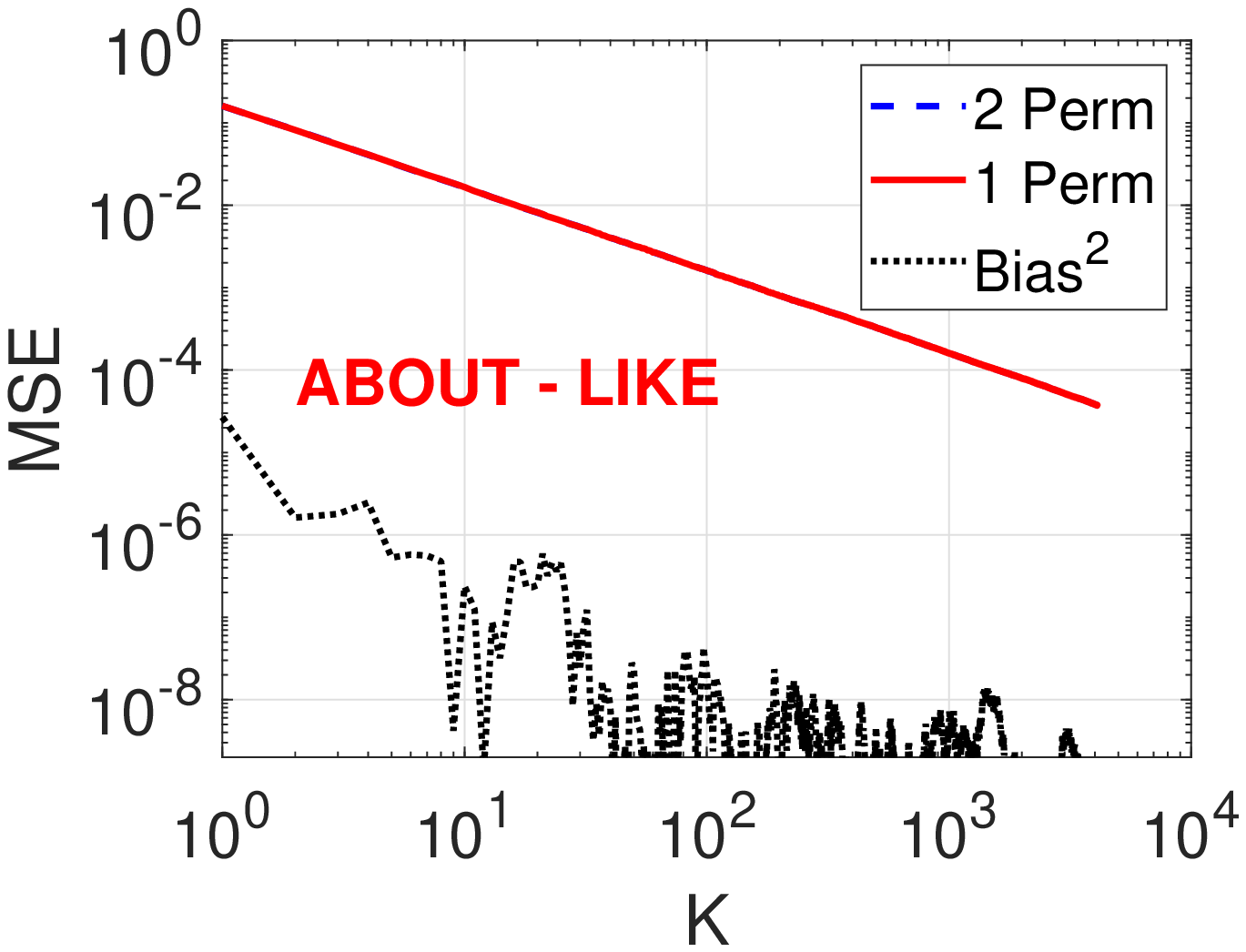}
    }
    \mbox{
    \includegraphics[width=2.1in]{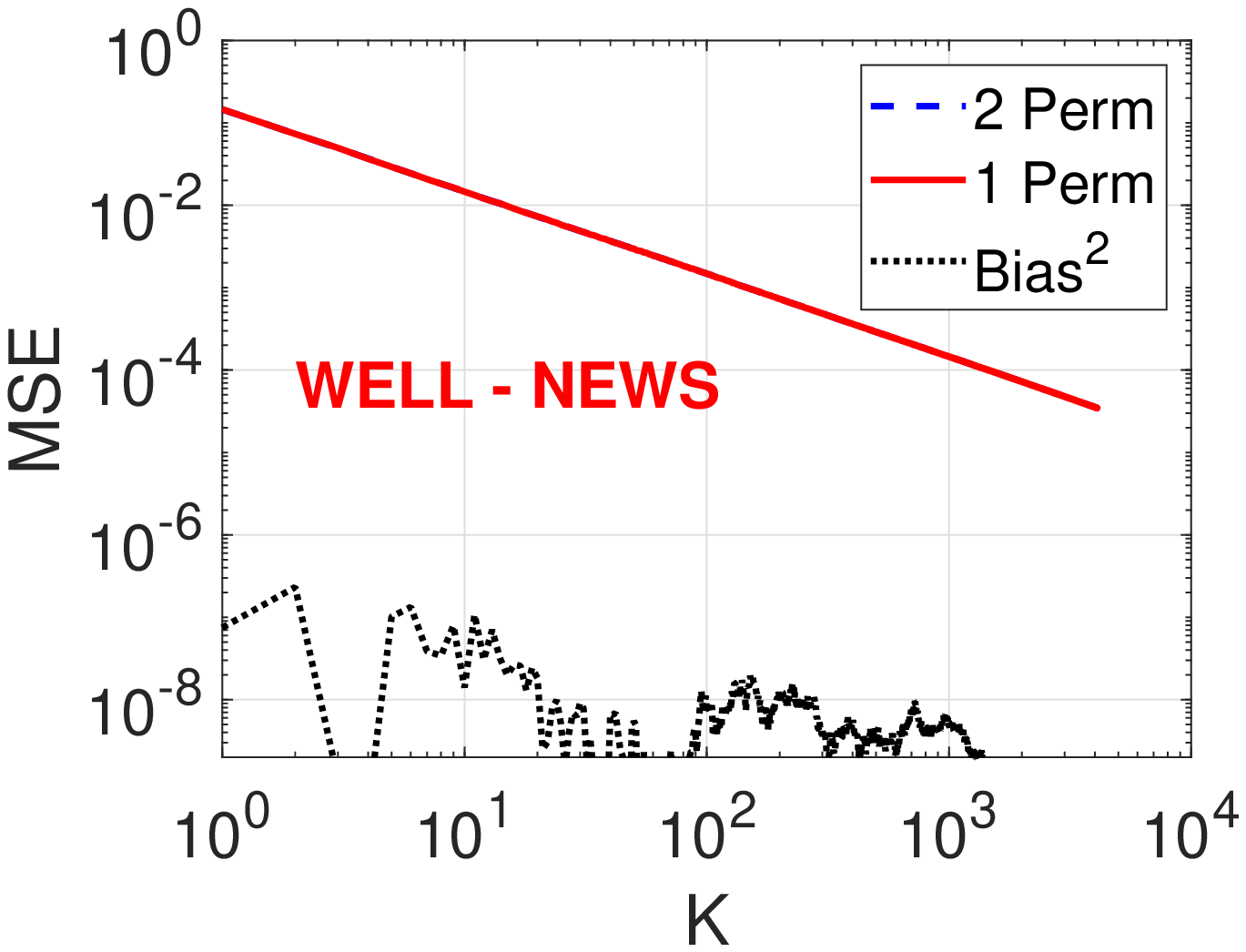}
    \includegraphics[width=2.1in]{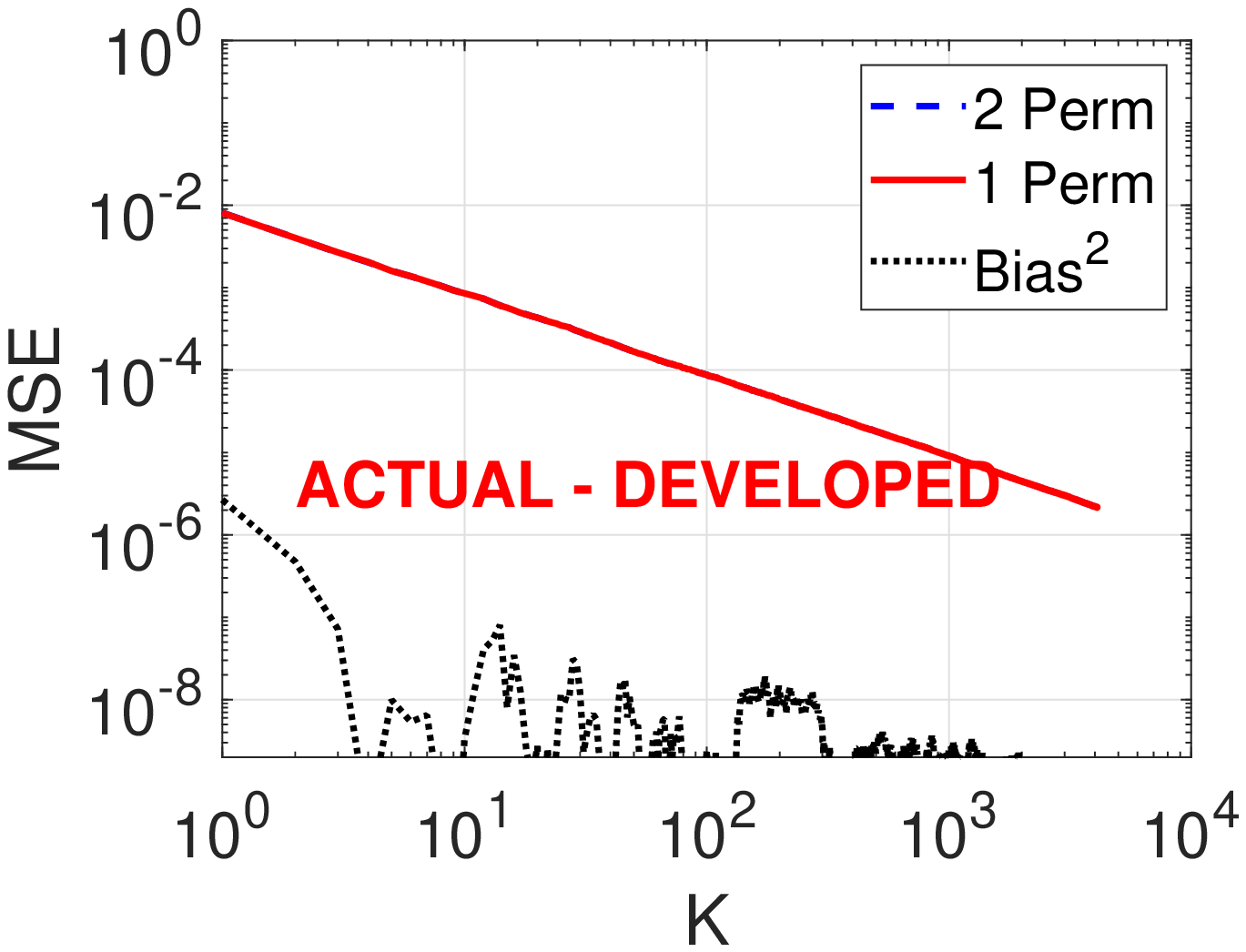}
    \includegraphics[width=2.1in]{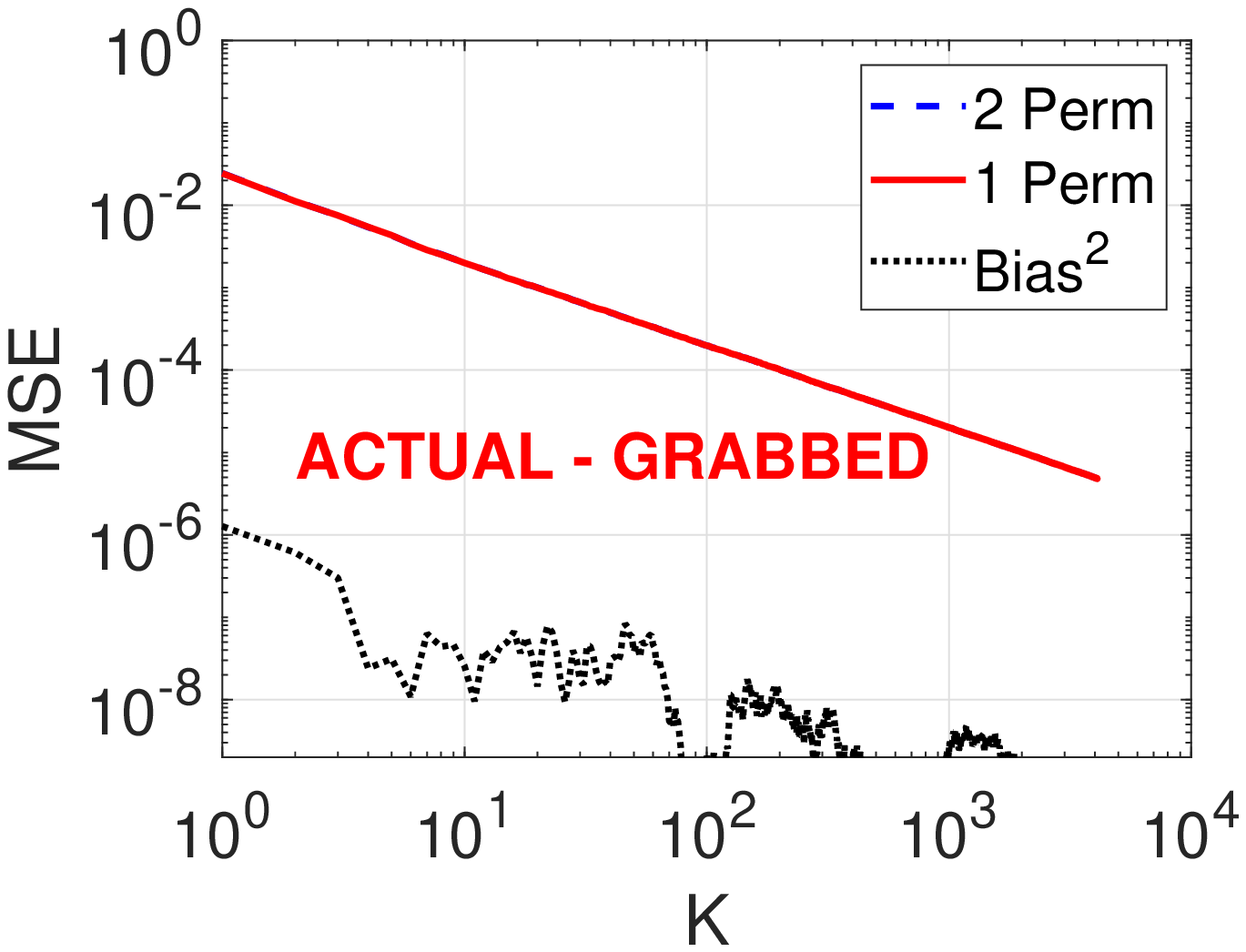}
    }
    \mbox{
    \includegraphics[width=2.1in]{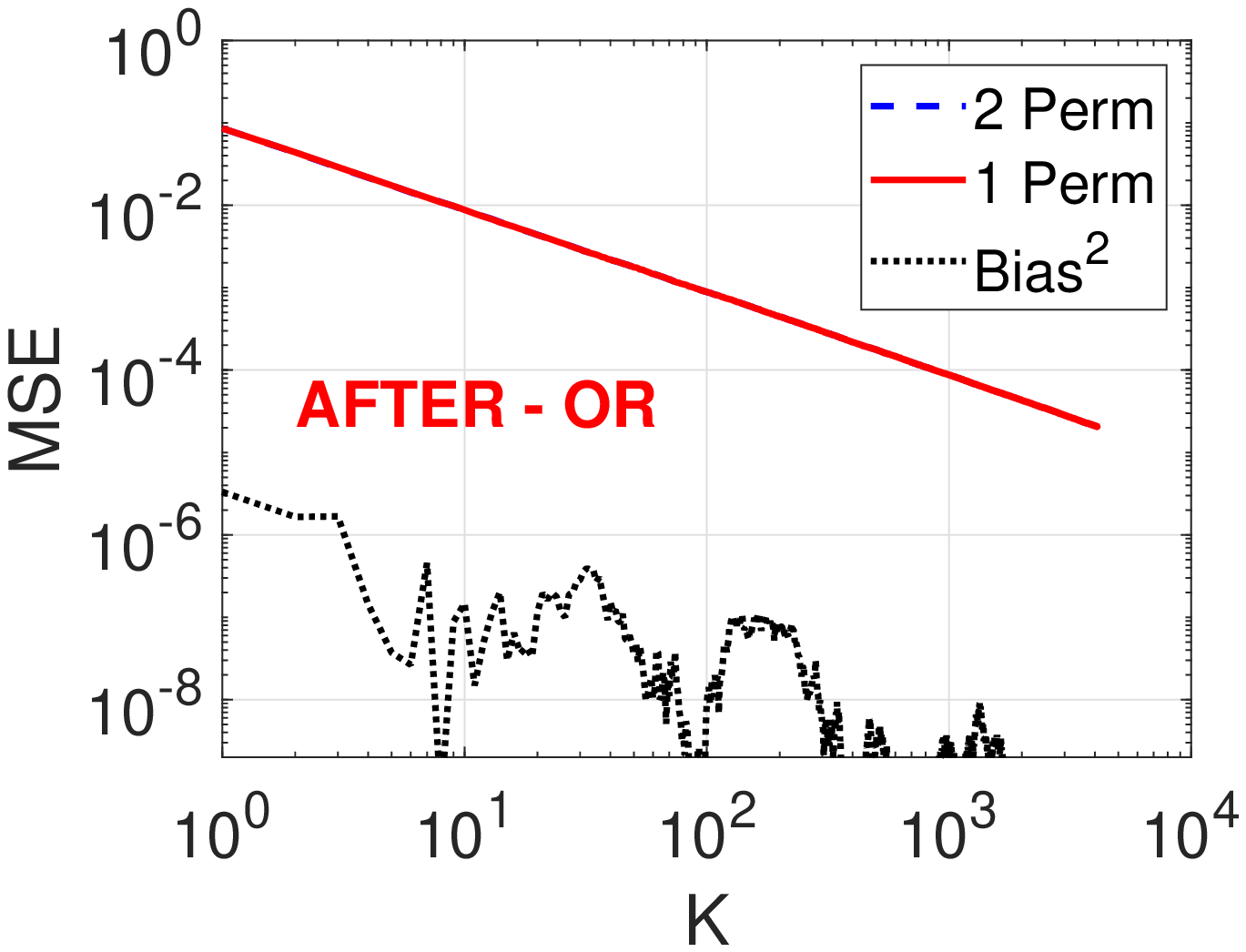}
    \includegraphics[width=2.1in]{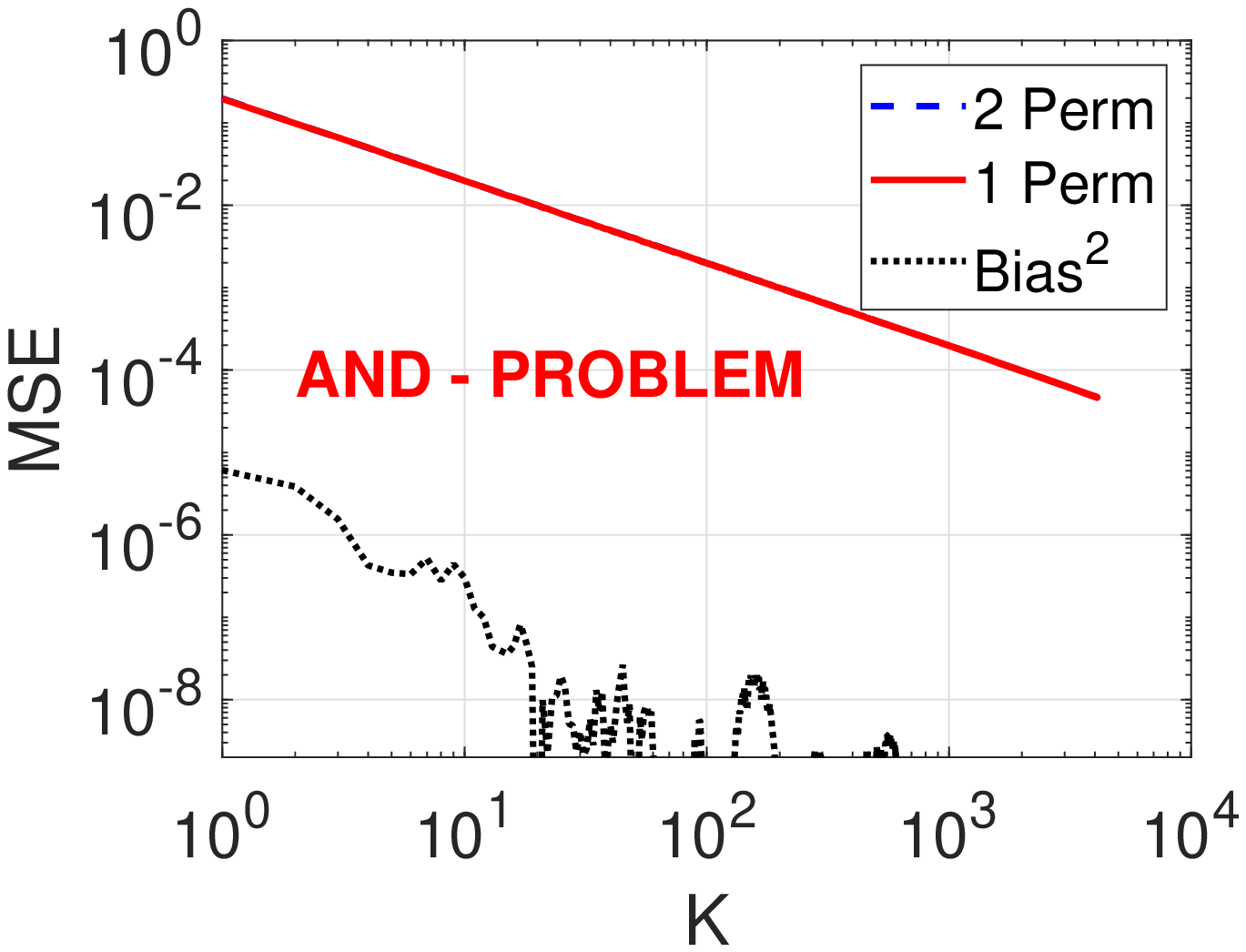}
    \includegraphics[width=2.1in]{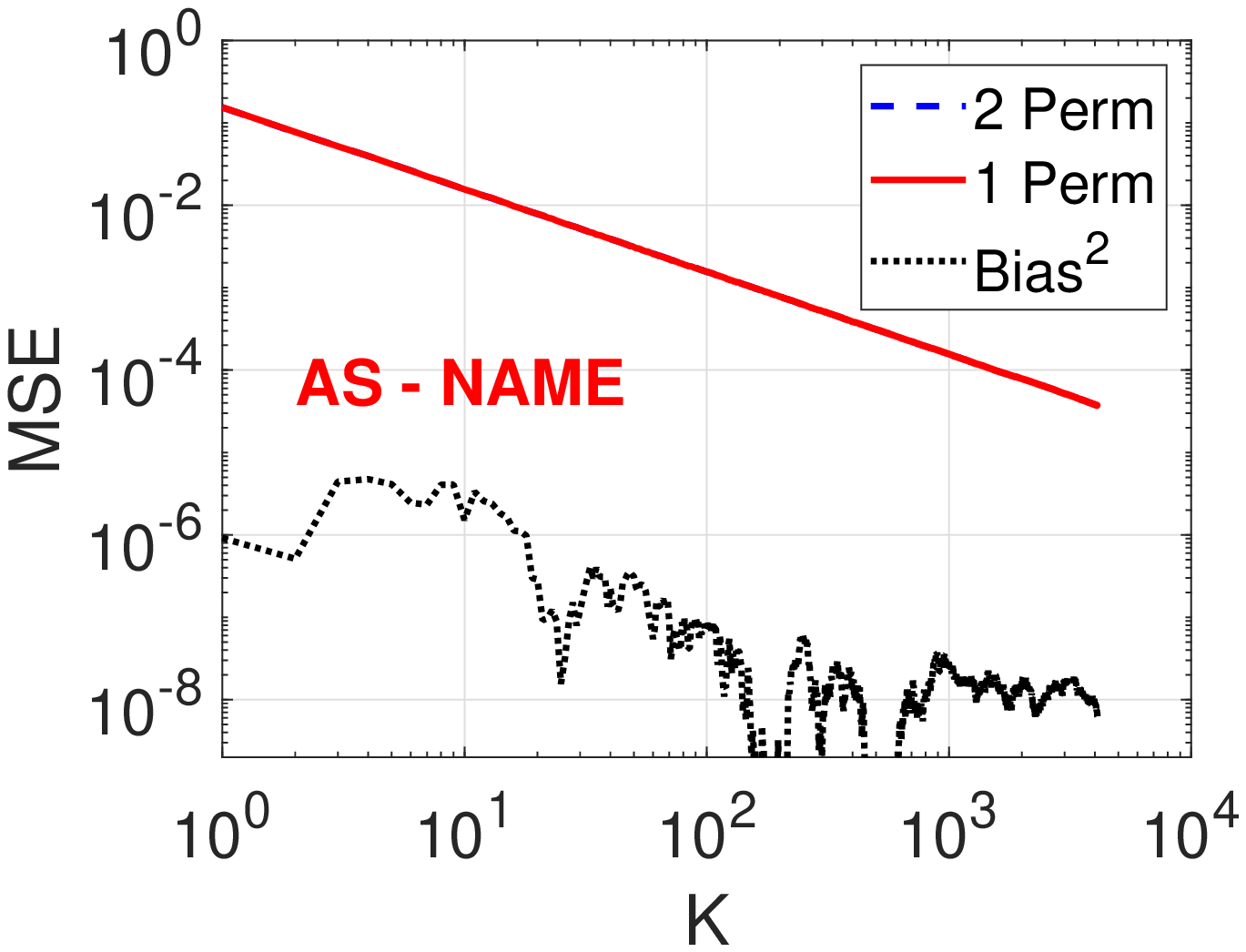}
    }
    \mbox{
    \includegraphics[width=2.1in]{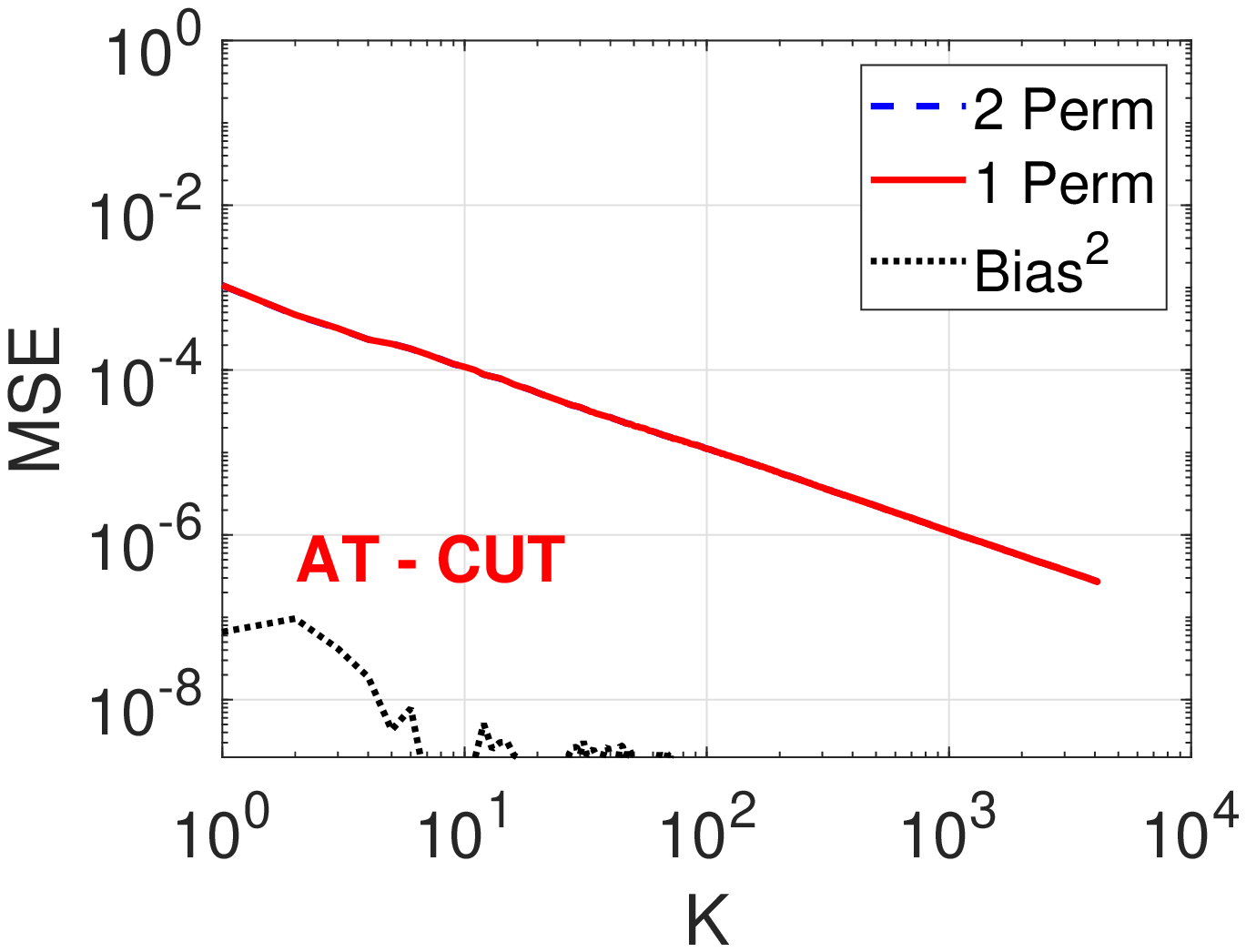}
    \includegraphics[width=2.1in]{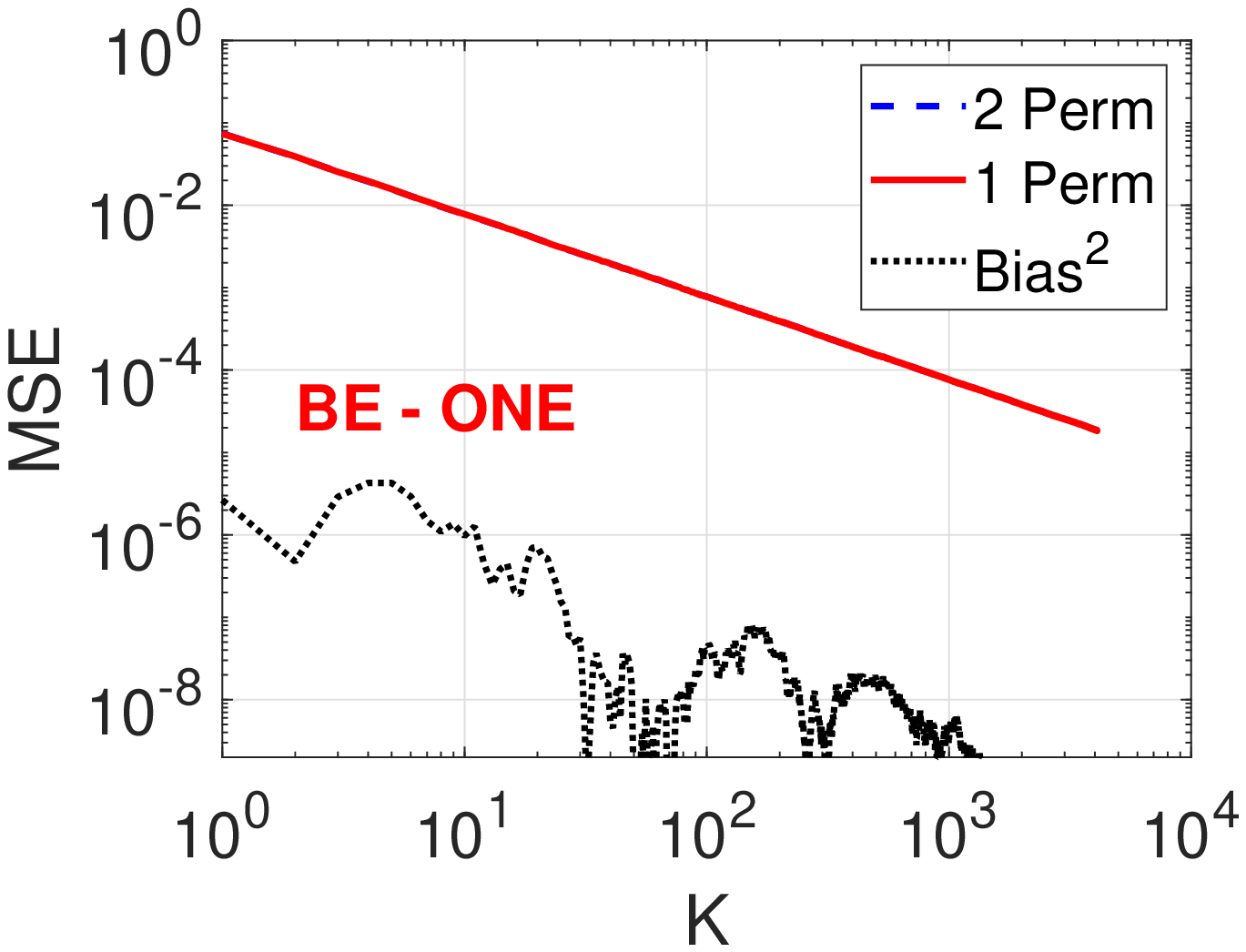}
    \includegraphics[width=2.1in]{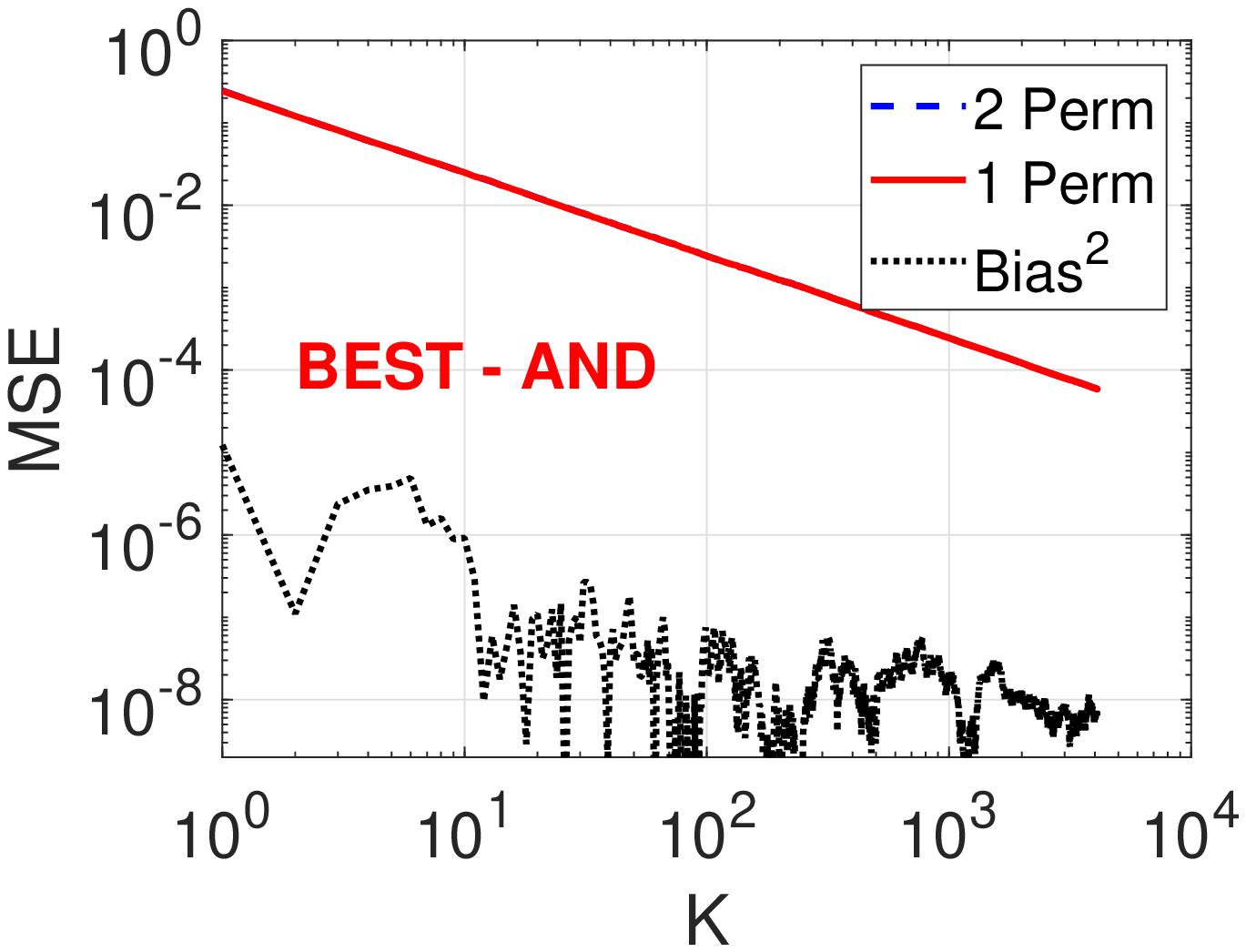}
    }
    \mbox{
    \includegraphics[width=2.1in]{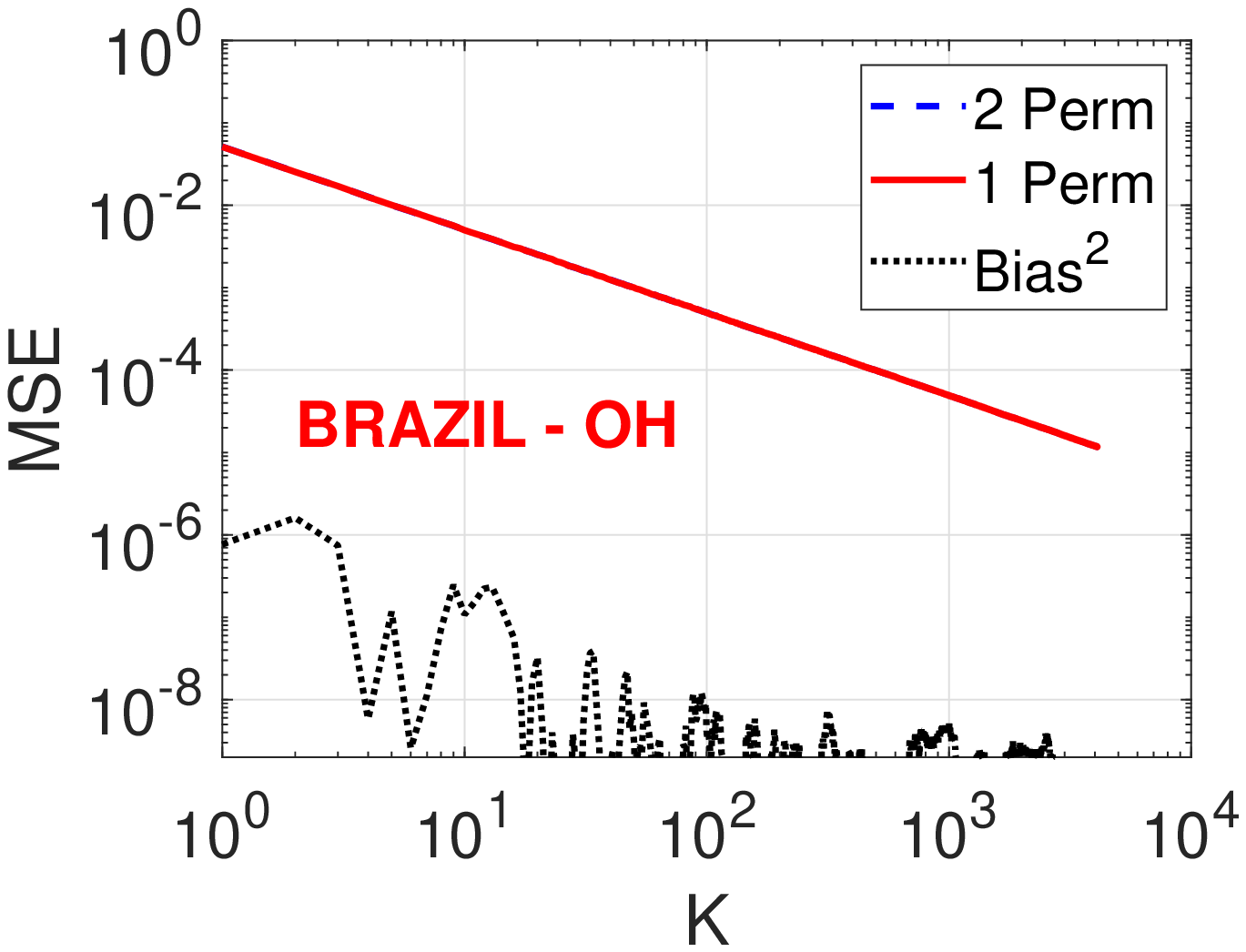}
    \includegraphics[width=2.1in]{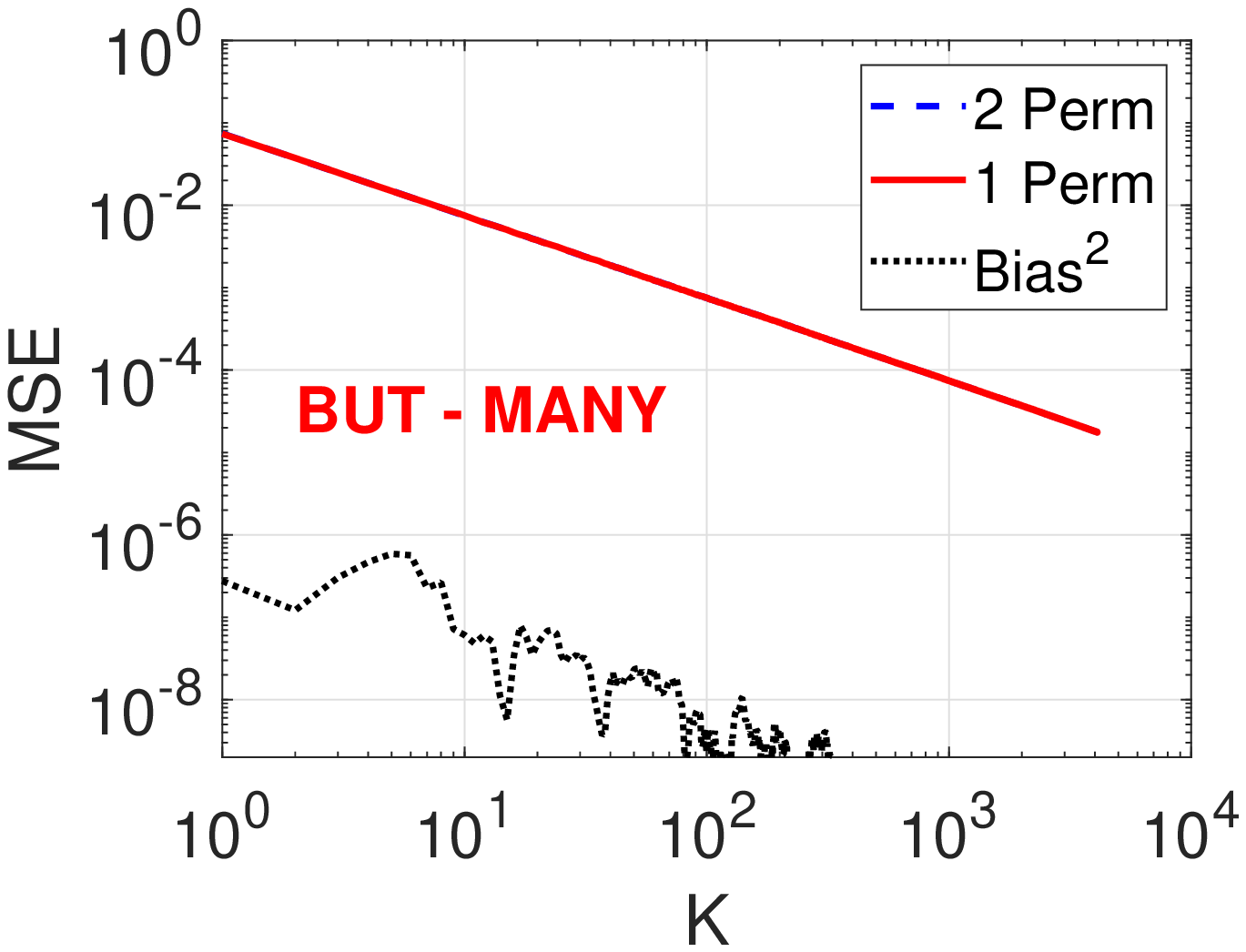}
    \includegraphics[width=2.1in]{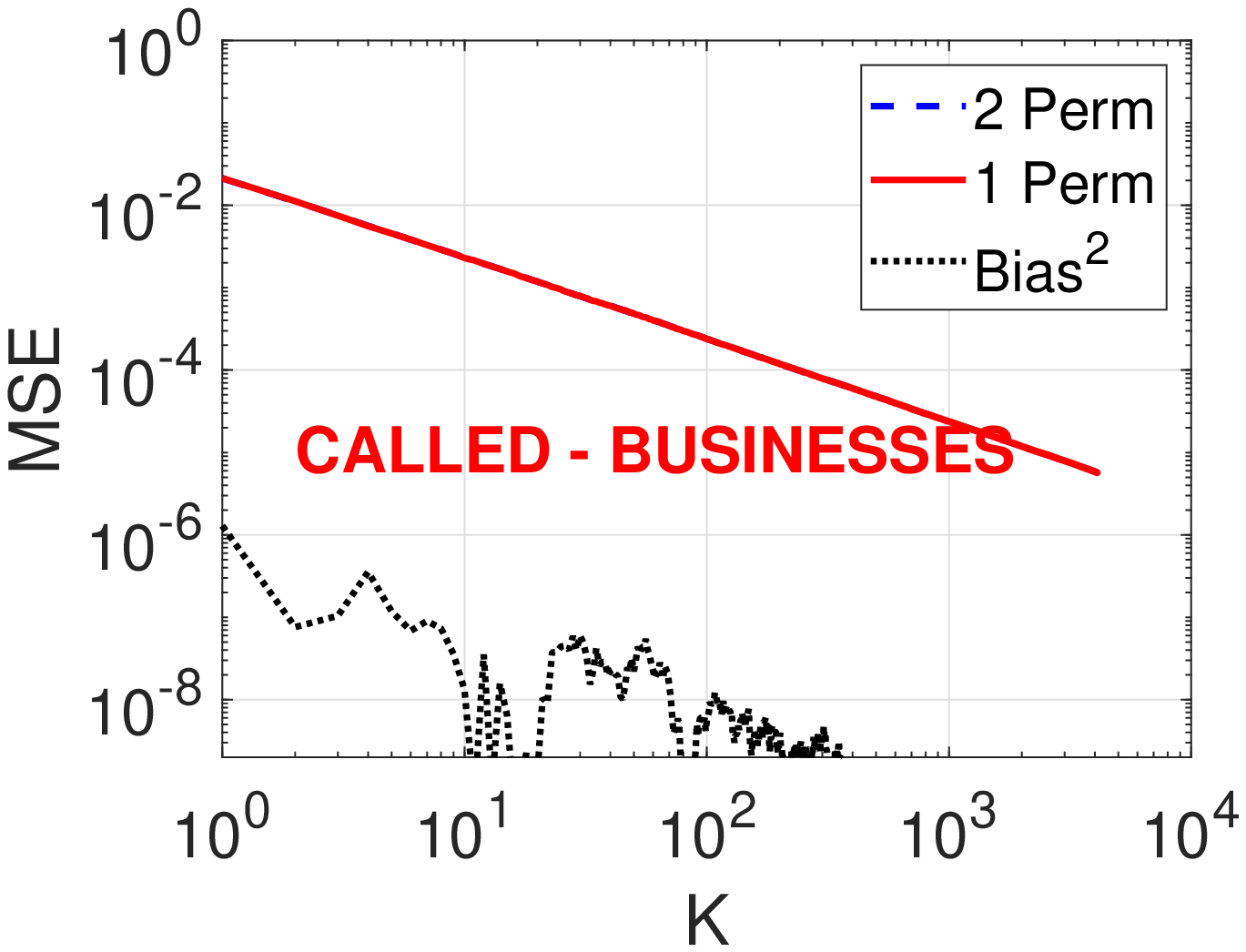}
    }

  \end{center}
  \vspace{-0.1in}
  \caption{Empirical MSEs of C-MinHash-$(\pi,\pi)$ (``1 Perm'', red, solid) vs. C-MinHash-$(\sigma,\pi)$ (``2 Perm'', blue, dashed) on various data pairs from the \textit{Words} dataset. We also report the empirical bias$^2$ for C-MinHash-$(\pi,\pi)$ to show that the bias is so small that it can be safely neglected. The empirical MSE curves for both estimators essentially overlap for all data pairs, for $K$ ranging from 1 to 4096. }
  \label{fig:word1}
\end{figure}

\begin{figure}[H]
  \begin{center}
   \mbox{
    \includegraphics[width=2.1in]{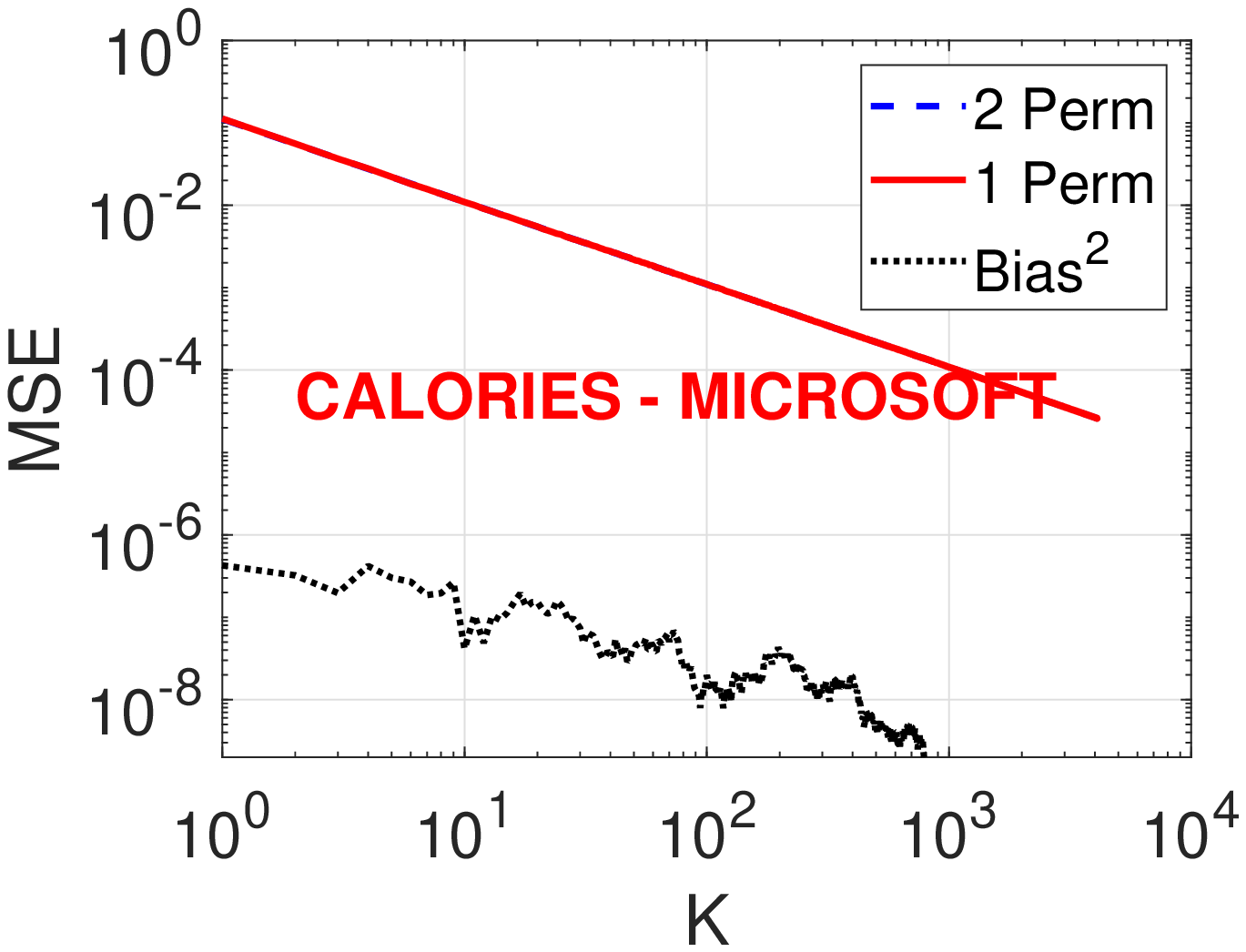}
    \includegraphics[width=2.1in]{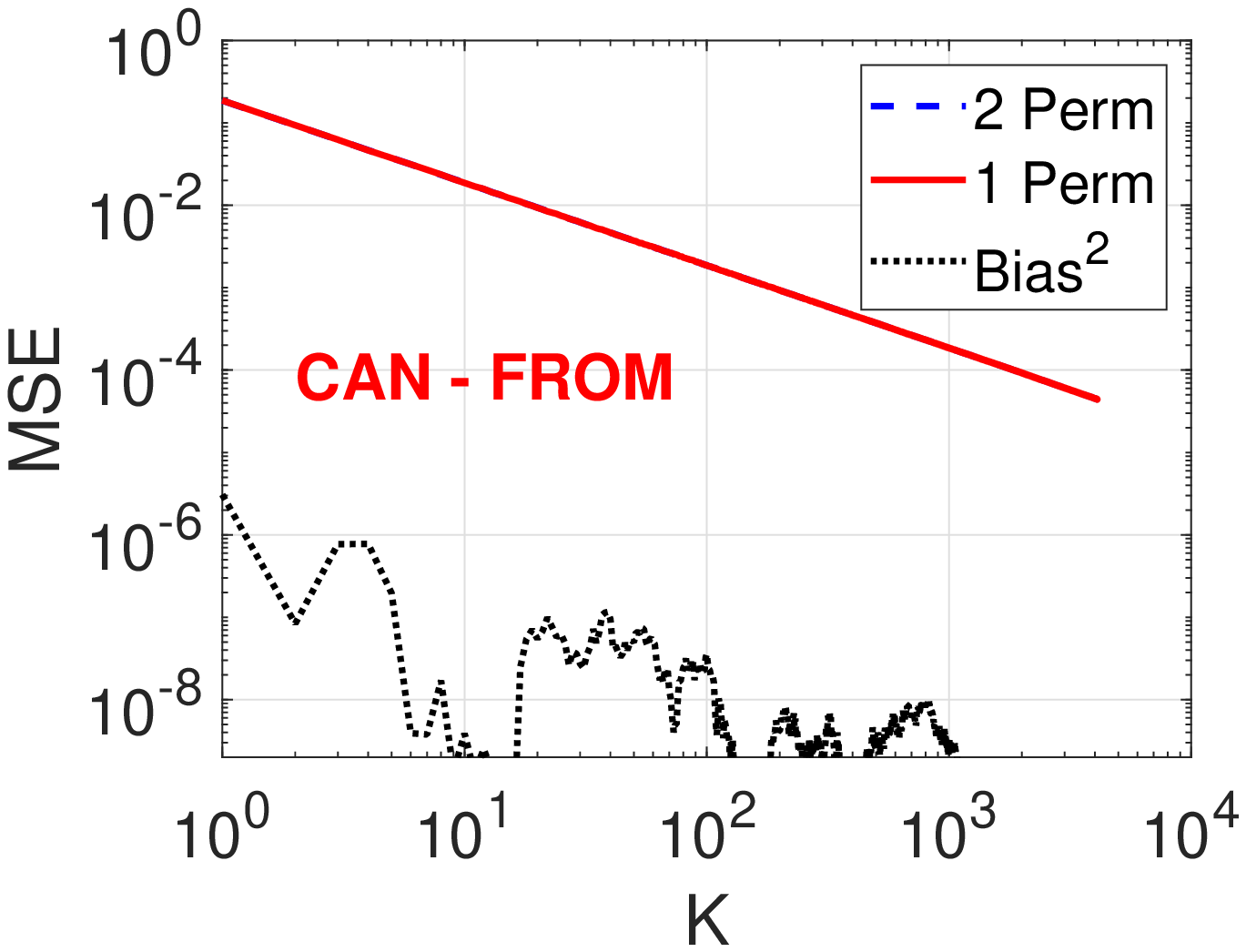}
    \includegraphics[width=2.1in]{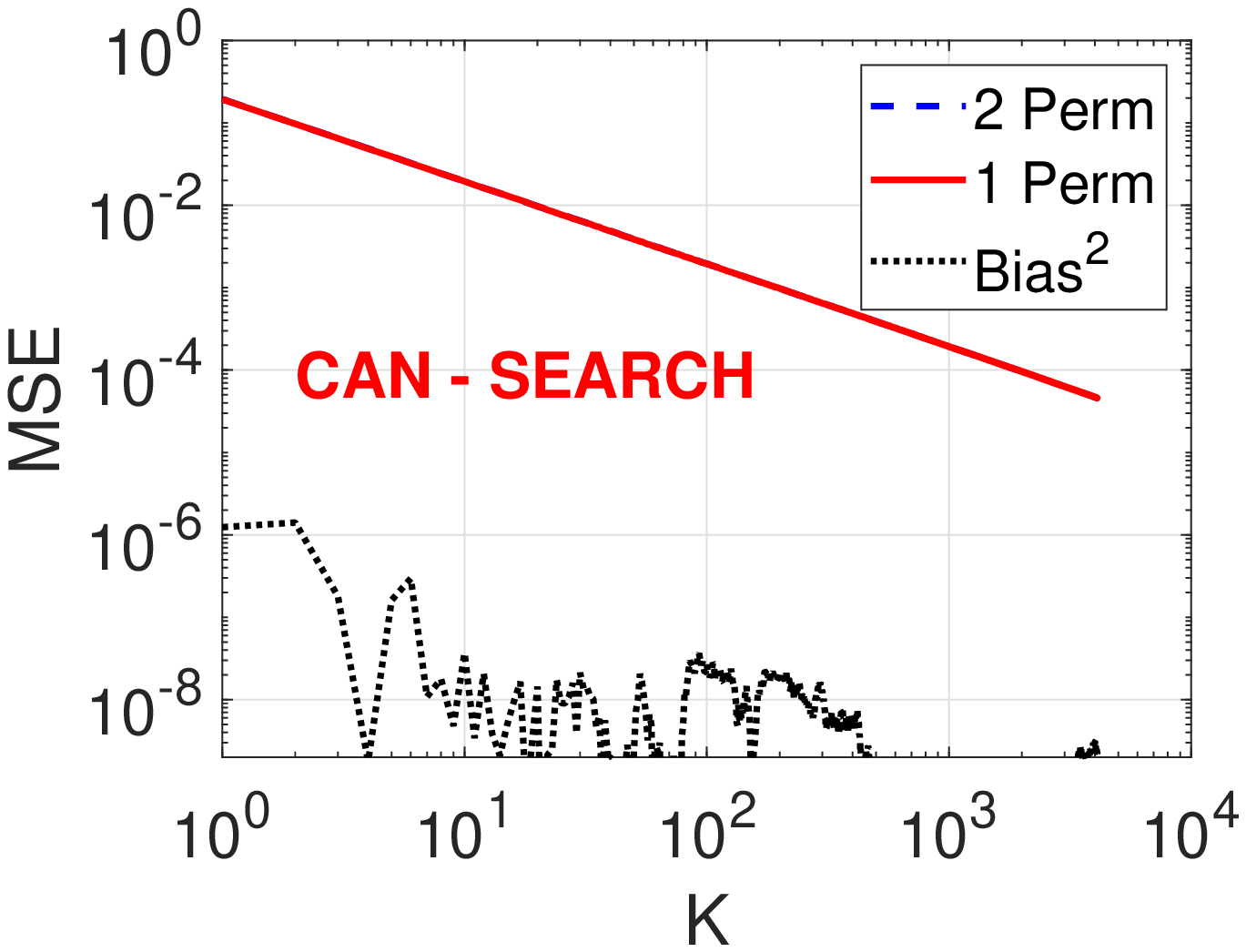}
    }
    \mbox{
    \includegraphics[width=2.1in]{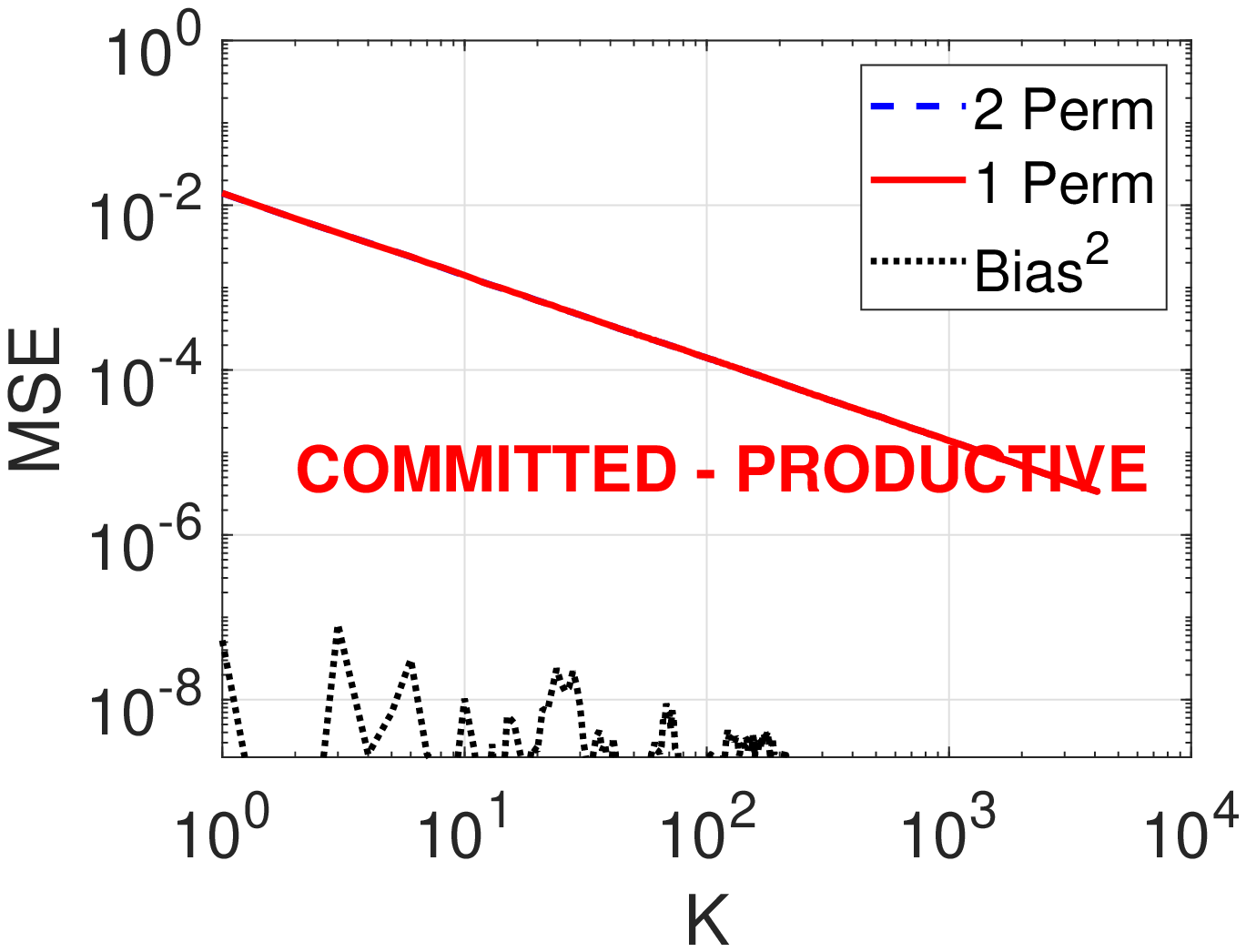}
    \includegraphics[width=2.1in]{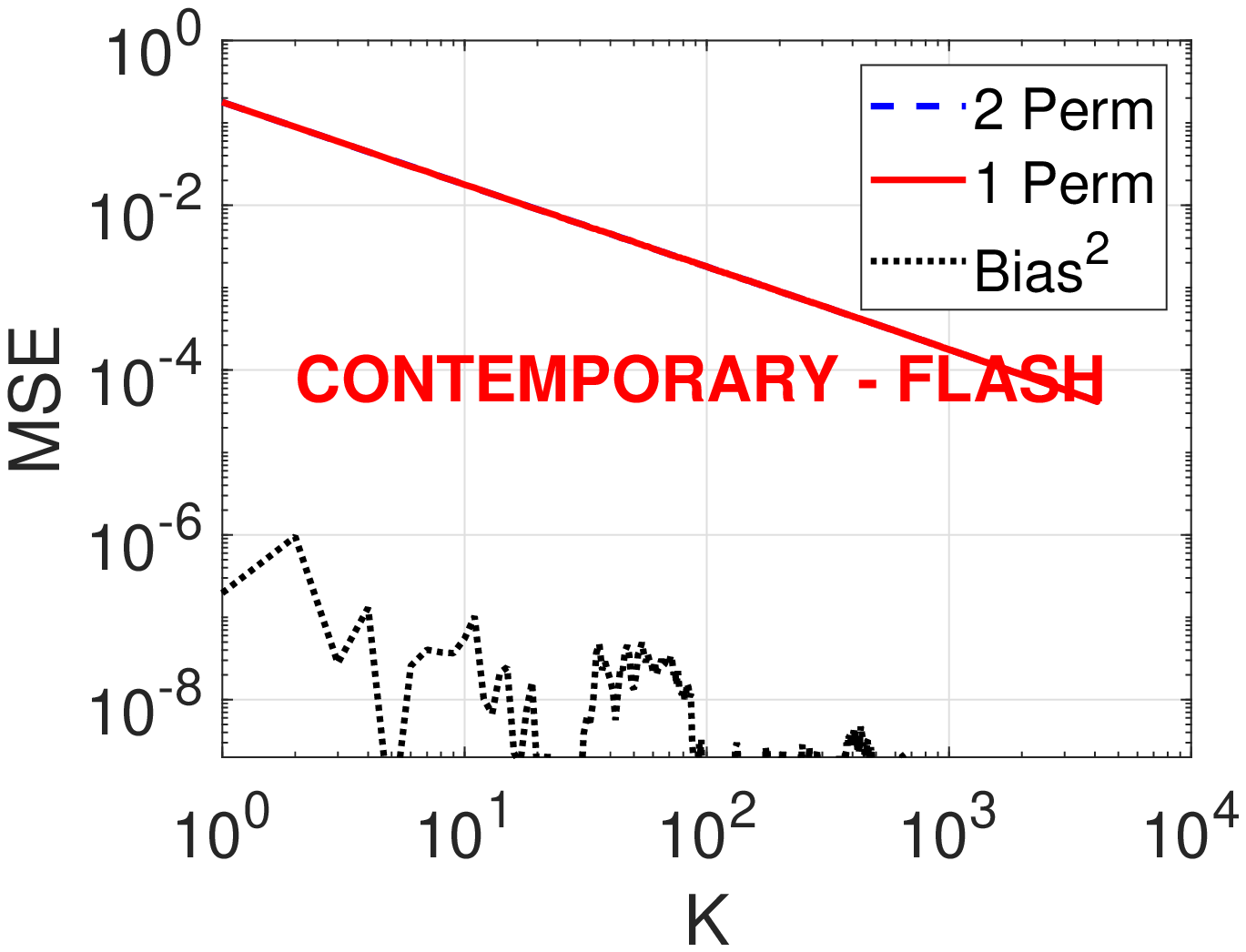}
    \includegraphics[width=2.1in]{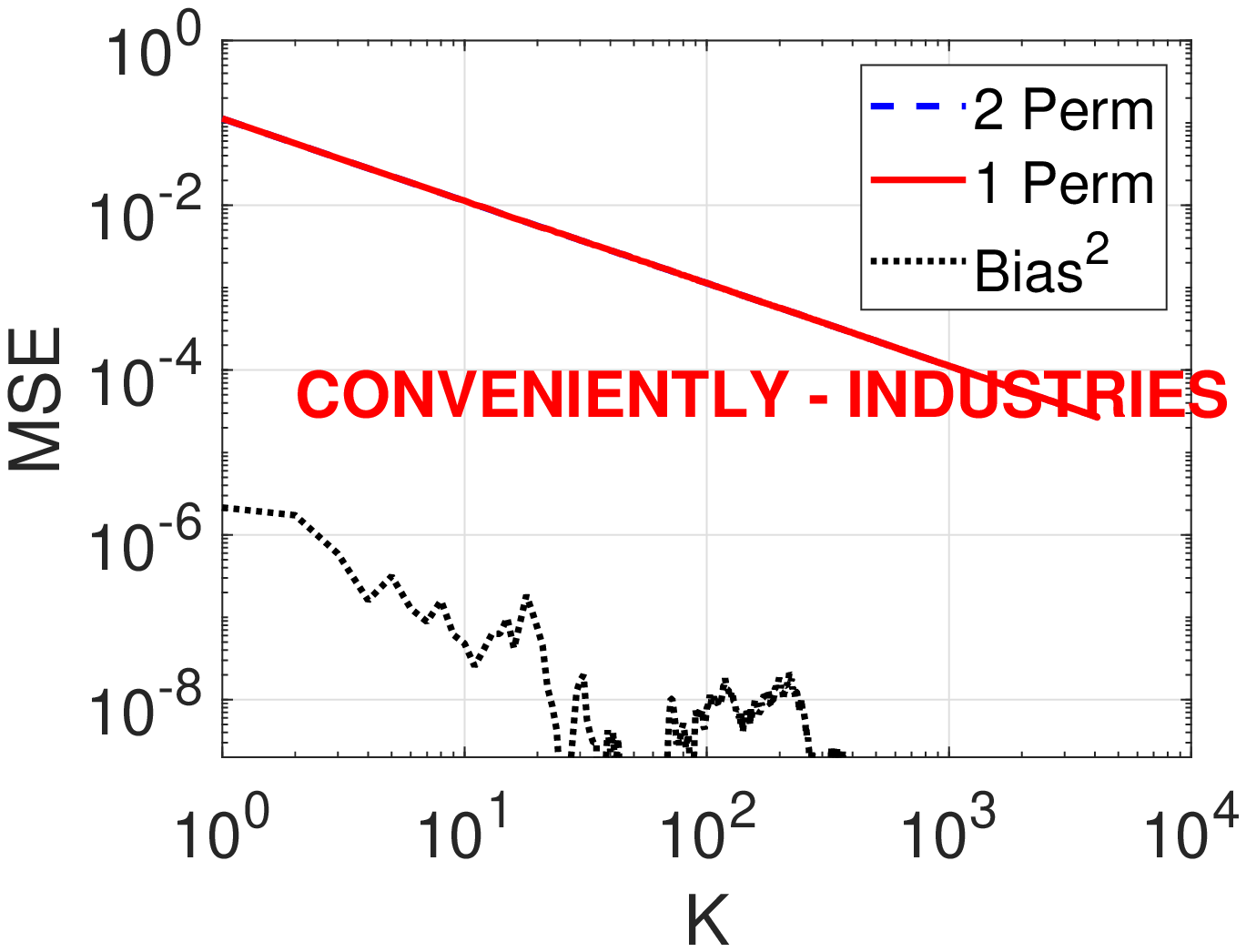}
    }
    \mbox{
    \includegraphics[width=2.1in]{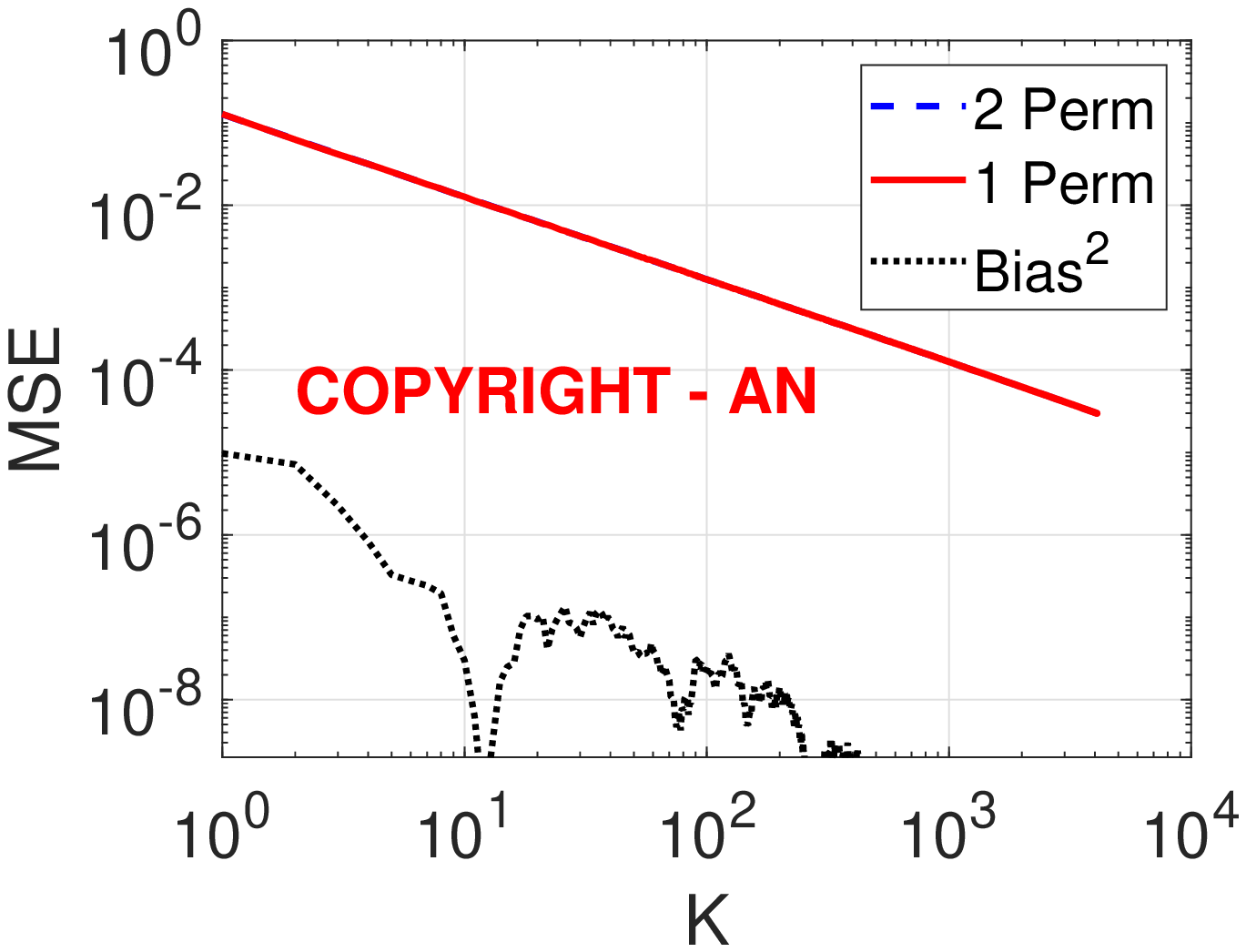}
    \includegraphics[width=2.1in]{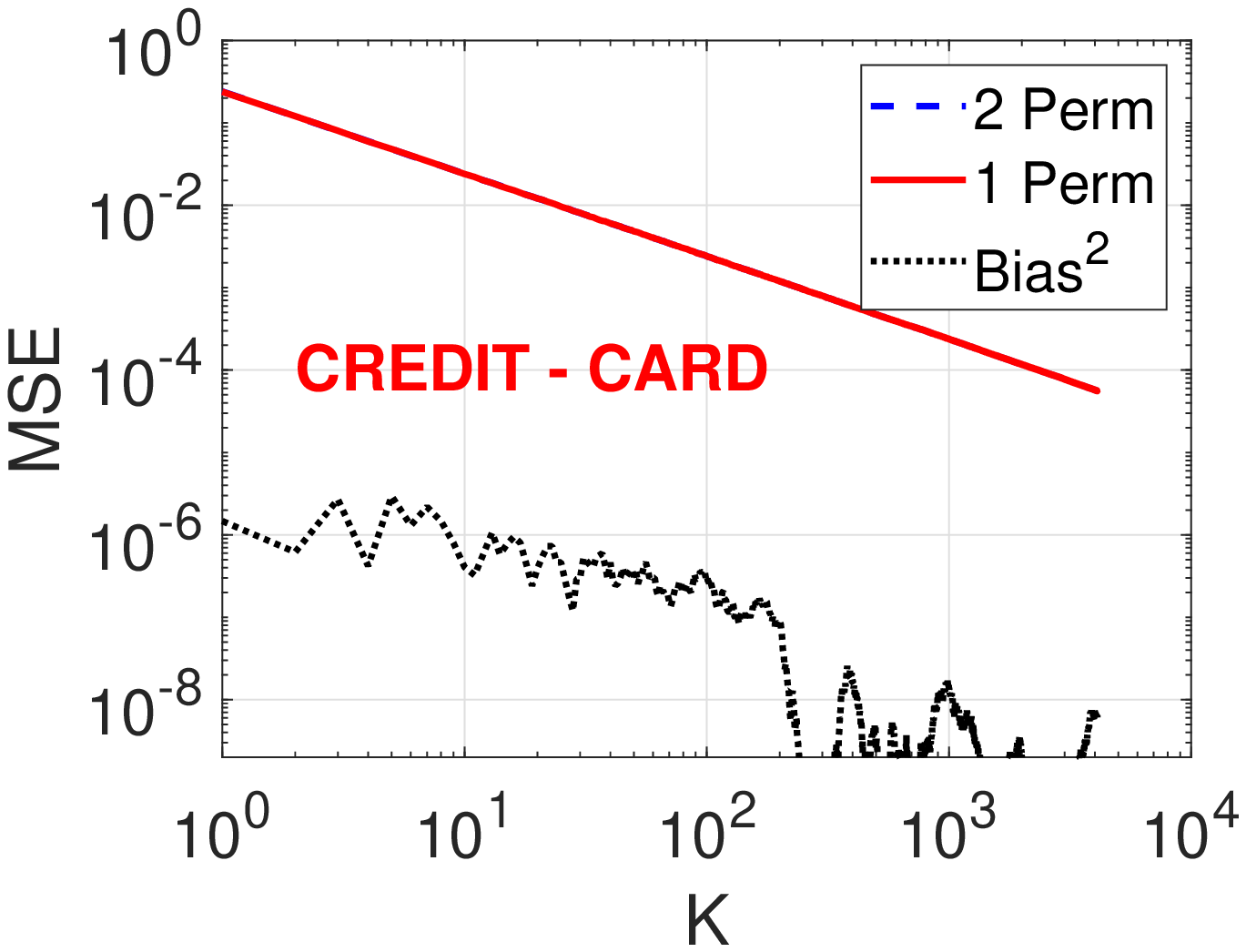}
    \includegraphics[width=2.1in]{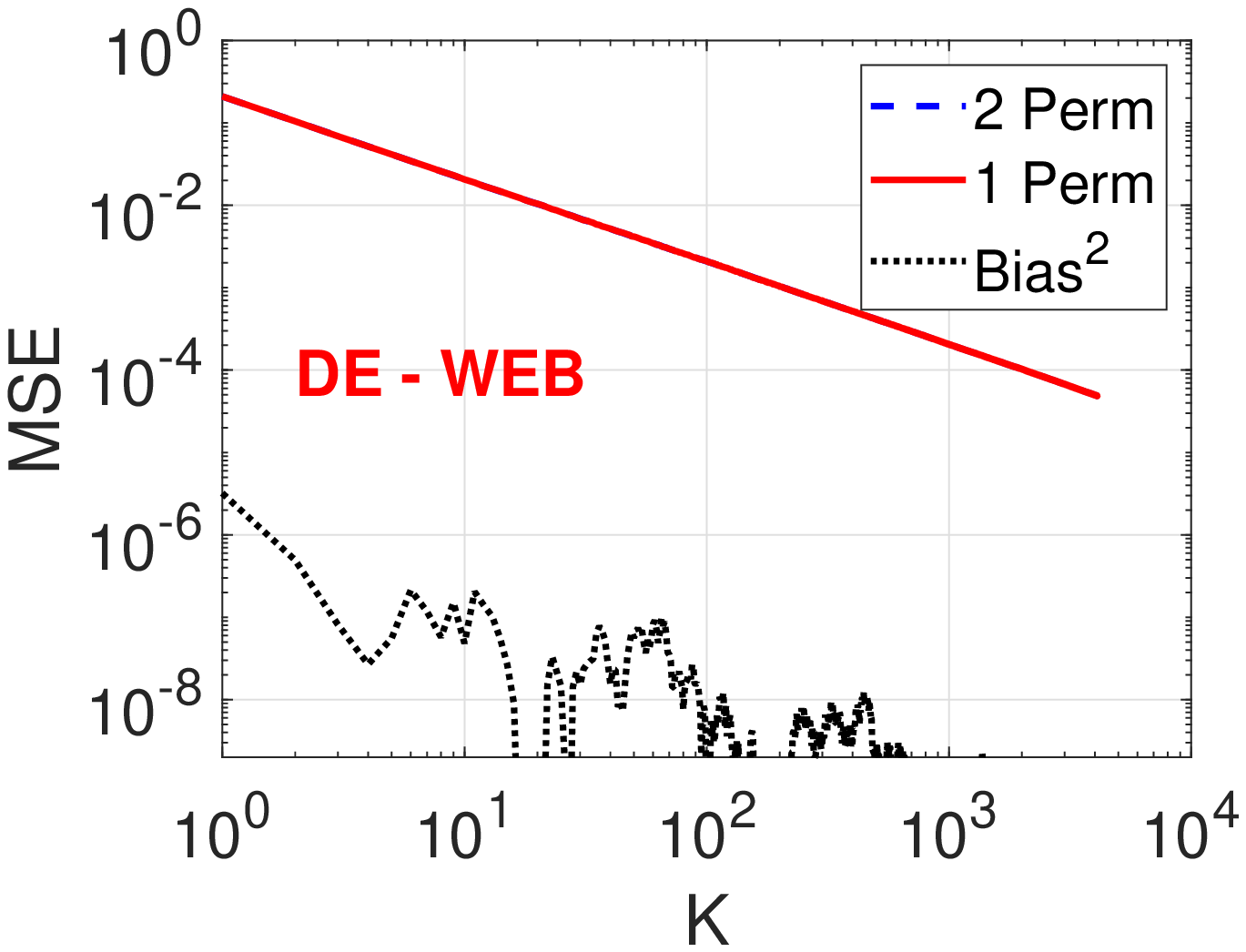}
    }
    \mbox{
    \includegraphics[width=2.1in]{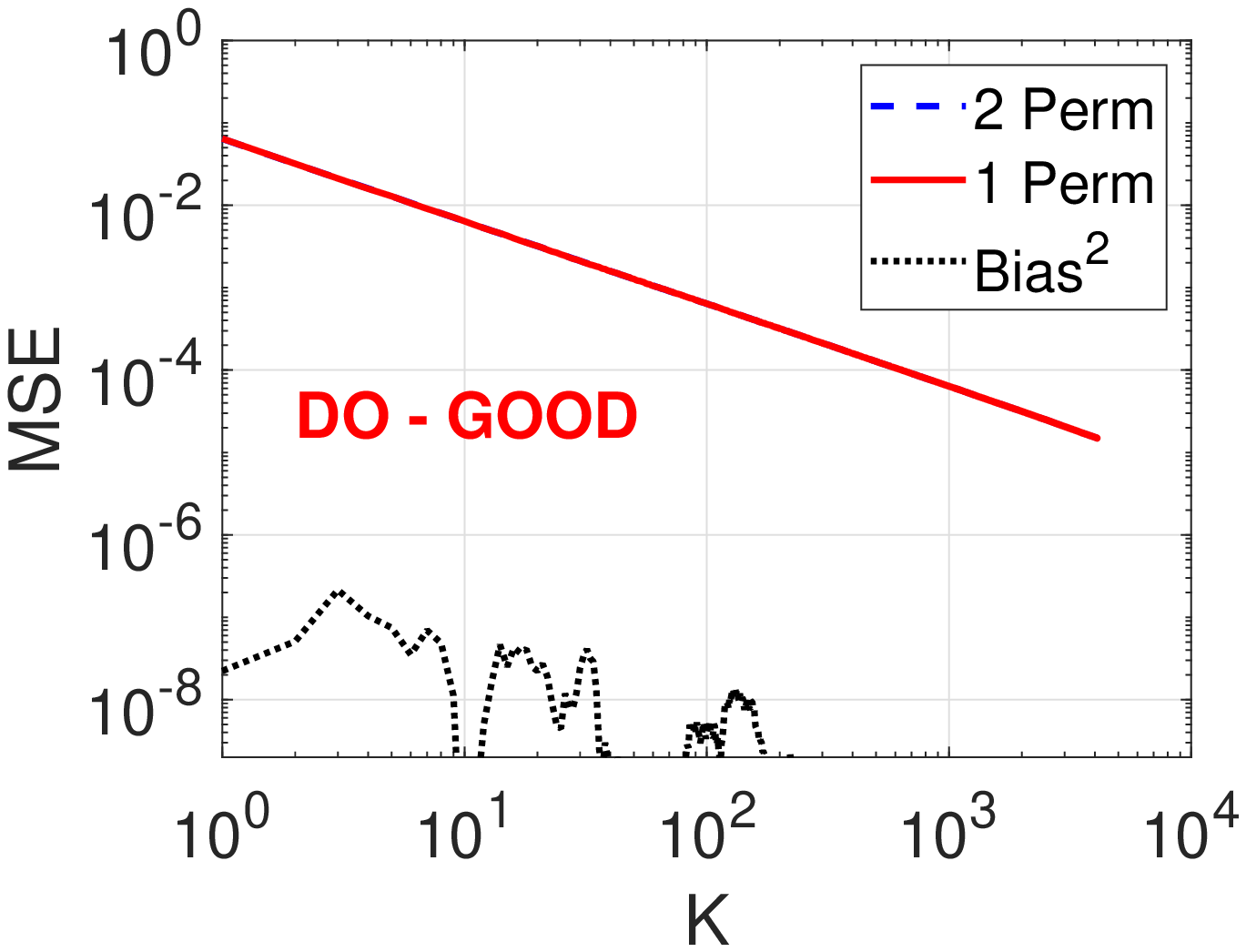}
    \includegraphics[width=2.1in]{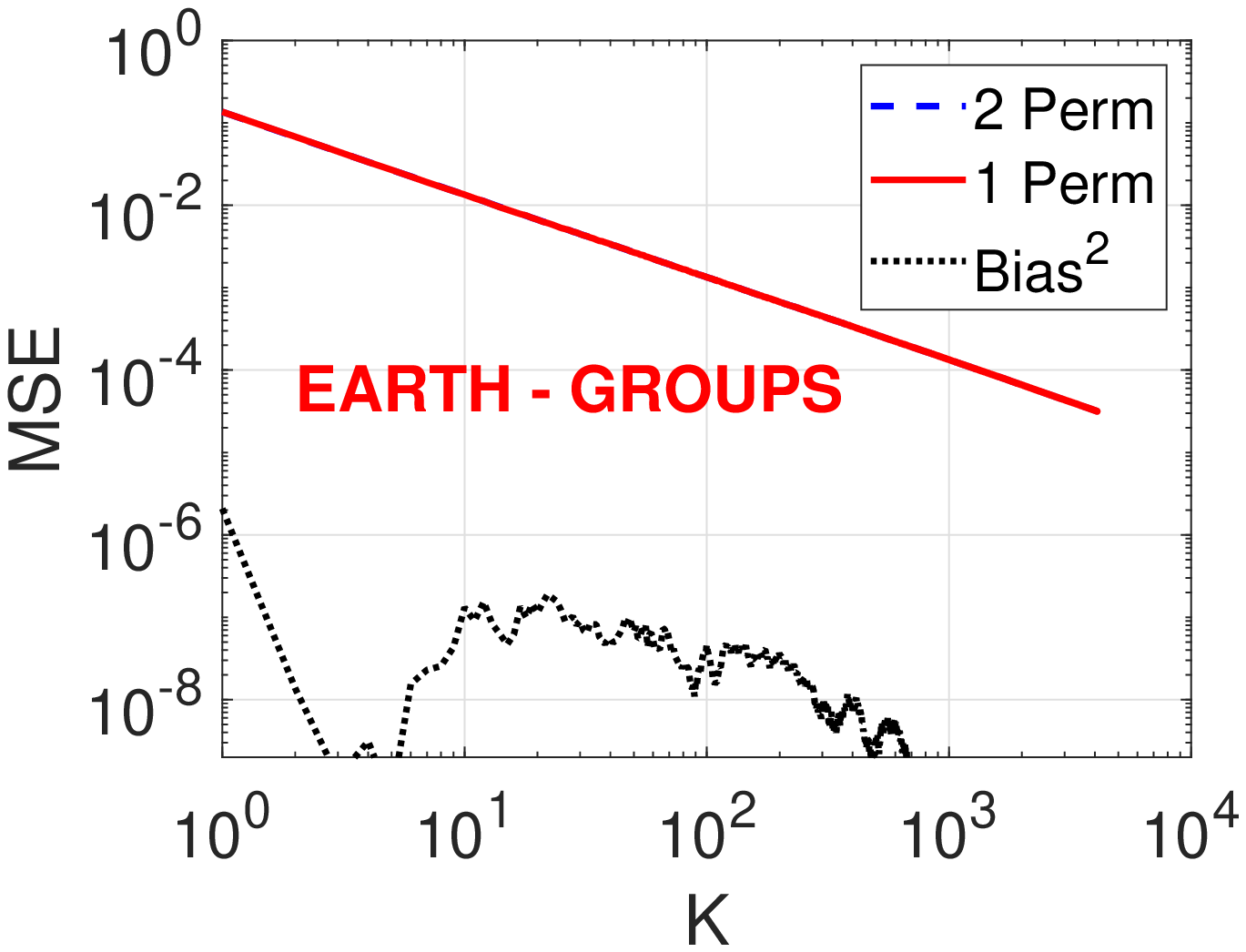}
    \includegraphics[width=2.1in]{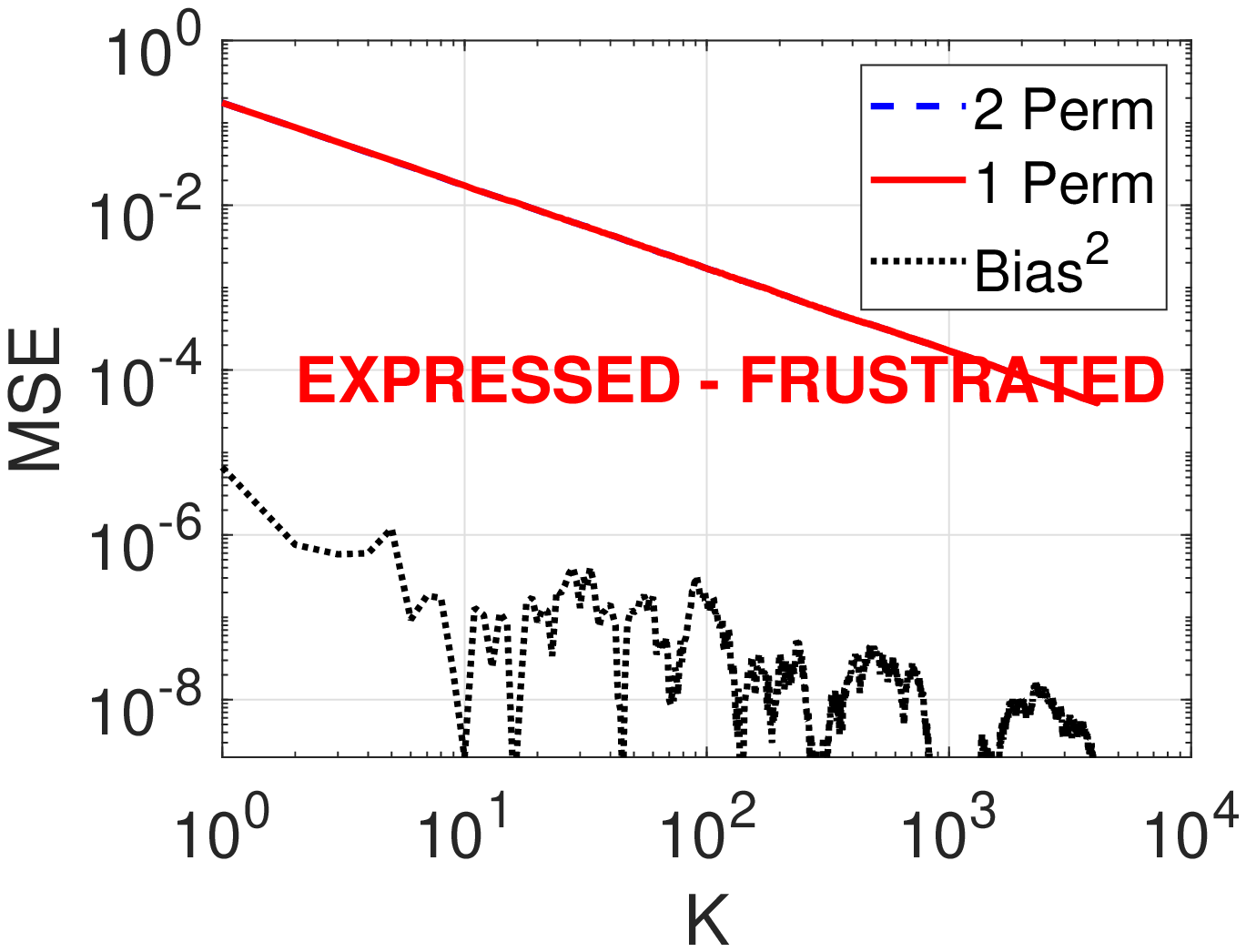}
    }
    \mbox{
    \includegraphics[width=2.1in]{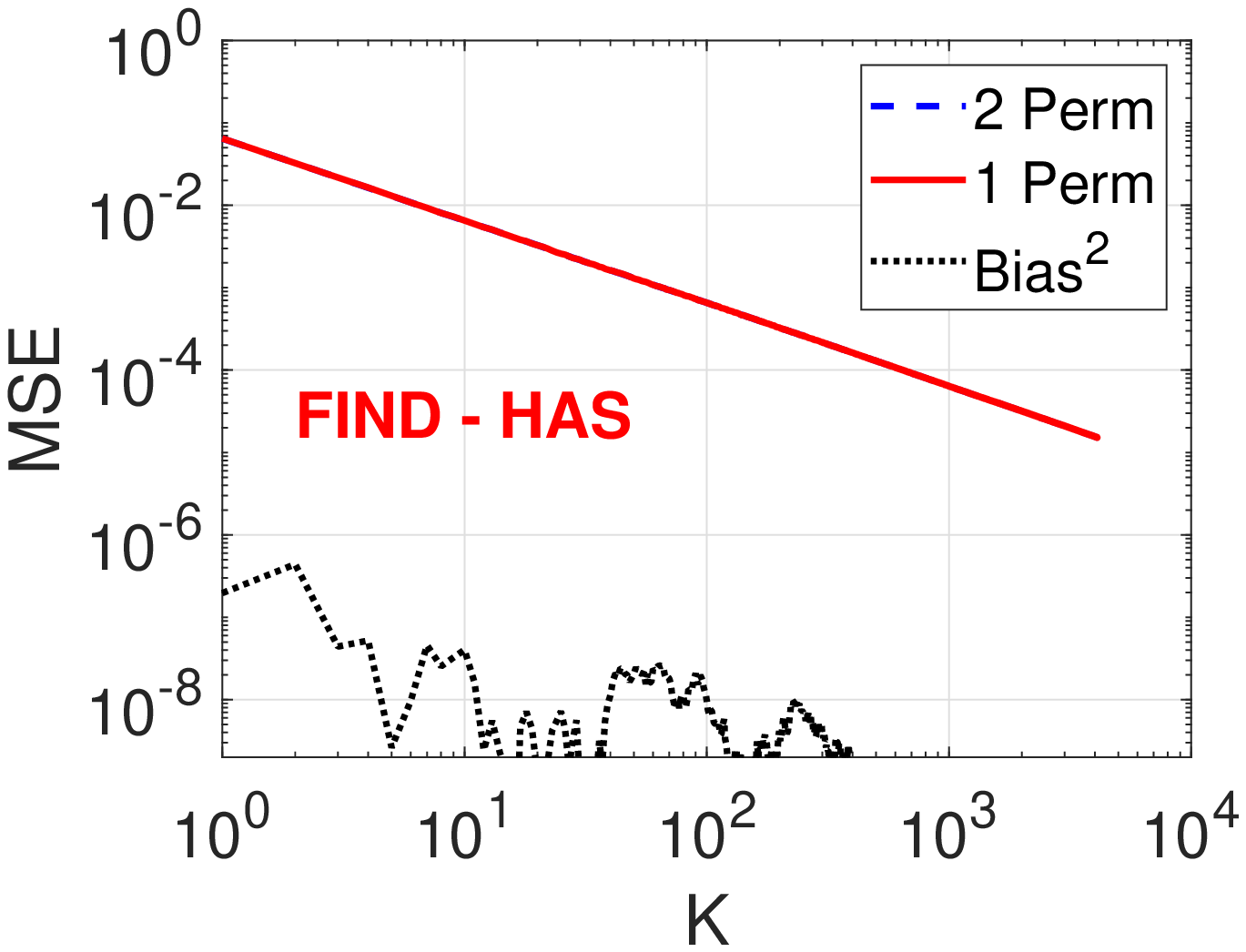}
    \includegraphics[width=2.1in]{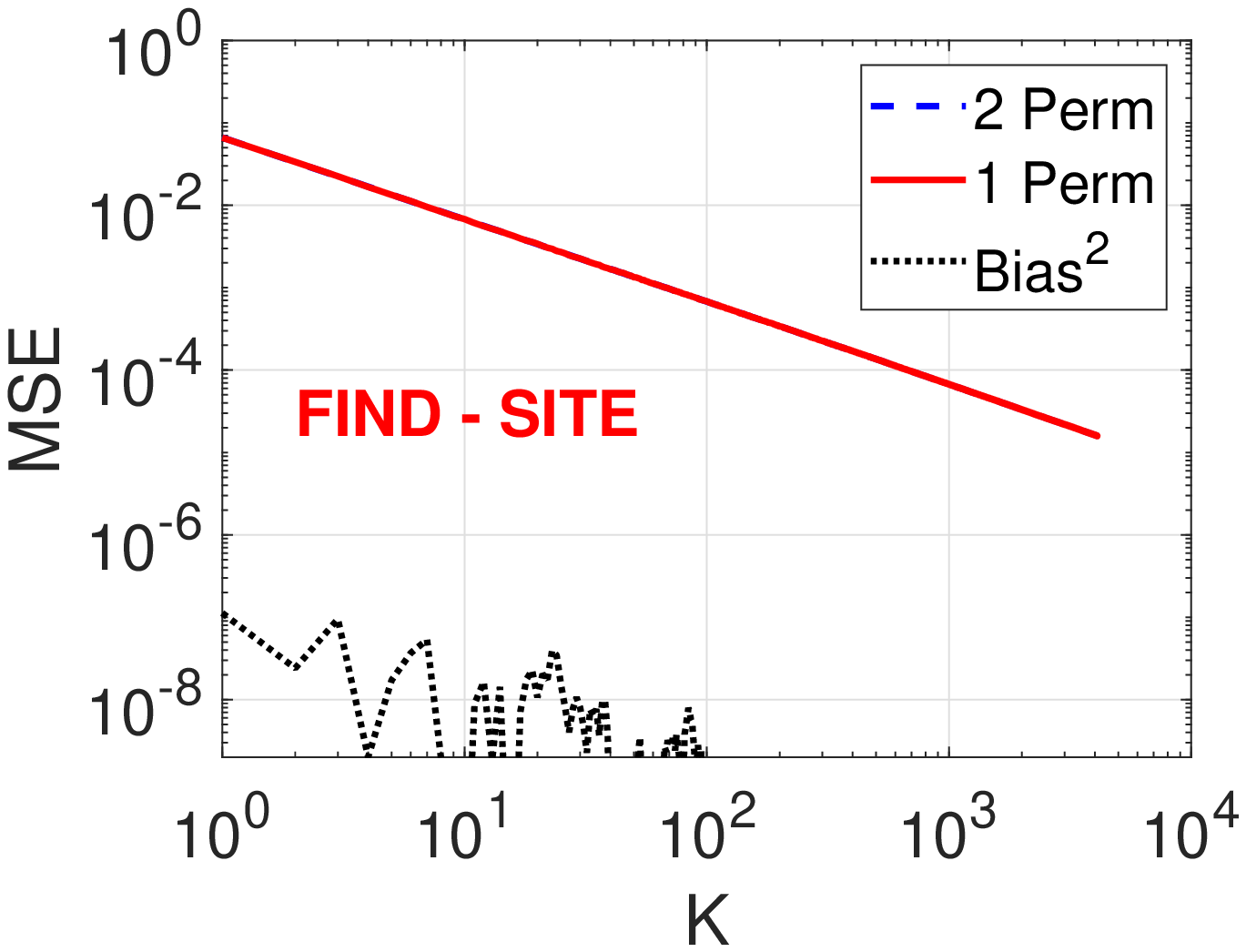}
    \includegraphics[width=2.1in]{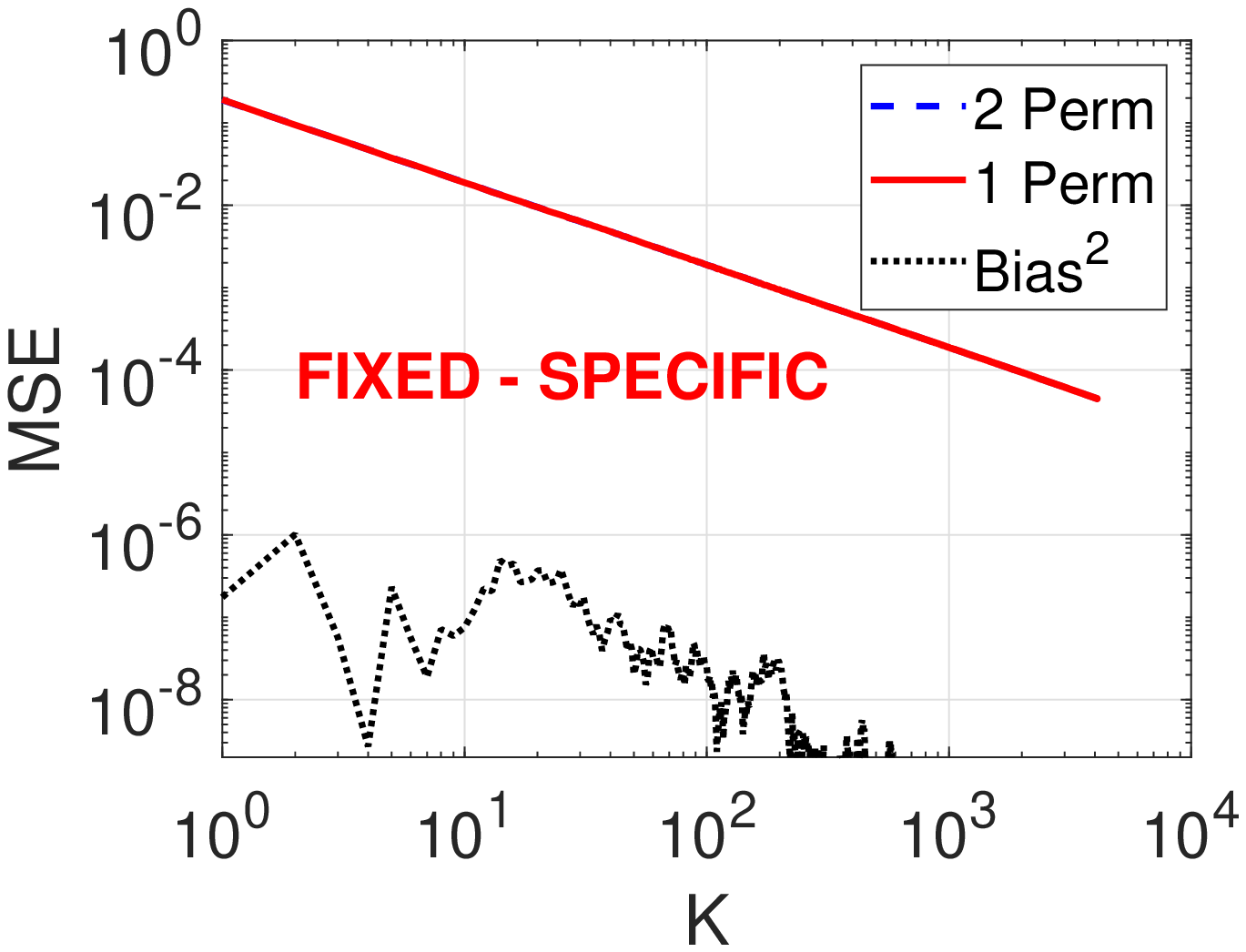}
    }

  \end{center}
  \vspace{-0.1in}
  \caption{Empirical MSEs of C-MinHash-$(\pi,\pi)$ (``1 Perm'', red, solid) vs. C-MinHash-$(\sigma,\pi)$ (``2 Perm'', blue, dashed) on various data pairs from the \textit{Words} dataset. We also report the empirical bias$^2$ for C-MinHash-$(\pi,\pi)$ to show that the bias is so small that it can be safely neglected. The empirical MSE curves for both estimators essentially overlap for all data pairs, for $K$ ranging from 1 to 4096. }
  \label{fig:word2}
\end{figure}

\begin{figure}[H]
  \begin{center}
   \mbox{
    \includegraphics[width=2.1in]{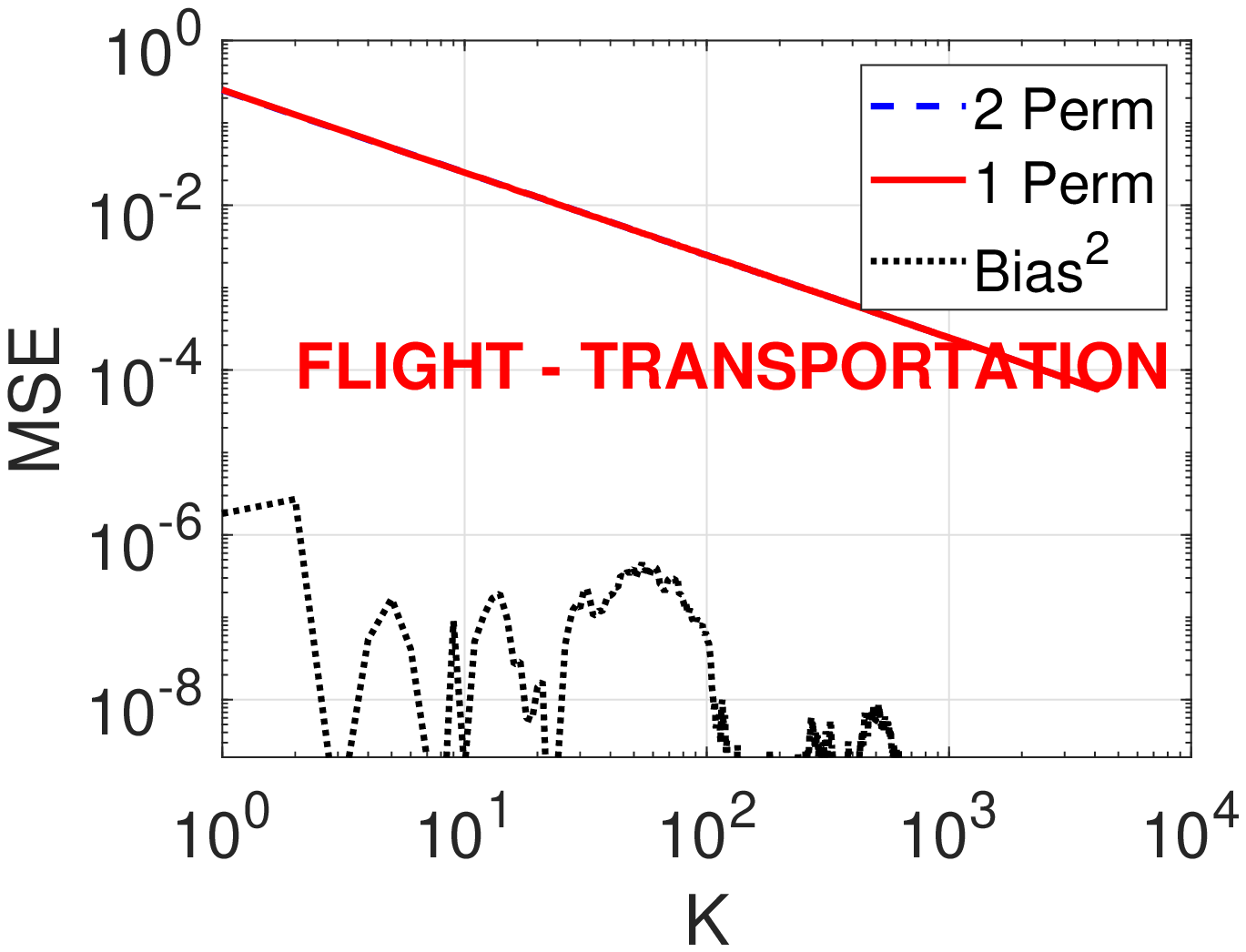}
    \includegraphics[width=2.1in]{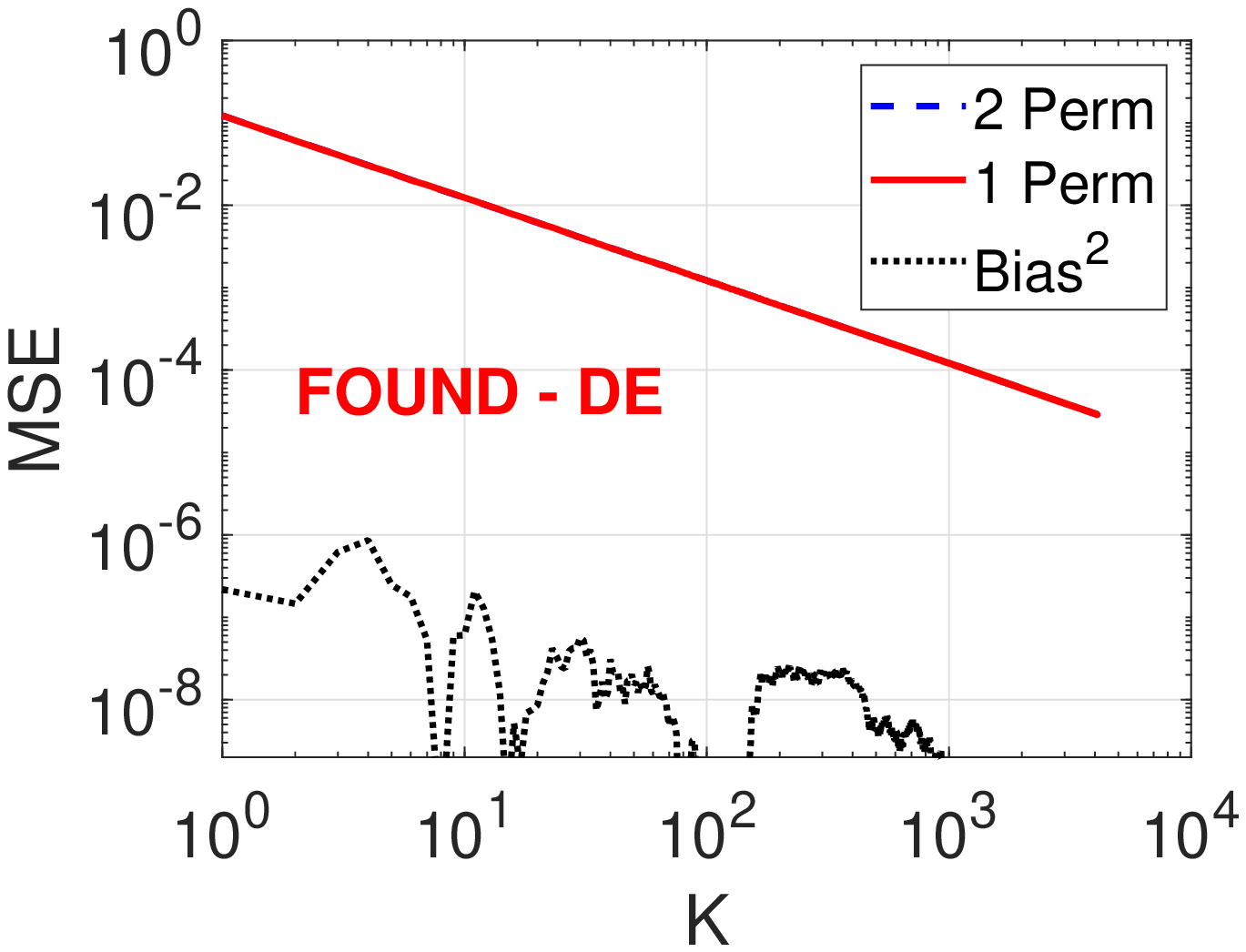}
    \includegraphics[width=2.1in]{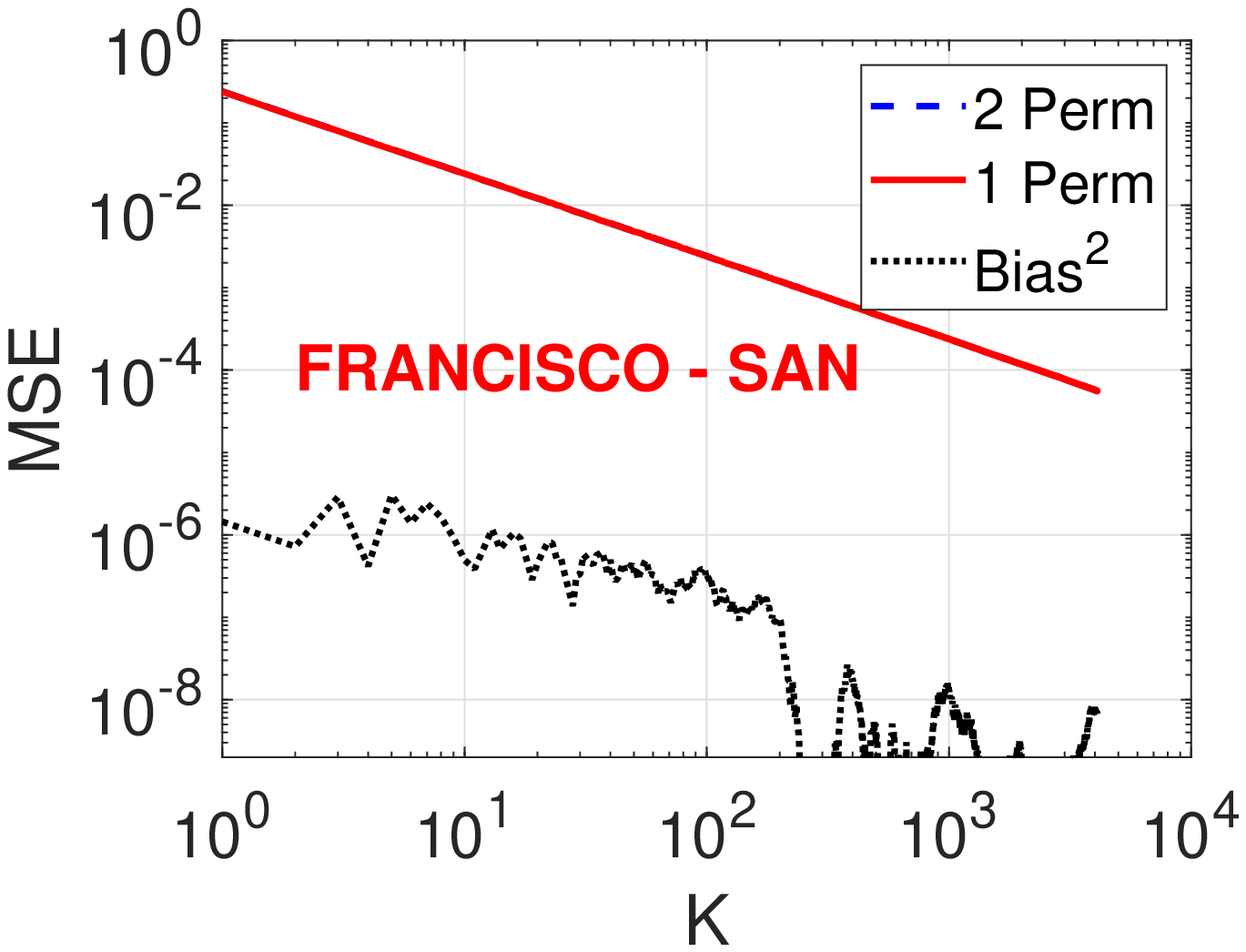}
    }
    \mbox{
    \includegraphics[width=2.1in]{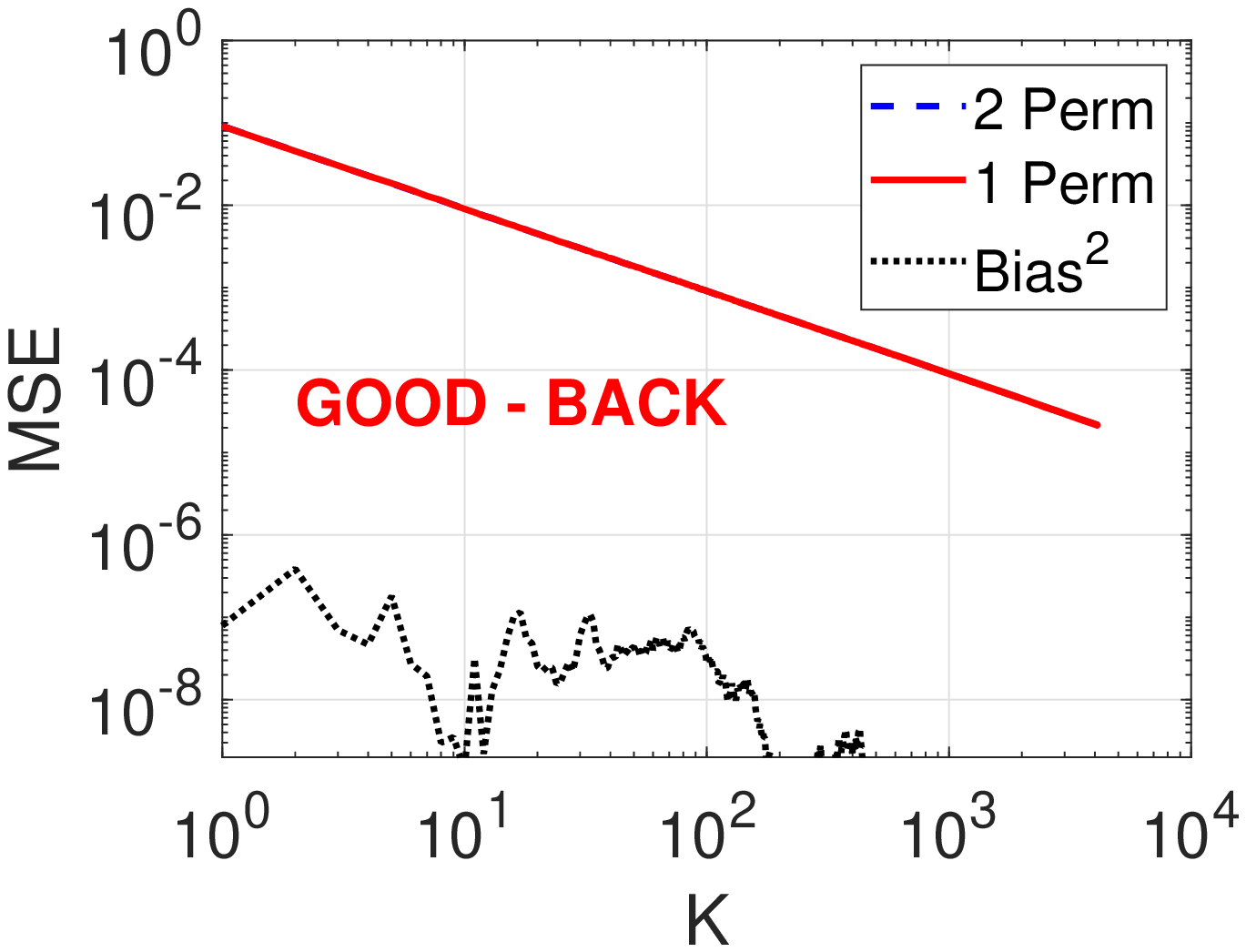}
    \includegraphics[width=2.1in]{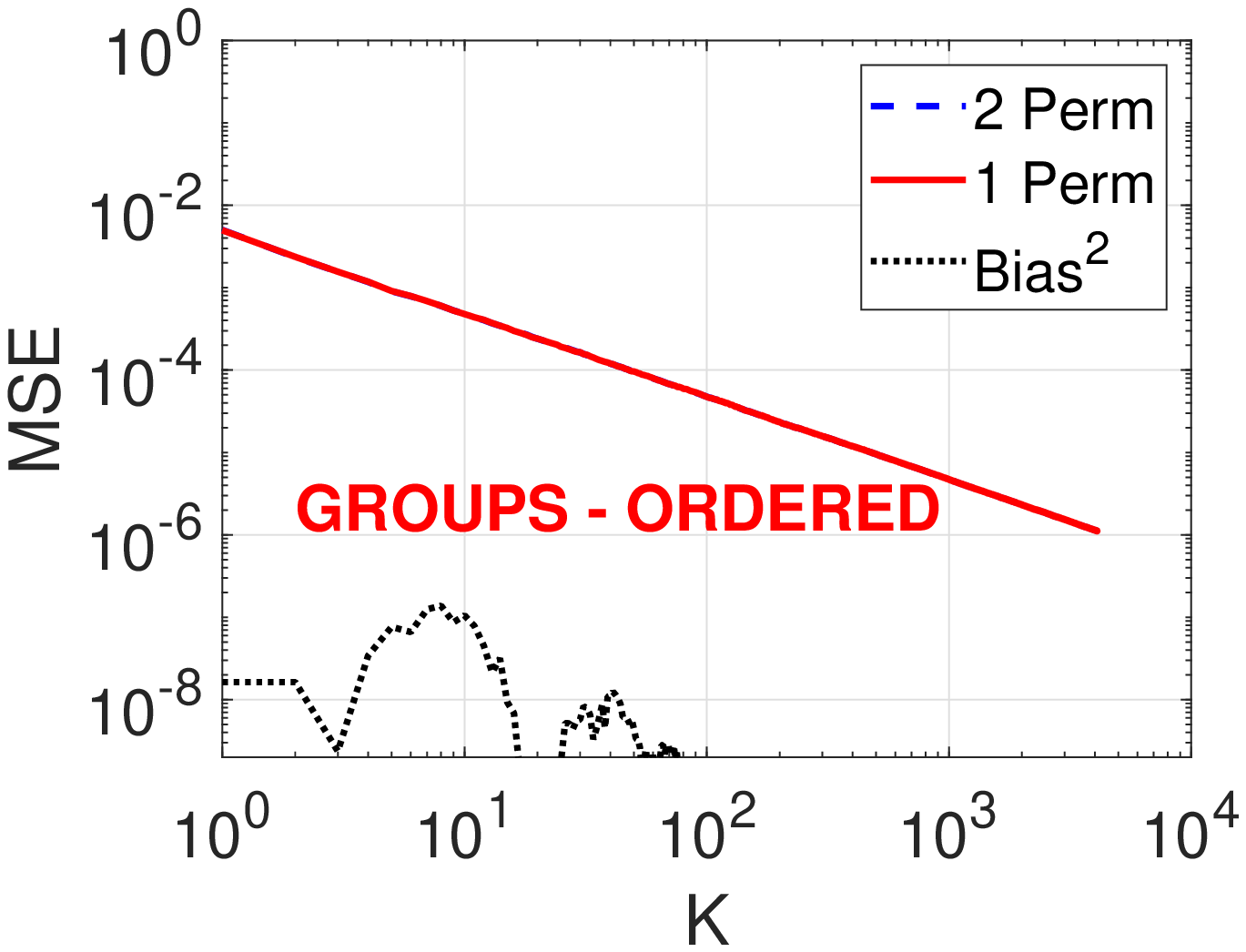}
    \includegraphics[width=2.1in]{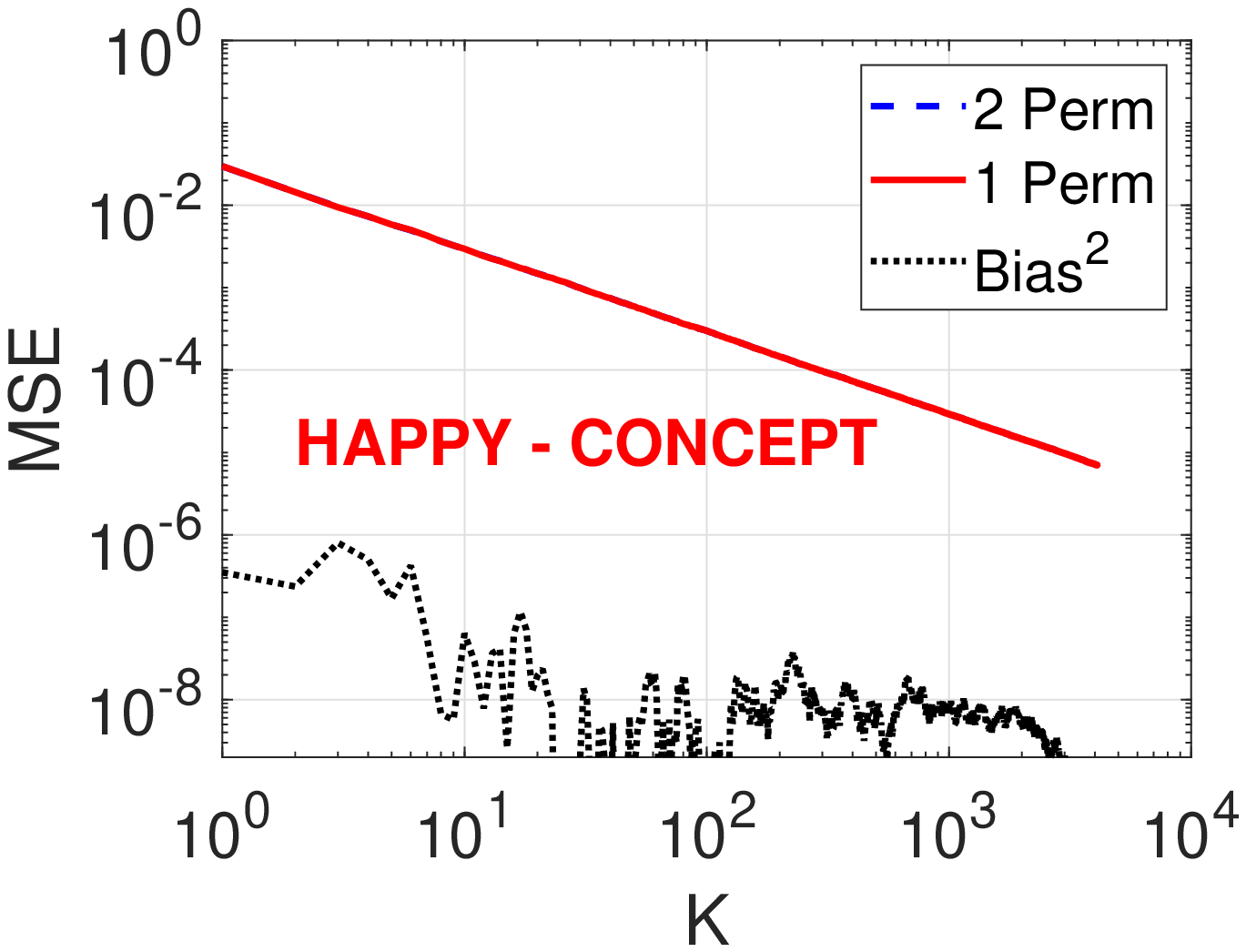}
    }
    \mbox{
    \includegraphics[width=2.1in]{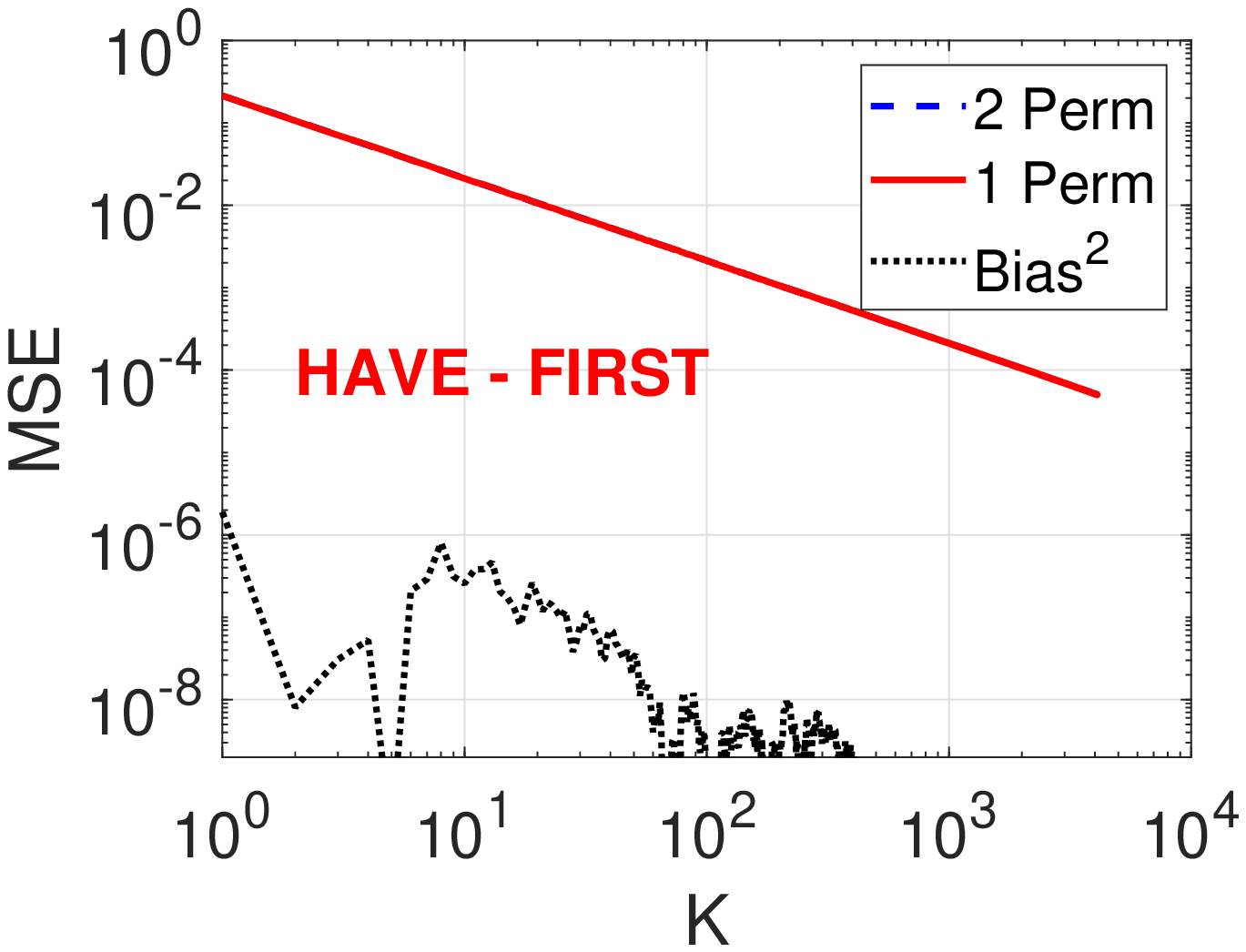}
    \includegraphics[width=2.1in]{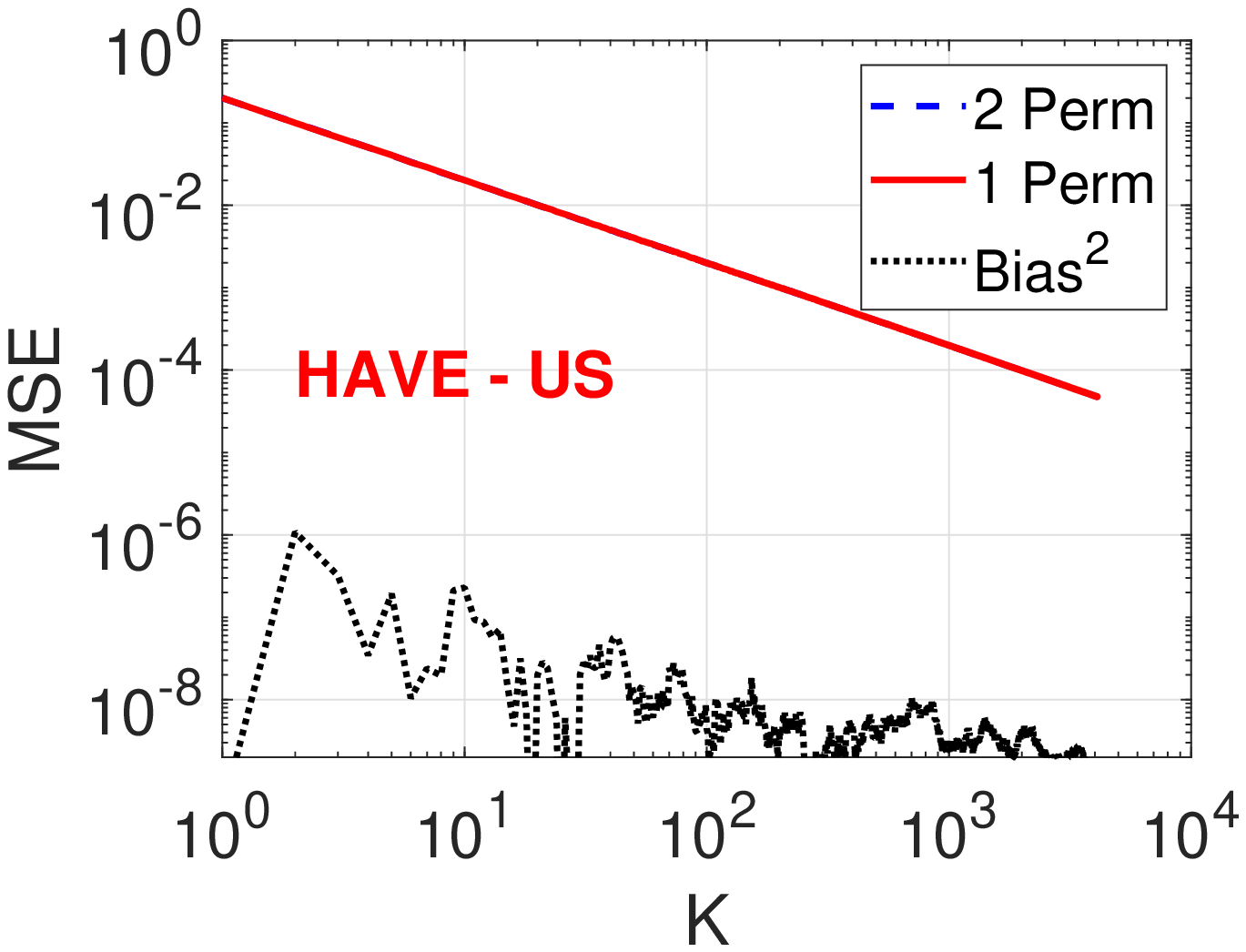}
    \includegraphics[width=2.1in]{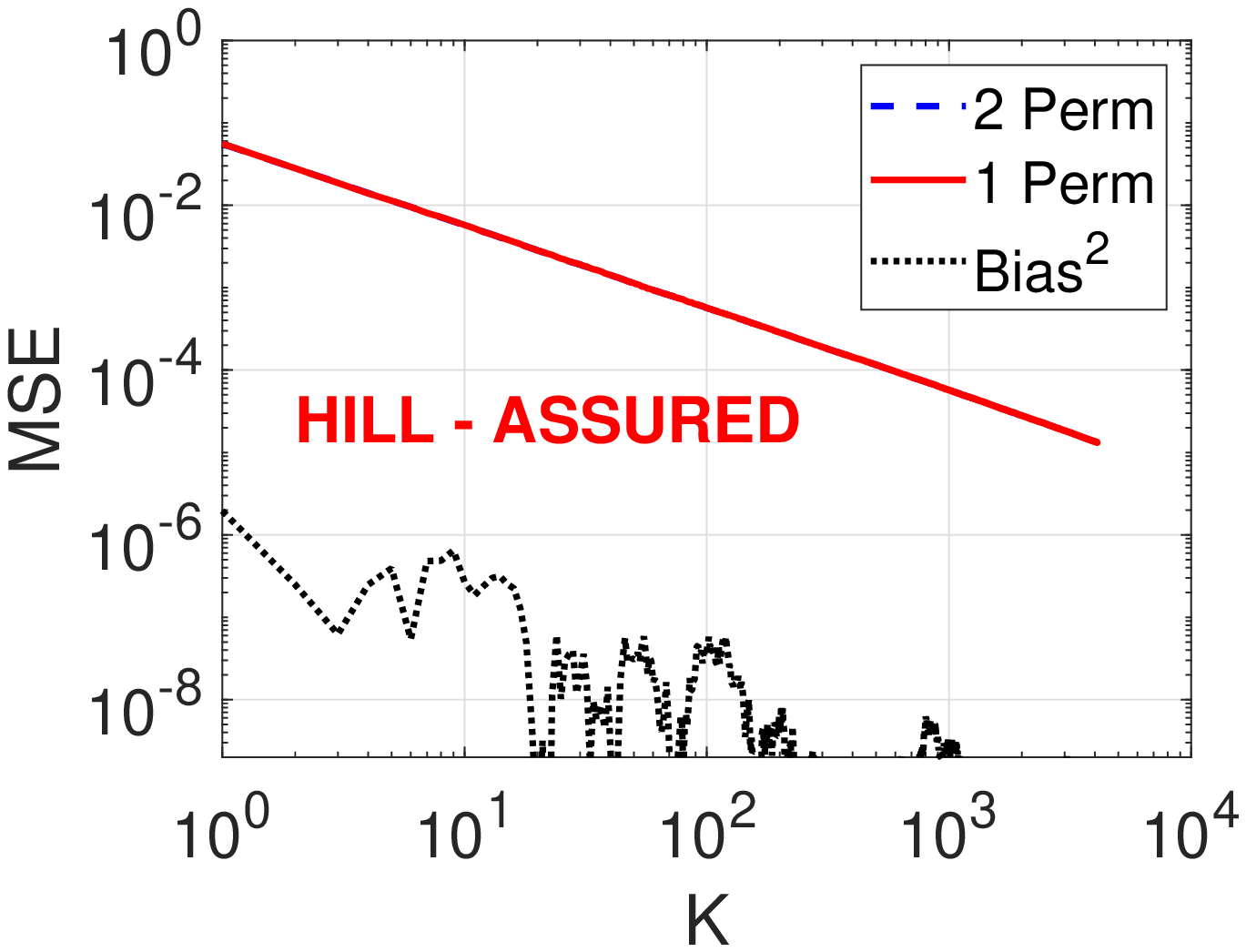}
    }
    \mbox{
    \includegraphics[width=2.1in]{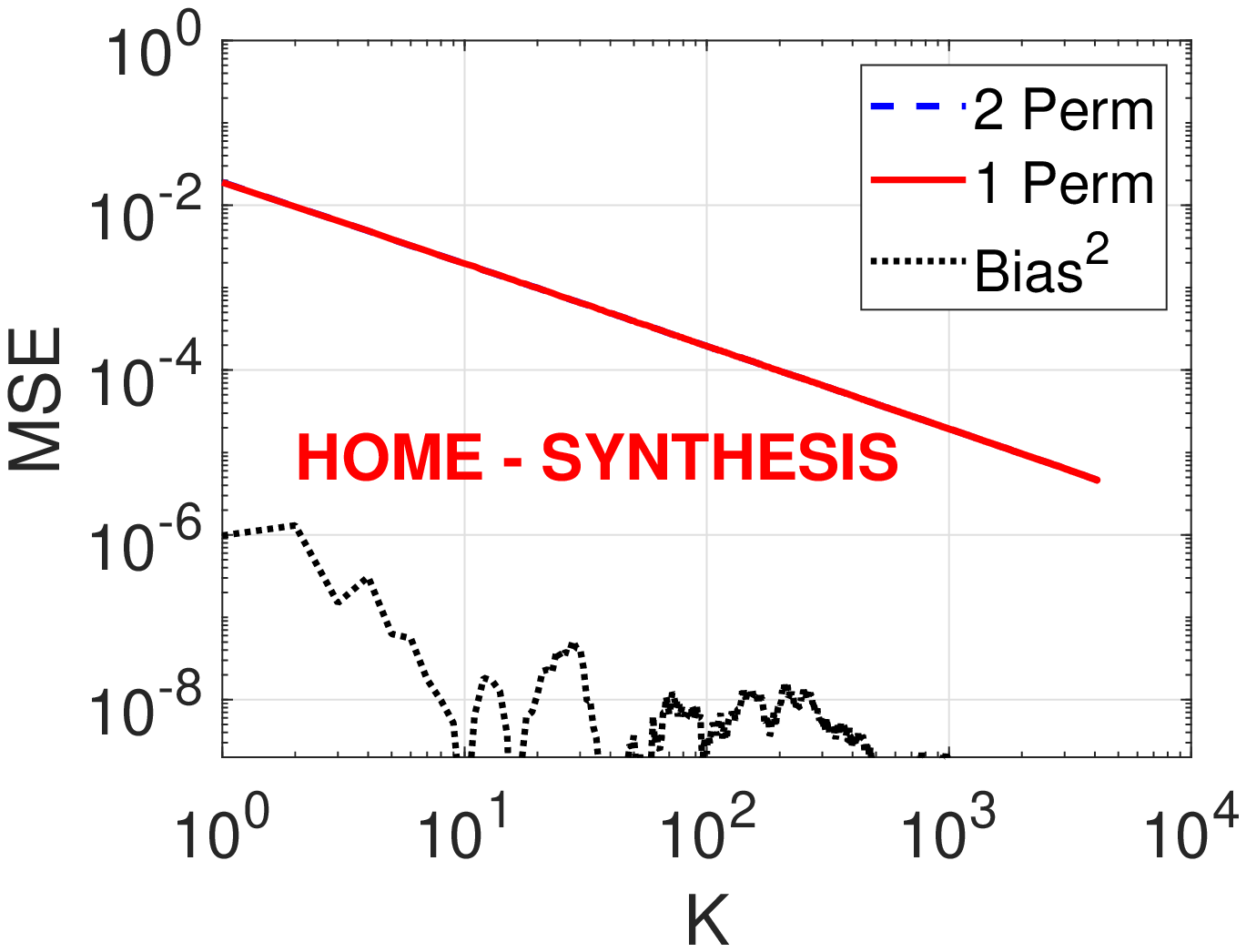}
    \includegraphics[width=2.1in]{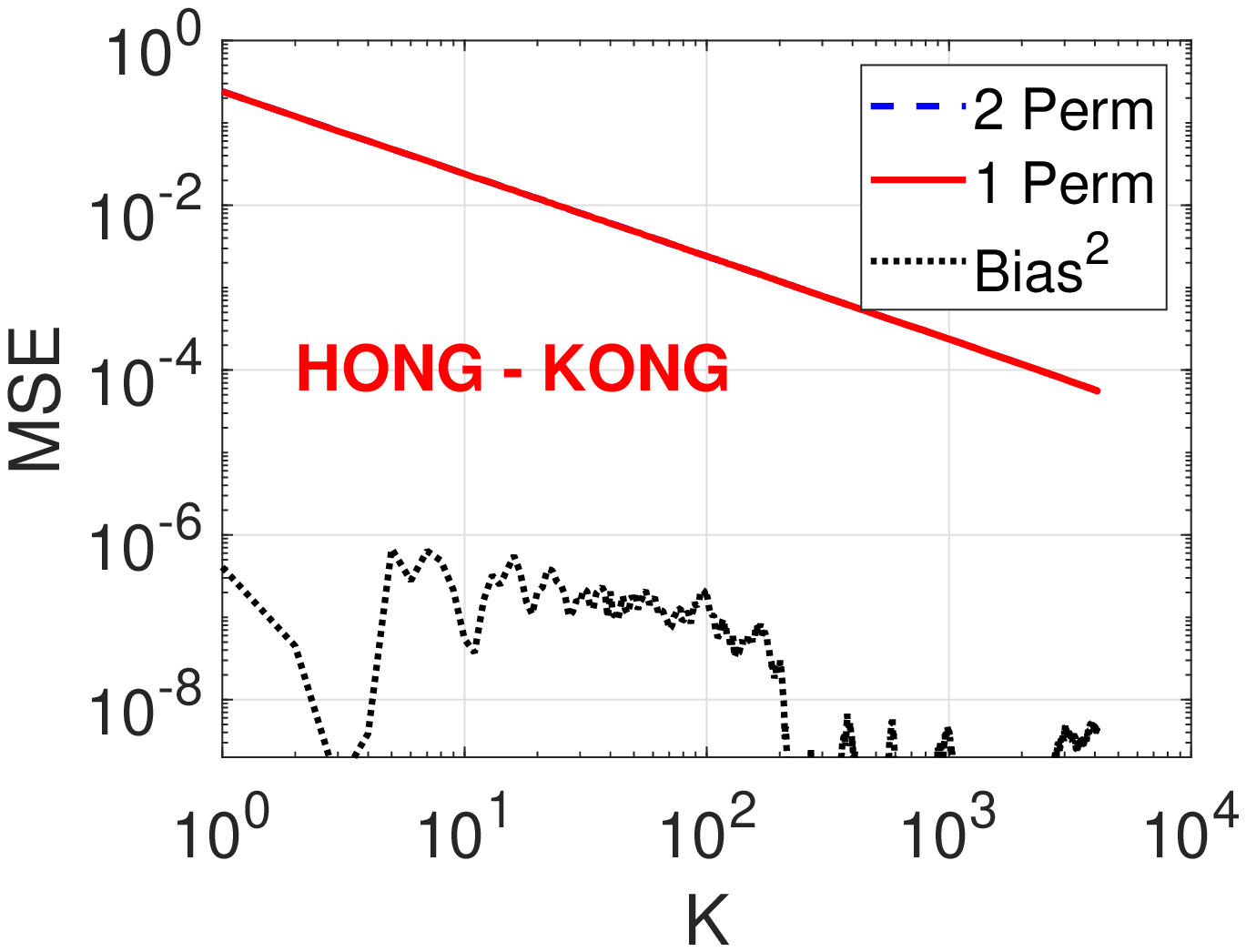}
    \includegraphics[width=2.1in]{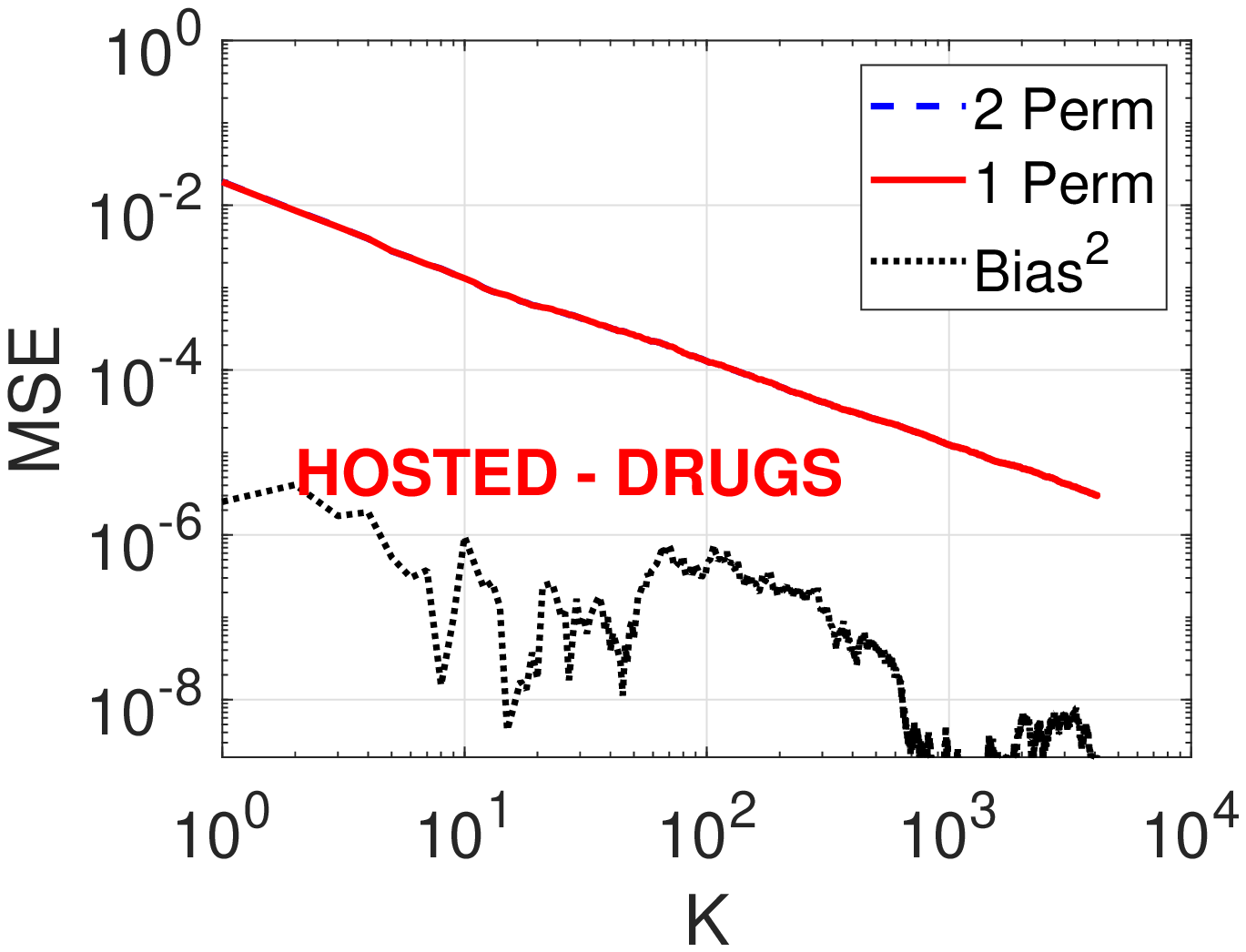}
    }
    \mbox{
    \includegraphics[width=2.1in]{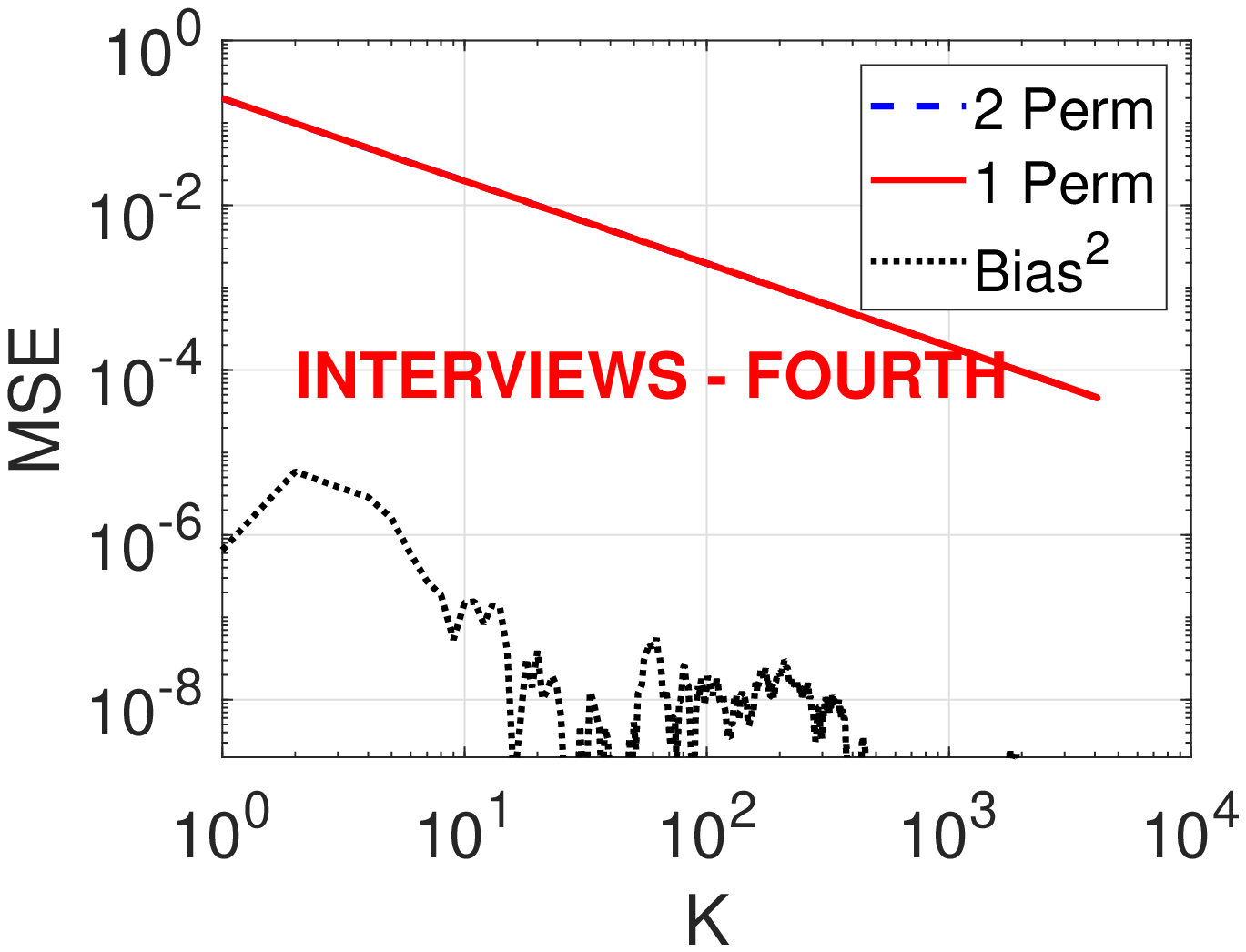}
    \includegraphics[width=2.1in]{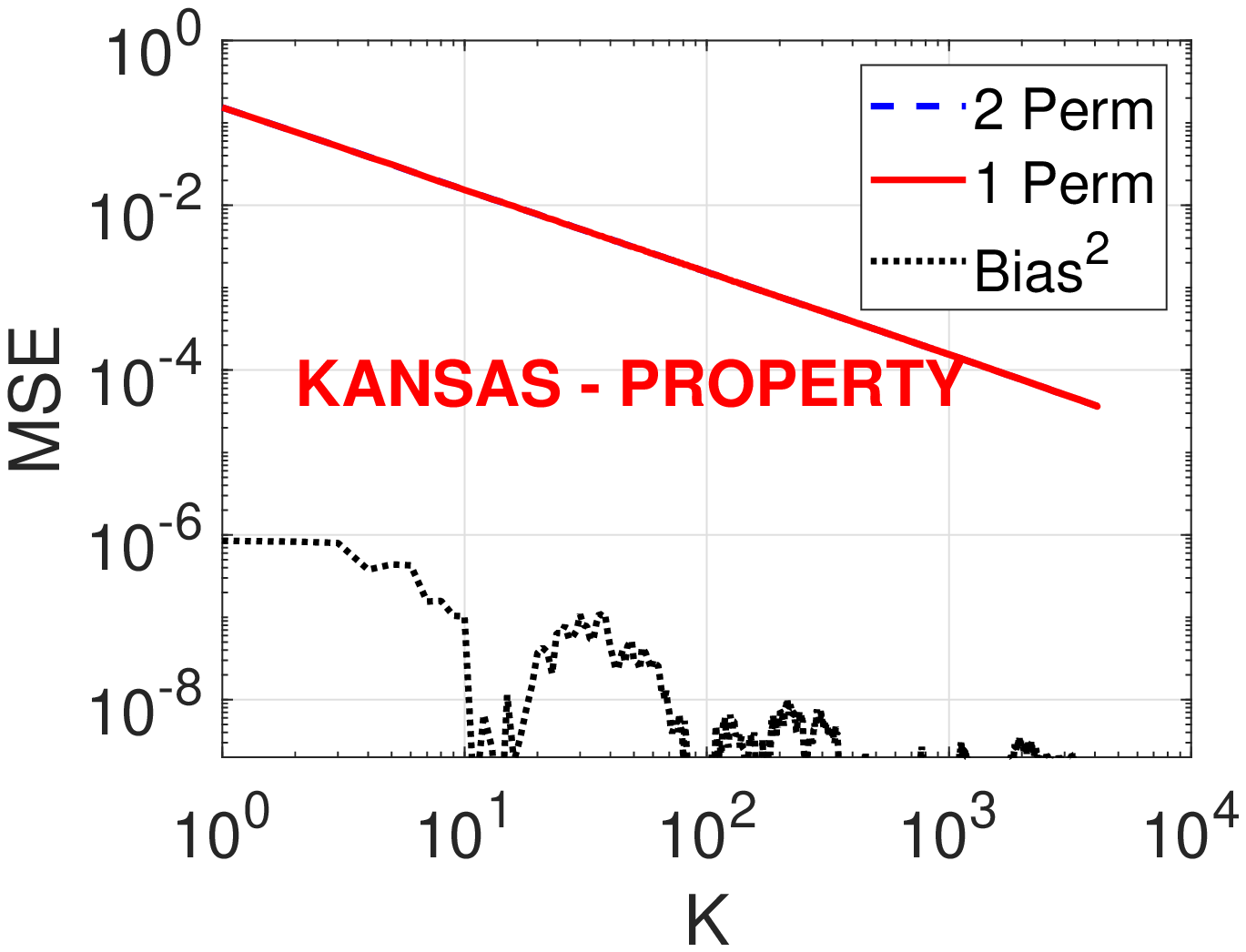}
    \includegraphics[width=2.1in]{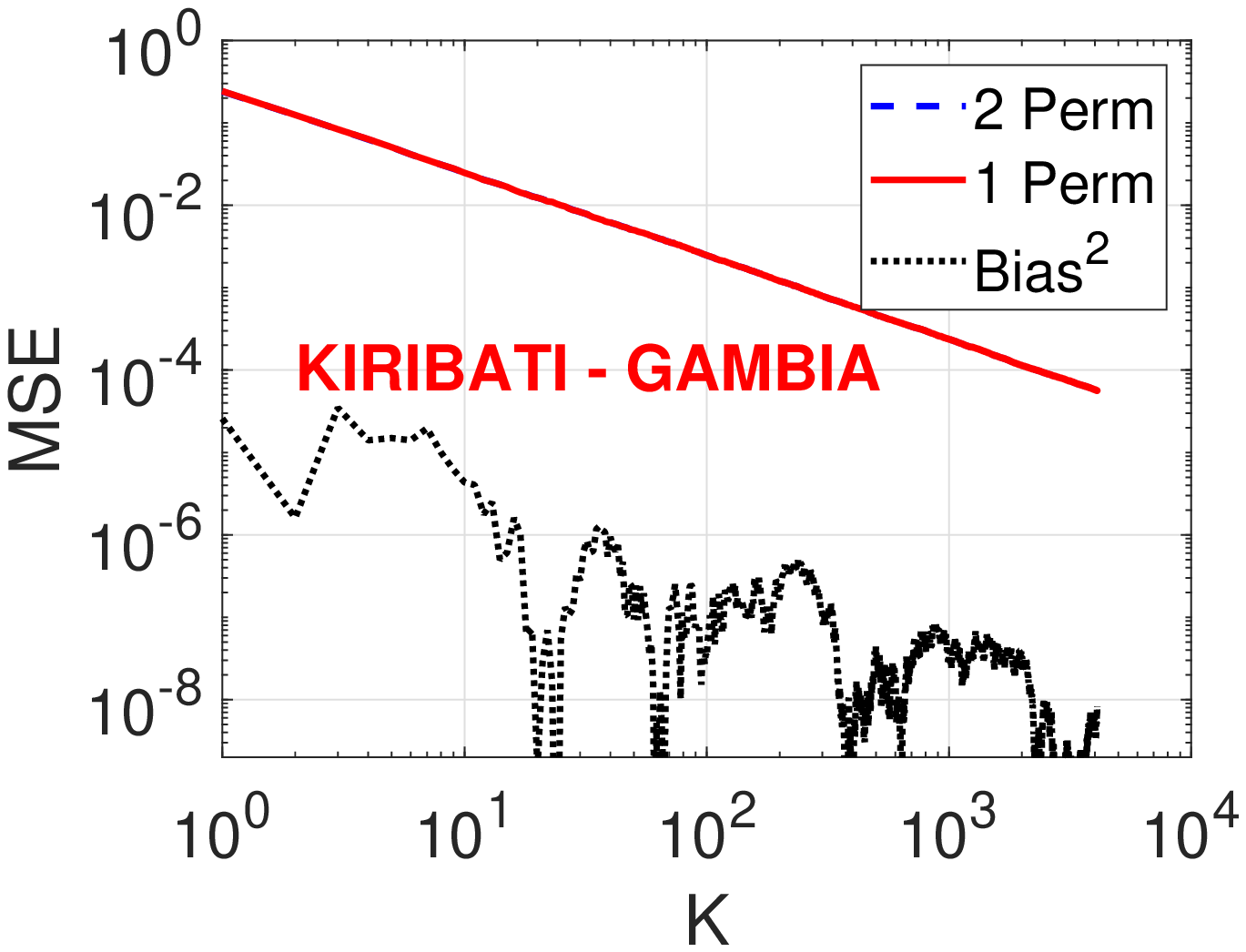}
    }

  \end{center}
  \vspace{-0.1in}
  \caption{Empirical MSEs of C-MinHash-$(\pi,\pi)$ (``1 Perm'', red, solid) vs. C-MinHash-$(\sigma,\pi)$ (``2 Perm'', blue, dashed) on various data pairs from the \textit{Words} dataset. We also report the empirical bias$^2$ for C-MinHash-$(\pi,\pi)$ to show that the bias is so small that it can be safely neglected. The empirical MSE curves for both estimators essentially overlap for all data pairs, for $K$ ranging from 1 to 4096. }
  \label{fig:word3}
\end{figure}

\begin{figure}[H]
  \begin{center}
   \mbox{
    \includegraphics[width=2.1in]{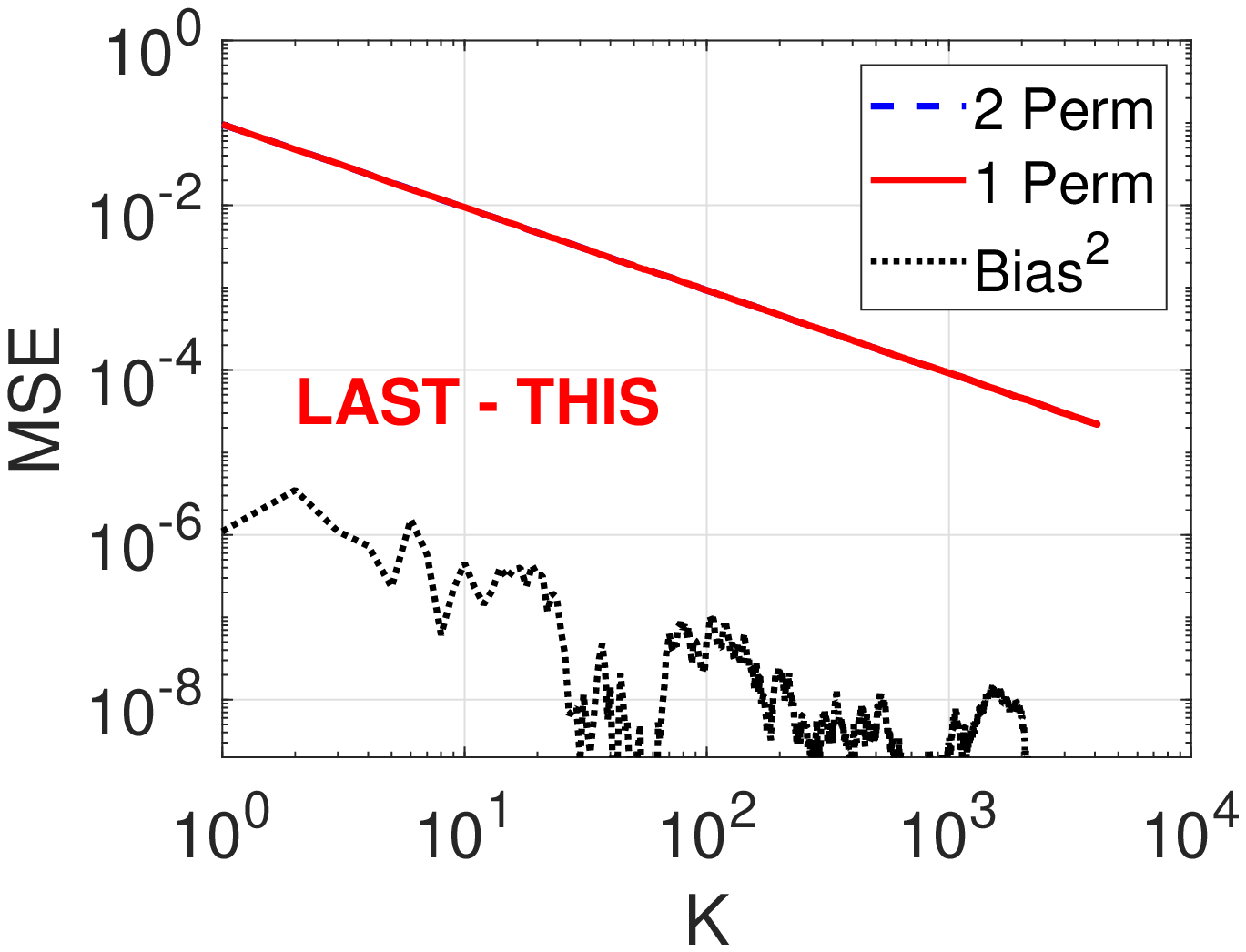}
    \includegraphics[width=2.1in]{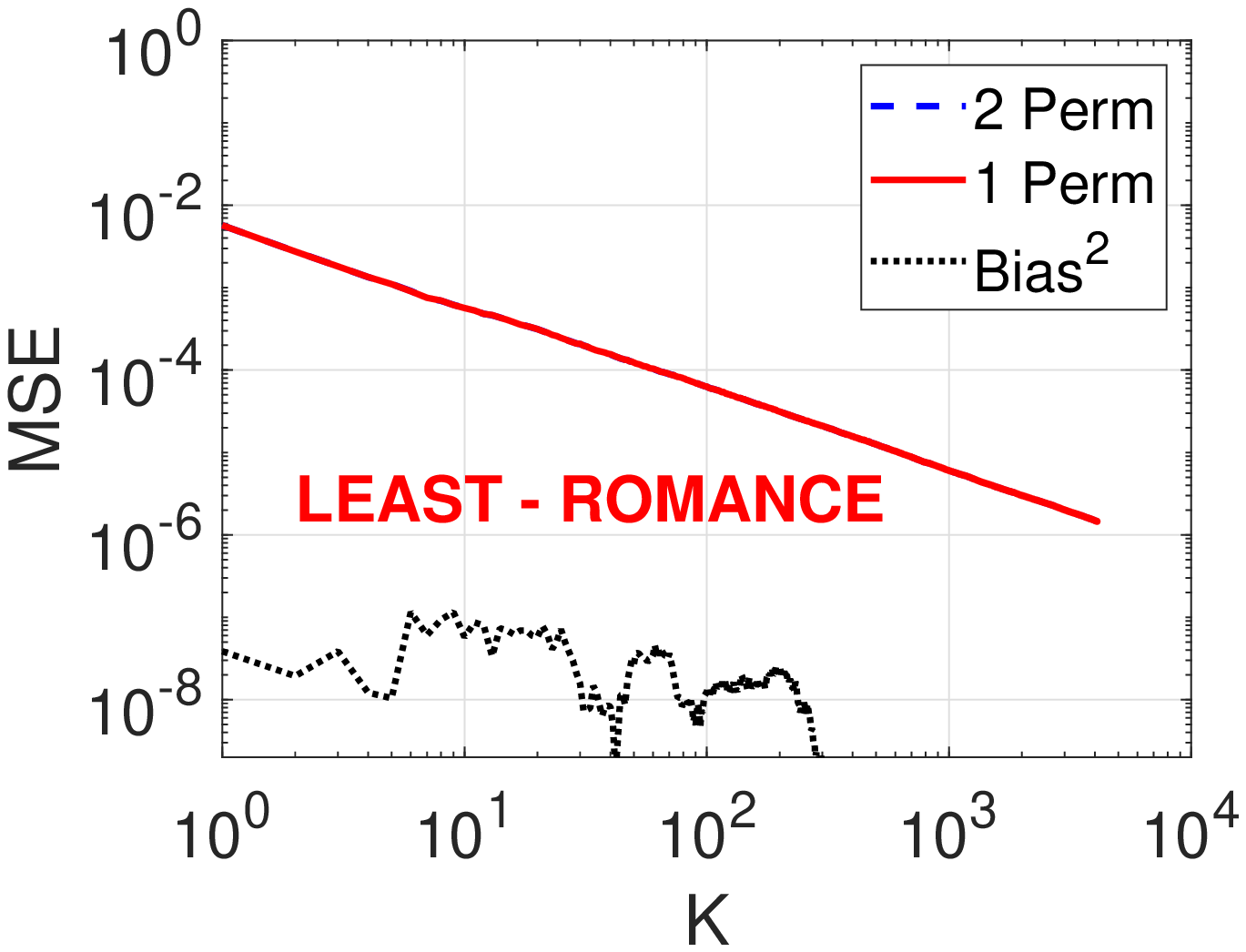}
    \includegraphics[width=2.1in]{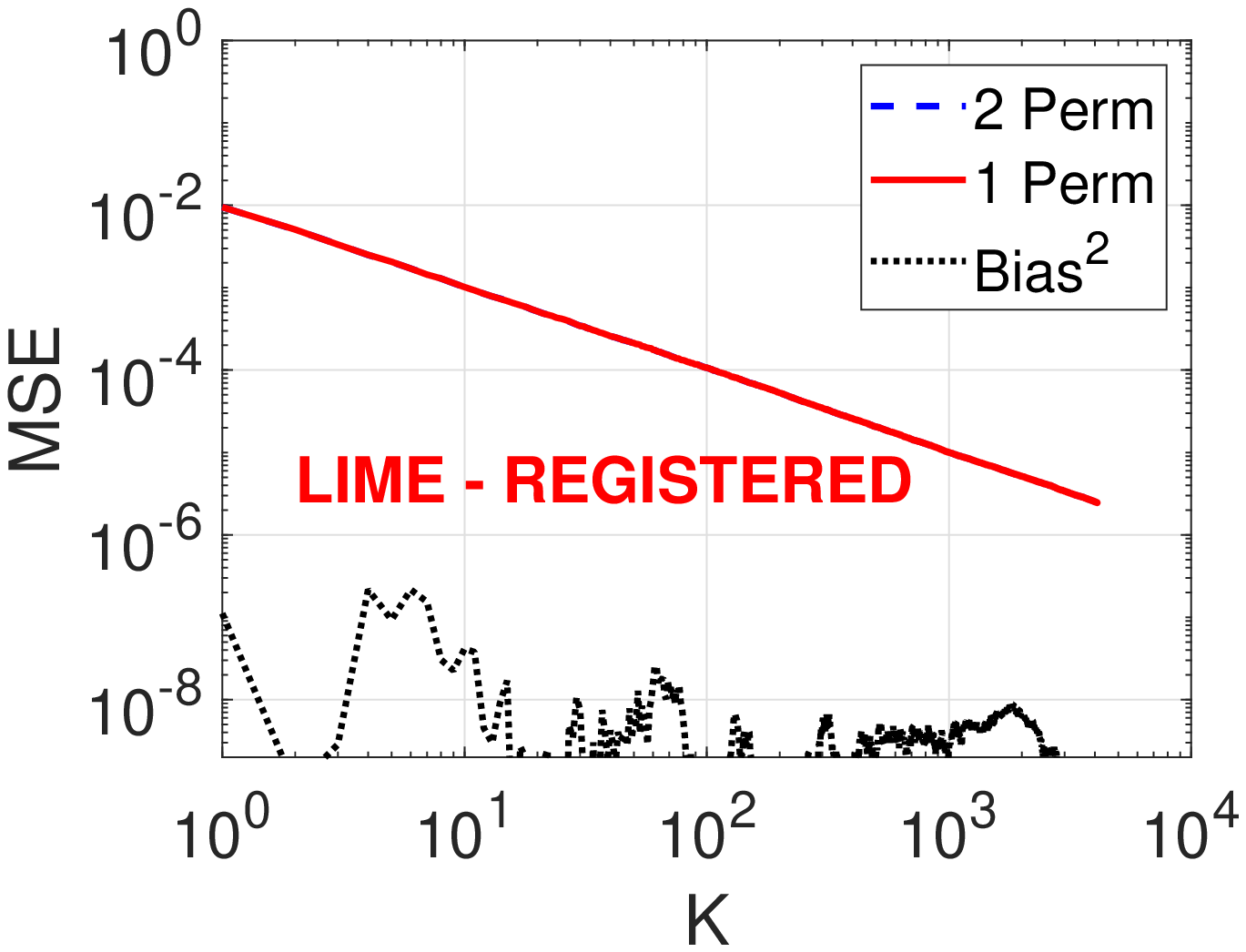}
    }
    \mbox{
    \includegraphics[width=2.1in]{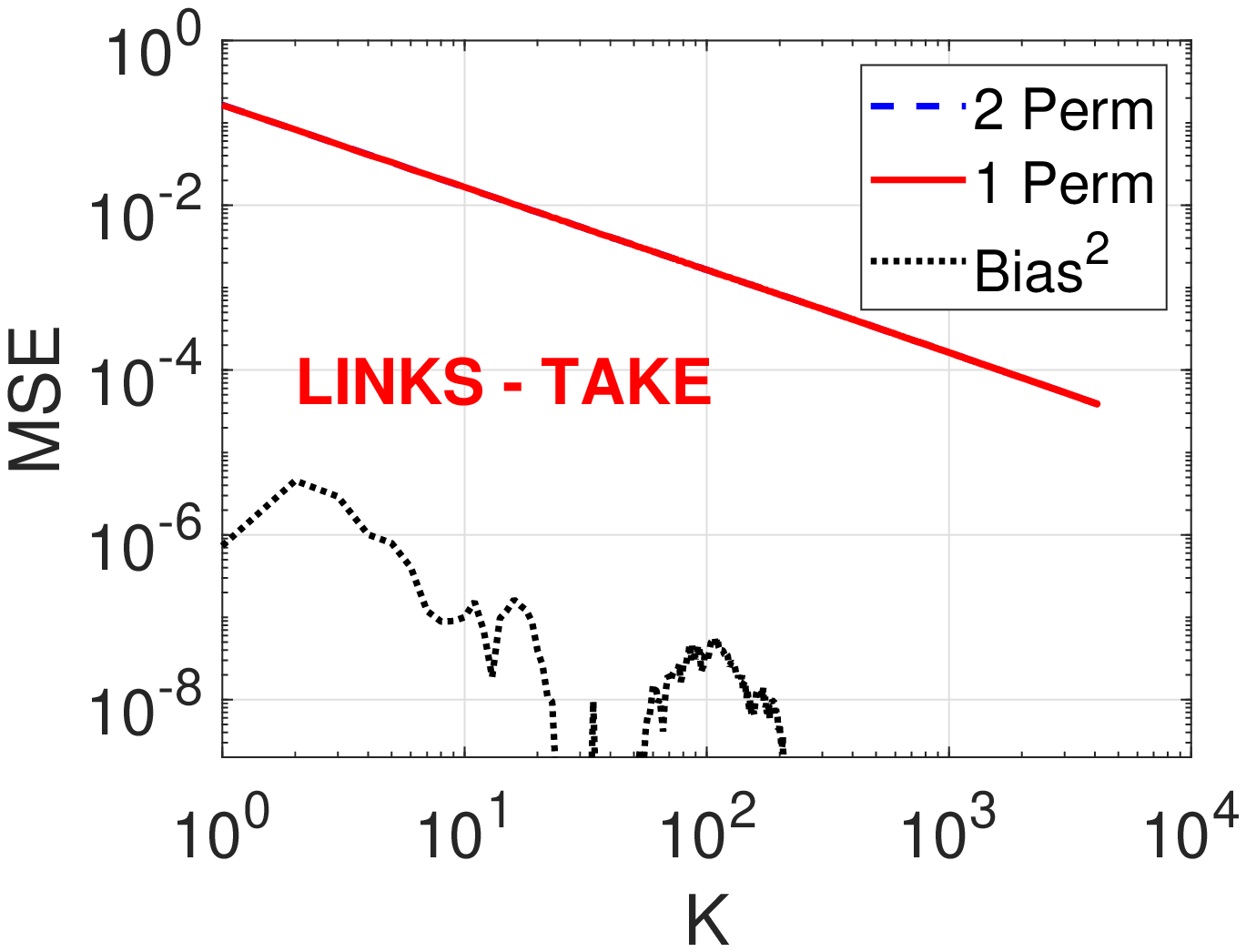}
    \includegraphics[width=2.1in]{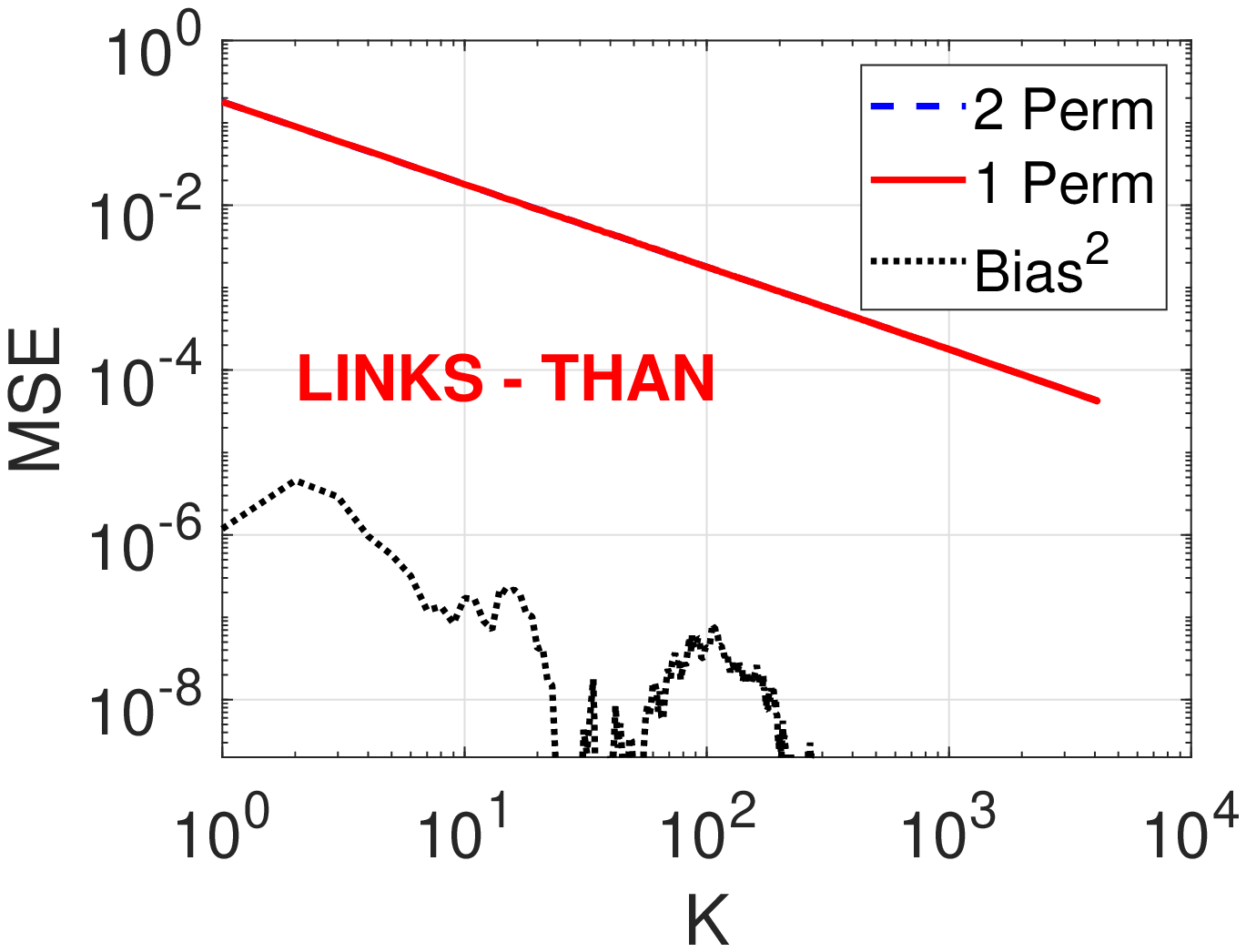}
    \includegraphics[width=2.1in]{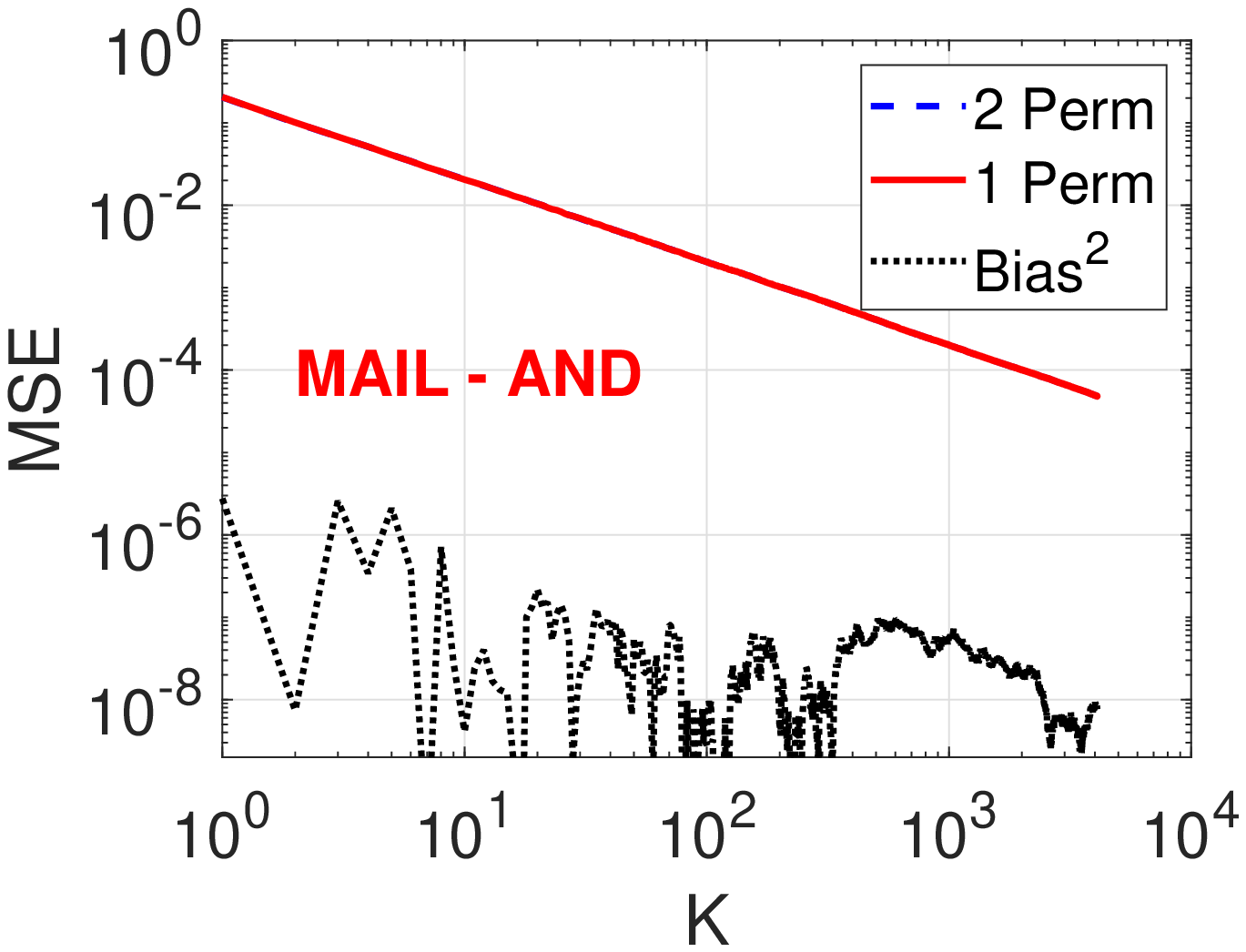}
    }
    \mbox{
    \includegraphics[width=2.1in]{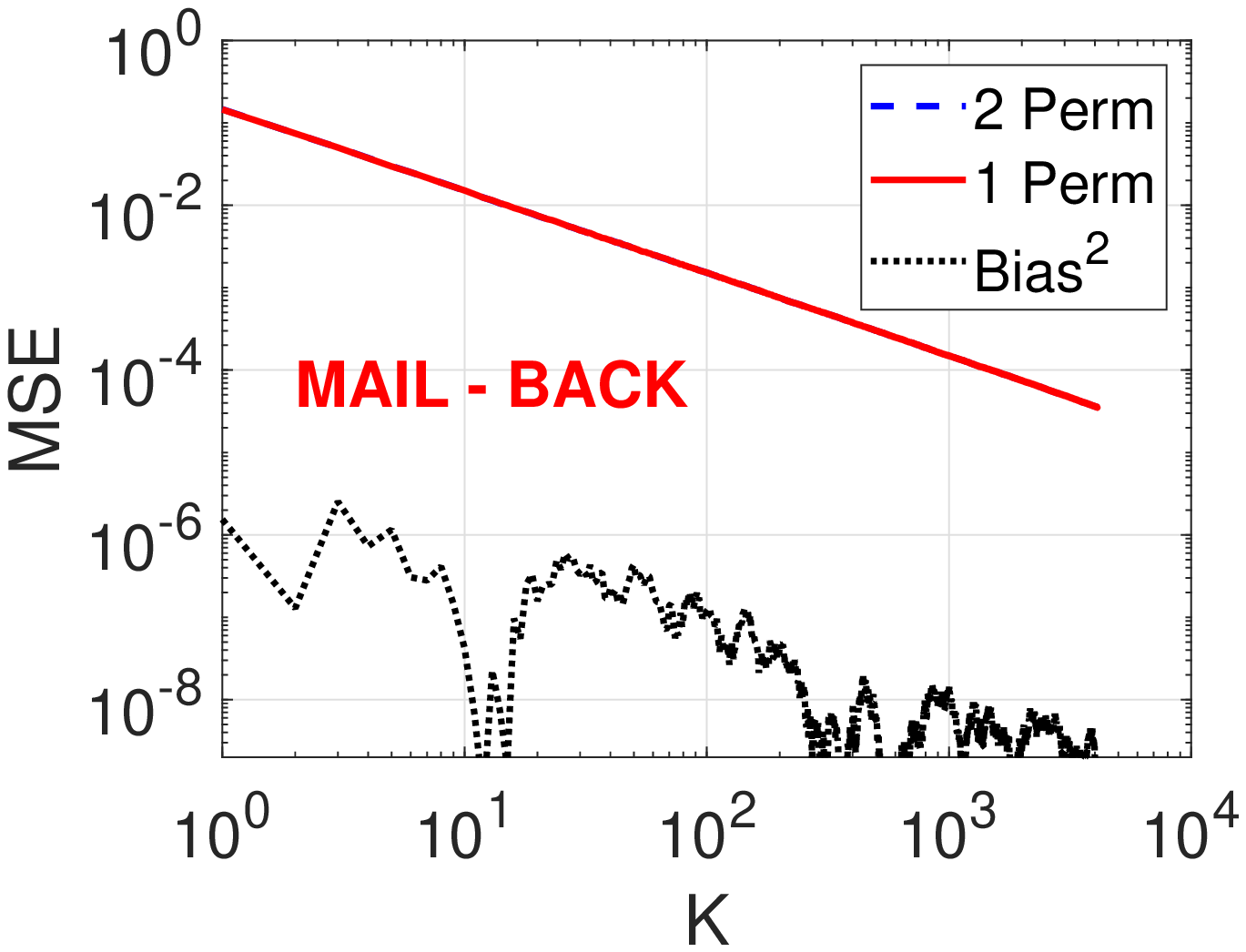}
    \includegraphics[width=2.1in]{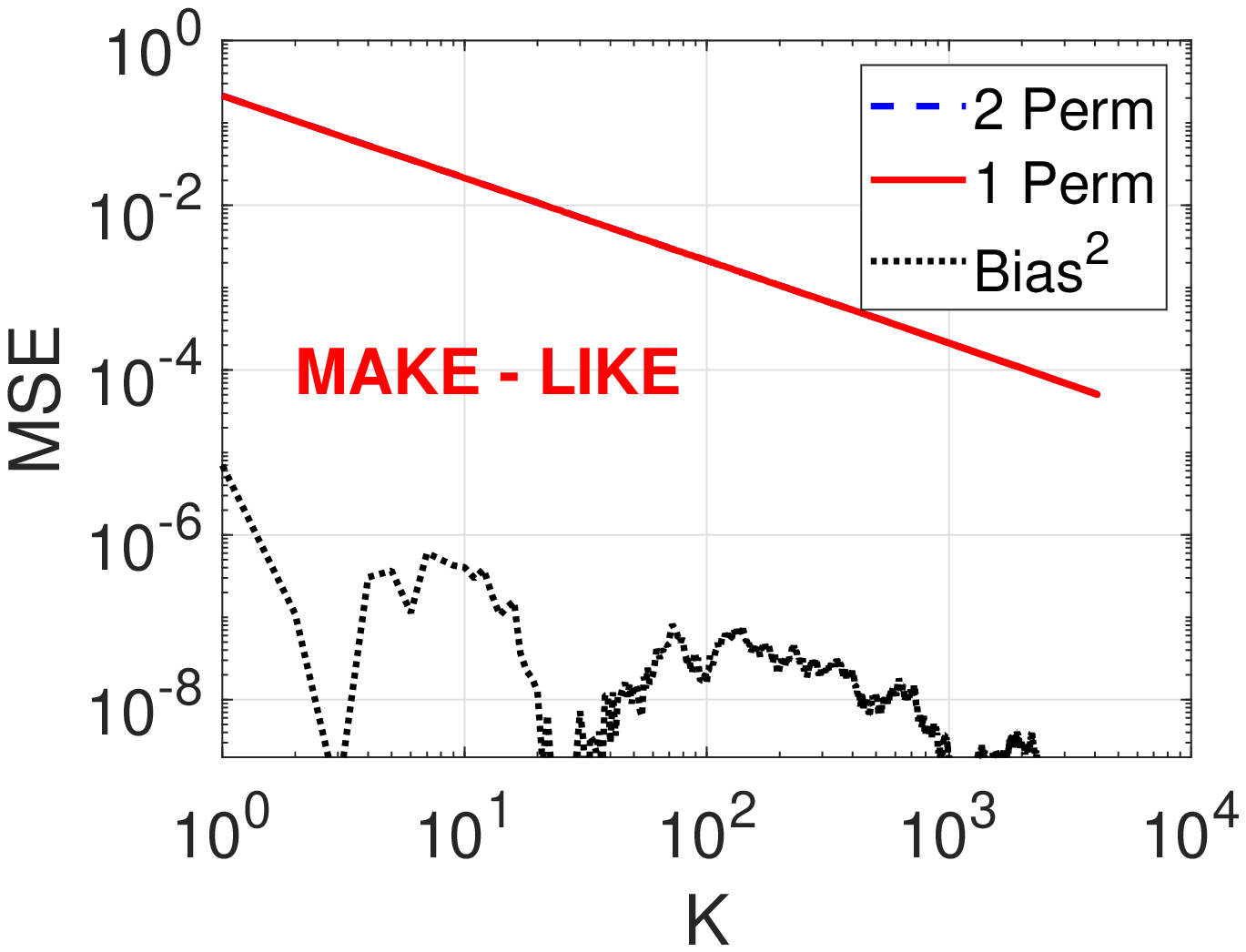}
    \includegraphics[width=2.1in]{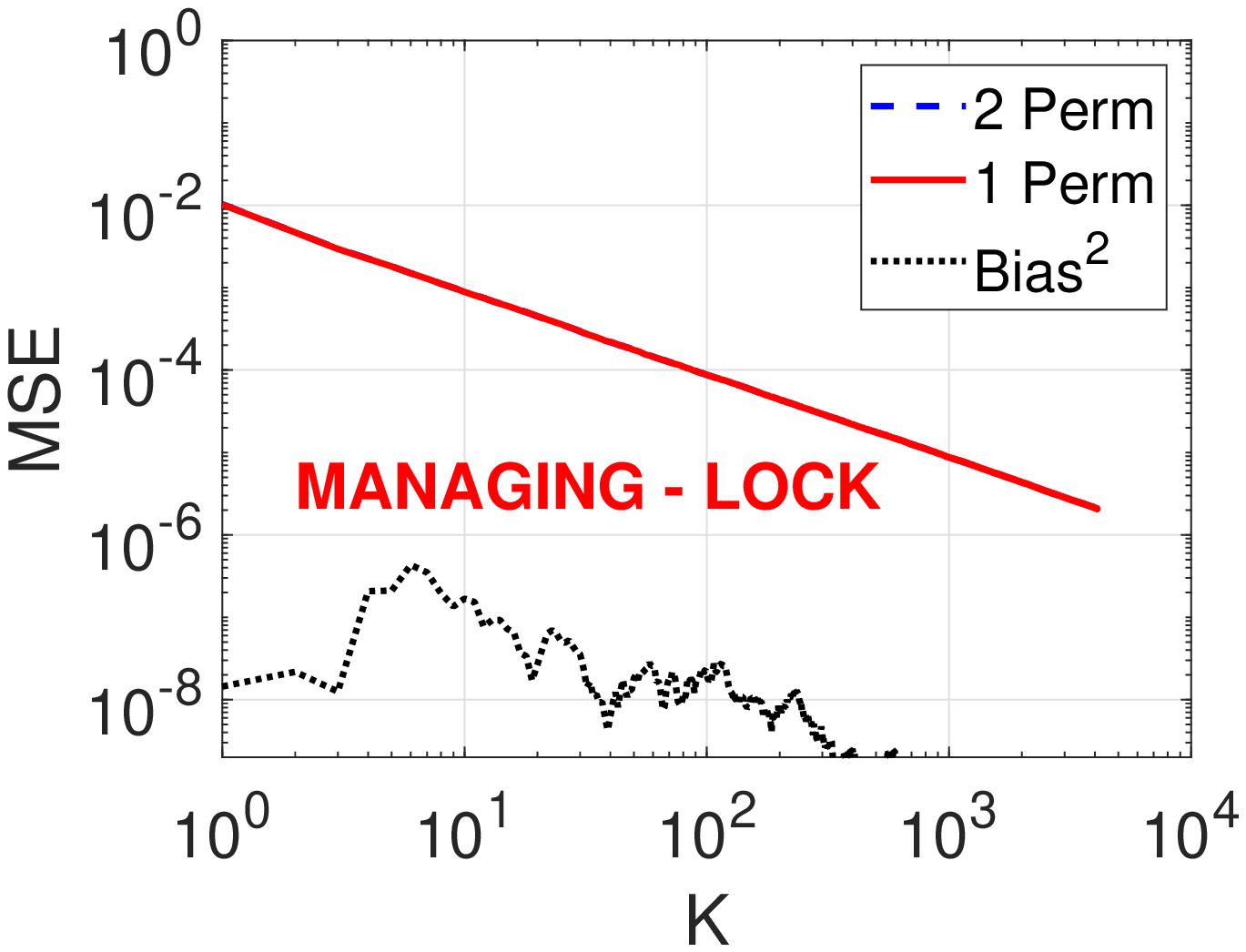}
    }
    \mbox{
    \includegraphics[width=2.1in]{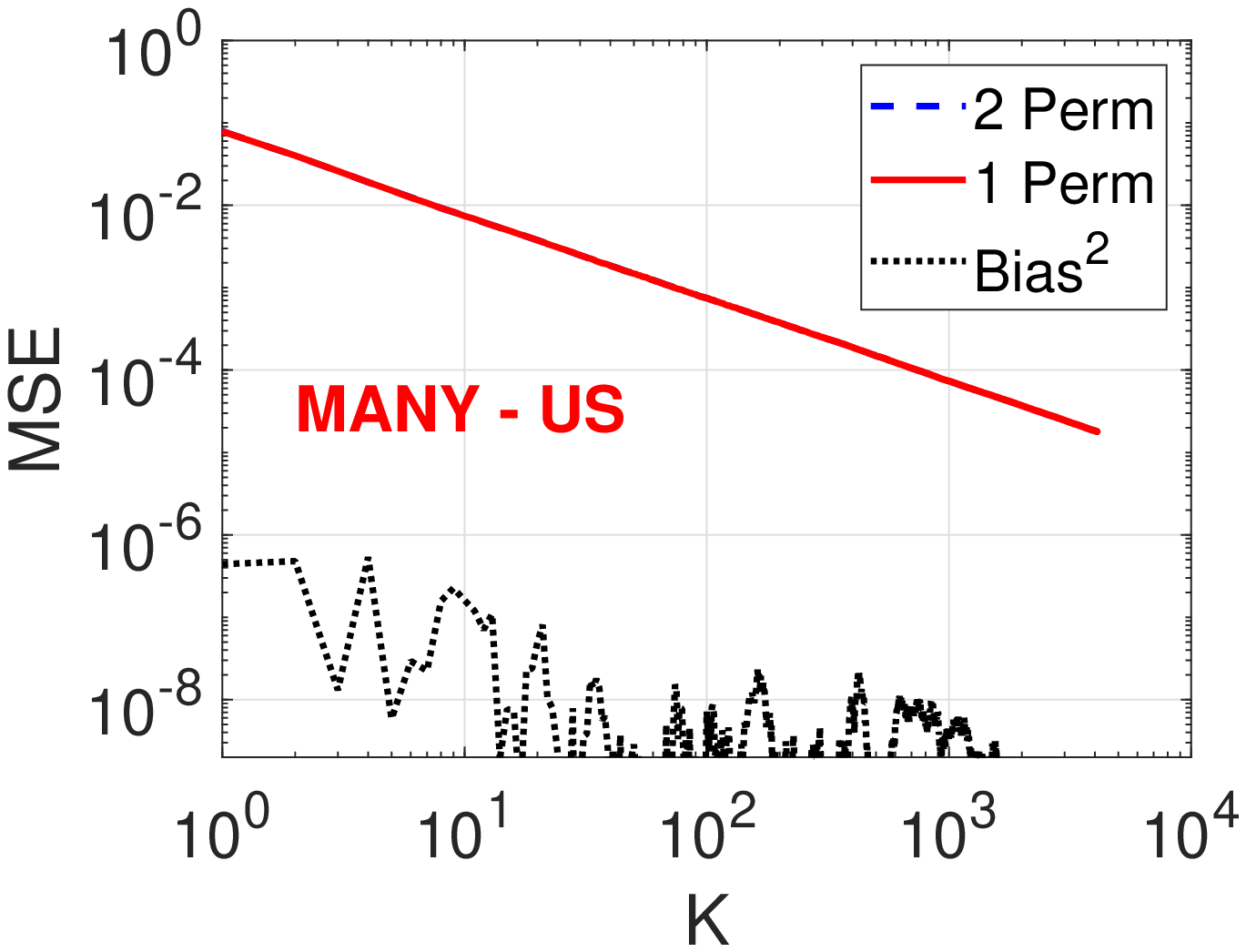}
    \includegraphics[width=2.1in]{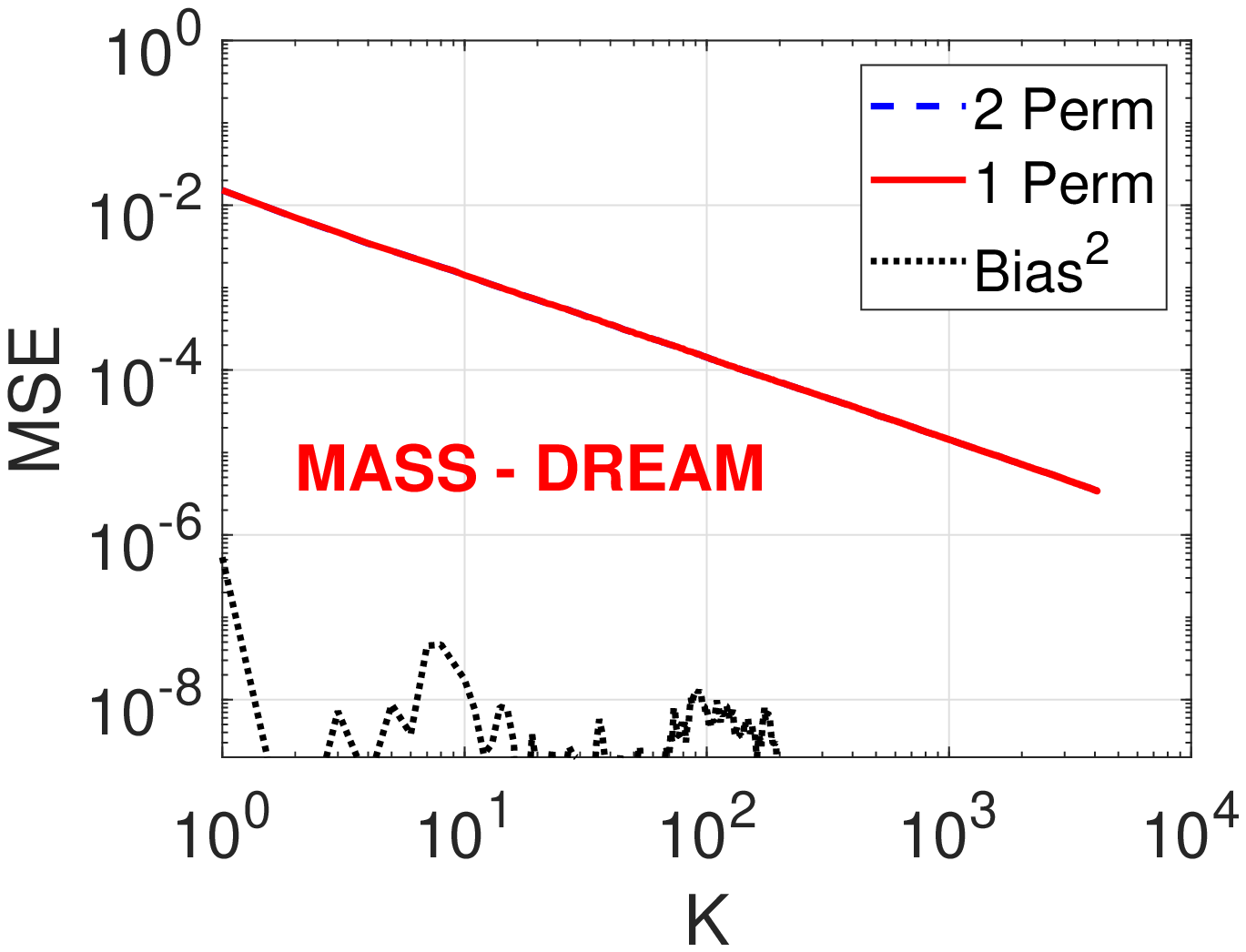}
    \includegraphics[width=2.1in]{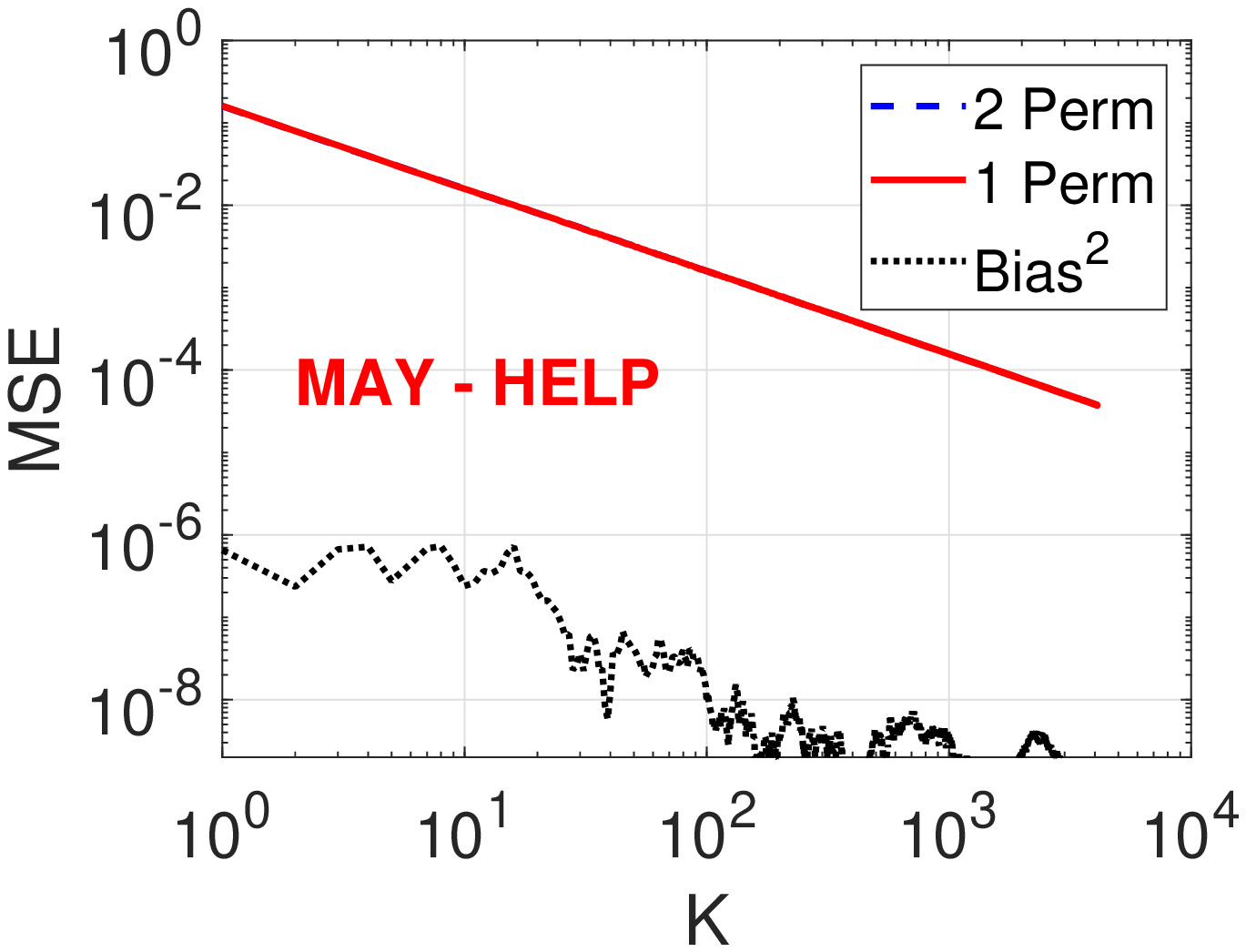}
    }
    \mbox{
    \includegraphics[width=2.1in]{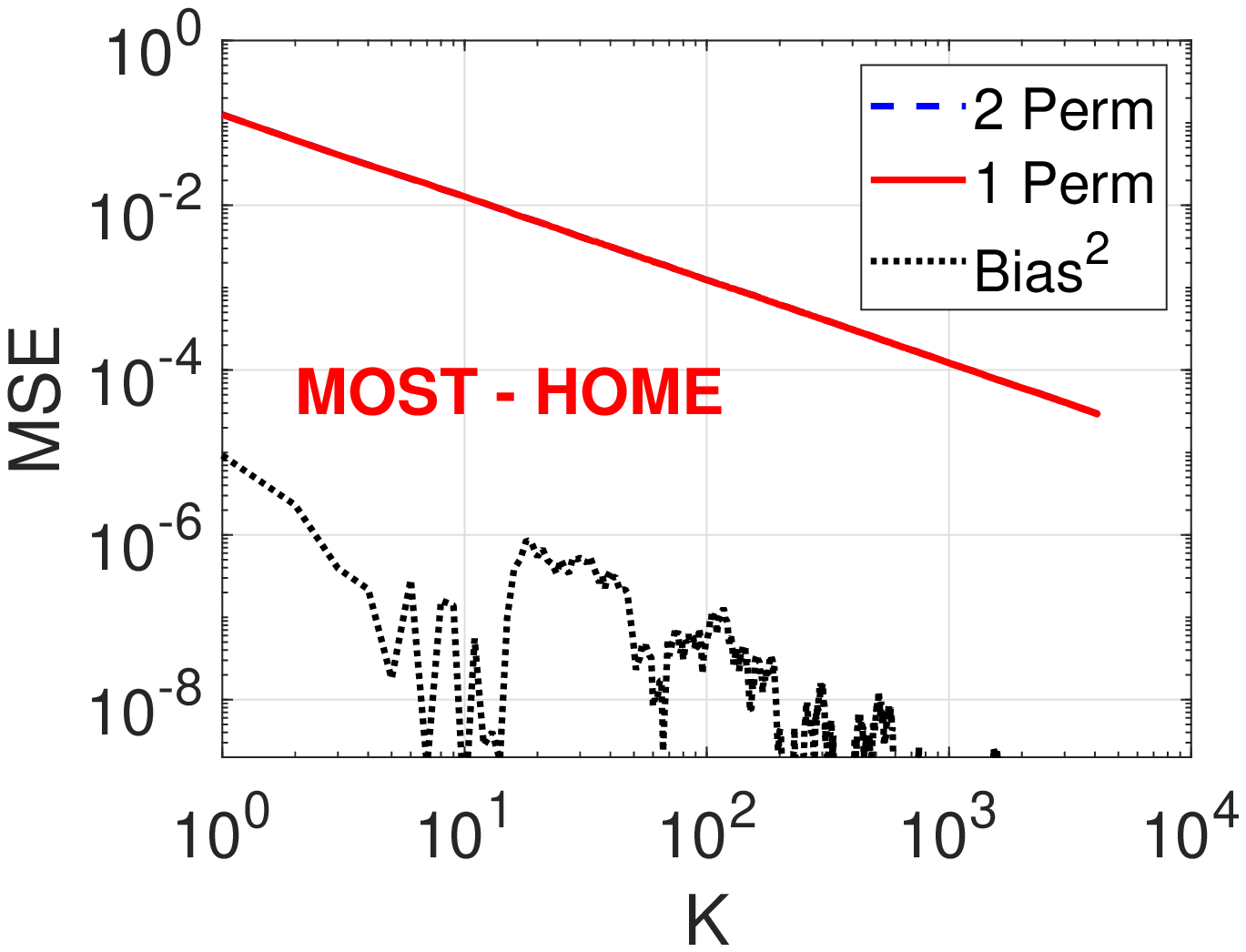}
    \includegraphics[width=2.1in]{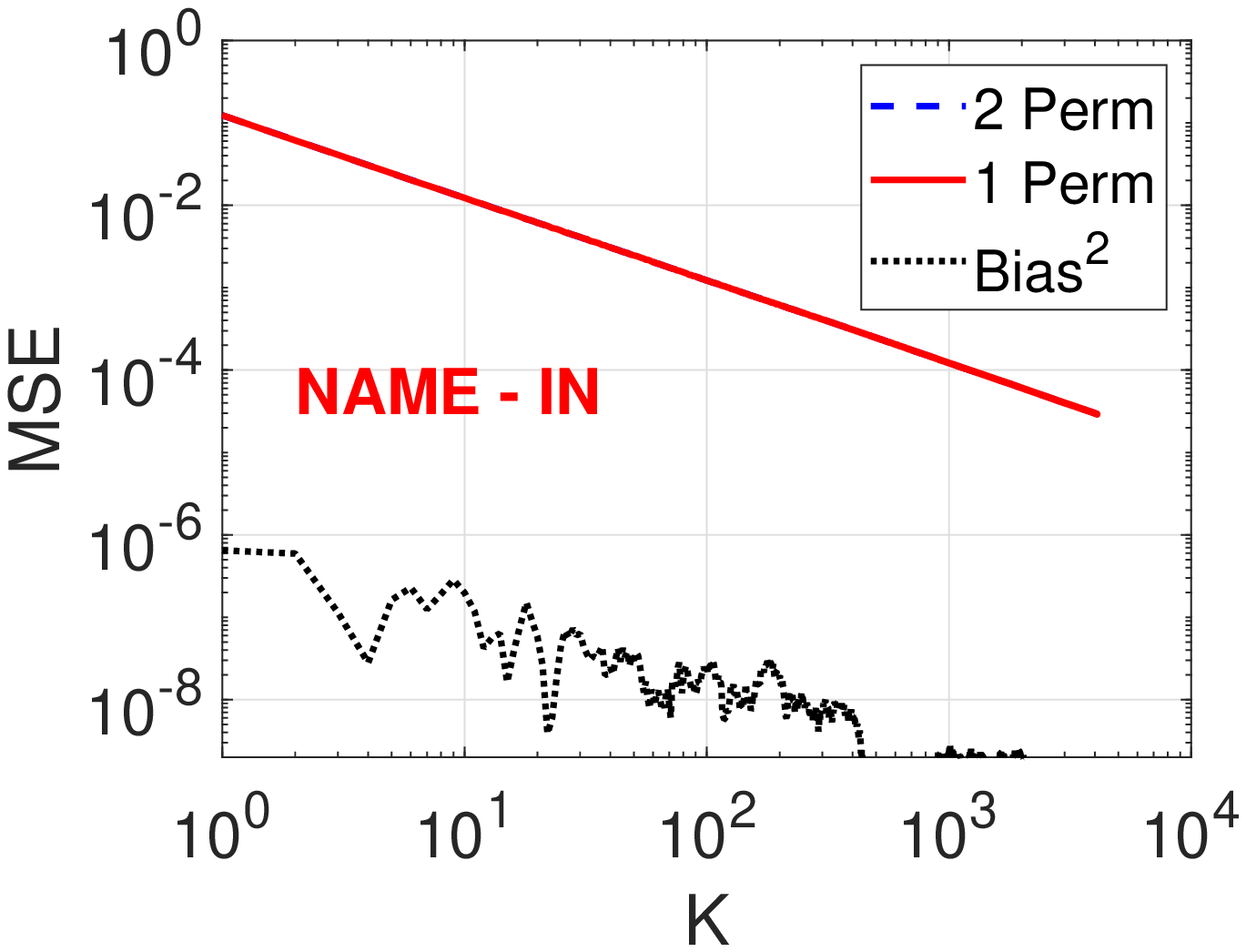}
    \includegraphics[width=2.1in]{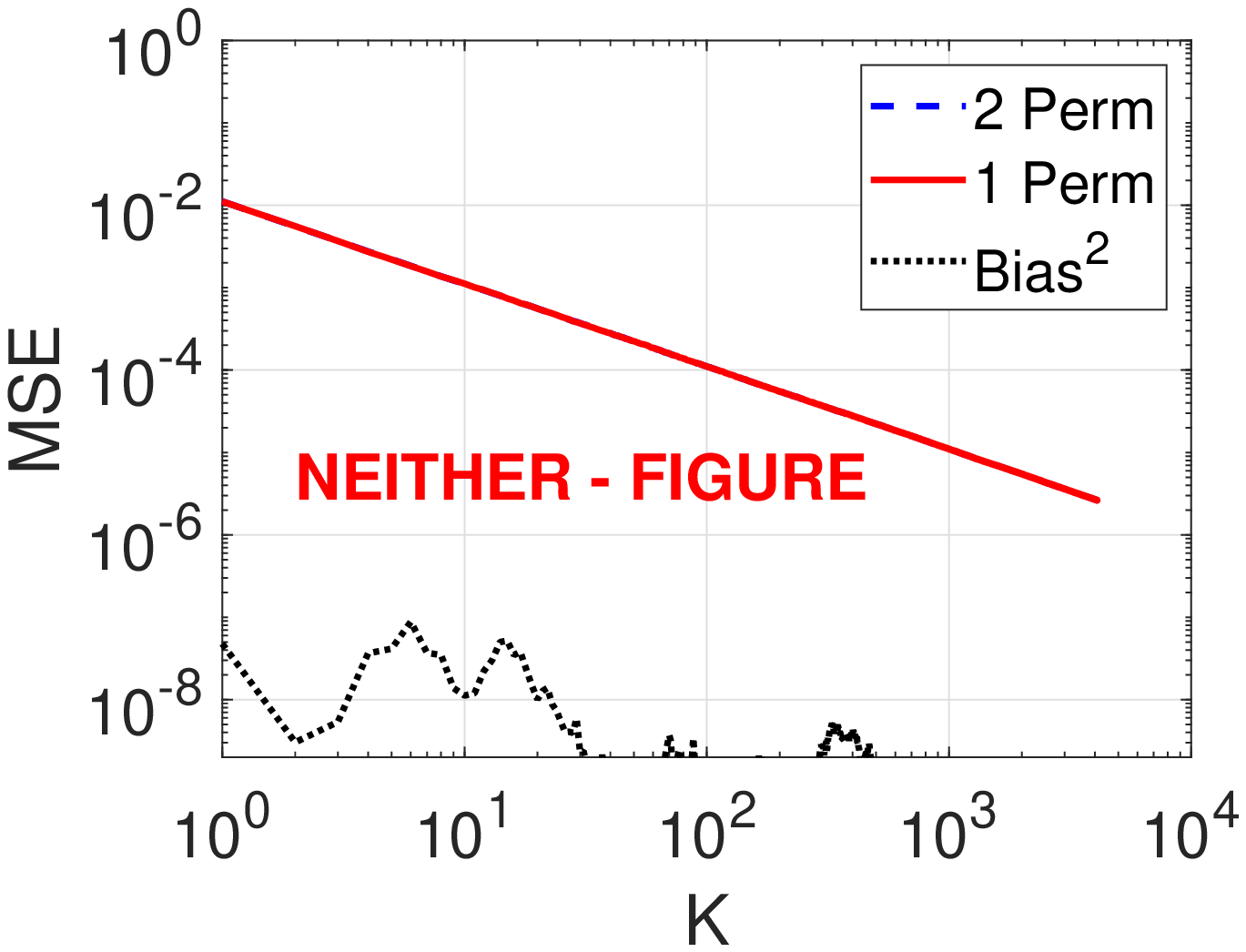}
    }

  \end{center}
  \vspace{-0.1in}
  \caption{Empirical MSEs of C-MinHash-$(\pi,\pi)$ (``1 Perm'', red, solid) vs. C-MinHash-$(\sigma,\pi)$ (``2 Perm'', blue, dashed) on various data pairs from the \textit{Words} dataset. We also report the empirical bias$^2$ for C-MinHash-$(\pi,\pi)$ to show that the bias is so small that it can be safely neglected. The empirical MSE curves for both estimators essentially overlap for all data pairs, for $K$ ranging from 1 to 4096. }
  \label{fig:word4}
\end{figure}

\begin{figure}[H]
  \begin{center}
   \mbox{
    \includegraphics[width=2.1in]{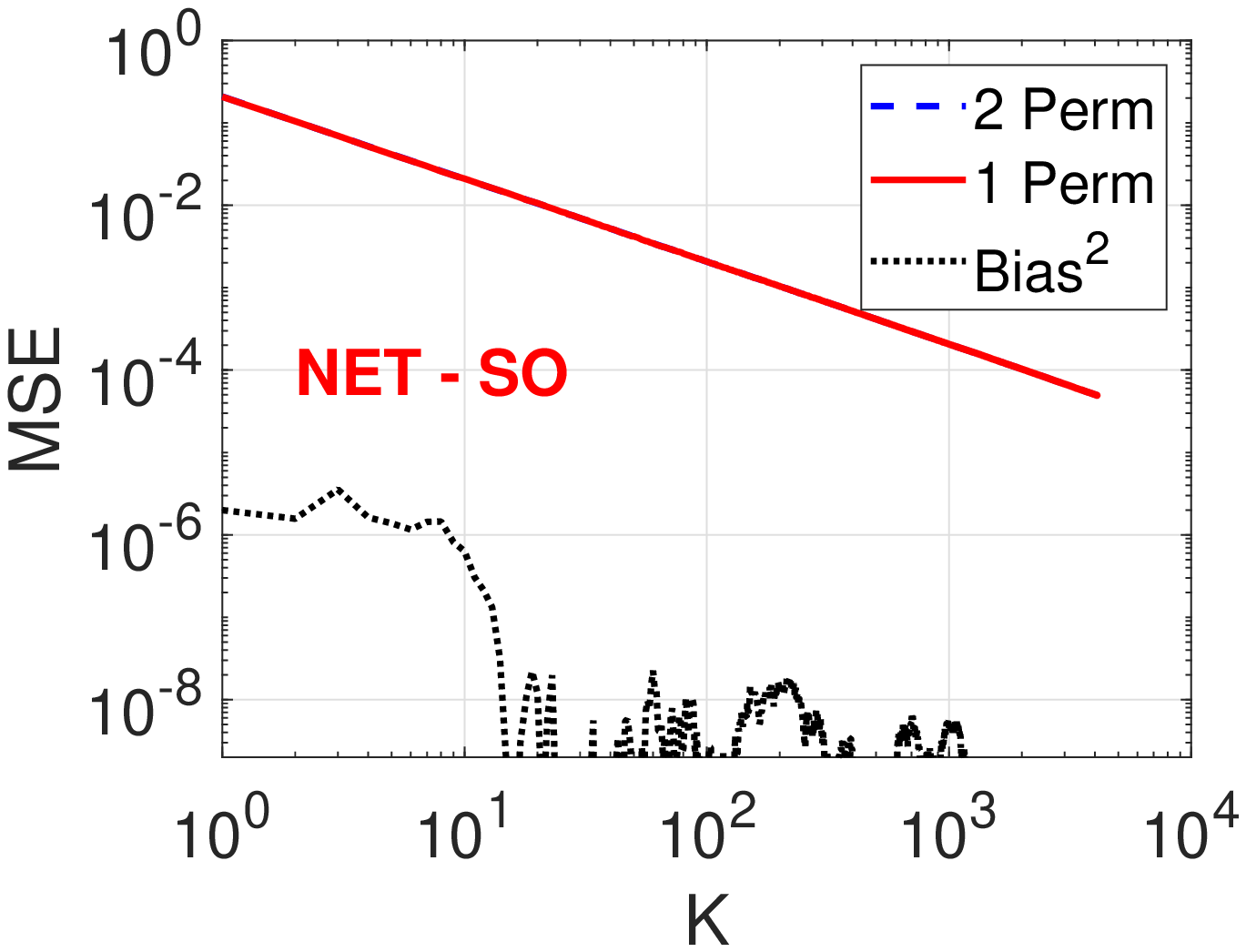}
    \includegraphics[width=2.1in]{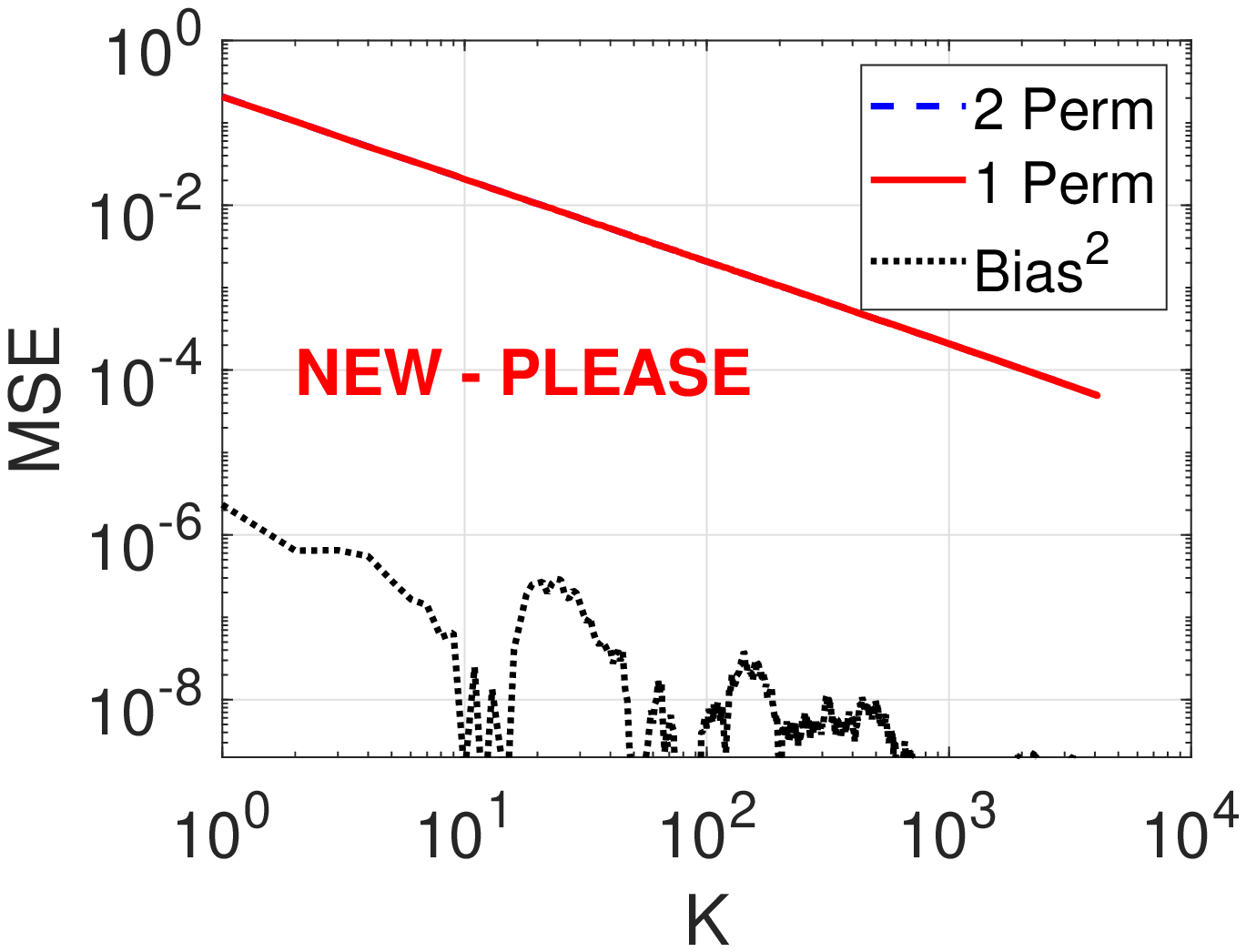}
    \includegraphics[width=2.1in]{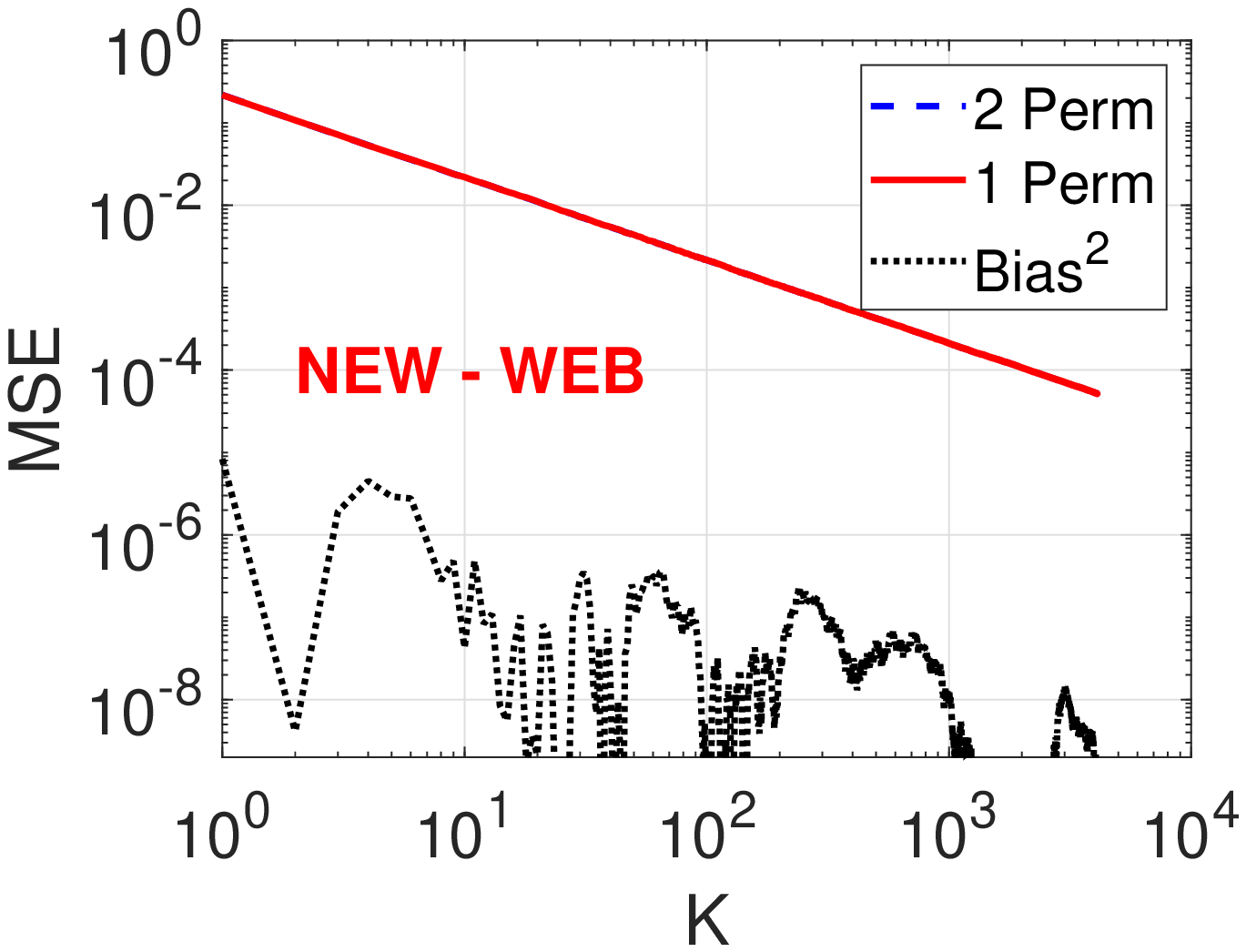}
    }
    \mbox{
    \includegraphics[width=2.1in]{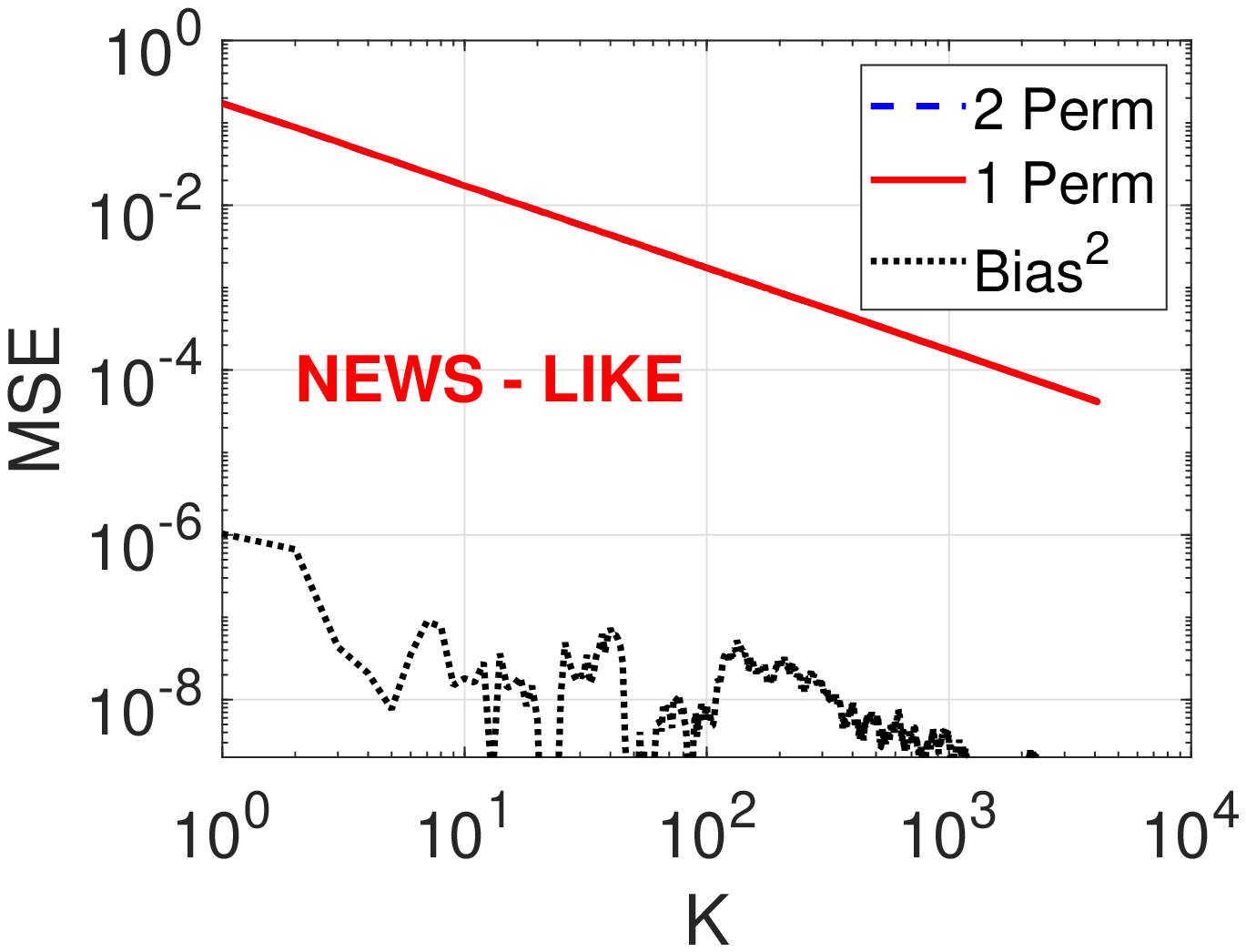}
    \includegraphics[width=2.1in]{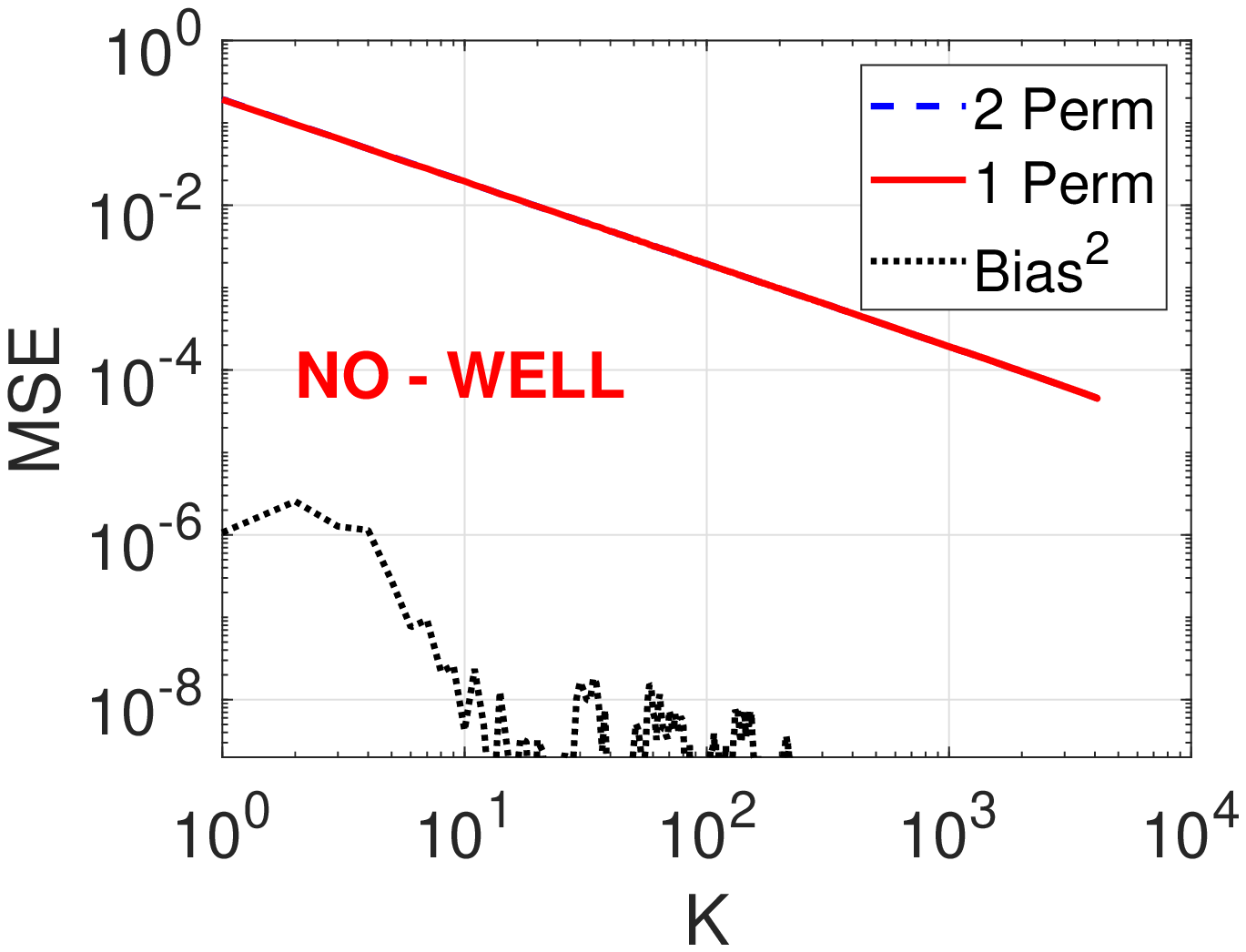}
    \includegraphics[width=2.1in]{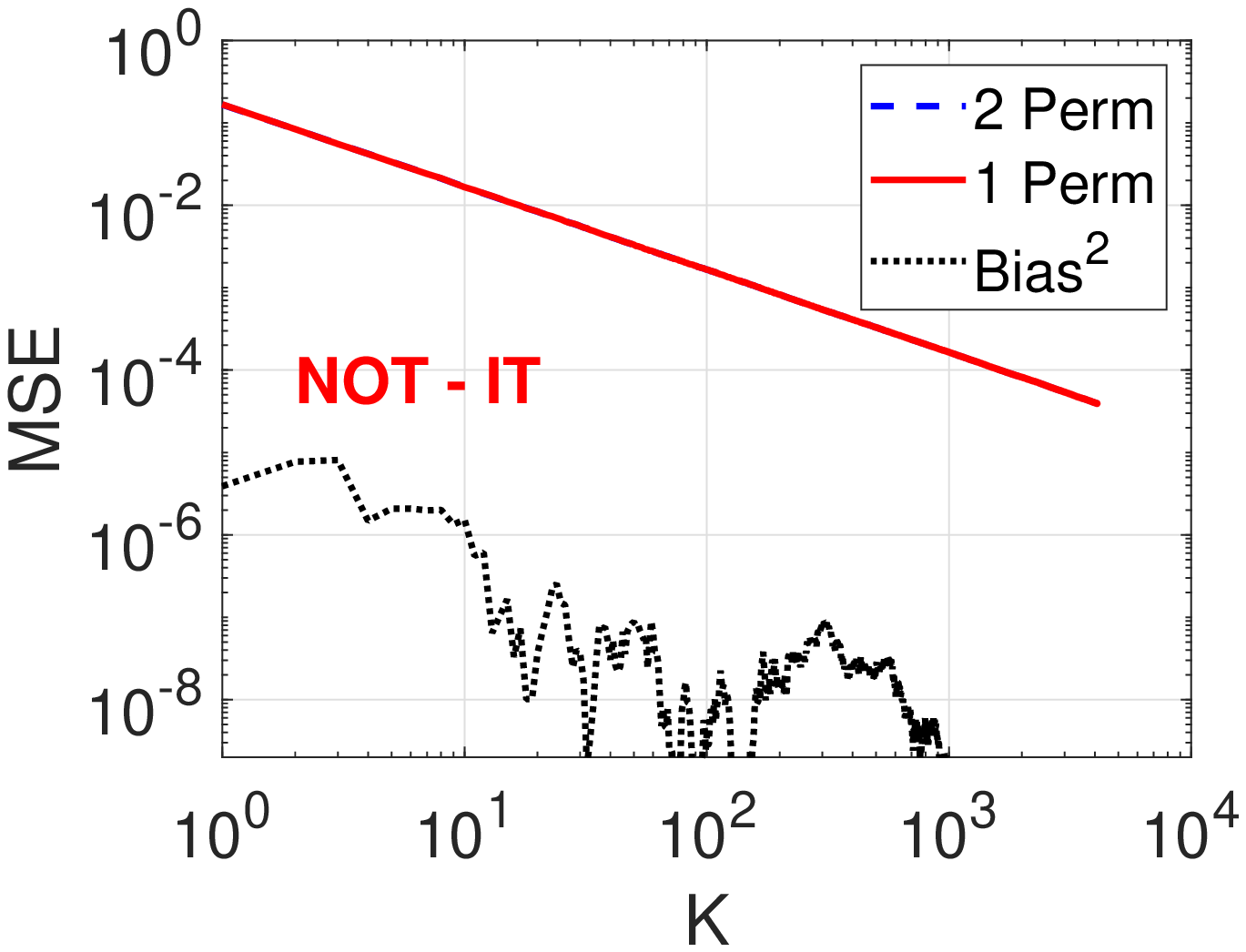}
    }
    \mbox{
    \includegraphics[width=2.1in]{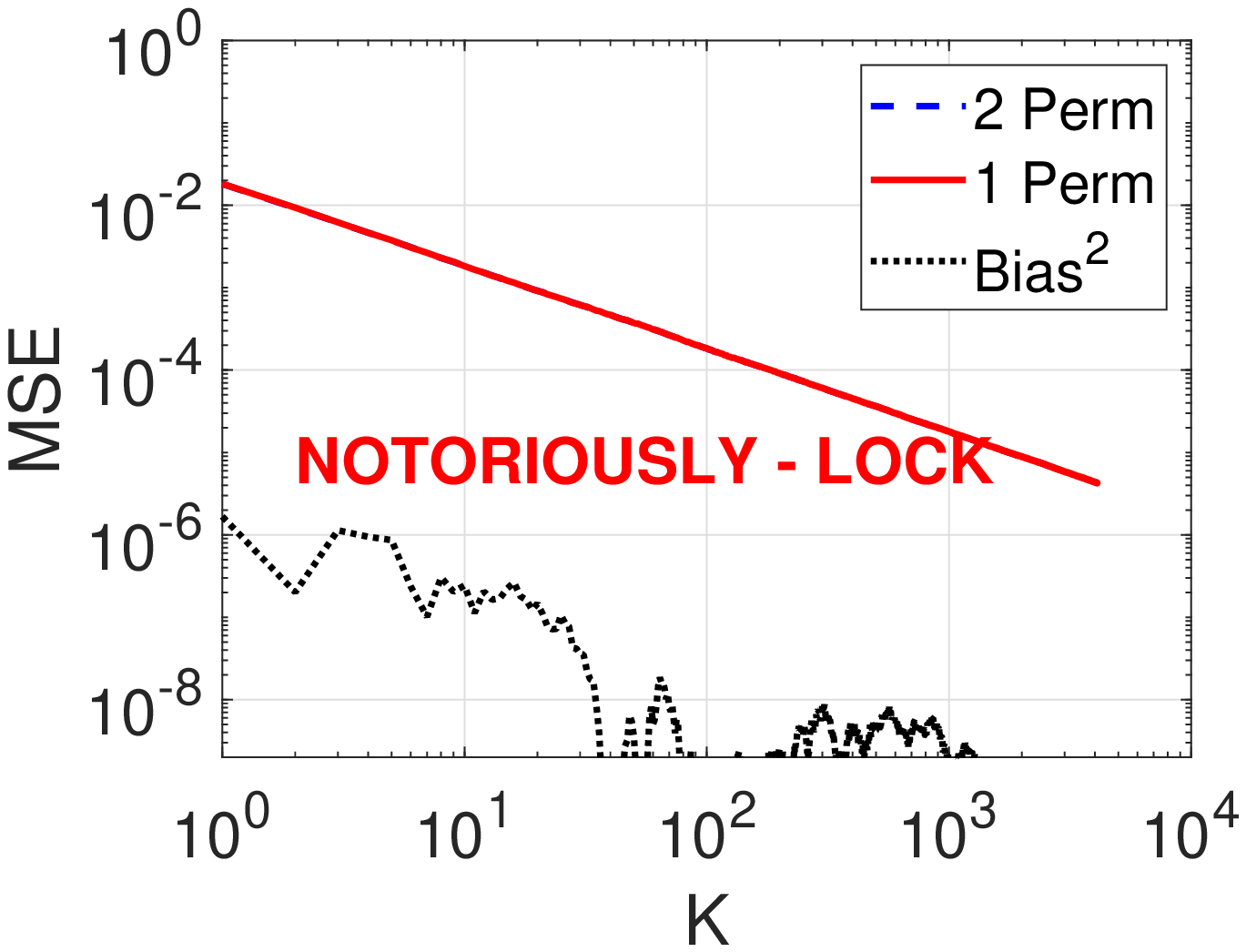}
    \includegraphics[width=2.1in]{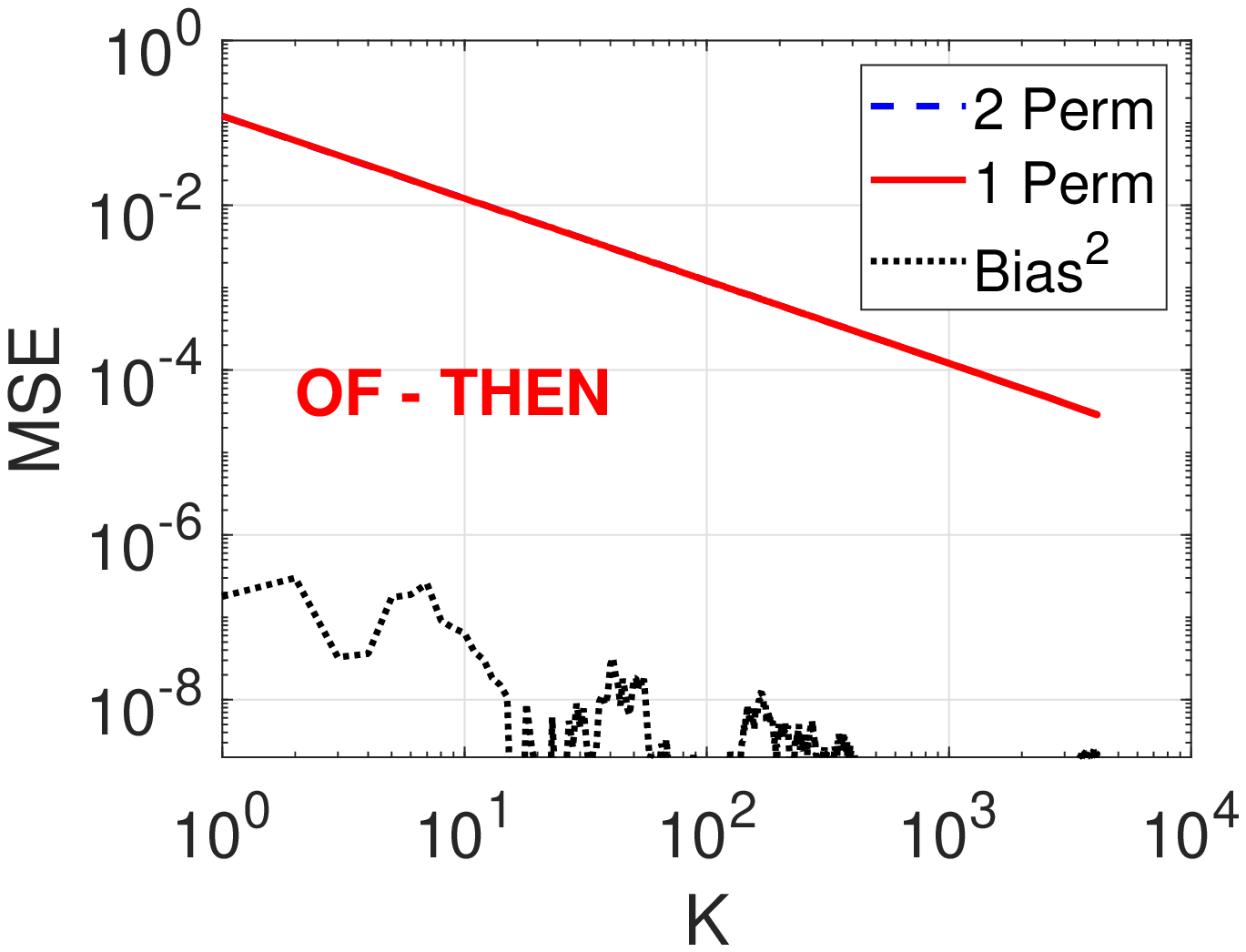}
    \includegraphics[width=2.1in]{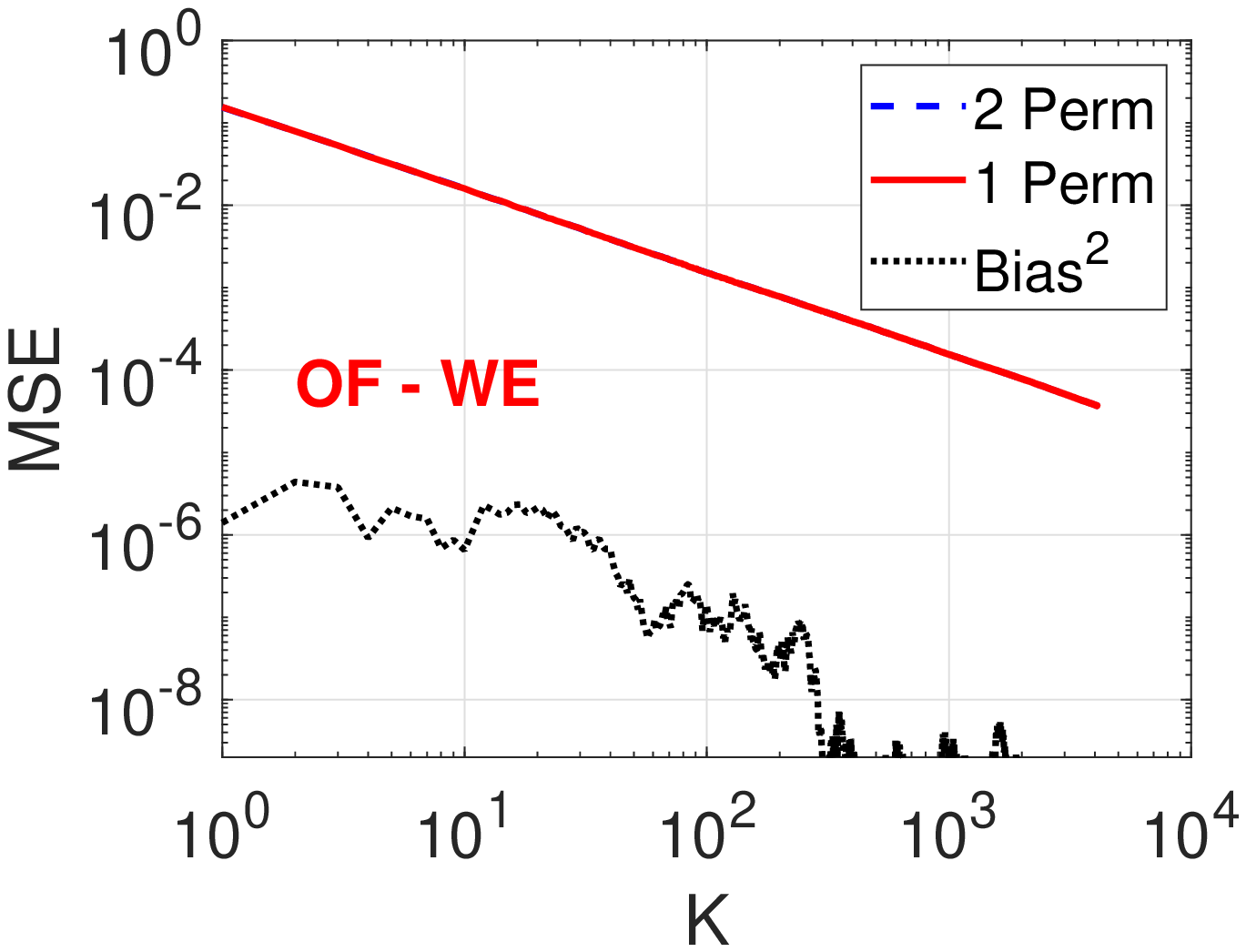}
    }
    \mbox{
    \includegraphics[width=2.1in]{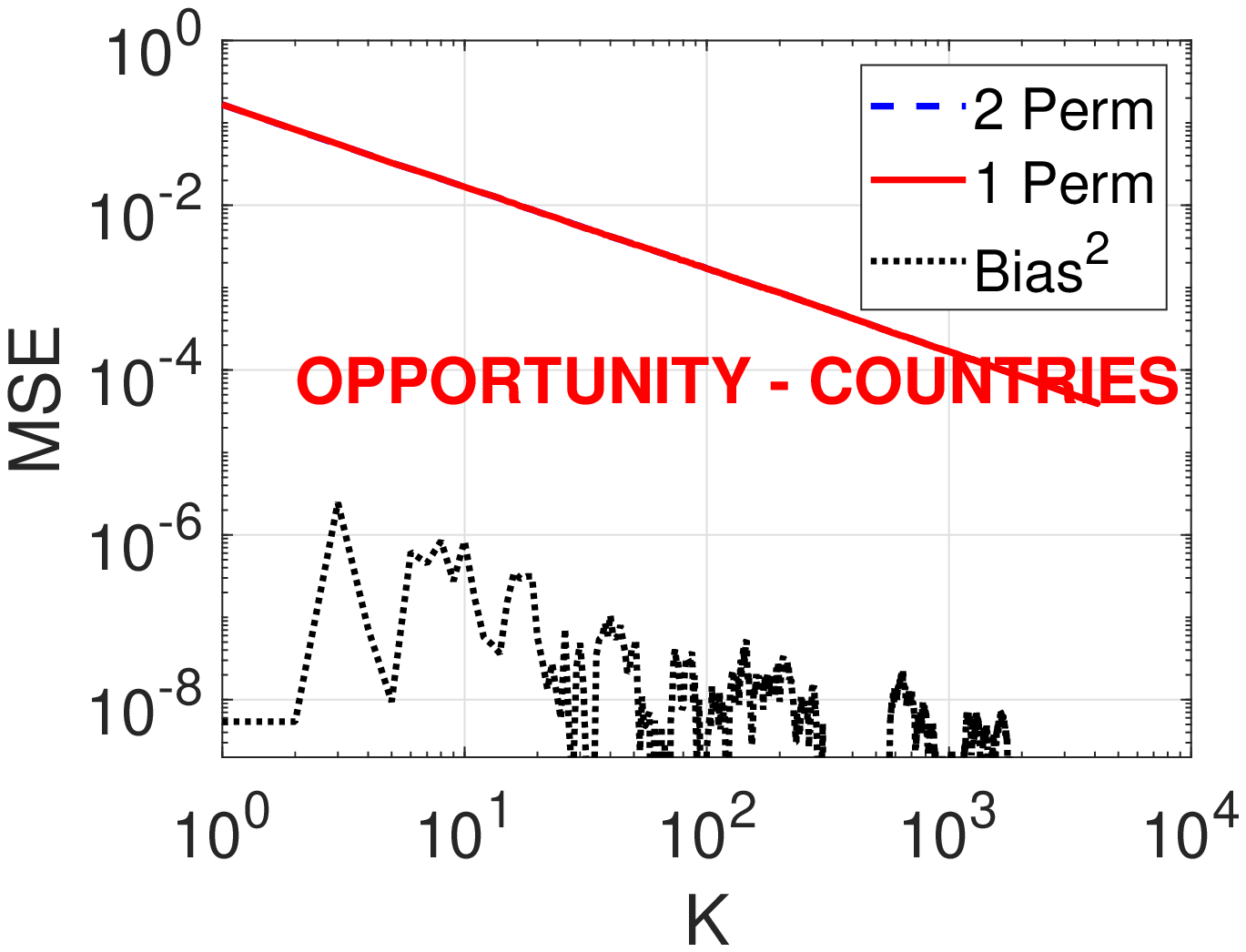}
    \includegraphics[width=2.1in]{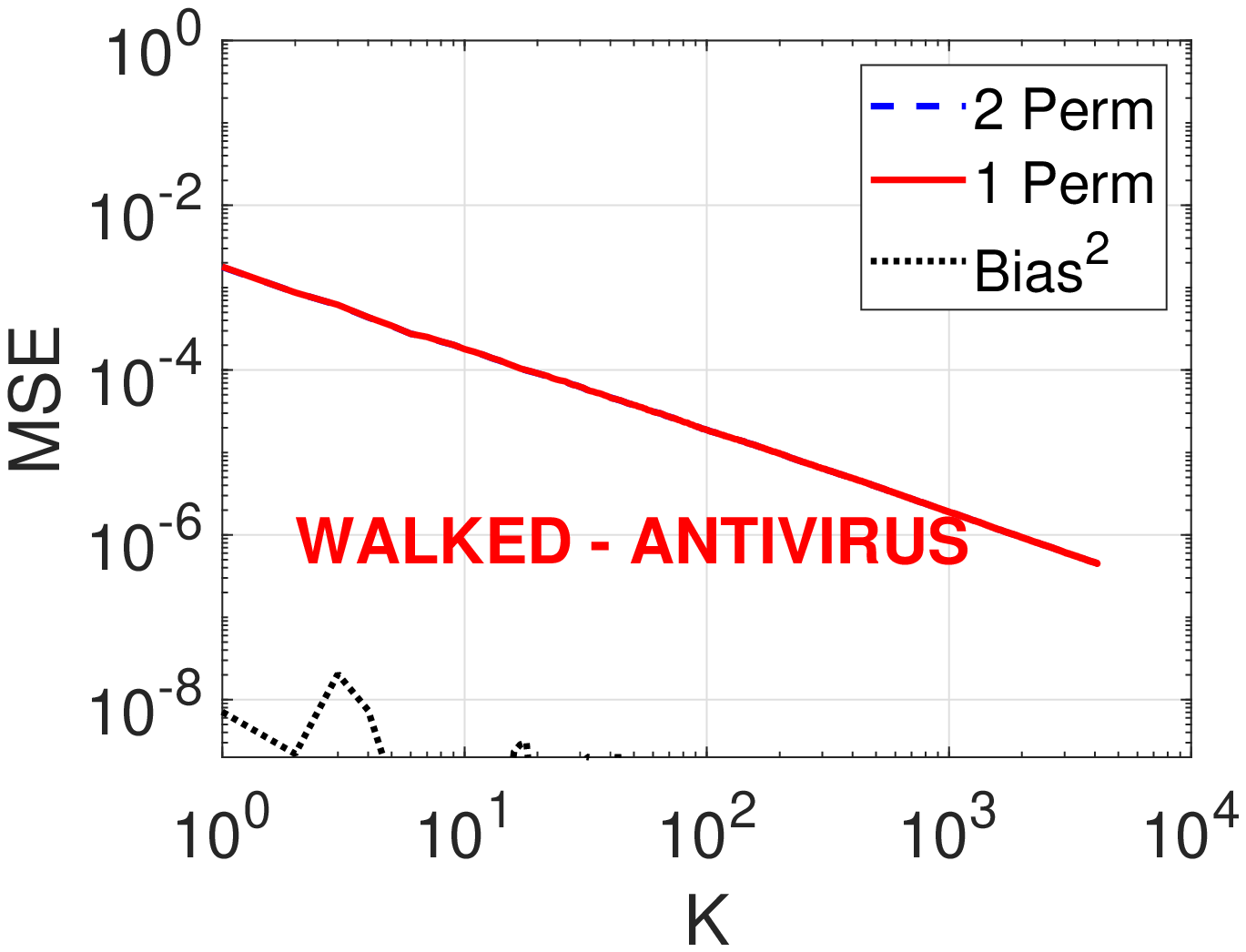}
    \includegraphics[width=2.1in]{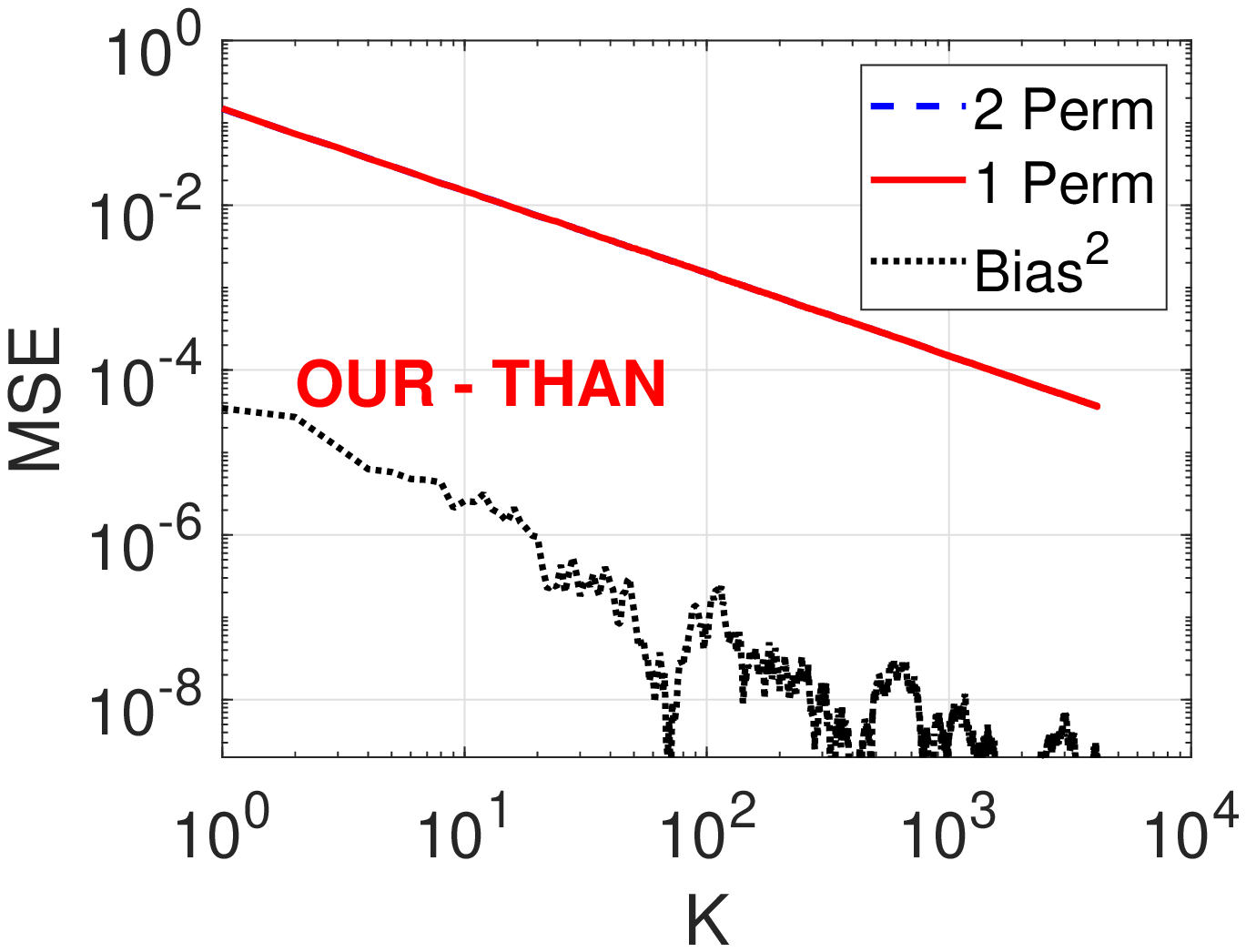}
    }
    \mbox{
    \includegraphics[width=2.1in]{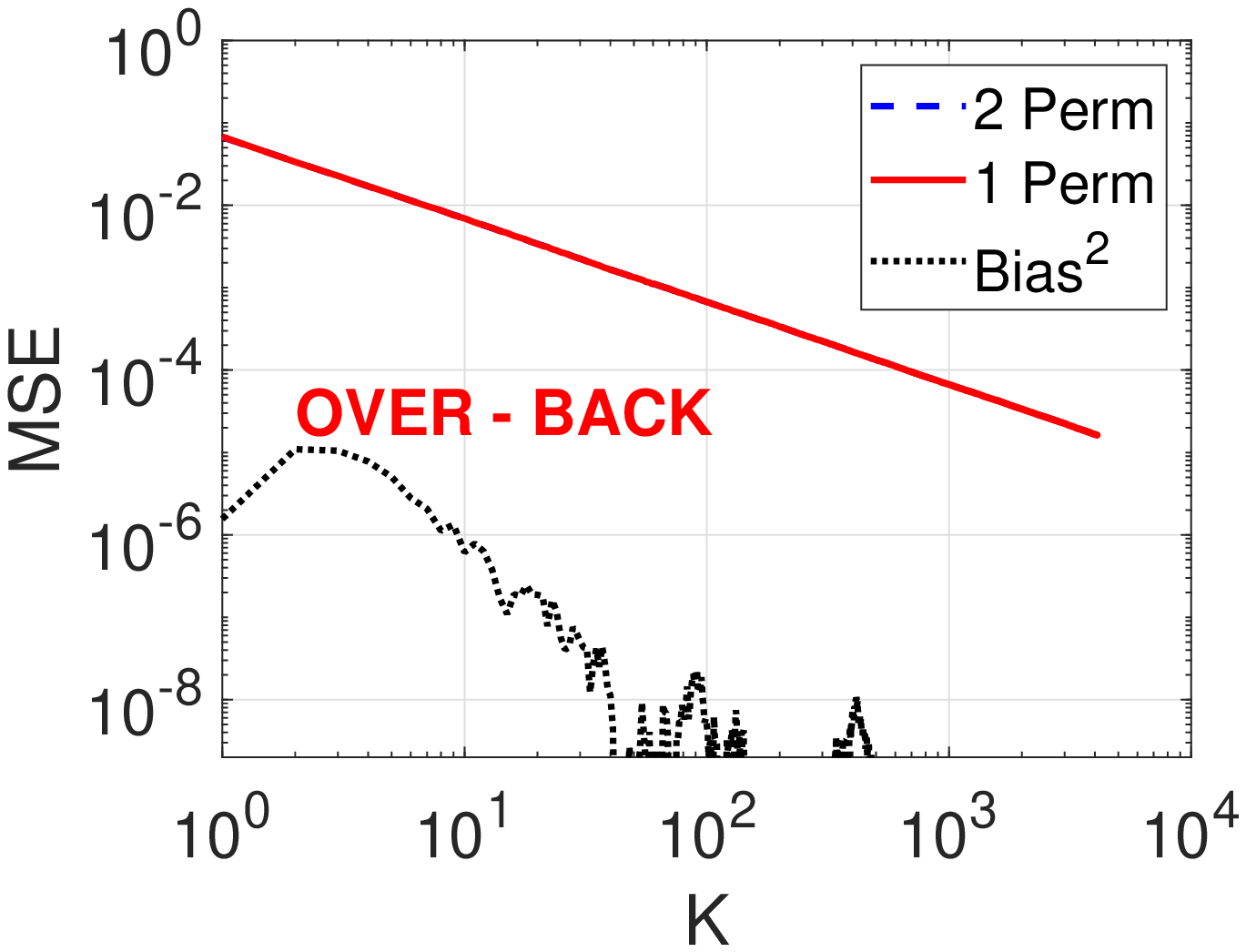}
    \includegraphics[width=2.1in]{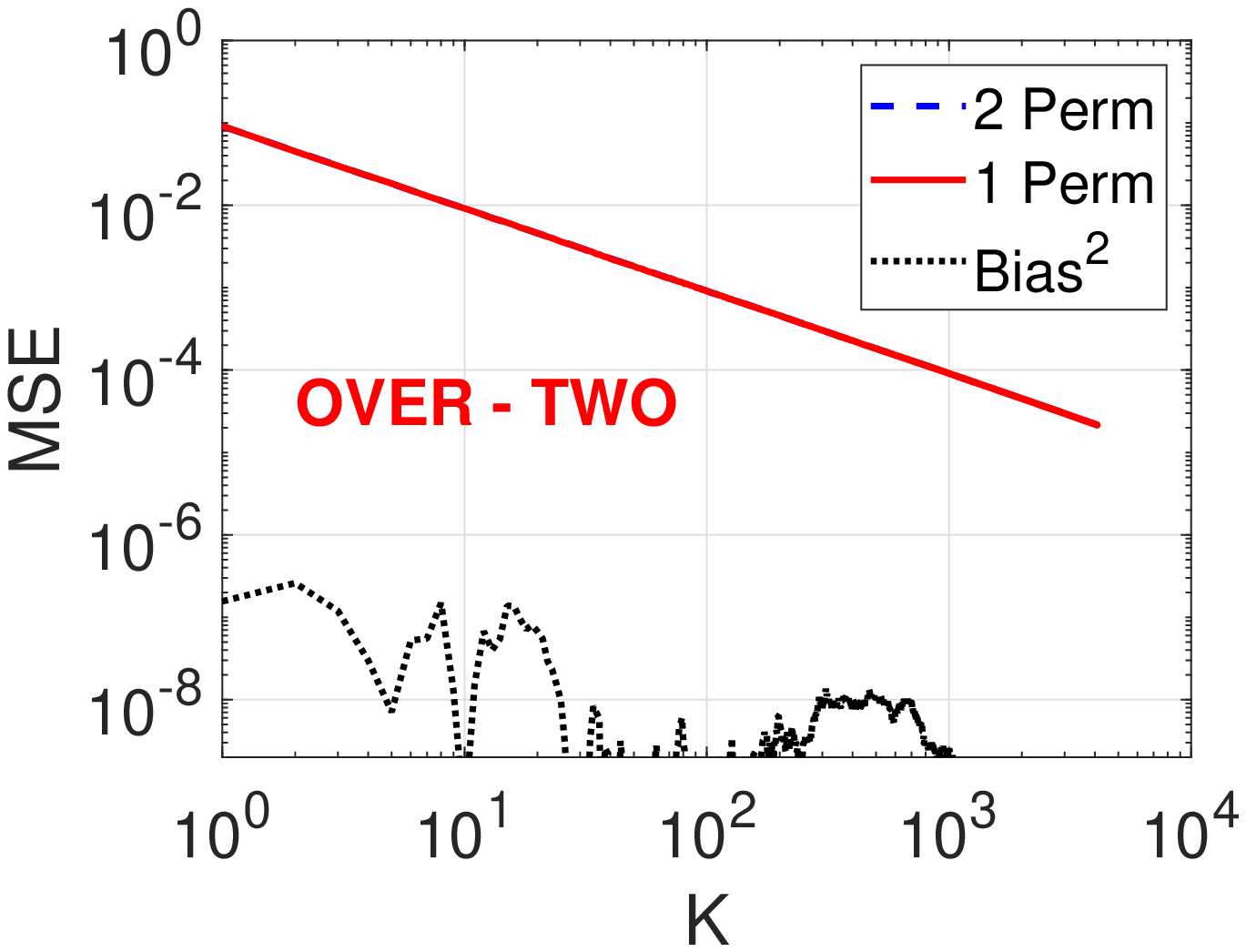}
    \includegraphics[width=2.1in]{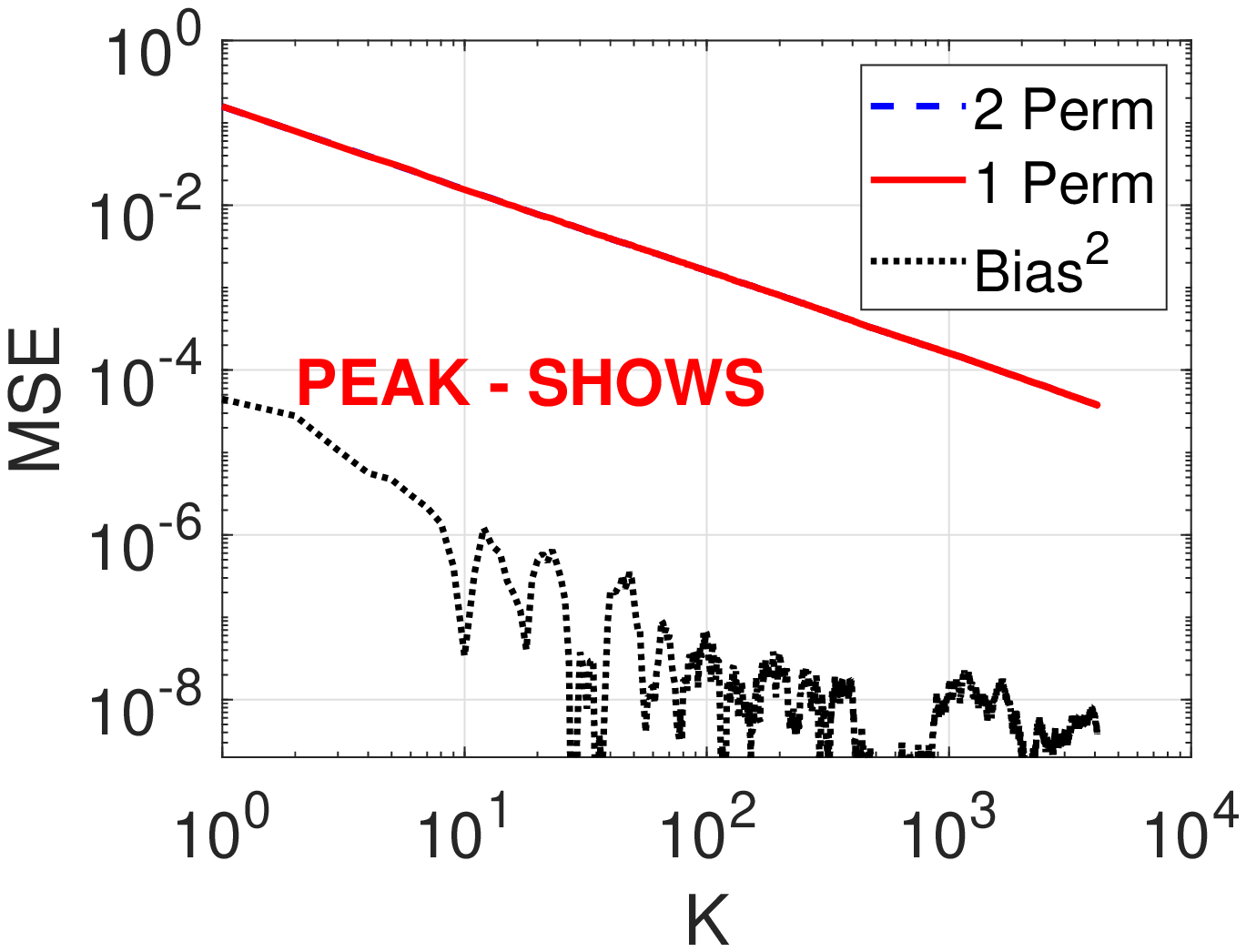}
    }

  \end{center}
  \vspace{-0.1in}
  \caption{Empirical MSEs of C-MinHash-$(\pi,\pi)$ (``1 Perm'', red, solid) vs. C-MinHash-$(\sigma,\pi)$ (``2 Perm'', blue, dashed) on various data pairs from the \textit{Words} dataset. We also report the empirical bias$^2$ for C-MinHash-$(\pi,\pi)$ to show that the bias is so small that it can be safely neglected. The empirical MSE curves for both estimators essentially overlap for all data pairs, for $K$ ranging from 1 to 4096. }
  \label{fig:word5}
\end{figure}

\begin{figure}[H]
  \begin{center}
   \mbox{
    \includegraphics[width=2.1in]{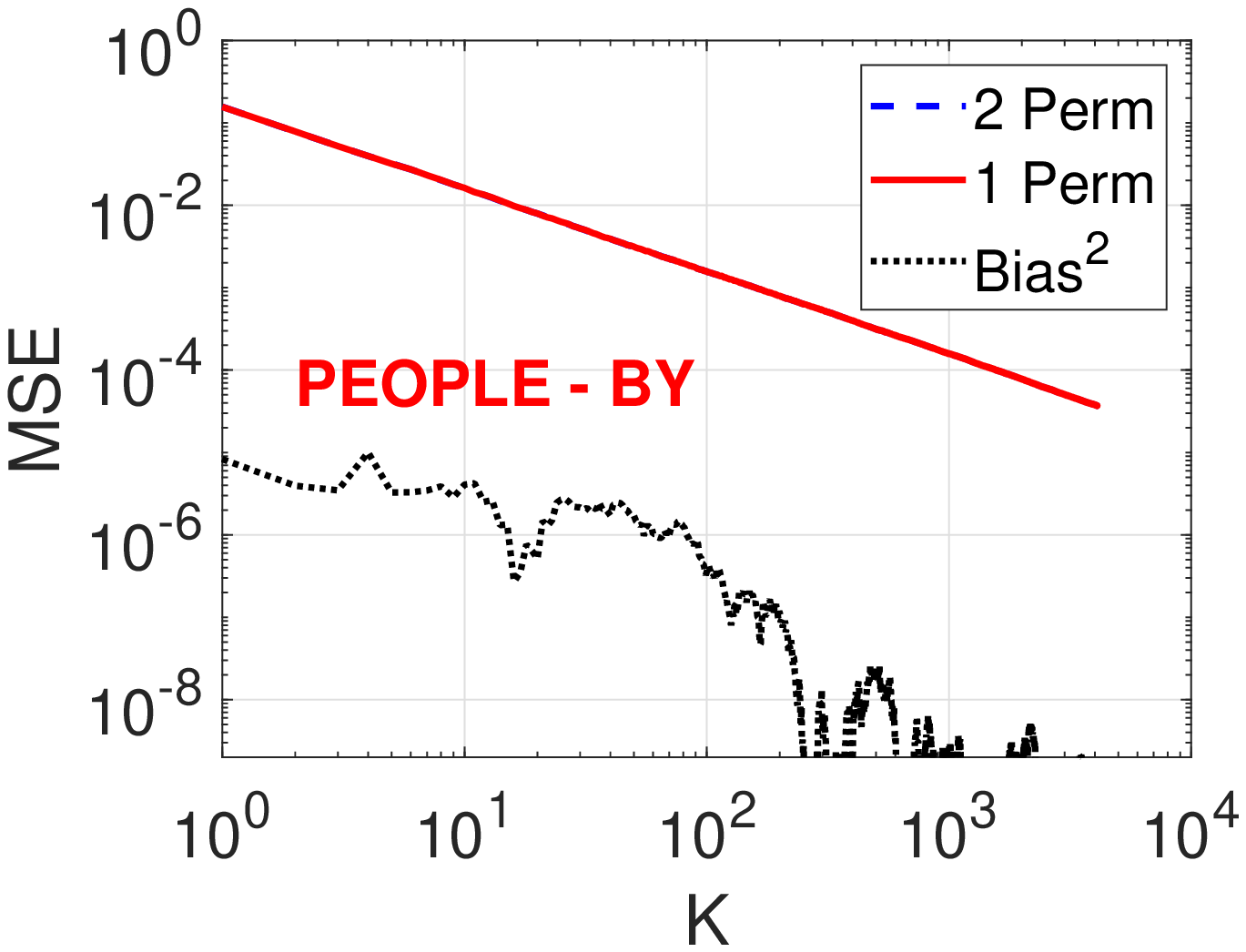}
    \includegraphics[width=2.1in]{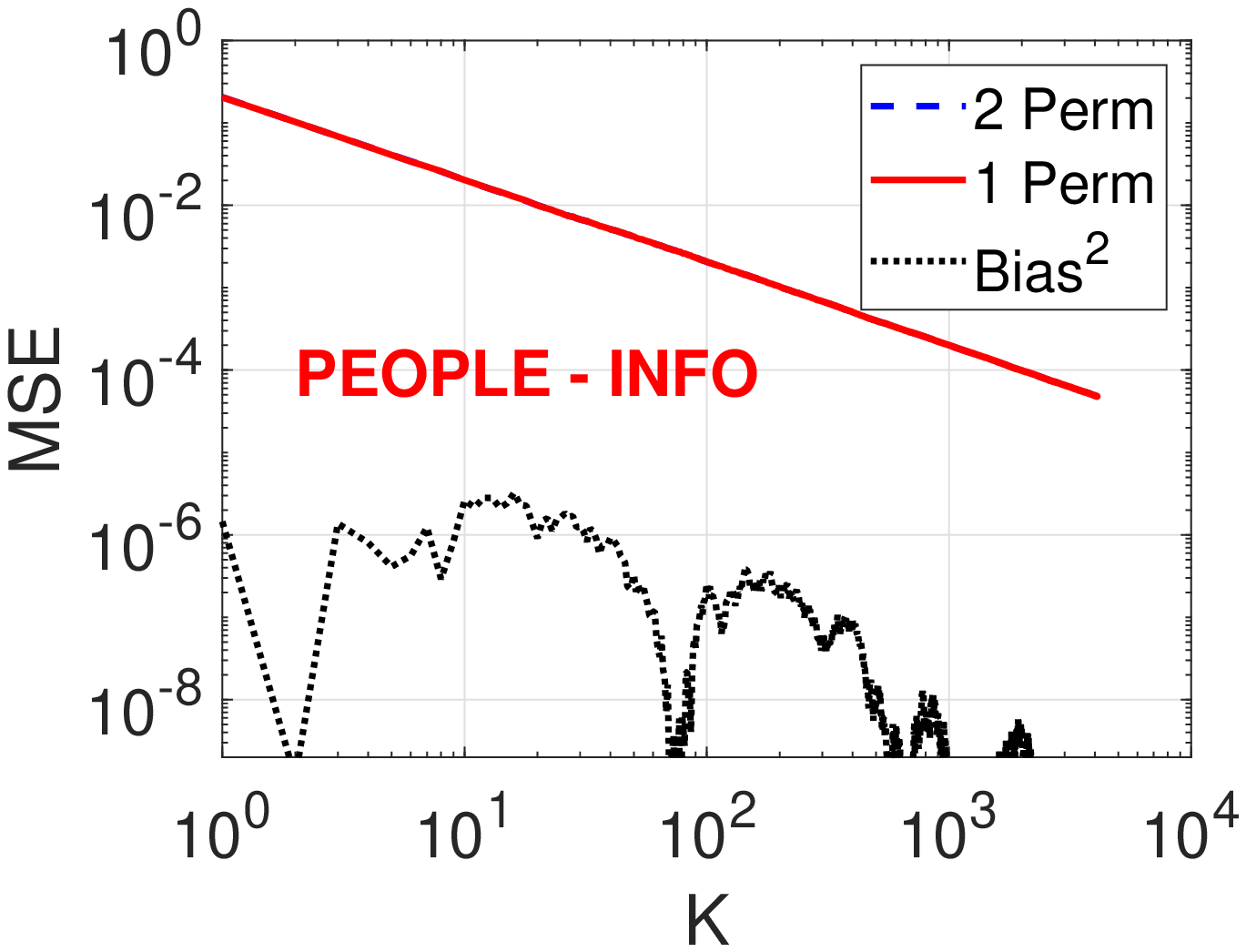}
    \includegraphics[width=2.1in]{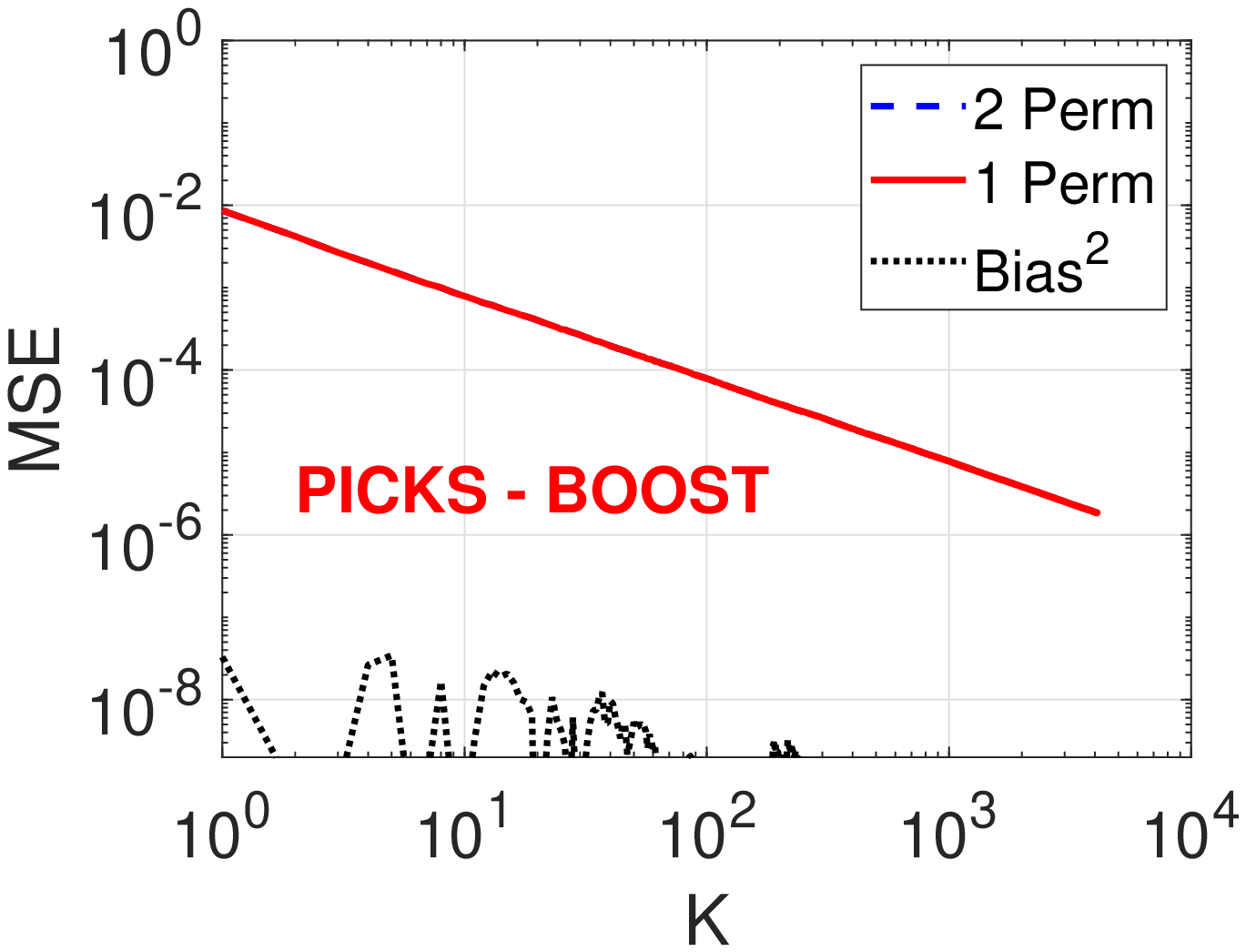}
    }
    \mbox{
    \includegraphics[width=2.1in]{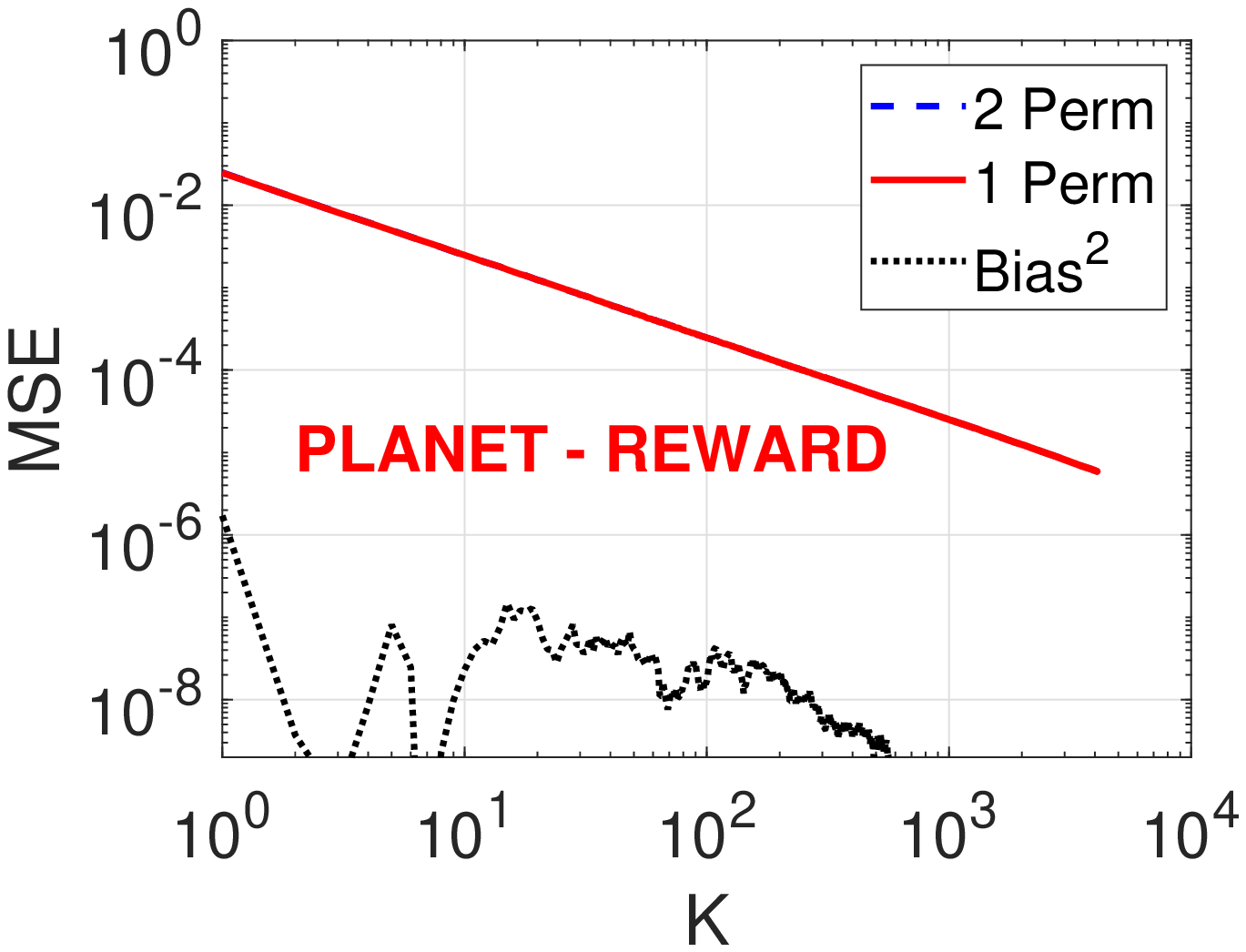}
    \includegraphics[width=2.1in]{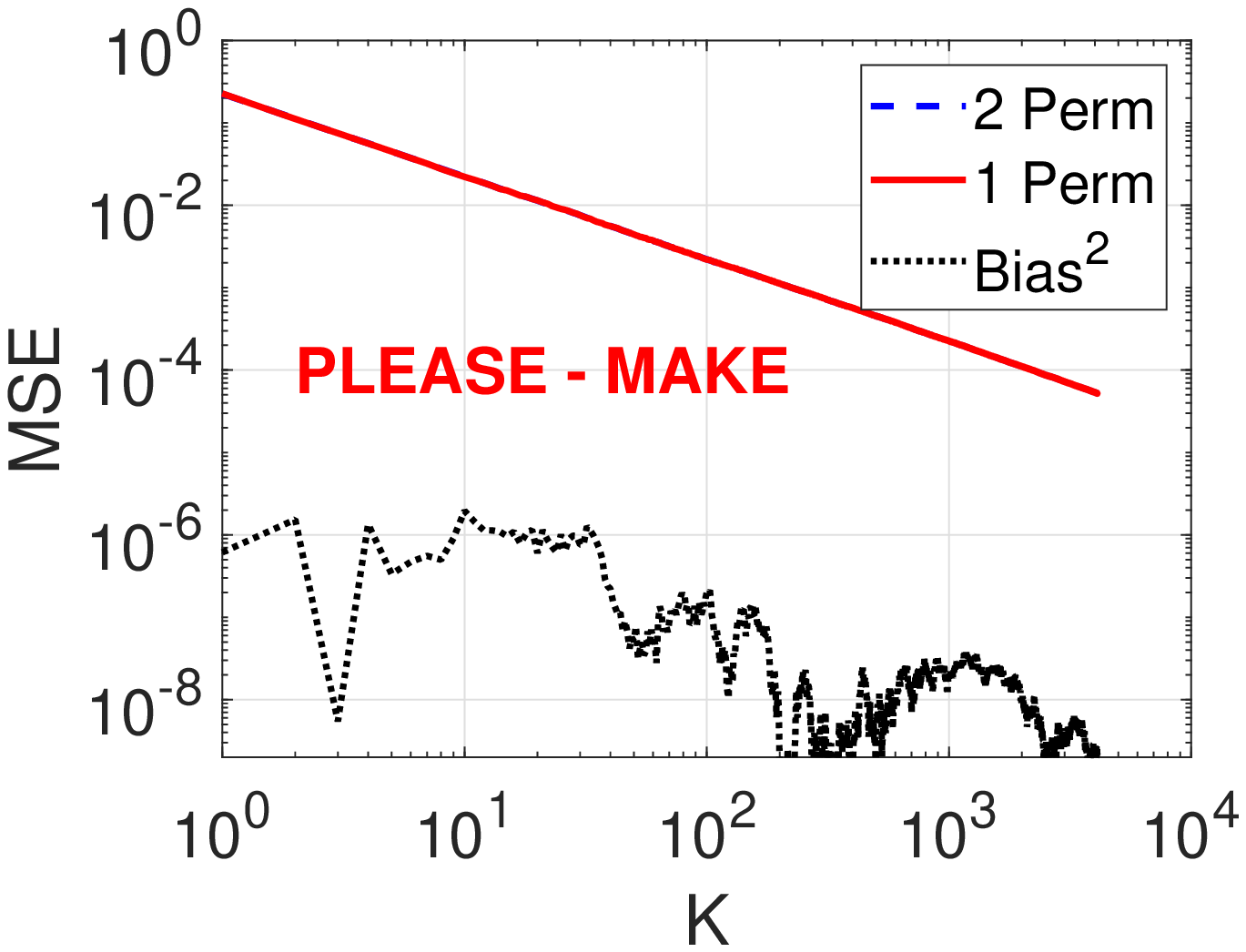}
    \includegraphics[width=2.1in]{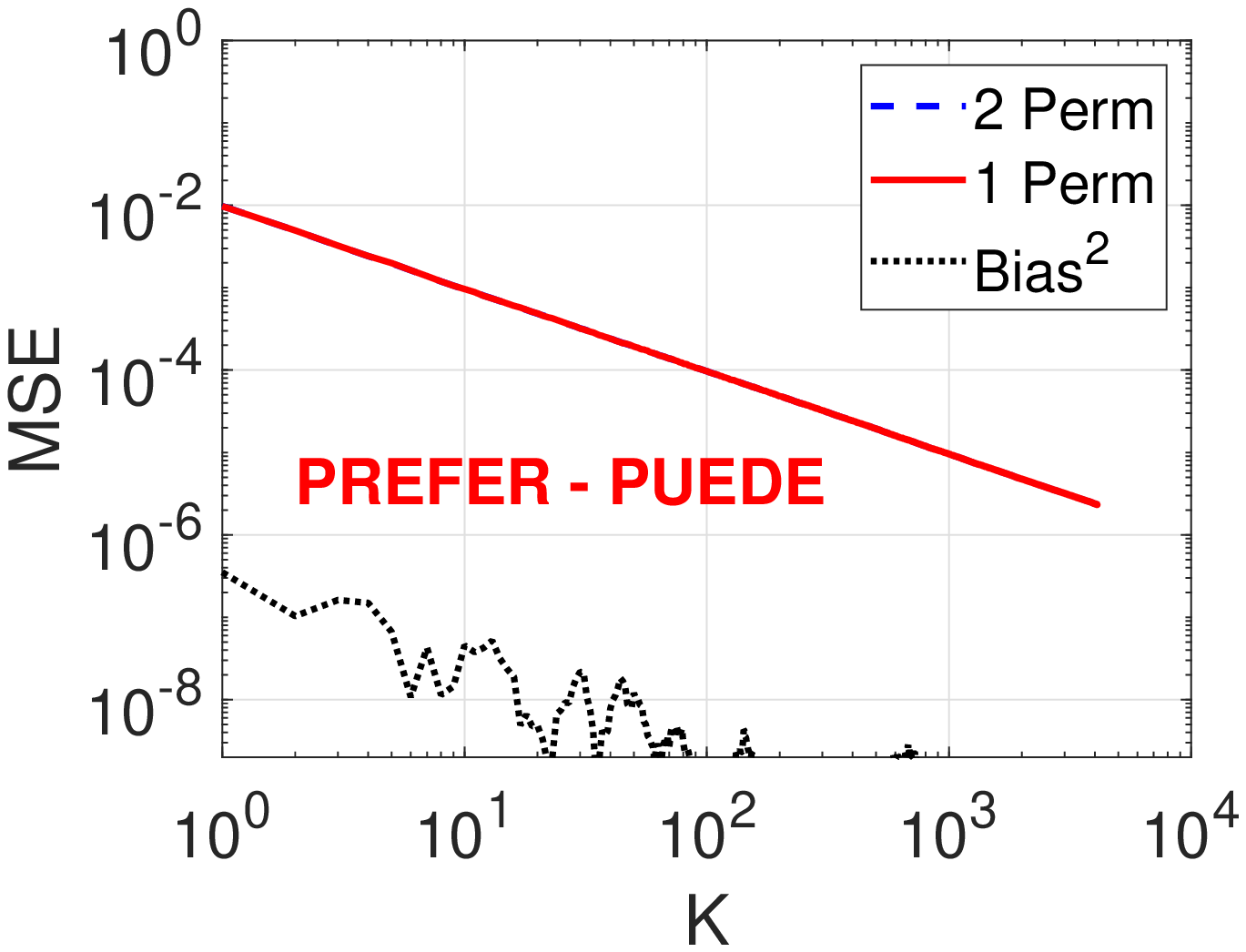}
    }
    \mbox{
    \includegraphics[width=2.1in]{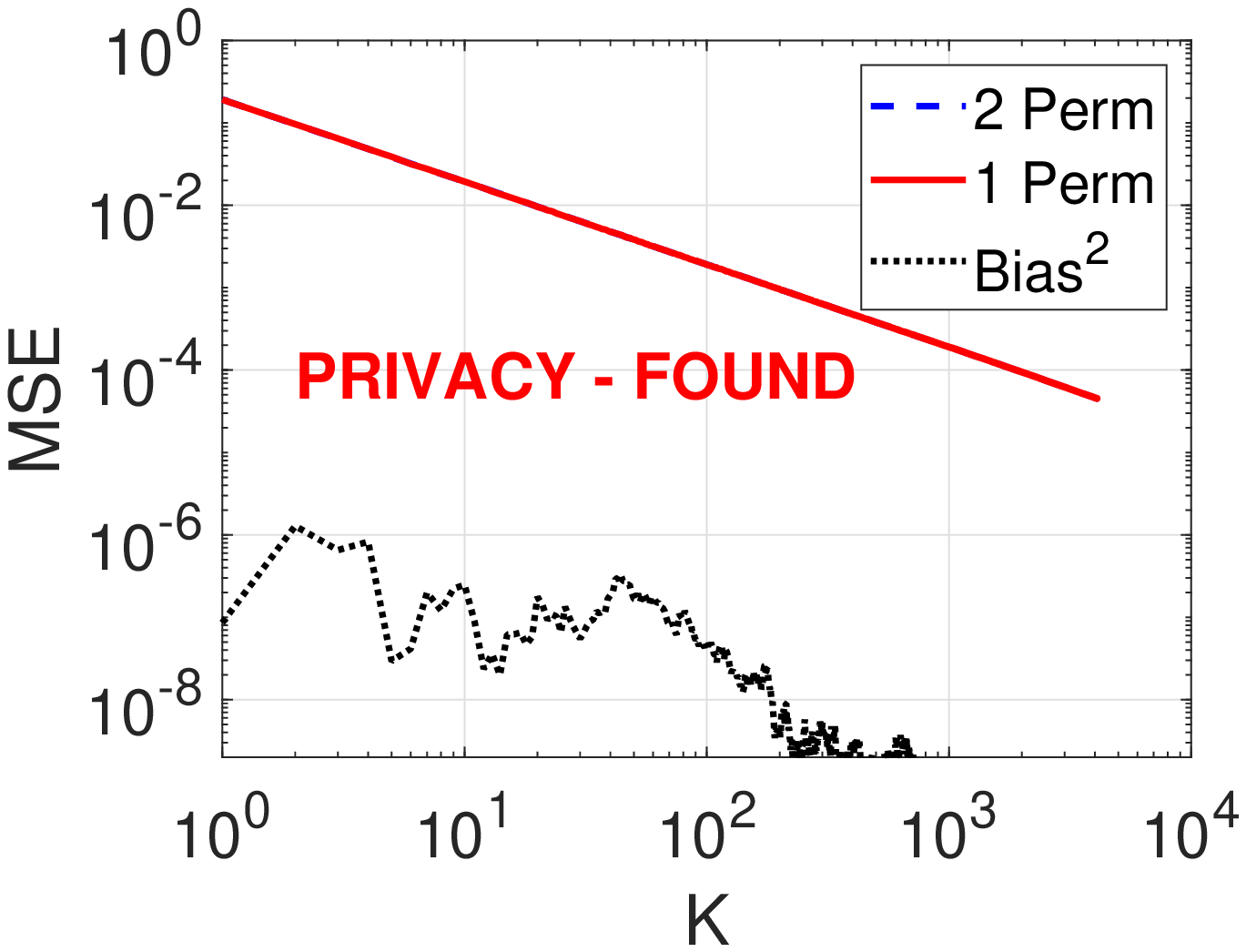}
    \includegraphics[width=2.1in]{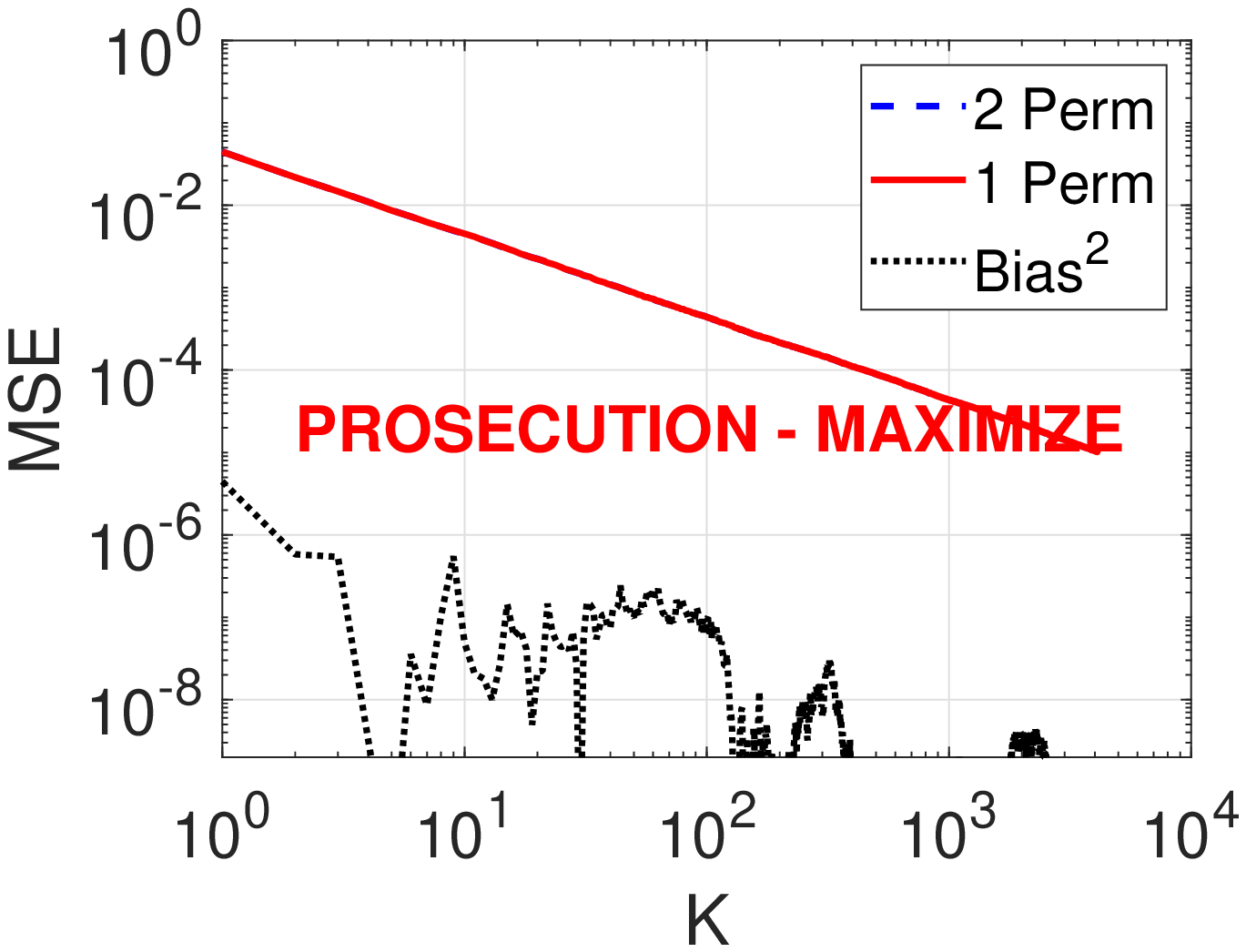}
    \includegraphics[width=2.1in]{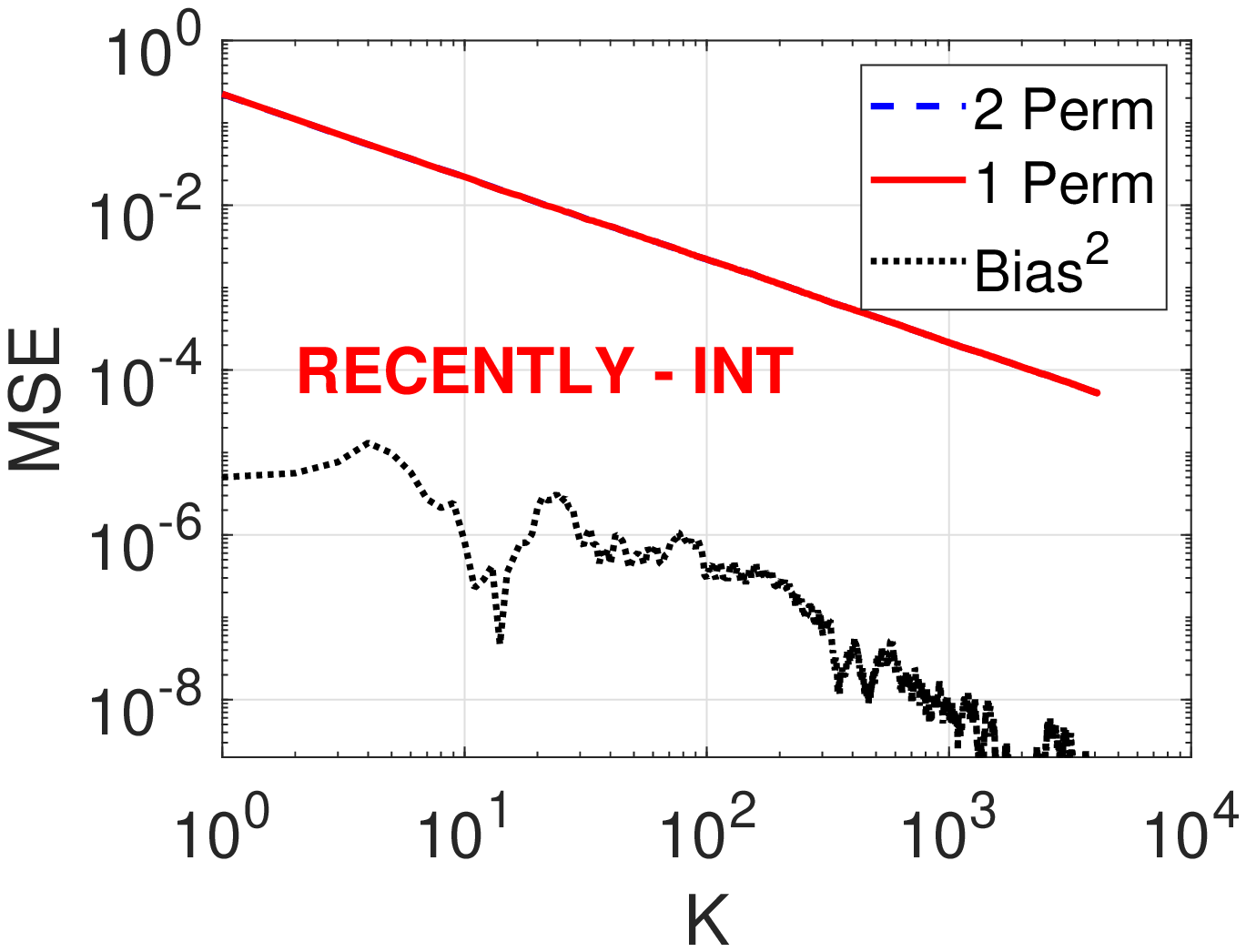}
    }
    \mbox{
    \includegraphics[width=2.1in]{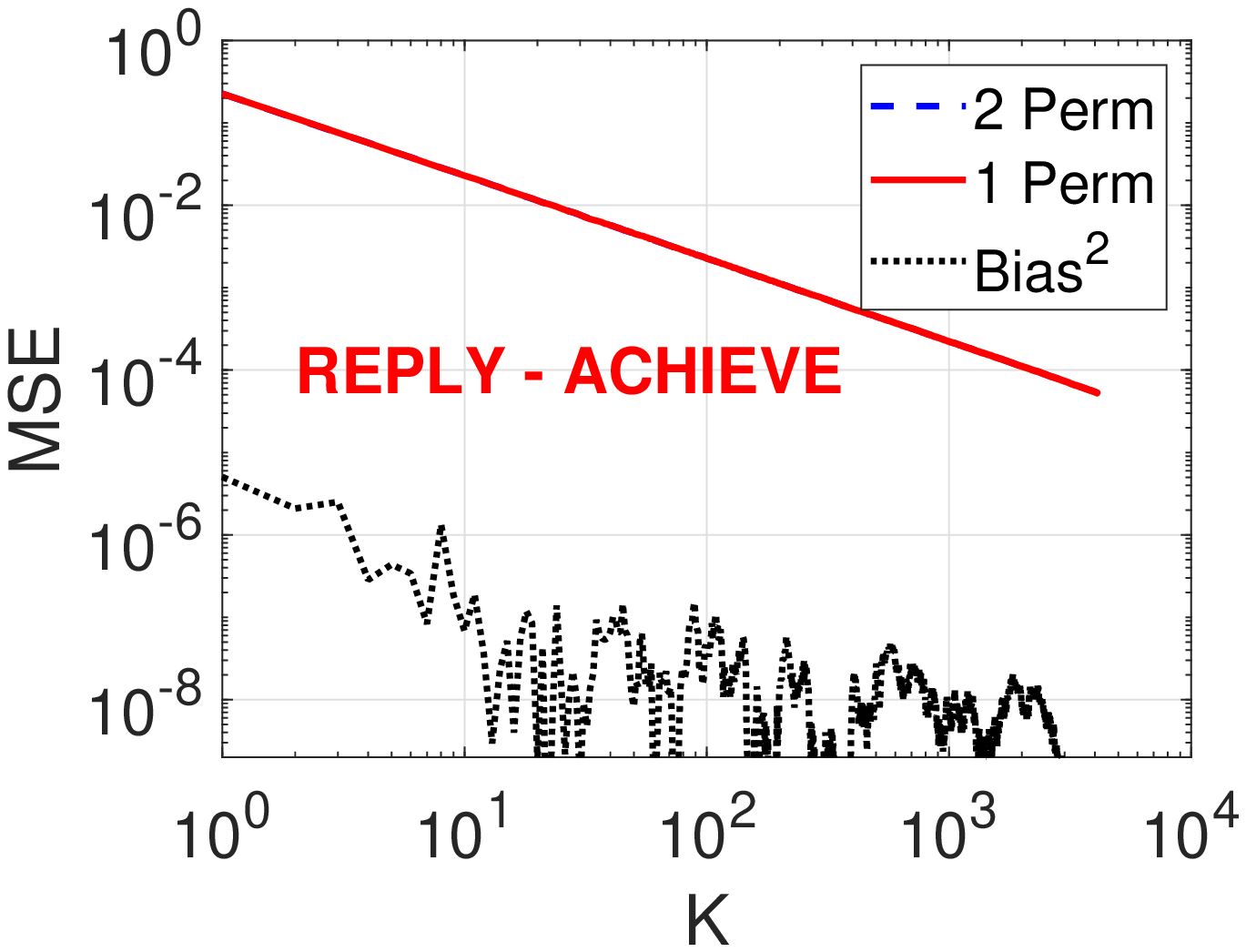}
    \includegraphics[width=2.1in]{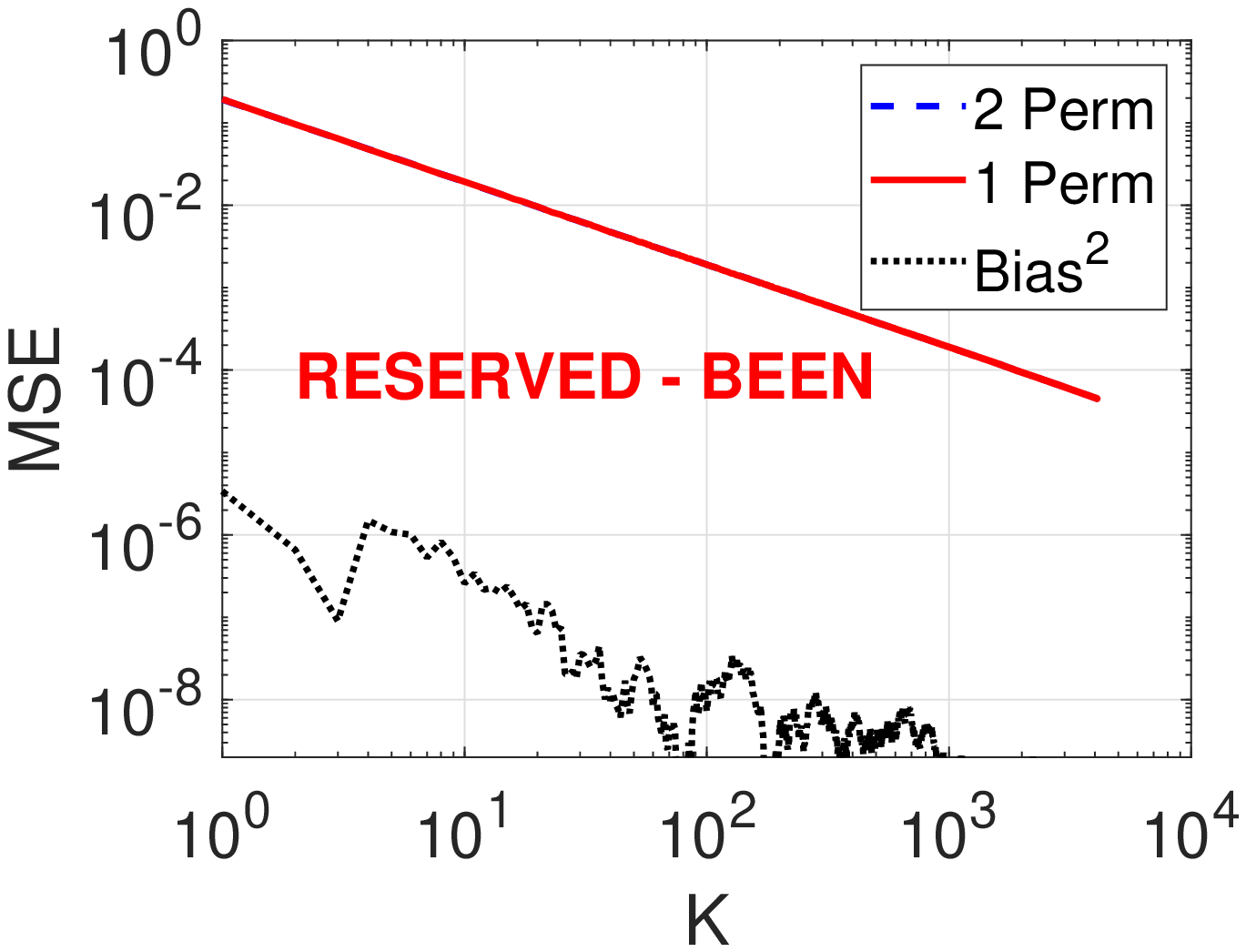}
    \includegraphics[width=2.1in]{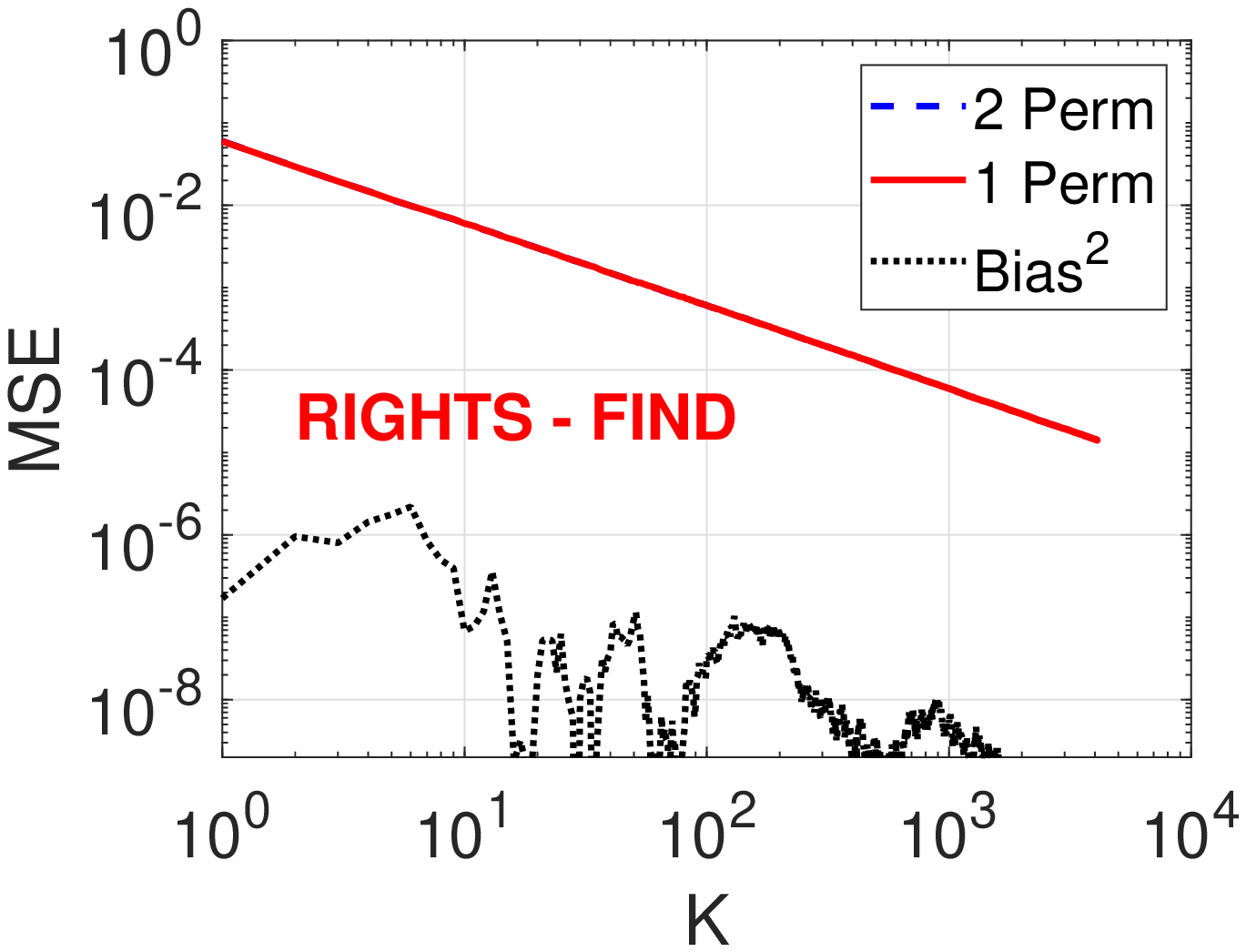}
    }
    \mbox{
    \includegraphics[width=2.1in]{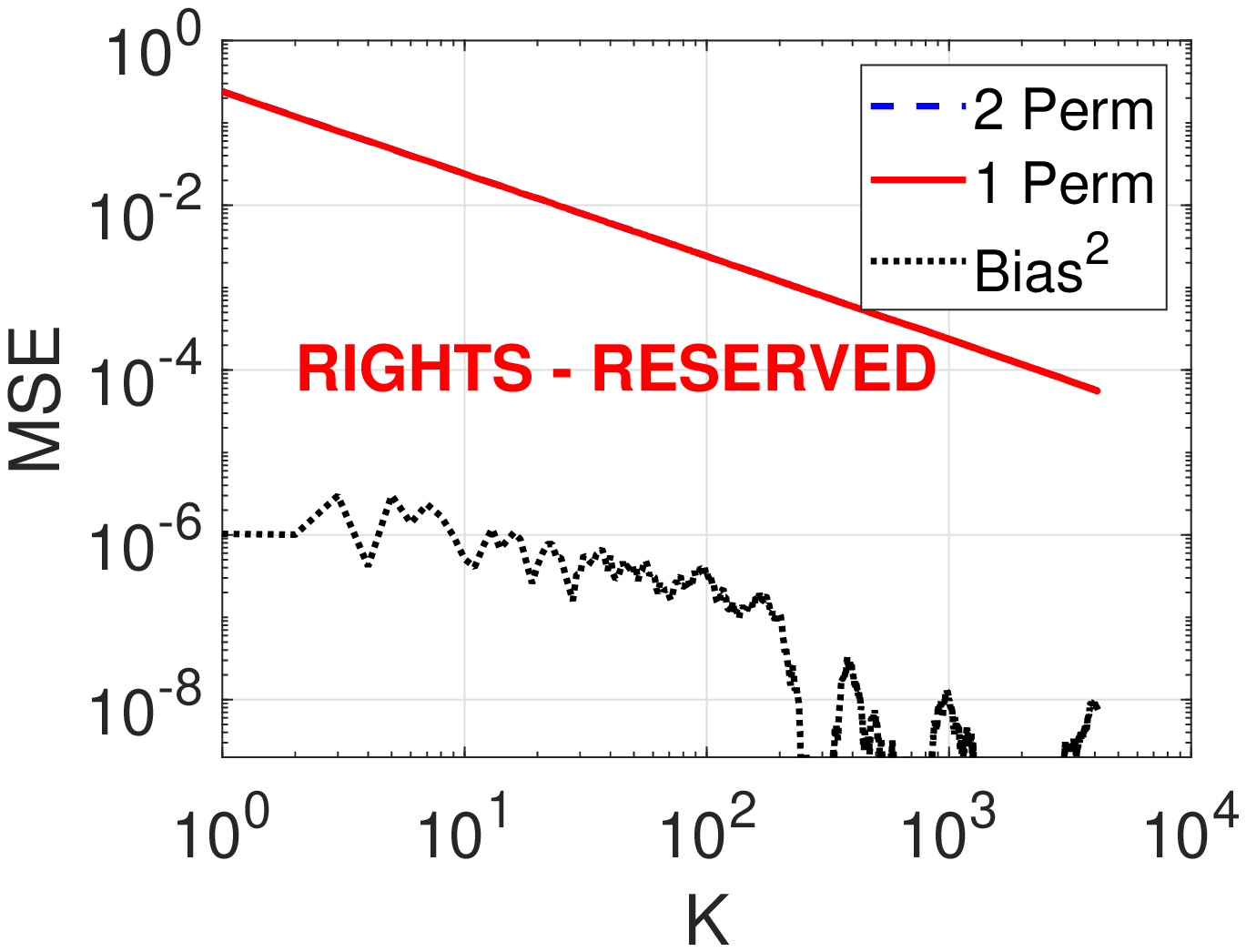}
    \includegraphics[width=2.1in]{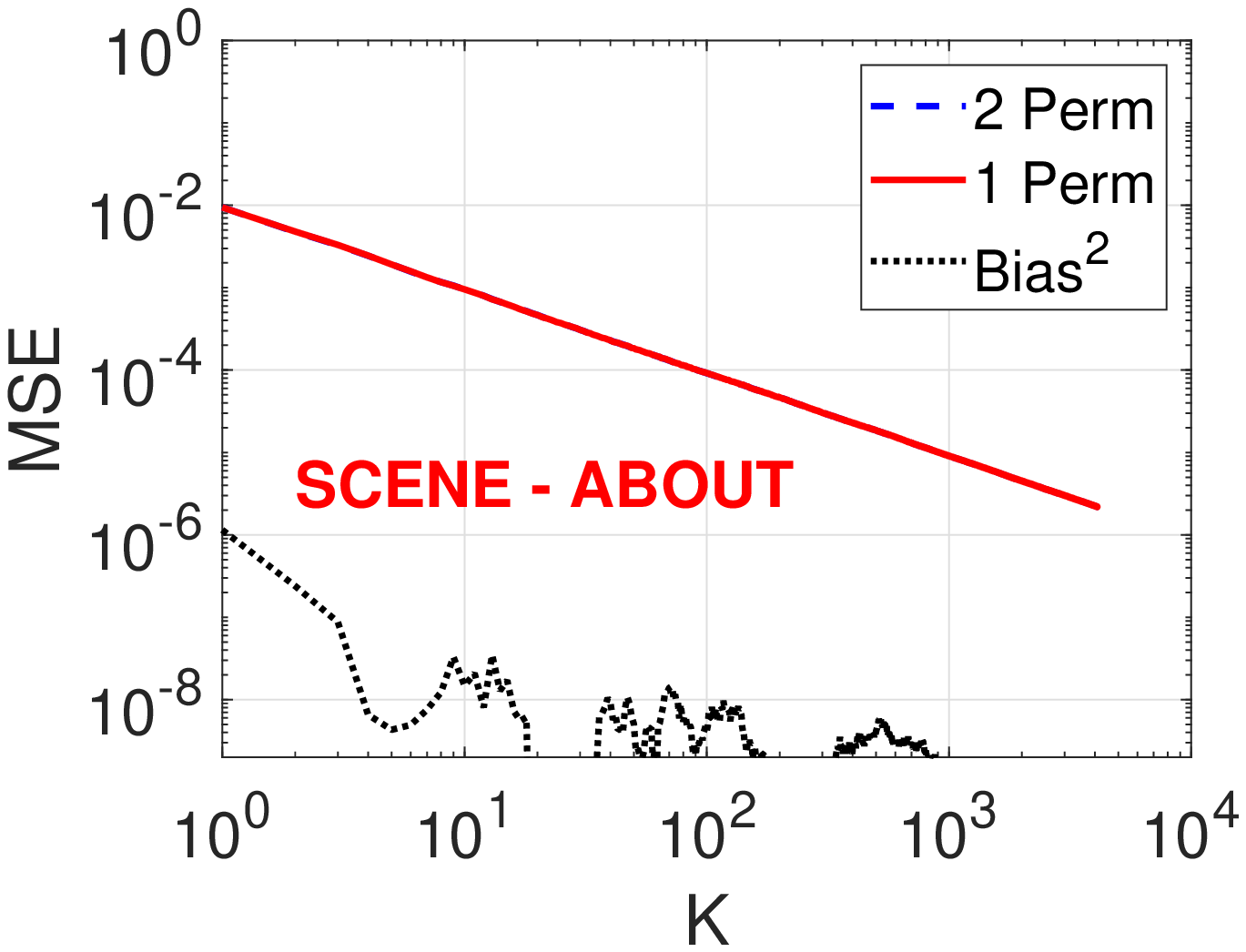}
    \includegraphics[width=2.1in]{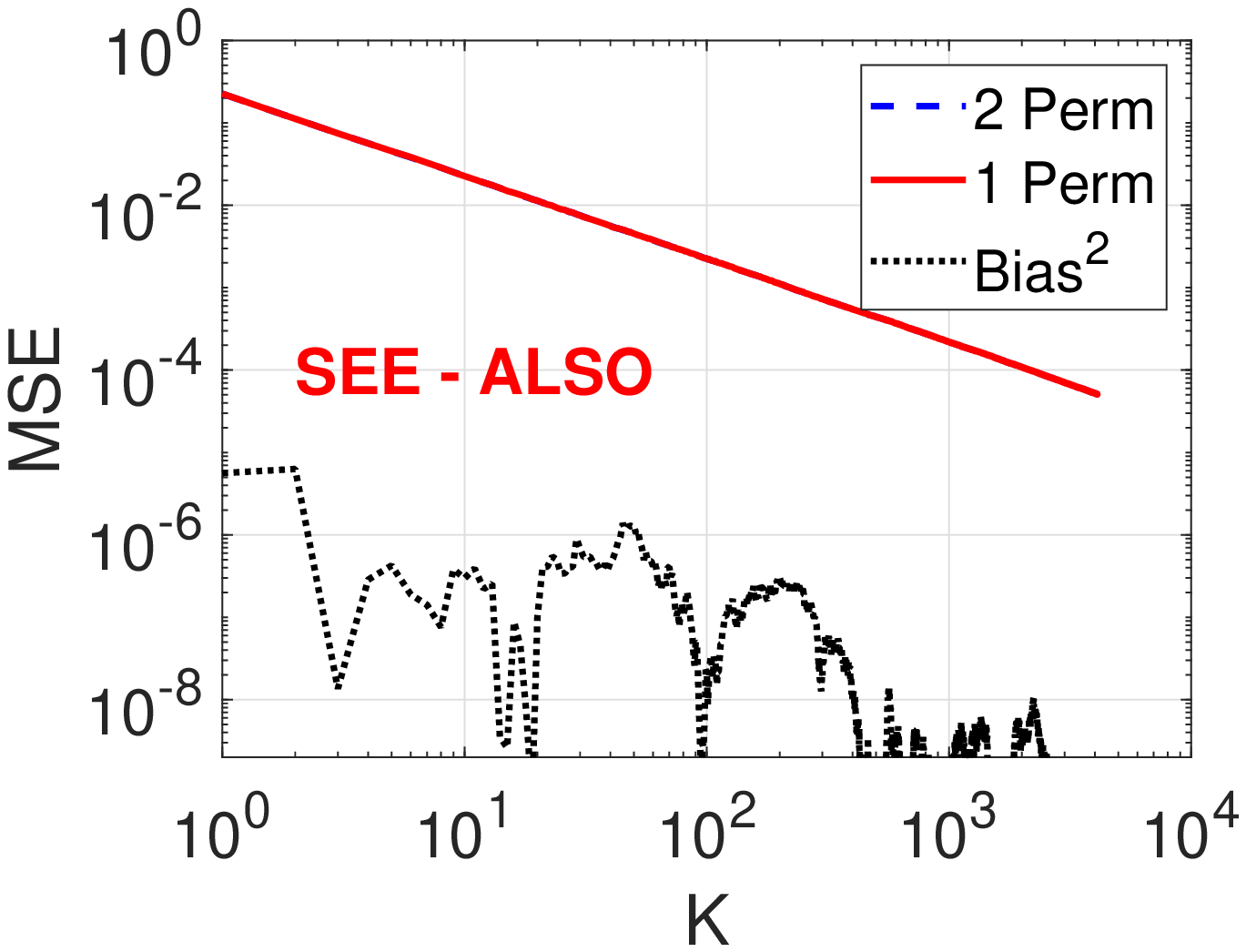}
    }

  \end{center}
  \vspace{-0.1in}
  \caption{Empirical MSEs of C-MinHash-$(\pi,\pi)$ (``1 Perm'', red, solid) vs. C-MinHash-$(\sigma,\pi)$ (``2 Perm'', blue, dashed) on various data pairs from the \textit{Words} dataset. We also report the empirical bias$^2$ for C-MinHash-$(\pi,\pi)$ to show that the bias is so small that it can be safely neglected. The empirical MSE curves for both estimators essentially overlap for all data pairs, for $K$ ranging from 1 to 4096. }
  \label{fig:word6}
\end{figure}

\begin{figure}[H]
  \begin{center}
   \mbox{
    \includegraphics[width=2.1in]{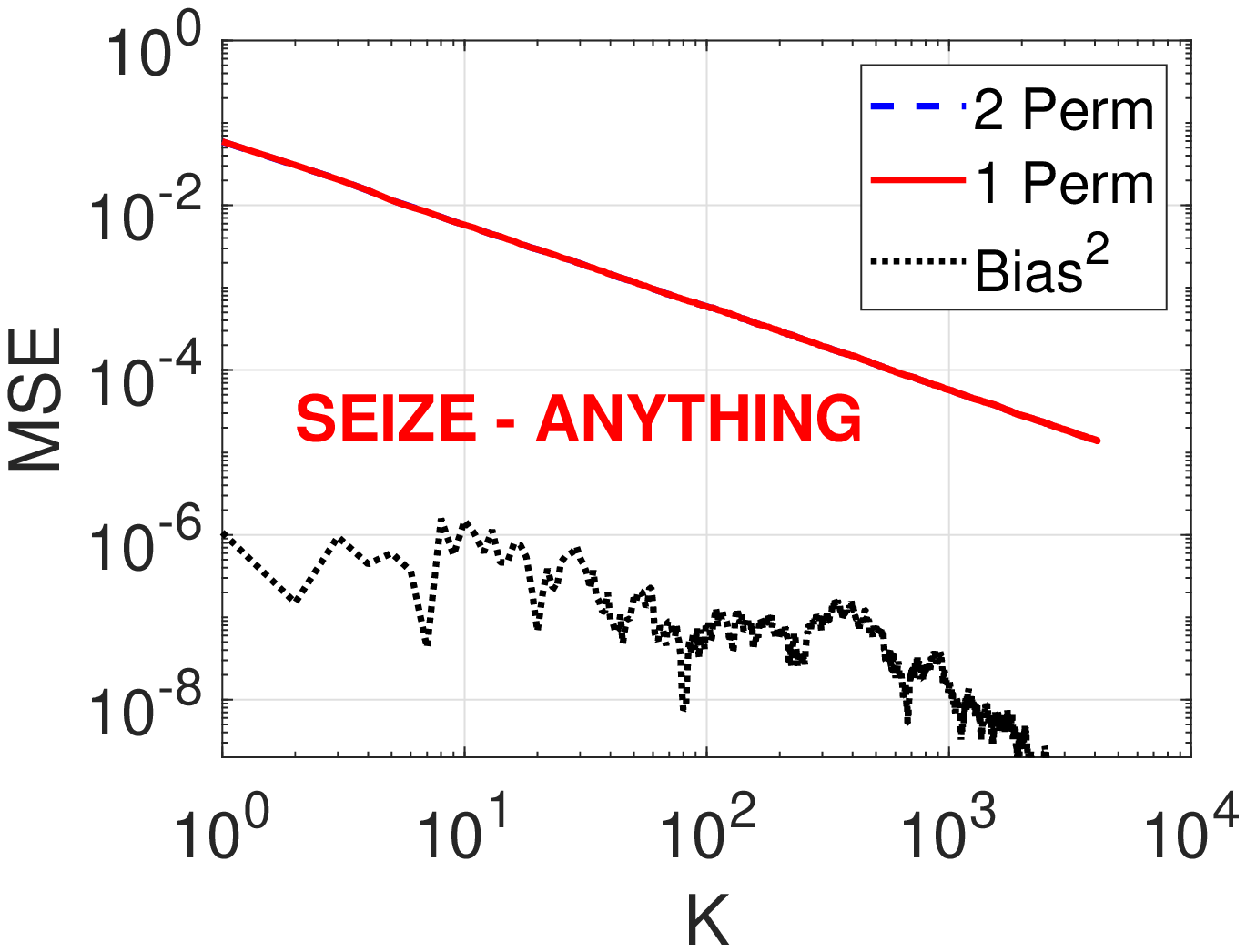}
    \includegraphics[width=2.1in]{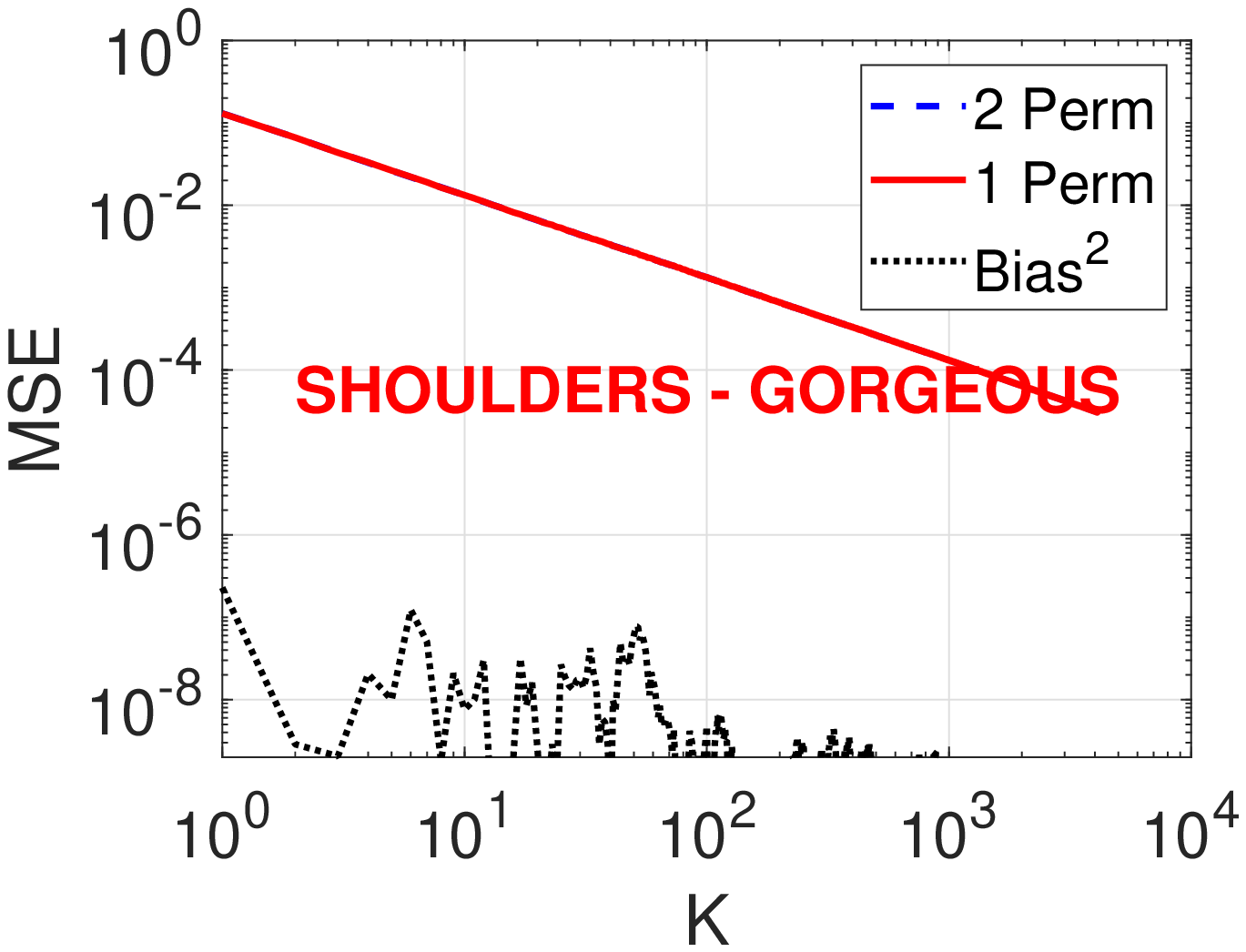}
    \includegraphics[width=2.1in]{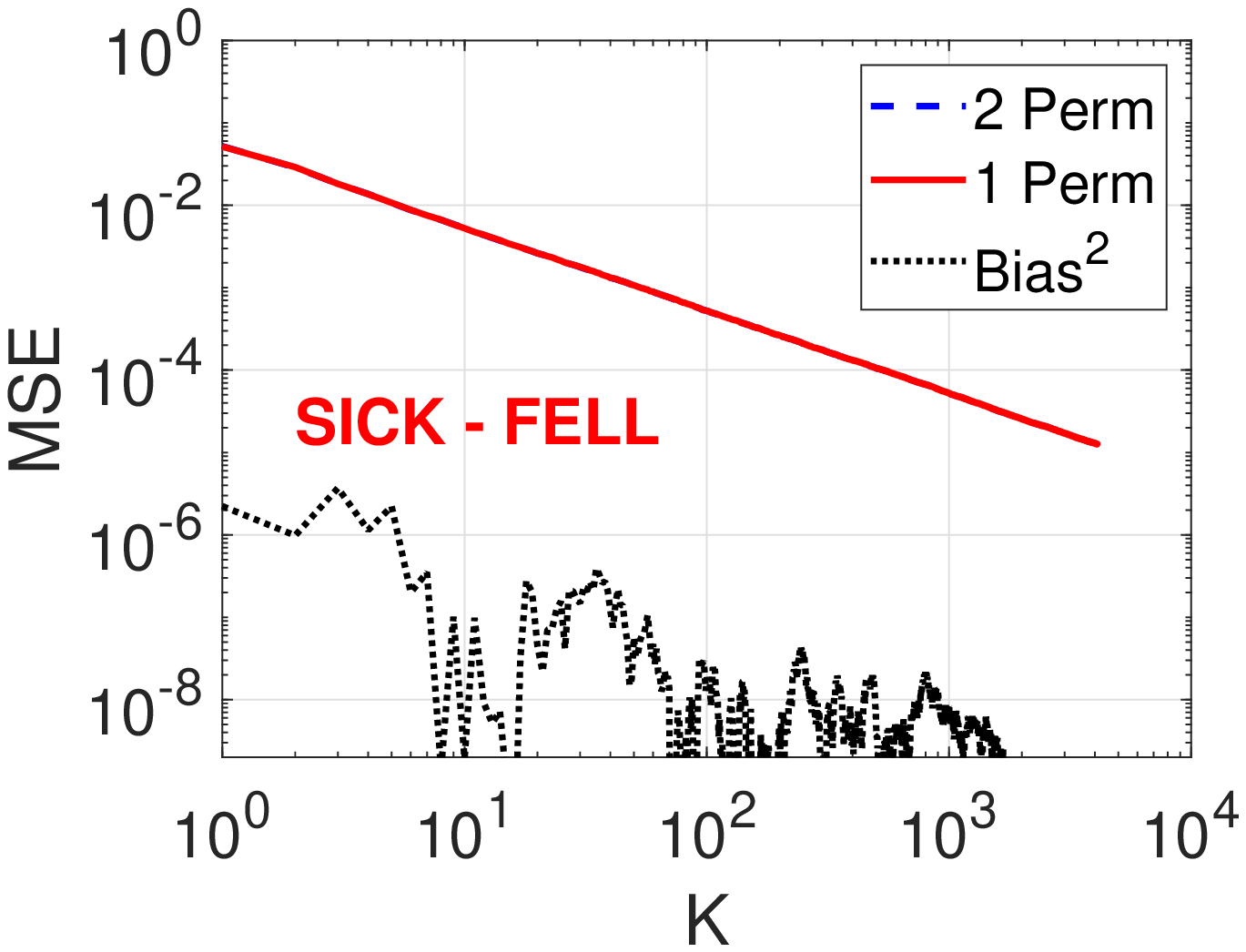}
    }
    \mbox{
    \includegraphics[width=2.1in]{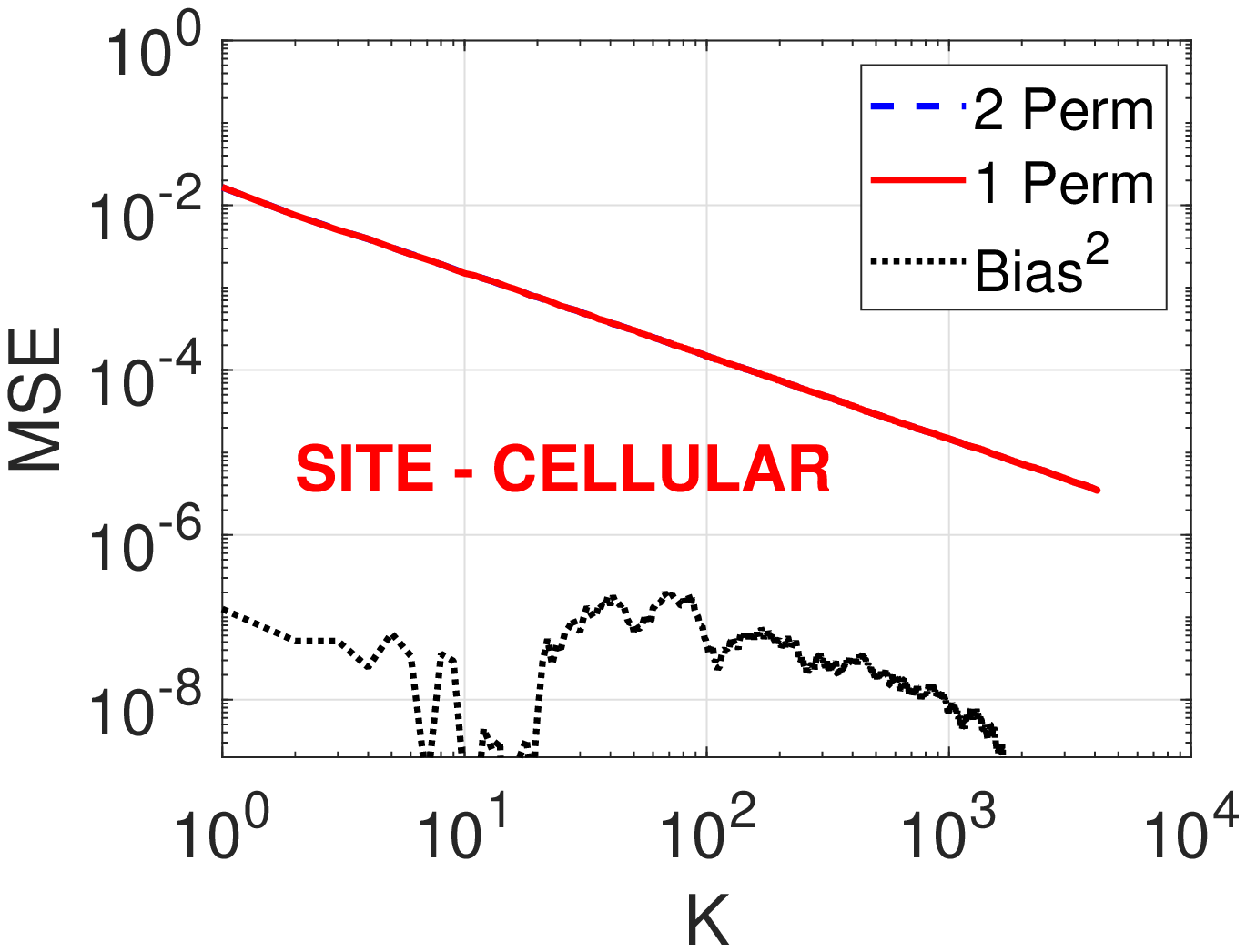}
    \includegraphics[width=2.1in]{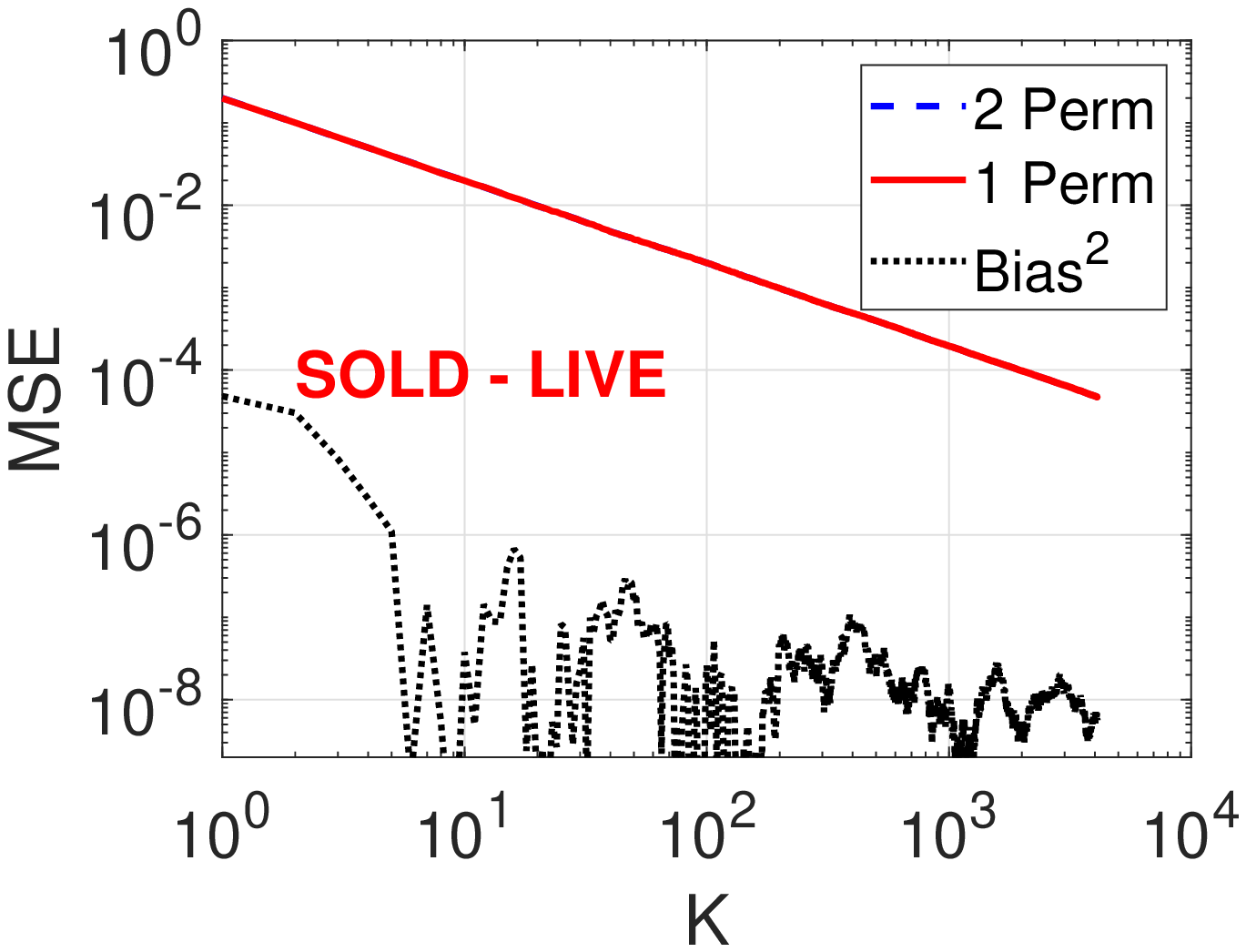}
    \includegraphics[width=2.1in]{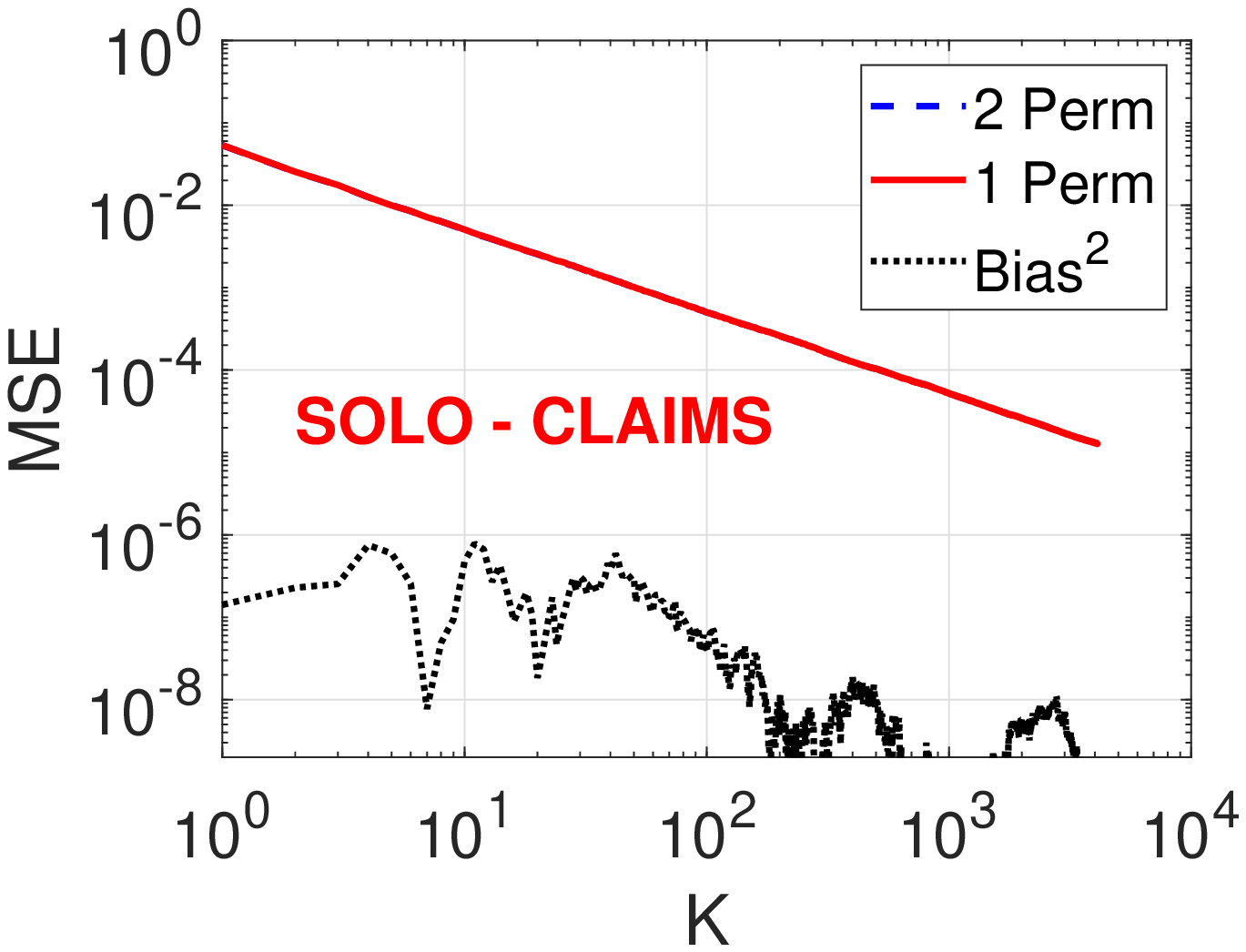}
    }
    \mbox{
    \includegraphics[width=2.1in]{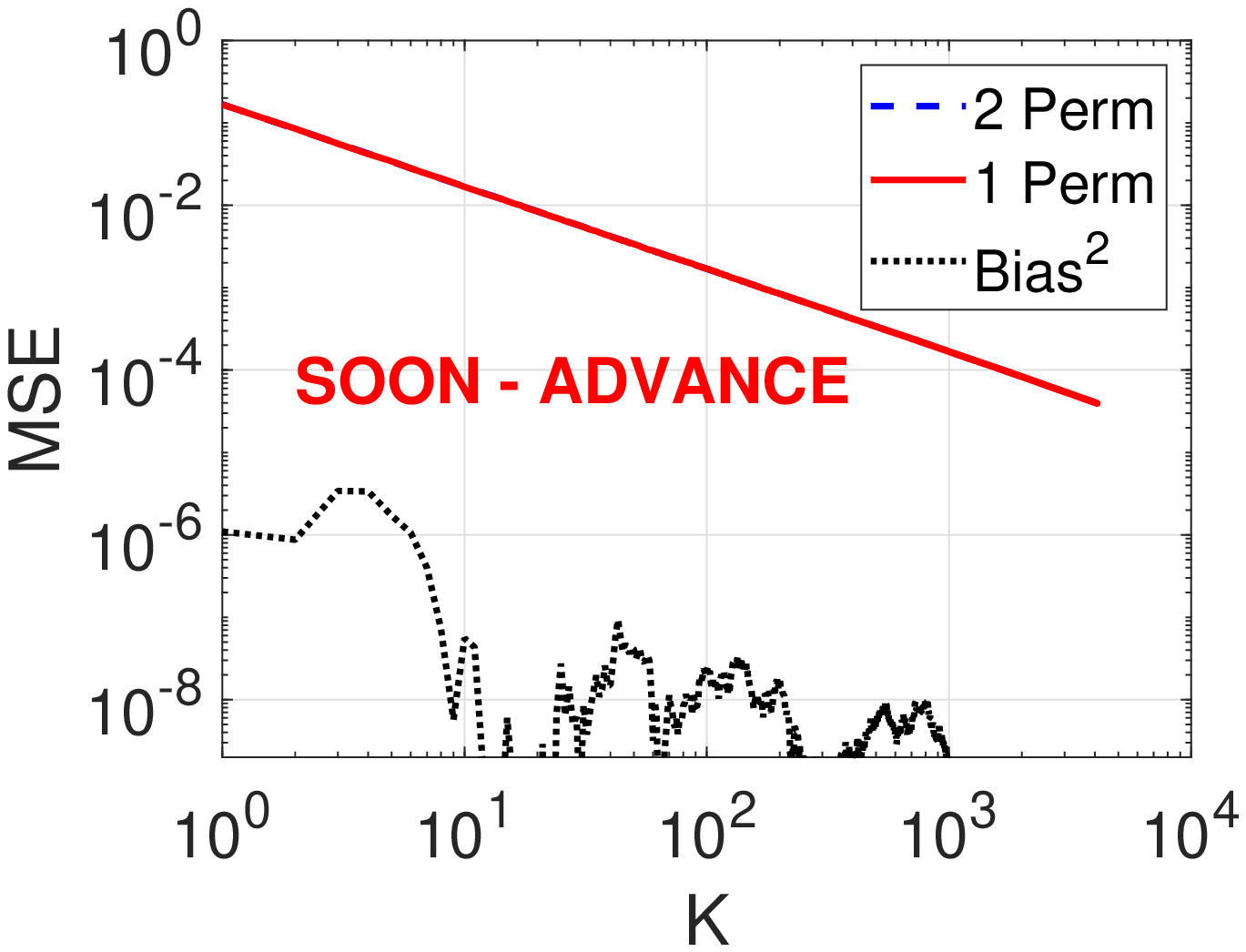}
    \includegraphics[width=2.1in]{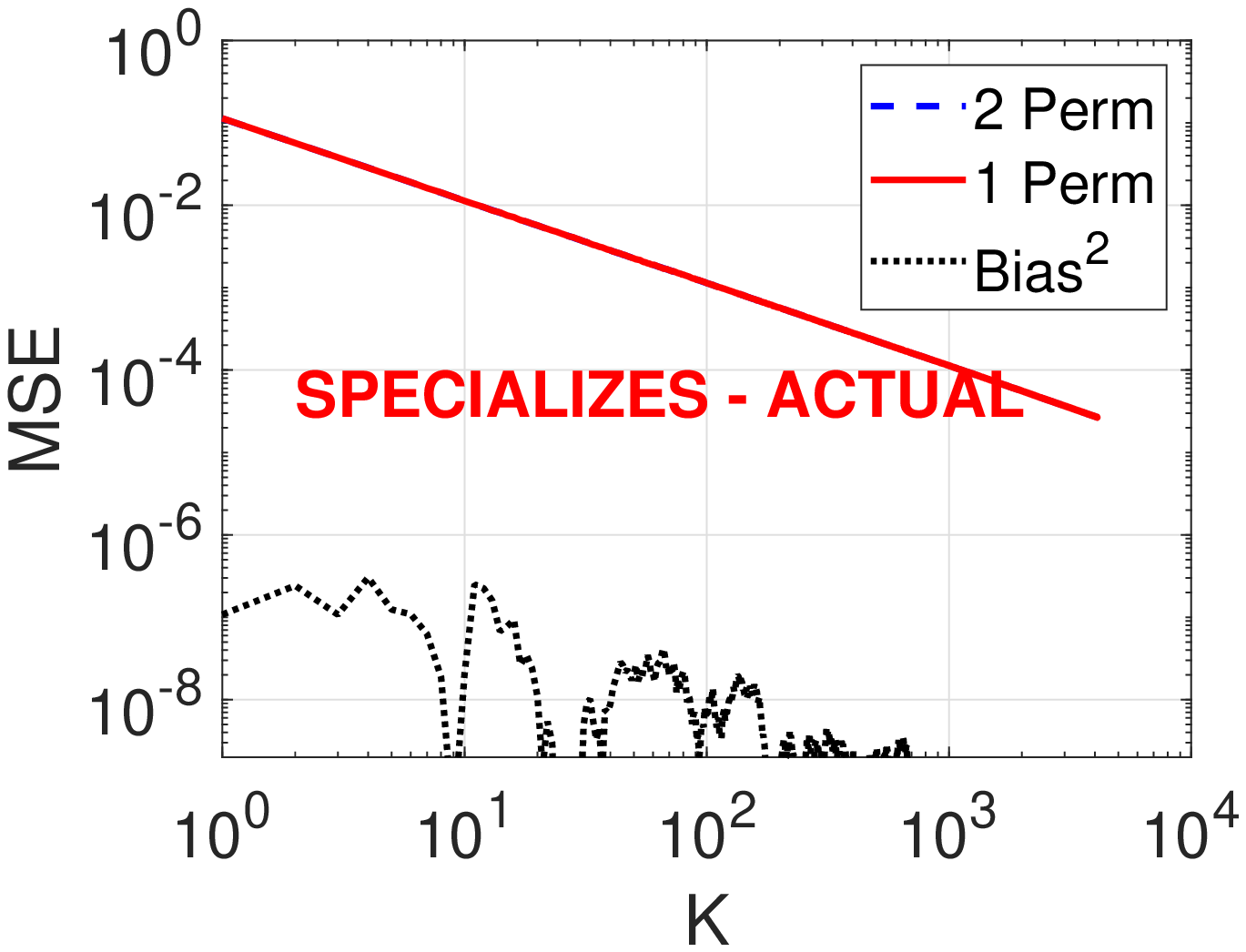}
    \includegraphics[width=2.1in]{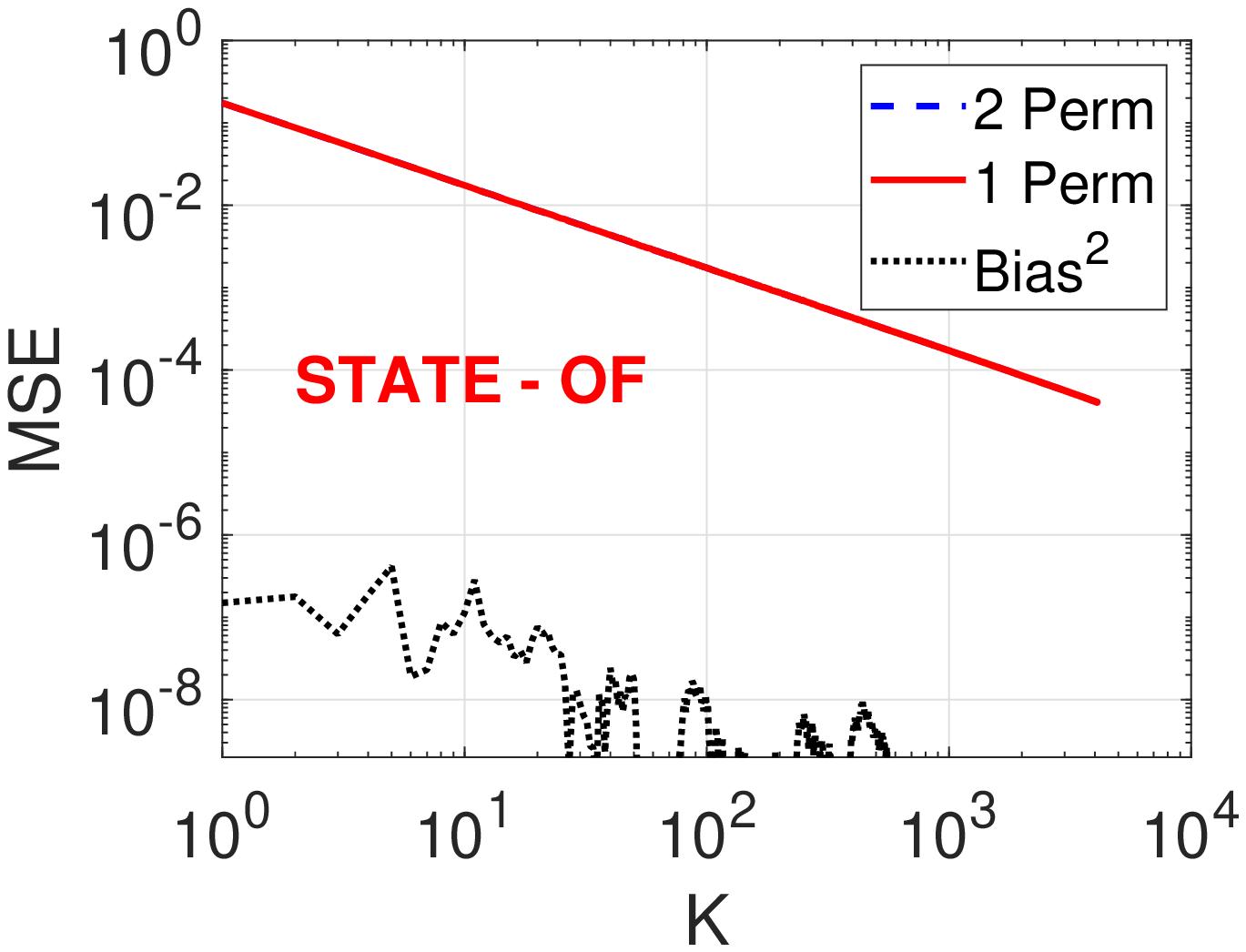}
    }
    \mbox{
    \includegraphics[width=2.1in]{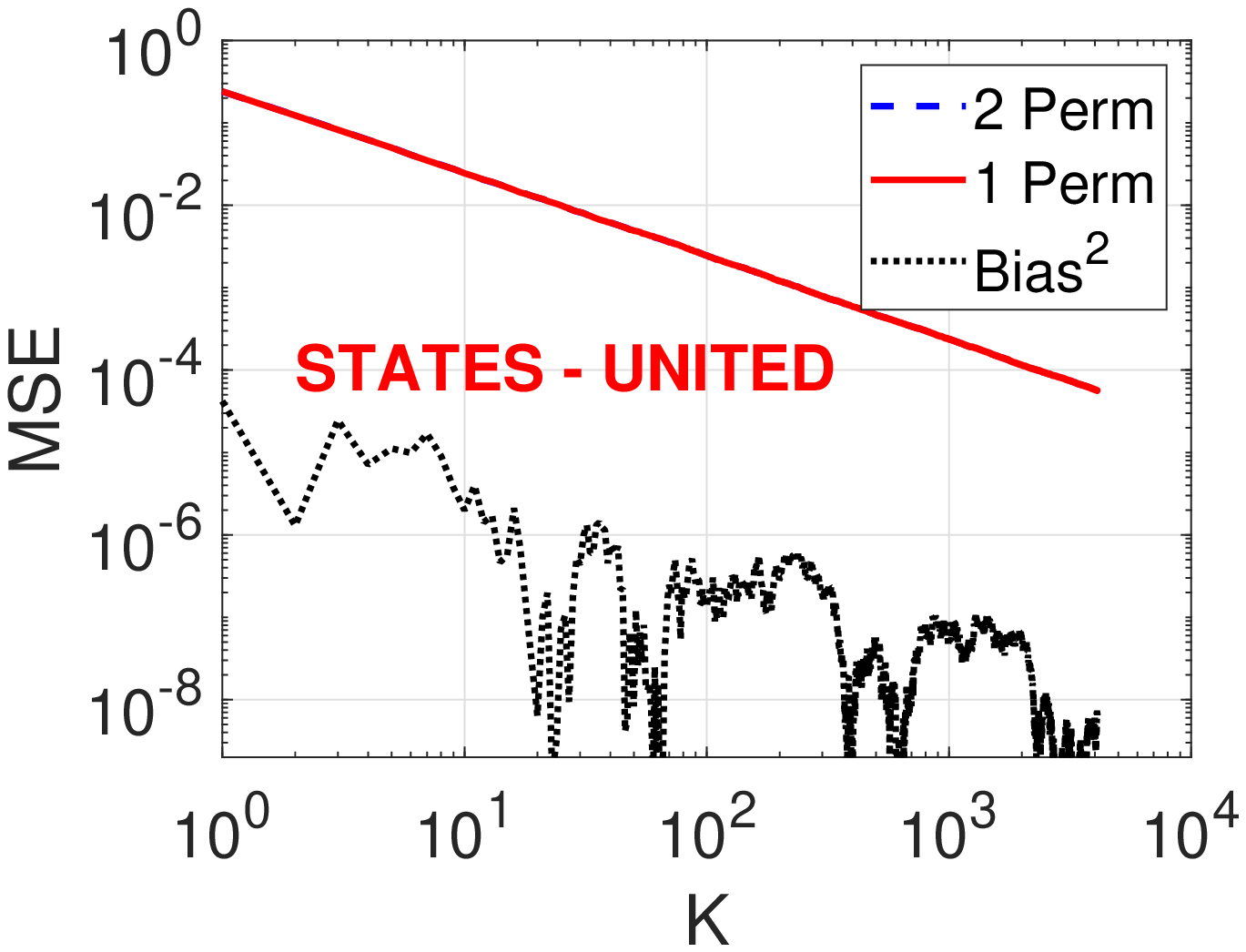}
    \includegraphics[width=2.1in]{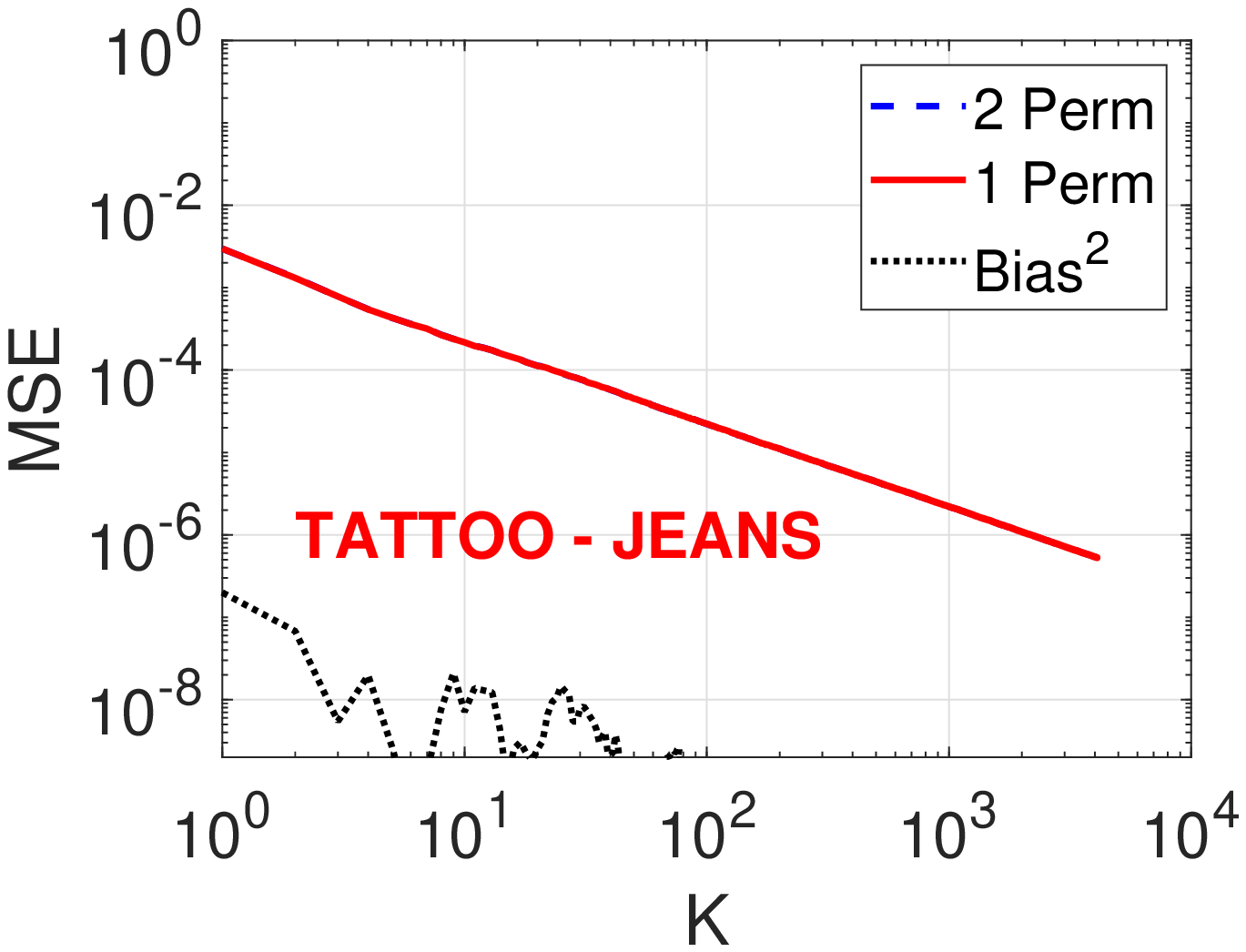}
    \includegraphics[width=2.1in]{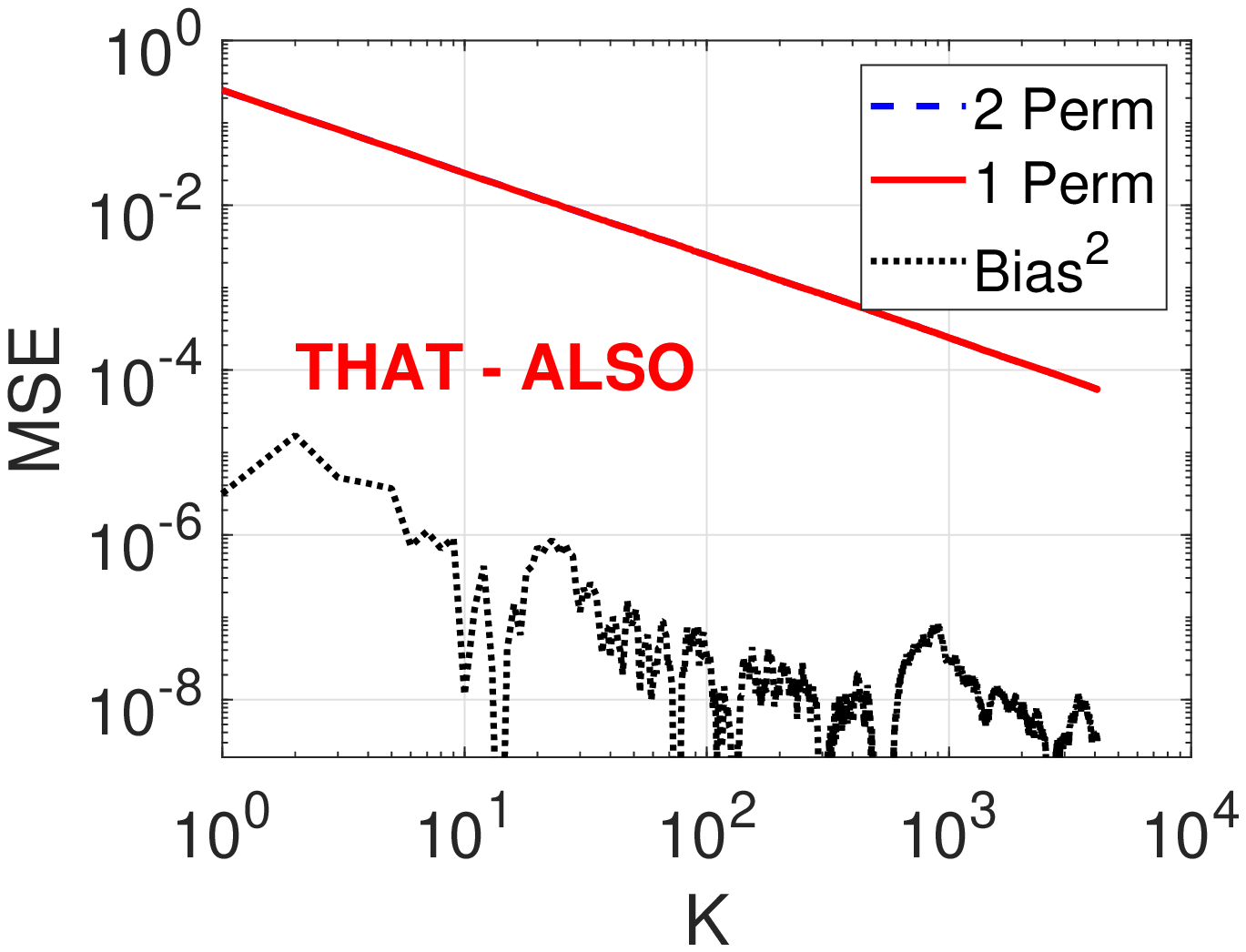}
    }
    \mbox{
    \includegraphics[width=2.1in]{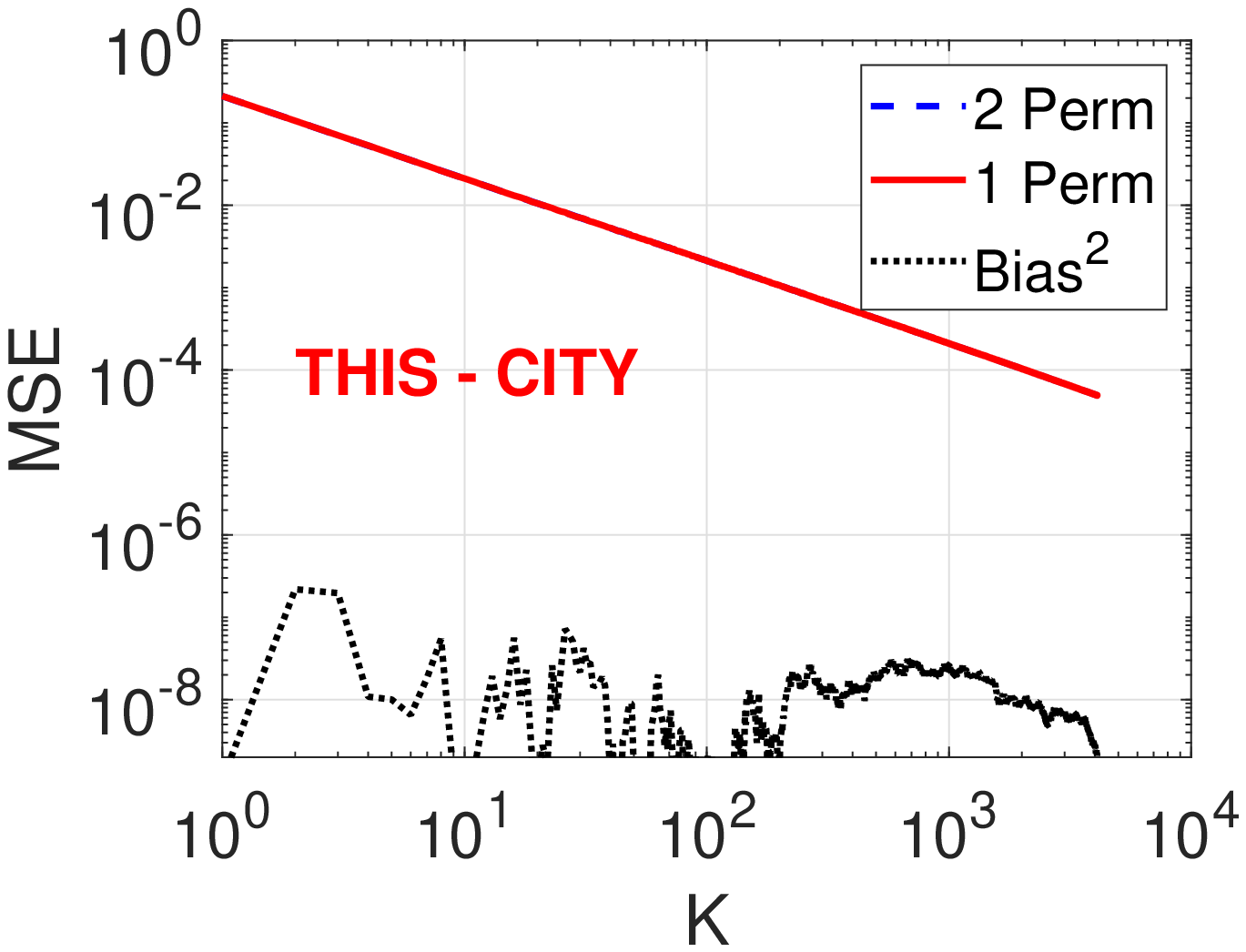}
    \includegraphics[width=2.1in]{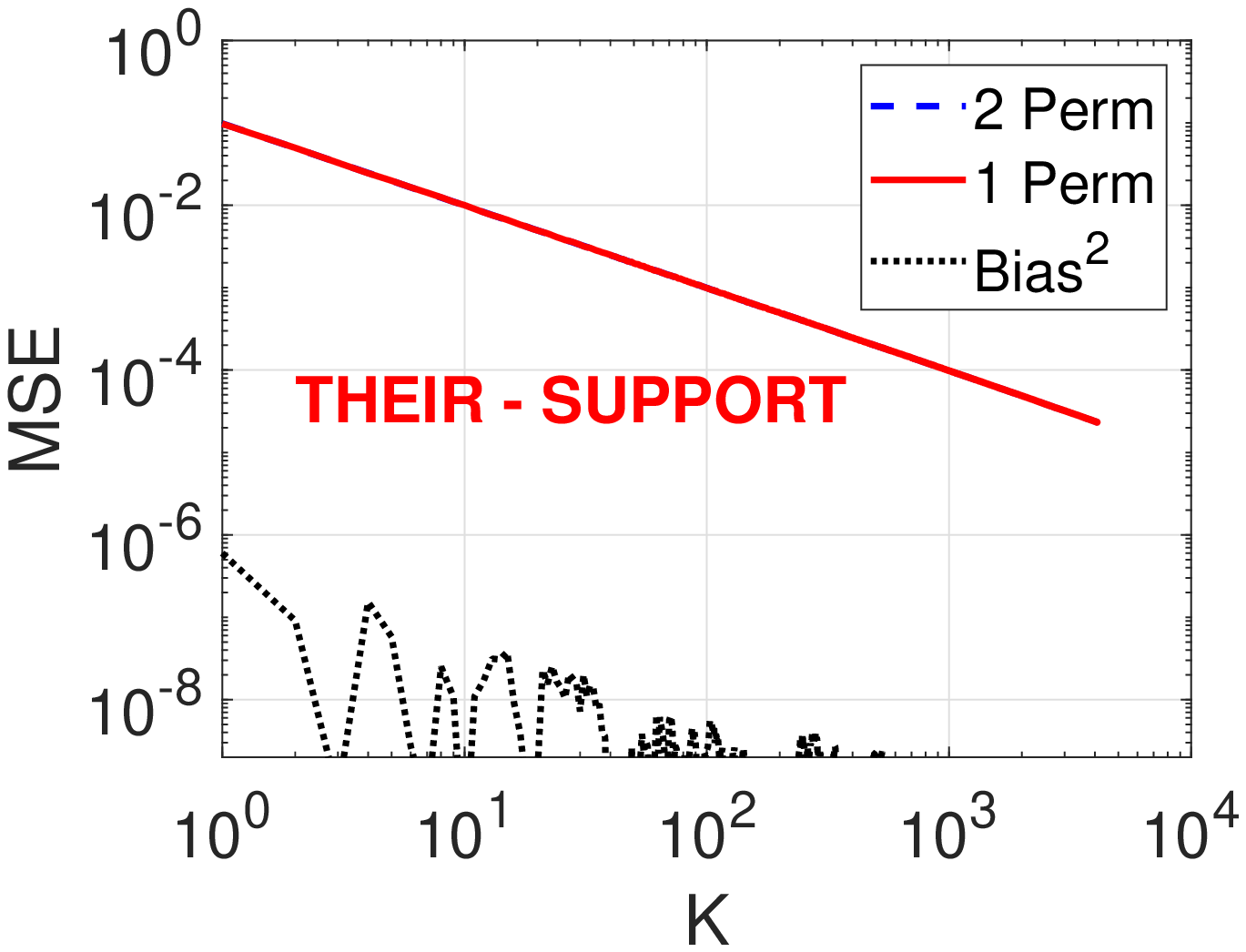}
    \includegraphics[width=2.1in]{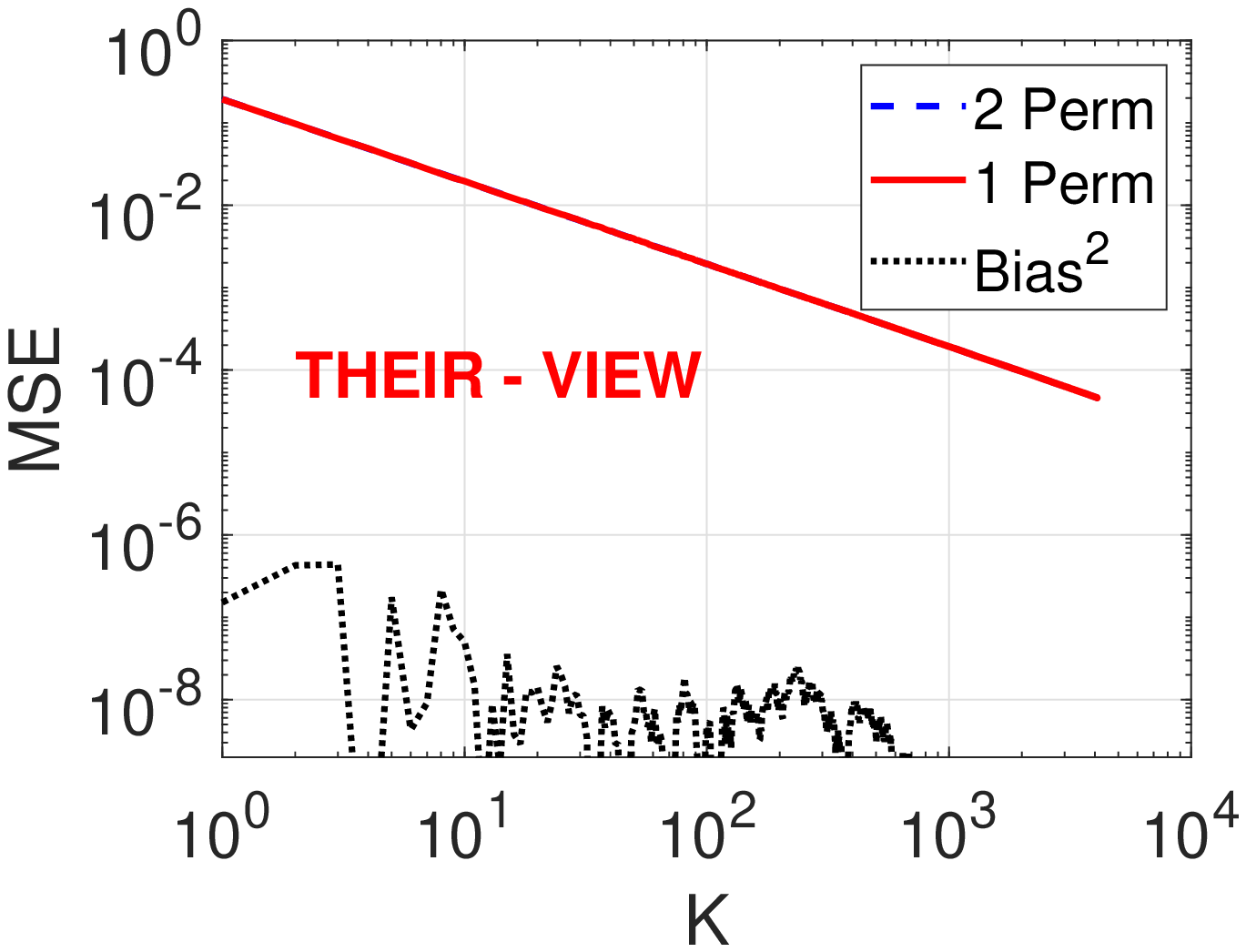}
    }

  \end{center}
  \vspace{-0.1in}
  \caption{Empirical MSEs of C-MinHash-$(\pi,\pi)$ (``1 Perm'', red, solid) vs. C-MinHash-$(\sigma,\pi)$ (``2 Perm'', blue, dashed) on various data pairs from the \textit{Words} dataset. We also report the empirical bias$^2$ for C-MinHash-$(\pi,\pi)$ to show that the bias is so small that it can be safely neglected. The empirical MSE curves for both estimators essentially overlap for all data pairs, for $K$ ranging from 1 to 4096. }
  \label{fig:word7}
\end{figure}

\begin{figure}[H]
  \begin{center}
   \mbox{
    \includegraphics[width=2.1in]{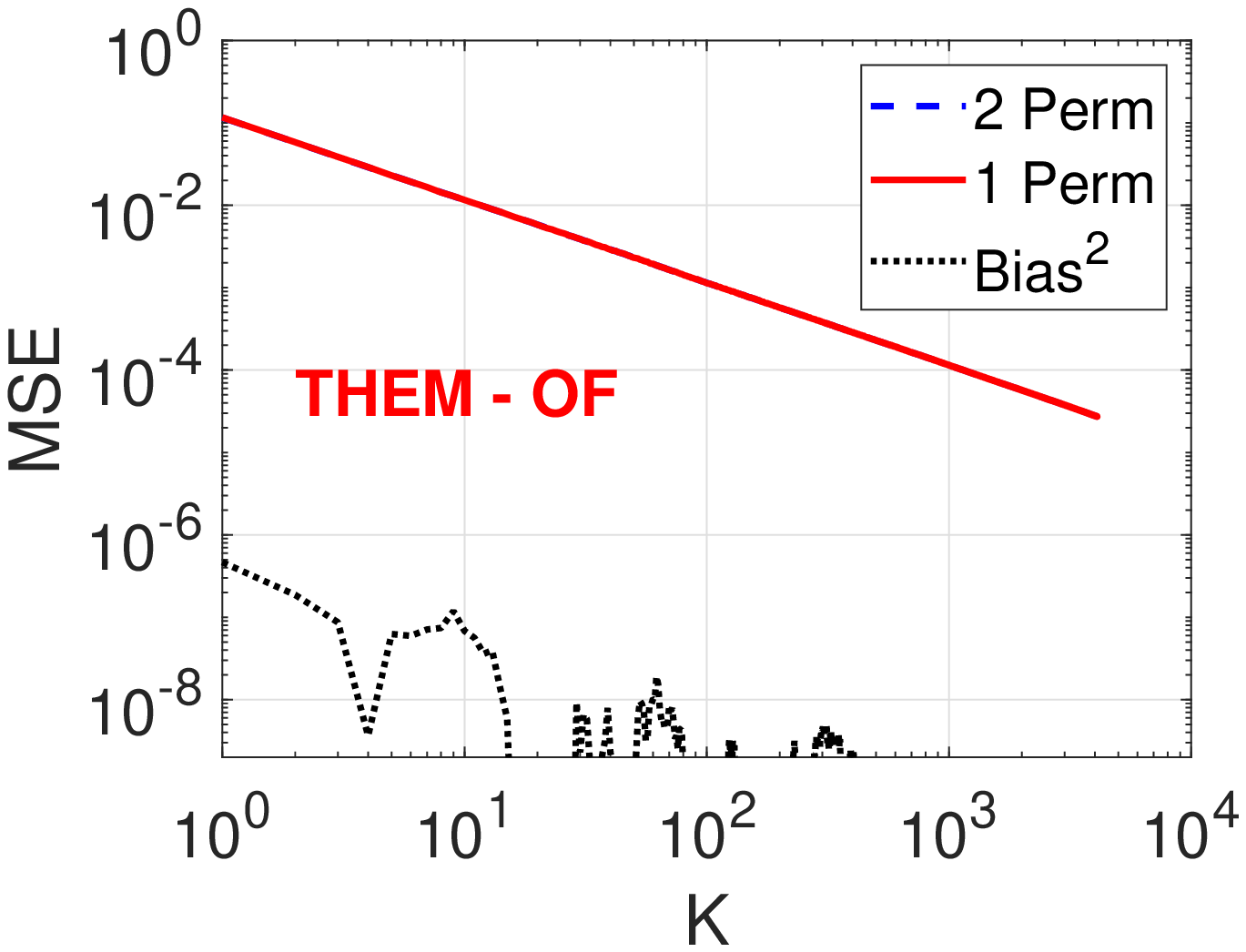}
    \includegraphics[width=2.1in]{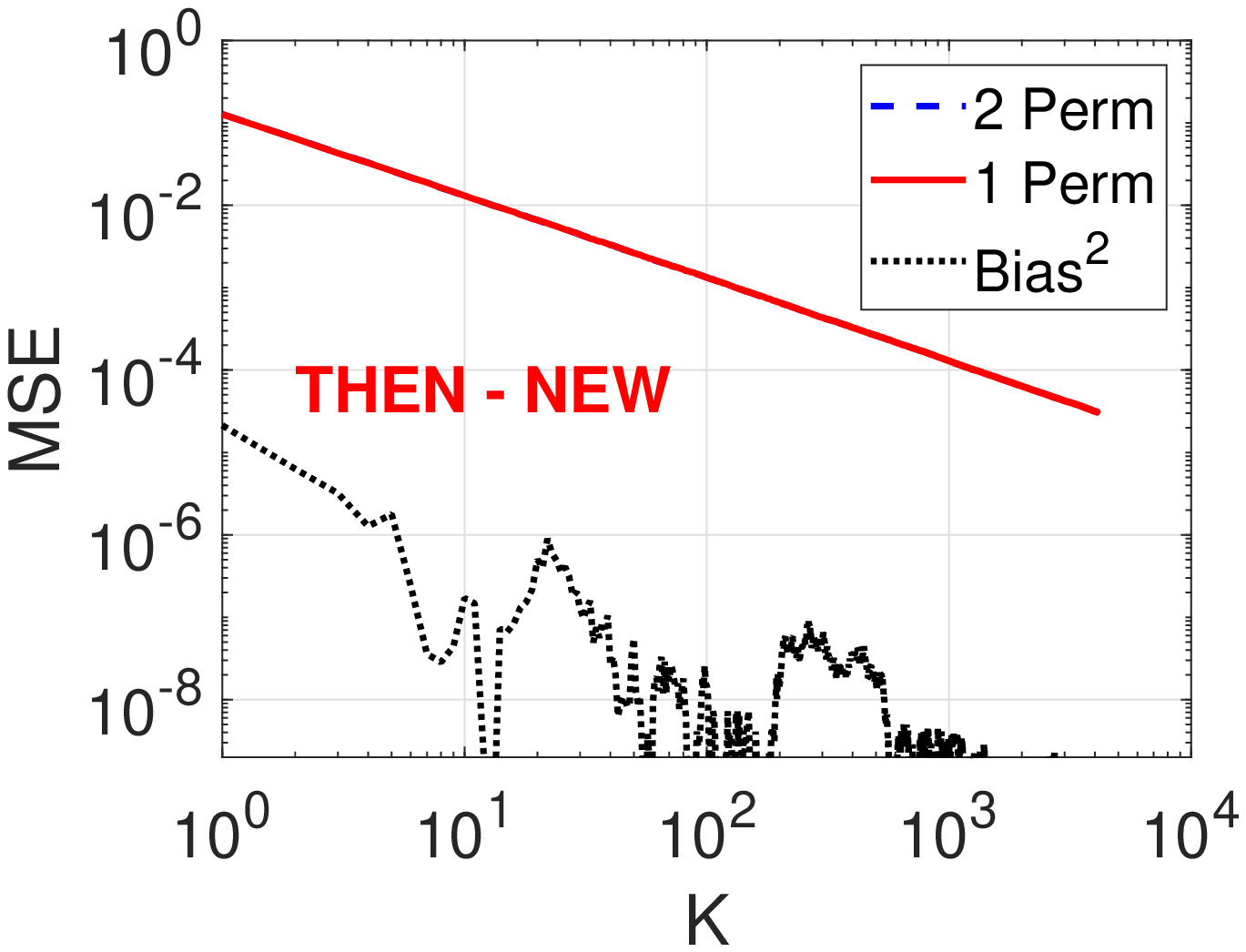}
    \includegraphics[width=2.1in]{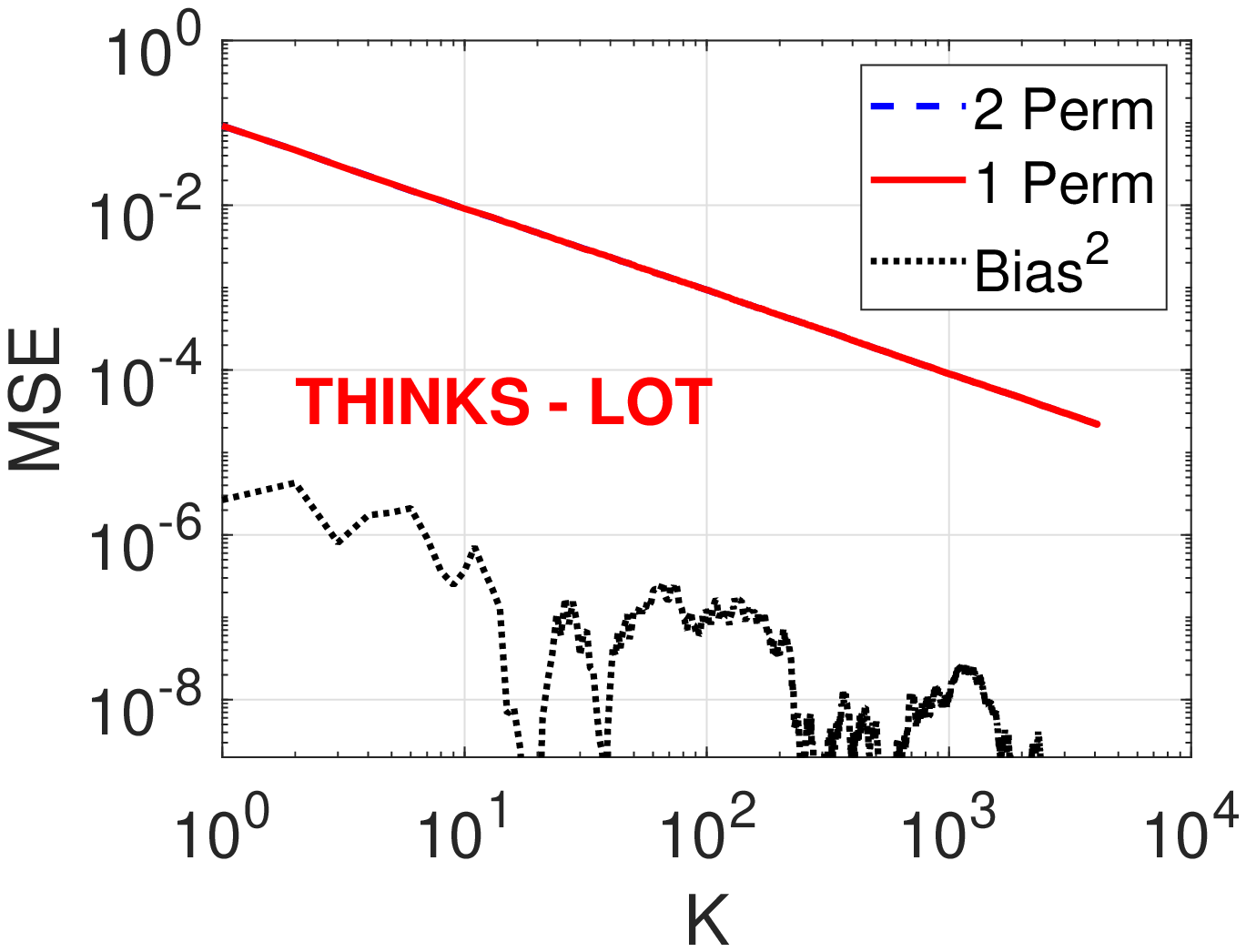}
    }
    \mbox{
    \includegraphics[width=2.1in]{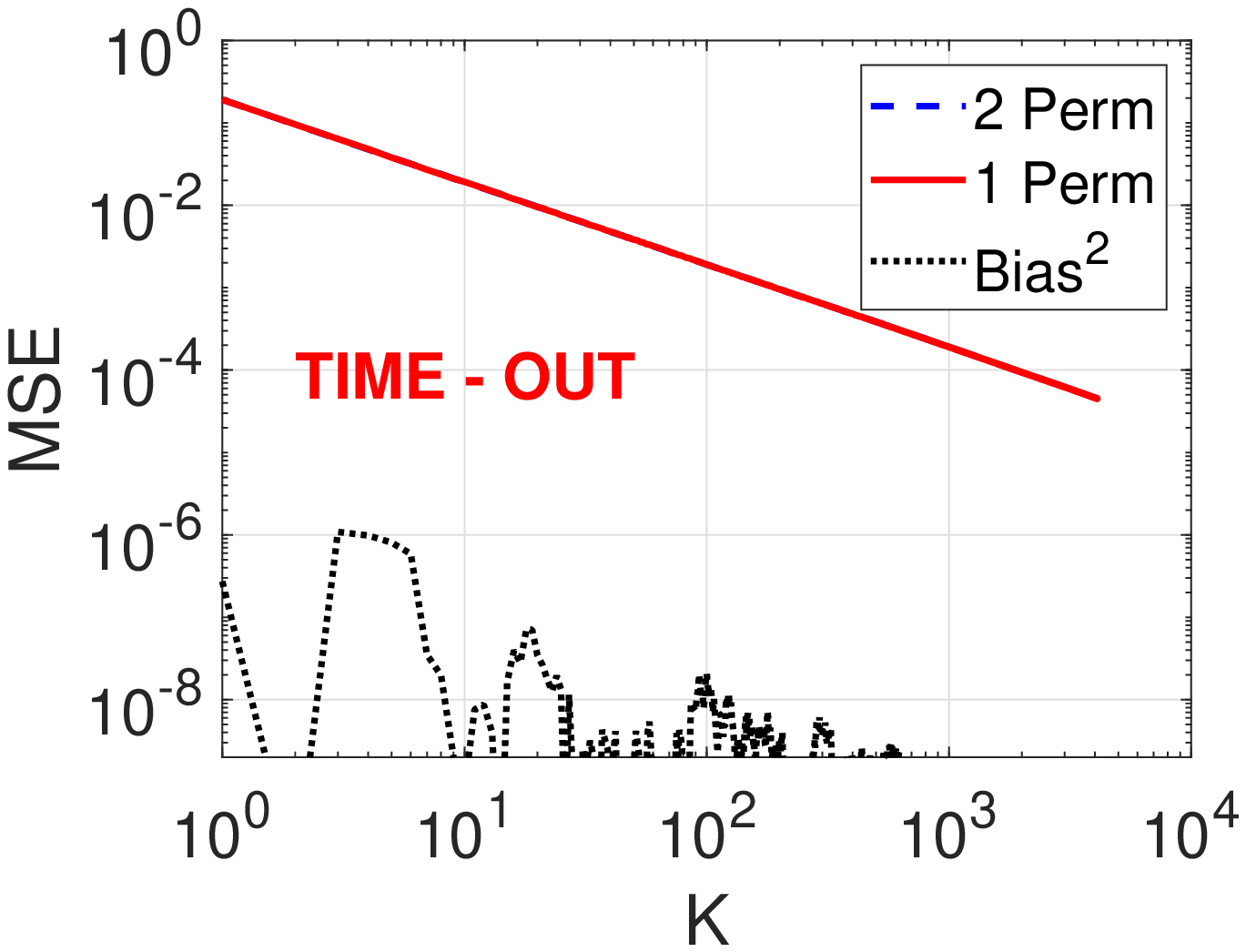}
    \includegraphics[width=2.1in]{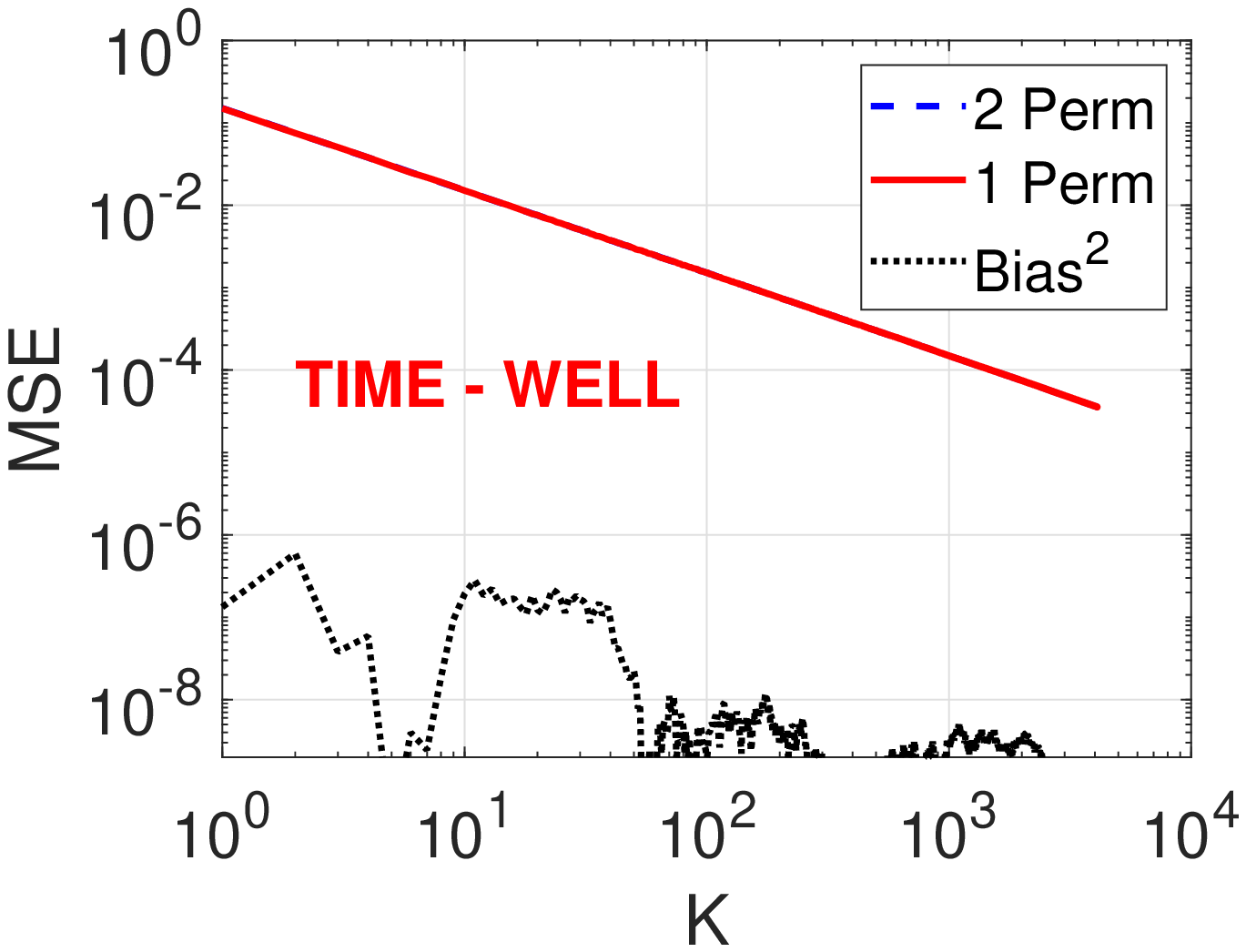}
    \includegraphics[width=2.1in]{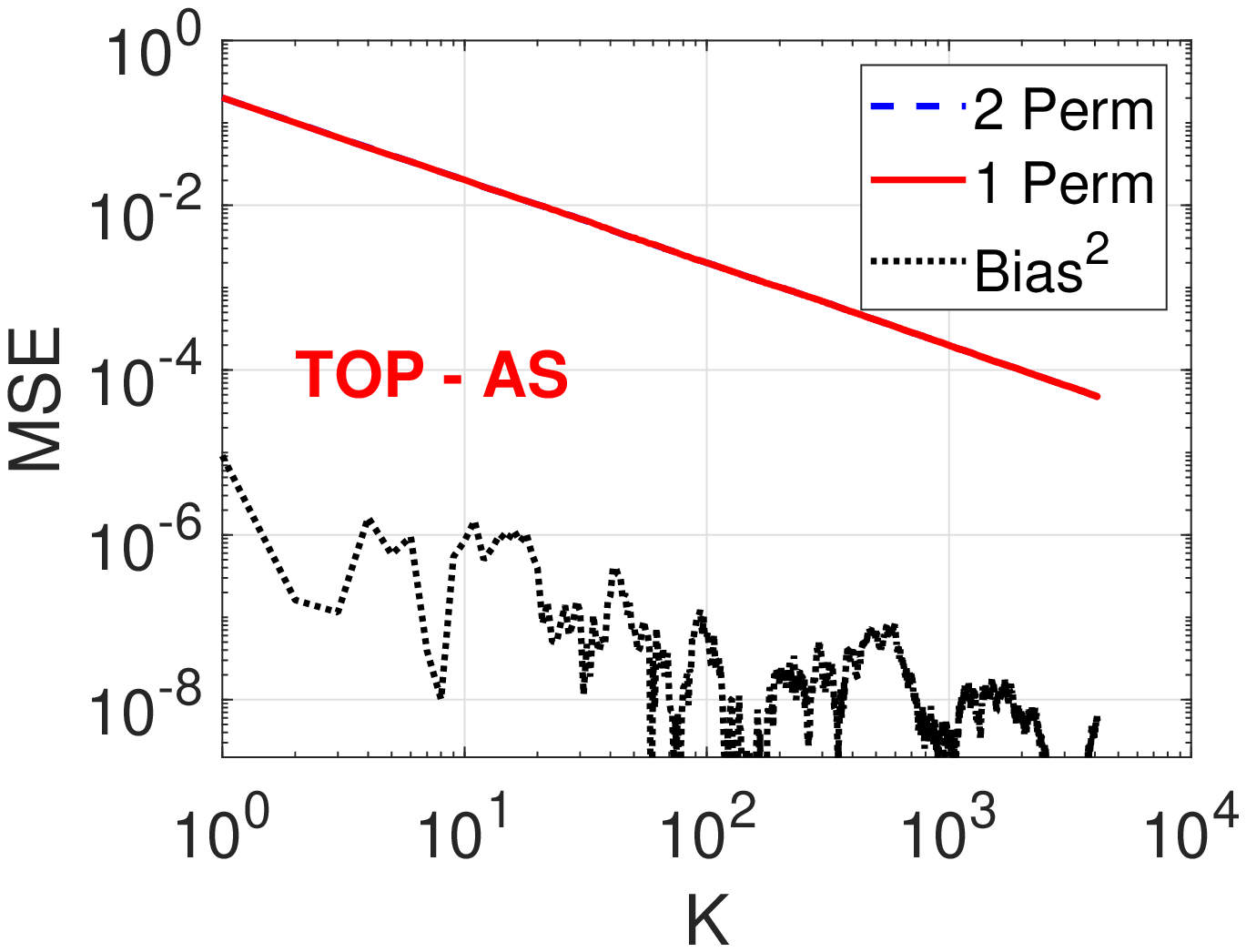}
    }
    \mbox{
    \includegraphics[width=2.1in]{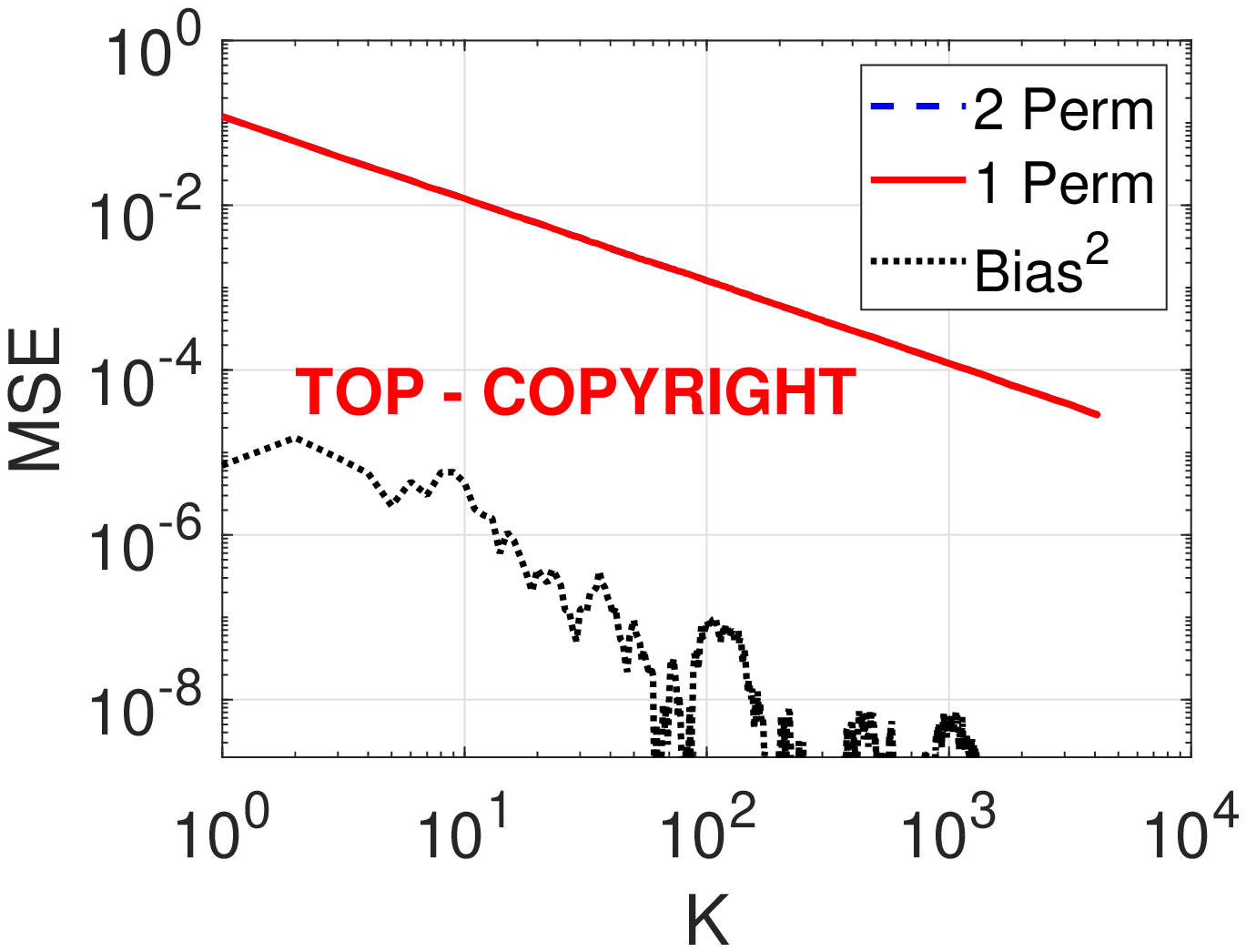}
    \includegraphics[width=2.1in]{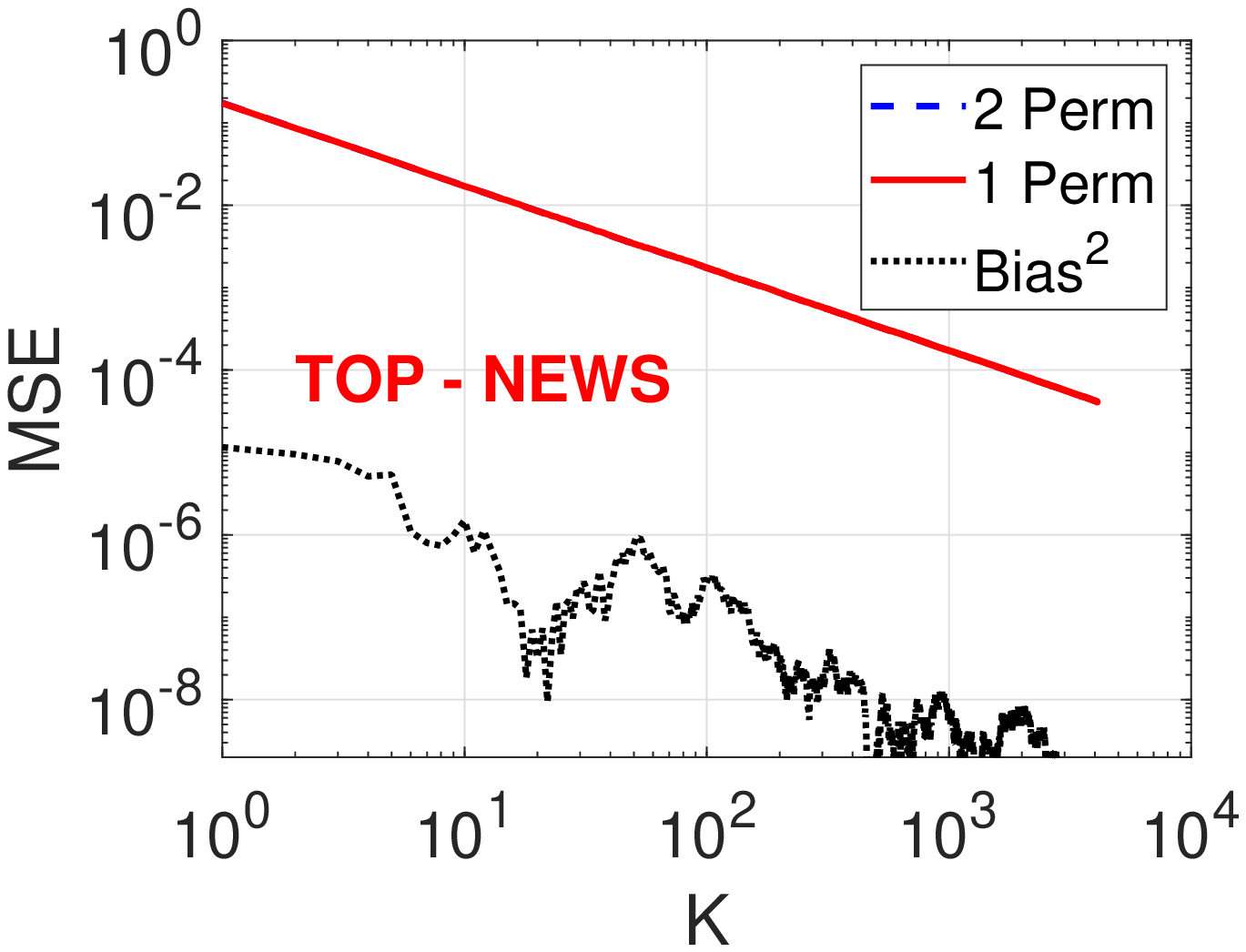}
    \includegraphics[width=2.1in]{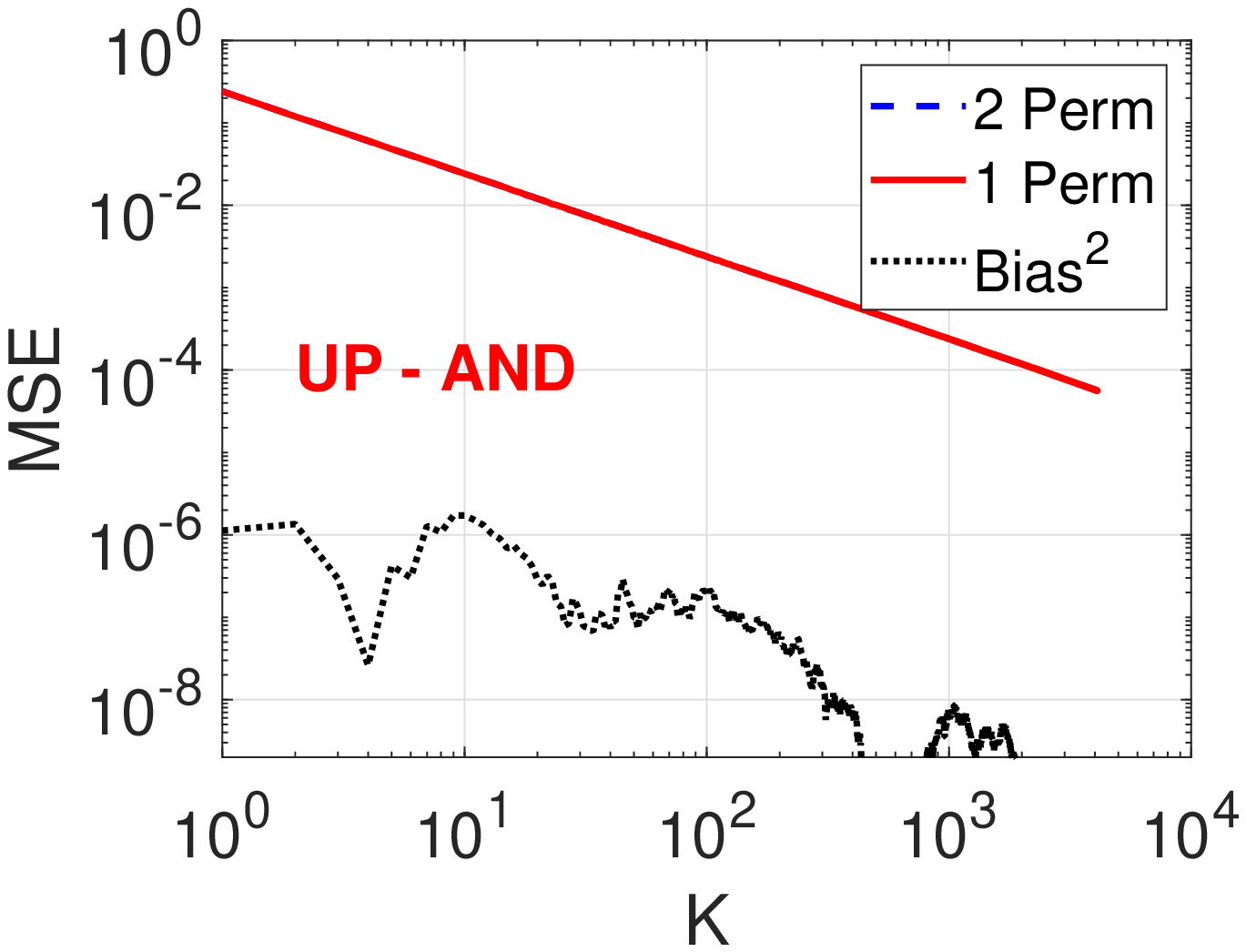}
    }
    \mbox{
    \includegraphics[width=2.1in]{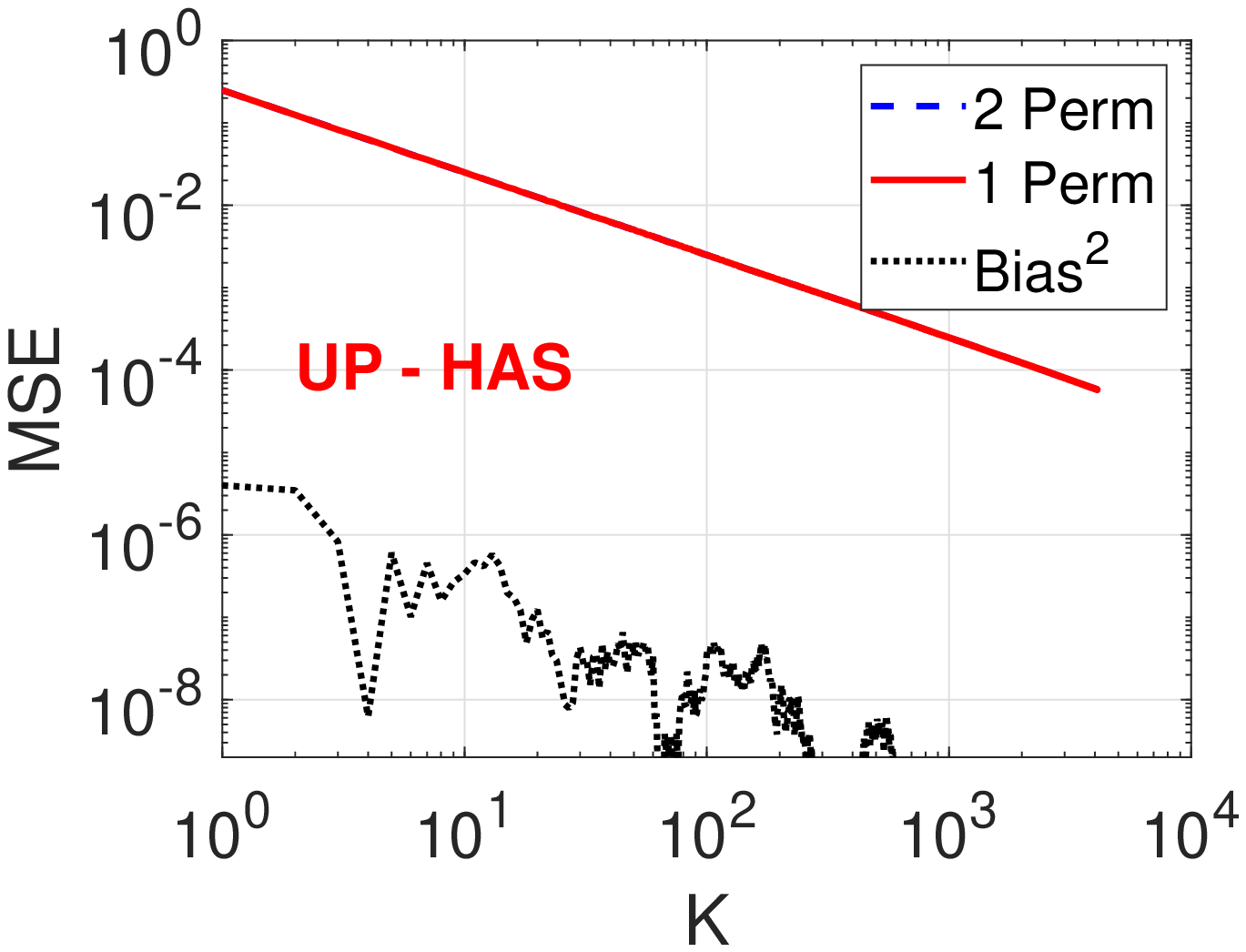}
    \includegraphics[width=2.1in]{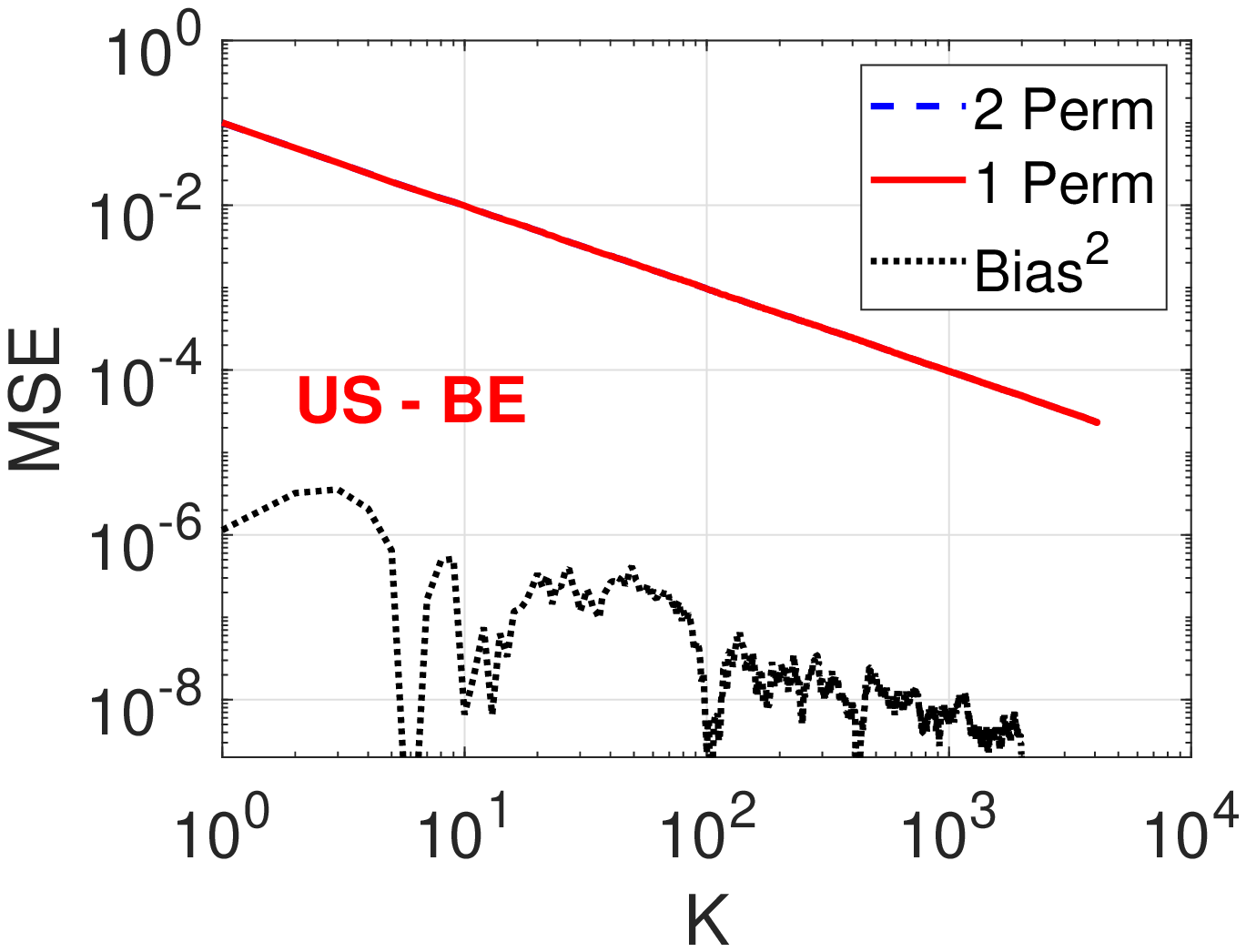}
    \includegraphics[width=2.1in]{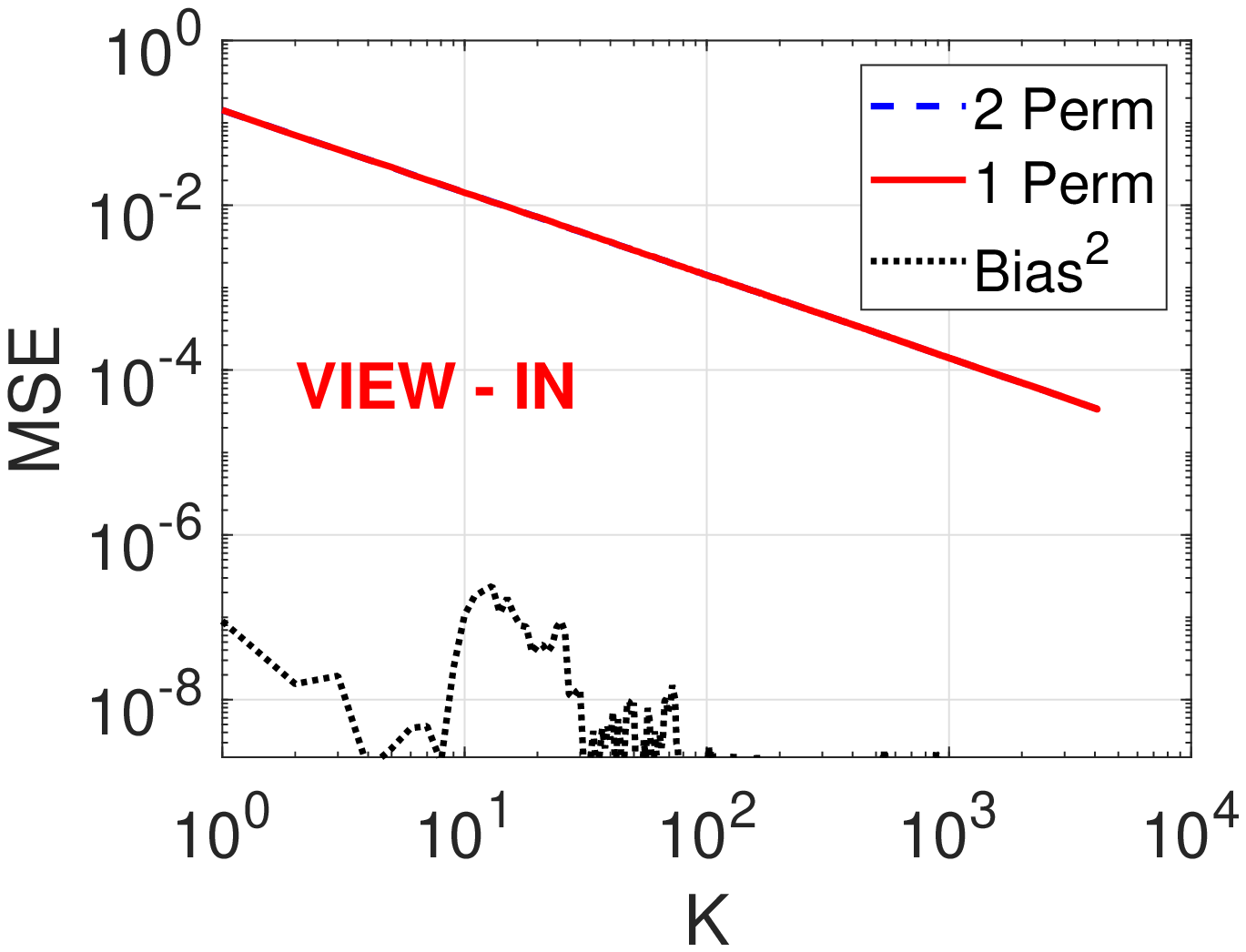}
    }
    \mbox{
    \includegraphics[width=2.1in]{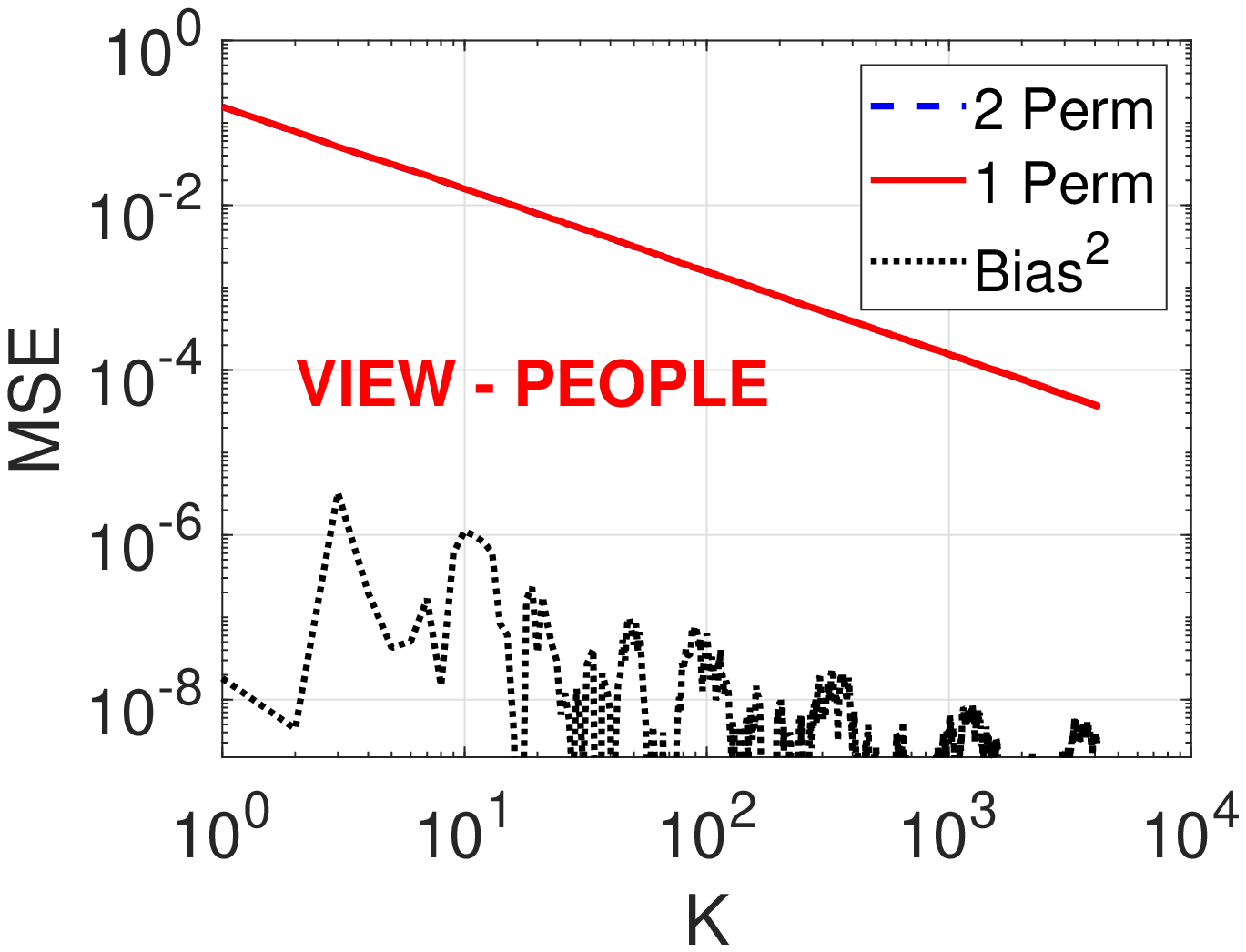}
    \includegraphics[width=2.1in]{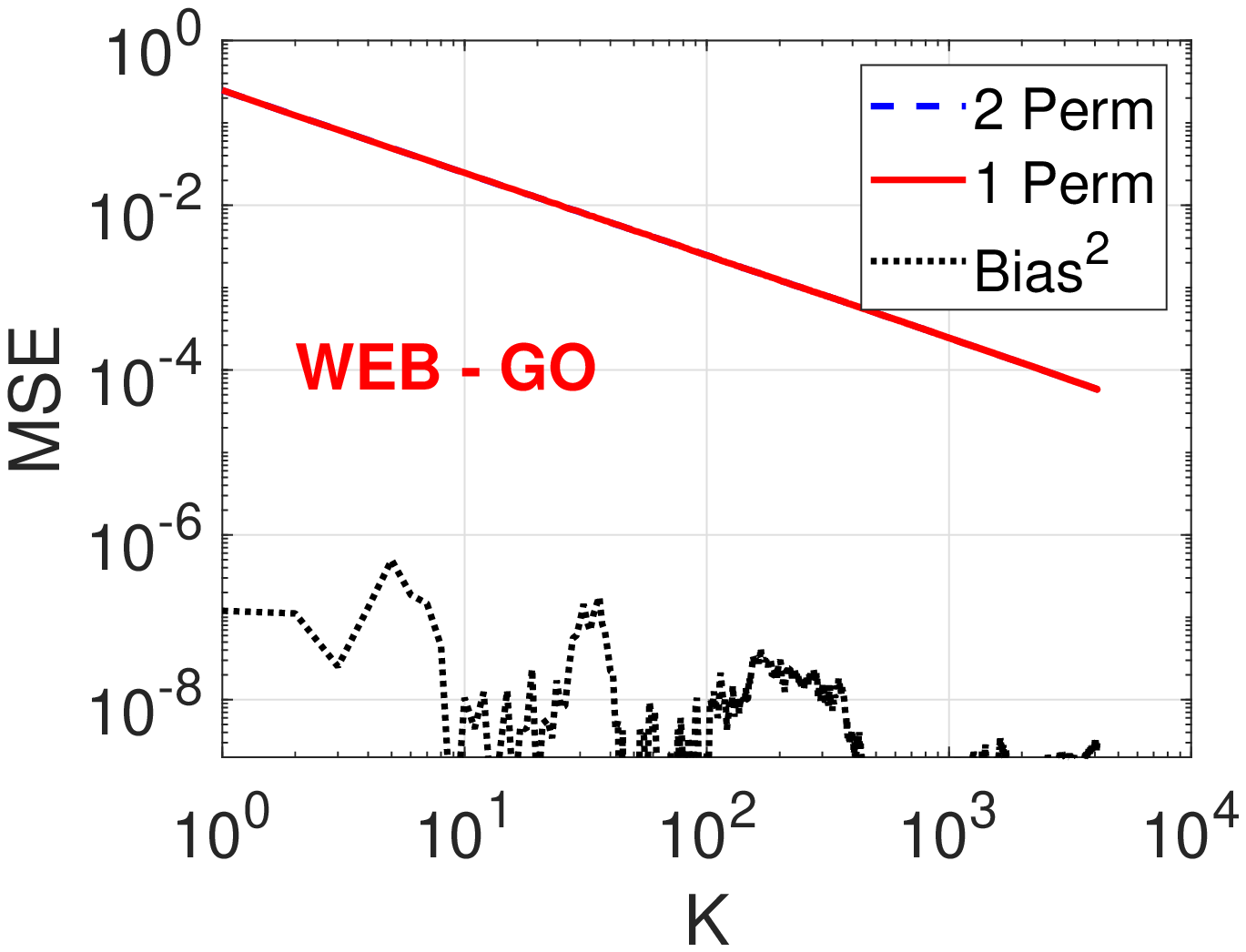}
    \includegraphics[width=2.1in]{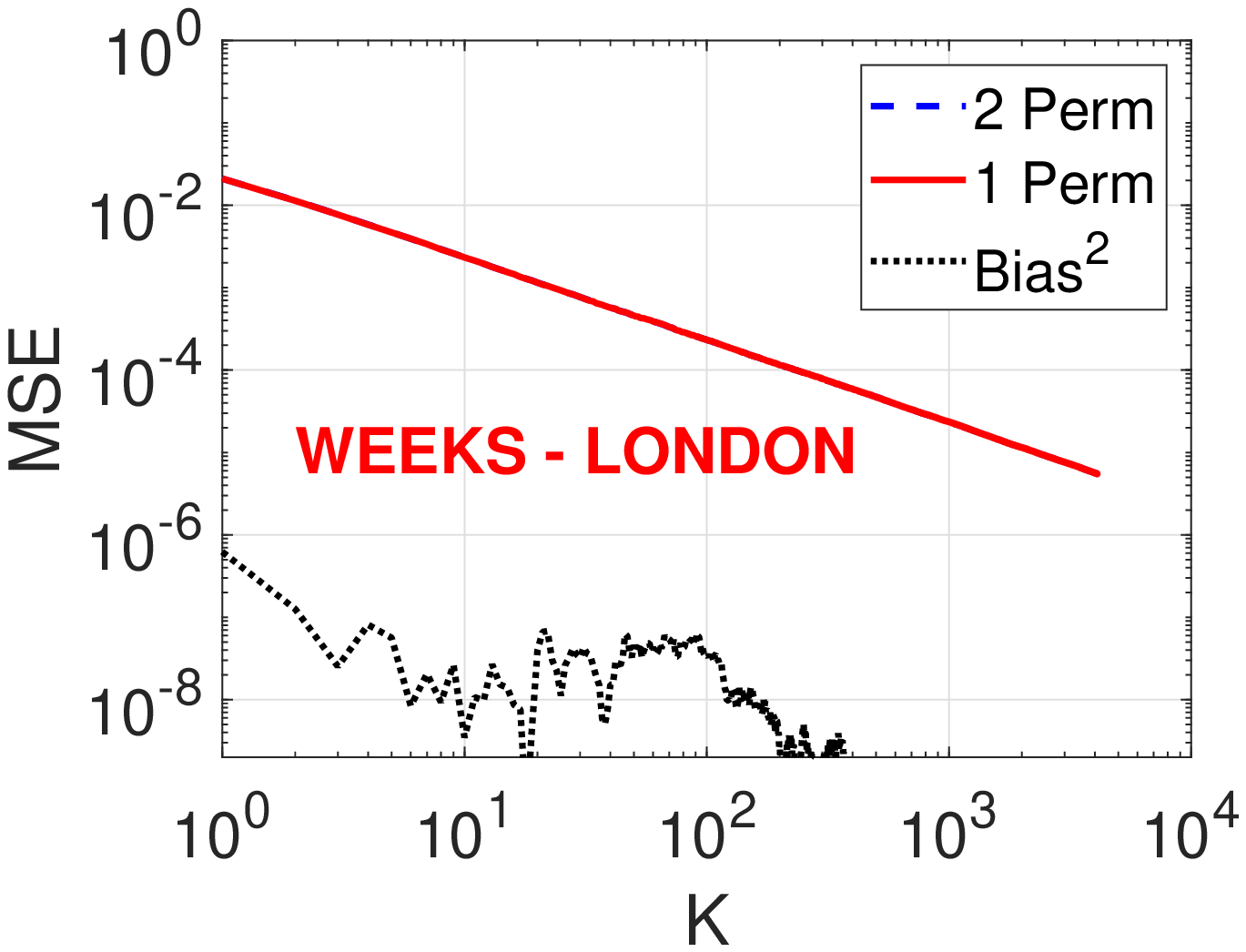}
    }

  \end{center}
  \vspace{-0.1in}
  \caption{Empirical MSEs of C-MinHash-$(\pi,\pi)$ (``1 Perm'', red, solid) vs. C-MinHash-$(\sigma,\pi)$ (``2 Perm'', blue, dashed) on various data pairs from the \textit{Words} dataset. We also report the empirical bias$^2$ for C-MinHash-$(\pi,\pi)$ to show that the bias is so small that it can be safely neglected. The empirical MSE curves for both estimators essentially overlap for all data pairs, for $K$ ranging from 1 to 4096. }
  \label{fig:word8}
\end{figure}

In Figures~\ref{fig:word1} - \ref{fig:word8}, we plot the empirical MSE along with the empirical bias$^2$ for $\hat J_{\pi,\pi}$, as well as the empirical MSE for $\hat J_{\sigma,\pi}$. Note that for $D$ this large, it is numerically difficult to evaluate the theoretical variance formulas in~\cite{CMH2Perm2021}. From the results in the Figures, we can observe
\begin{itemize}
    \item For all the data pairs, the MSE of C-MinHash-$(\pi,\pi)$ estimator overlaps with the empirical MSE of C-MinHash-$(\sigma,\pi)$ estimator for all $K$ from 1 up to 4096.

    \item The bias$^2$ is several orders of magnitudes  smaller than the MSE, in all data pairs. This verifies that the bias of $\hat J_{\pi,\pi}$ is extremely small in practice and can be safely neglected.
\end{itemize}

We have many more plots on more data pairs. Nevertheless, we believe the current set of experiments on this ``Words'' dataset should be sufficient to verify that, the proposed C-MinHash-$(\pi,\pi)$ could give indistinguishable Jaccard estimation accuracy in practice compared~with~C-MinHash-$(\sigma,\pi)$.

\newpage

\begin{table}[H]
\fontsize{8}{9.7}\selectfont
\caption{120 selected word pairs from the \textit{Words} dataset~\citep{Proc:Li_Church_EMNLP05}. For each pair, we report the density $\tilde d$ (number of non-zero entries divided by $D=2^{16}$) for each word as well as the Jaccard similarity~$J$.\vspace{0.1in} }
\label{tab:word pairs}

\hspace{-0.15in}
\begin{tabular}{l|ccc??l|ccc}
\toprule \hline
& \bm{$\tilde d_1$} & \bm{$\tilde d_2$} & \bm{$J$}    &                       & \bm{$\tilde d_1$} & \bm{$\tilde d_2$} & \bm{$J$  }  \\ \hline
ABOUT - INTO              & 0.302 & 0.125 & 0.258  & NEW - WEB               & 0.291 & 0.194 & 0.224  \\
ABOUT - LIKE              & 0.302 & 0.140 & 0.281  & NEWS - LIKE             & 0.168 & 0.140 & 0.172  \\
ACTUAL - DEVELOPED        & 0.017 & 0.030 & 0.071  & NO - WELL               & 0.220 & 0.120 & 0.244  \\
ACTUAL - GRABBED          & 0.017 & 0.002 & 0.016  & NOT - IT                & 0.281 & 0.295 & 0.437  \\
AFTER - OR                & 0.103 & 0.356 & 0.220  & NOTORIOUSLY - LOCK      & 0.0006 & 0.006 & 0.004  \\
AND - PROBLEM             & 0.554 & 0.044 & 0.070  & OF - THEN               & 0.570 & 0.104 & 0.168  \\
AS - NAME                 & 0.280 & 0.144 & 0.204  & OF - WE                 & 0.570 & 0.226 & 0.361  \\
AT - CUT                  & 0.374 & 0.242 & 0.052  & OPPORTUNITY - COUNTRIES & 0.029 & 0.024 & 0.066  \\
BE - ONE                  & 0.323 & 0.221 & 0.403  & OUR - THAN              & 0.244 & 0.125 & 0.245  \\
BEST - AND                & 0.136 & 0.554 & 0.228  & OVER - BACK             & 0.148 & 0.160 & 0.233  \\
BRAZIL - OH               & 0.010 & 0.031 & 0.019  & OVER - TWO              & 0.148 & 0.121 & 0.289  \\
BUT - MANY                & 0.167 & 0.116 & 0.340  & PEAK - SHOWS            & 0.006 & 0.033 & 0.026  \\
CALLED - BUSINESSES       & 0.016 & 0.018 & 0.043  & PEOPLE - BY             & 0.121 & 0.425 & 0.228  \\
CALORIES - MICROSOFT      & 0.002 & 0.045 & 0.0003 & PEOPLE - INFO           & 0.121 & 0.138 & 0.117  \\
CAN - FROM                & 0.243 & 0.326 & 0.444  & PICKS - BOOST           & 0.007 & 0.005 & 0.007  \\
CAN - SEARCH              & 0.243 & 0.214 & 0.237  & PLANET - REWARD         & 0.013 & 0.003 & 0.018  \\
COMMITTED - PRODUCTIVE    & 0.013 & 0.004 & 0.029  & PLEASE - MAKE           & 0.168 & 0.141 & 0.195  \\
CONTEMPORARY - FLASH      & 0.011 & 0.021 & 0.013  & PREFER - PUEDE          & 0.010 & 0.003 & 0.0001 \\
CONVENIENTLY - INDUSTRIES & 0.003 & 0.011 & 0.009  & PRIVACY - FOUND         & 0.126 & 0.136 & 0.053  \\
COPYRIGHT - AN            & 0.218 & 0.290 & 0.209  & PROSECUTION - MAXIMIZE  & 0.002 & 0.003 & 0.006  \\
CREDIT - CARD             & 0.046 & 0.041 & 0.285  & RECENTLY - INT          & 0.028 & 0.007 & 0.014  \\
DE - WEB                  & 0.117 & 0.194 & 0.091  & REPLY - ACHIEVE         & 0.013 & 0.012 & 0.023  \\
DO - GOOD                 & 0.174 & 0.102 & 0.276  & RESERVED - BEEN         & 0.172 & 0.141 & 0.108  \\
EARTH - GROUPS            & 0.021 & 0.035 & 0.056  & RIGHTS - FIND           & 0.187 & 0.144 & 0.166  \\
EXPRESSED - FRUSTRATED    & 0.010 & 0.002 & 0.024  & RIGHTS - RESERVED       & 0.187 & 0.172 & 0.877  \\
FIND - HAS                & 0.144 & 0.228 & 0.214  & SCENE - ABOUT           & 0.012 & 0.301 & 0.029  \\
FIND - SITE               & 0.144 & 0.275 & 0.212  & SEE - ALSO              & 0.138 & 0.166 & 0.291  \\
FIXED - SPECIFIC          & 0.011 & 0.039 & 0.054  & SEIZE - ANYTHING        & 0.0007 & 0.037 & 0.012  \\
FLIGHT - TRANSPORTATION   & 0.011 & 0.018 & 0.040  & SHOULDERS - GORGEOUS    & 0.003 & 0.004 & 0.028  \\
FOUND - DE                & 0.136 & 0.117 & 0.039  & SICK - FELL             & 0.008 & 0.008 & 0.085  \\
FRANCISCO - SAN           & 0.025 & 0.049 & 0.476  & SITE - CELLULAR         & 0.275 & 0.006 & 0.010  \\
GOOD - BACK               & 0.102 & 0.160 & 0.220  & SOLD - LIVE             & 0.018 & 0.064 & 0.055  \\
GROUPS - ORDERED          & 0.035 & 0.011 & 0.034  & SOLO - CLAIMS           & 0.010 & 0.012 & 0.007  \\
HAPPY - CONCEPT           & 0.029 & 0.013 & 0.054  & SOON - ADVANCE          & 0.040 & 0.017 & 0.057  \\
HAVE - FIRST              & 0.267 & 0.151 & 0.320  & SPECIALIZES - ACTUAL    & 0.003 & 0.017 & 0.008  \\
HAVE - US                 & 0.267 & 0.284 & 0.349  & STATE - OF             & 0.101 & 0.570 & 0.165  \\
HILL - ASSURED            & 0.020 & 0.004 & 0.011  & STATES - UNITED         & 0.061 & 0.062 & 0.591  \\
HOME - SYNTHESIS          & 0.365 & 0.002 & 0.003  & TATTOO - JEANS          & 0.002 & 0.004 & 0.035  \\
HONG - KONG               & 0.014 & 0.014 & 0.925  & THAT - ALSO             & 0.301 & 0.166 & 0.376  \\
HOSTED - DRUGS            & 0.016 & 0.013 & 0.013  & THIS - CITY              & 0.423 & 0.123 & 0.132  \\
INTERVIEWS - FOURTH       & 0.012 & 0.011 & 0.031  & THEIR - SUPPORT         & 0.165 & 0.117 & 0.189  \\
KANSAS - PROPERTY         & 0.017 & 0.045 & 0.052  & THEIR - VIEW            & 0.165 & 0.103 & 0.151  \\
KIRIBATI - GAMBIA         & 0.003 & 0.003 & 0.712  & THEM - OF               & 0.112 & 0.570 & 0.187  \\
LAST - THIS               & 0.135 & 0.423 & 0.221  & THEN - NEW              & 0.104 & 0.291 & 0.192  \\
LEAST - ROMANCE           & 0.046 & 0.007 & 0.019  & THINKS - LOT            & 0.007 & 0.040 & 0.079  \\
LIME - REGISTERED         & 0.002 & 0.030 & 0.004  & TIME - OUT              & 0.189 & 0.191 & 0.366  \\
LINKS - TAKE              & 0.191 & 0.105 & 0.134  & TIME - WELL             & 0.189 & 0.120 & 0.299  \\
LINKS - THAN              & 0.191 & 0.125 & 0.141  & TOP - AS                & 0.140 & 0.280 & 0.217  \\
MAIL - AND                & 0.160 & 0.554 & 0.192  & TOP - COPYRIGHT         & 0.140 & 0.218 & 0.149  \\
MAIL - BACK               & 0.160 & 0.160 & 0.132  & TOP - NEWS              & 0.140 & 0.168 & 0.192  \\
MAKE - LIKE               & 0.141 & 0.140 & 0.297  & UP - AND                & 0.200 & 0.554 & 0.334  \\
MANAGING - LOCK           & 0.010 & 0.006 & 0.010  & UP - HAS                & 0.200 & 0.228 & 0.312  \\
MANY - US                 & 0.116 & 0.284 & 0.210  & US - BE                 & 0.284 & 0.323 & 0.335  \\
MASS - DREAM              & 0.016 & 0.017 & 0.048  & VIEW - IN               & 0.103 & 0.540 & 0.153  \\
MAY - HELP                & 0.184 & 0.156 & 0.206  & VIEW - PEOPLE           & 0.103 & 0.121 & 0.138  \\
MOST - HOME               & 0.141 & 0.365 & 0.207  & WALKED - ANTIVIRUS      & 0.006 & 0.002 & 0.002  \\
NAME - IN                 & 0.144 & 0.540 & 0.207  & WEB - GO                & 0.194 & 0.111 & 0.138  \\
NEITHER - FIGURE          & 0.011 & 0.016 & 0.085  & WELL - INFO             & 0.120 & 0.138 & 0.110  \\
NET - SO                  & 0.101 & 0.154 & 0.112  & WELL - NEWS             & 0.120 & 0.168 & 0.161  \\
NEW - PLEASE              & 0.291 & 0.168 & 0.205  & WEEKS - LONDON          & 0.028 & 0.032 & 0.050 \\
\hline \bottomrule
\end{tabular}
\end{table}

\begin{figure}[H]
  \begin{center}
   \mbox{\hspace{-0.1in}
    \includegraphics[width=2.1in]{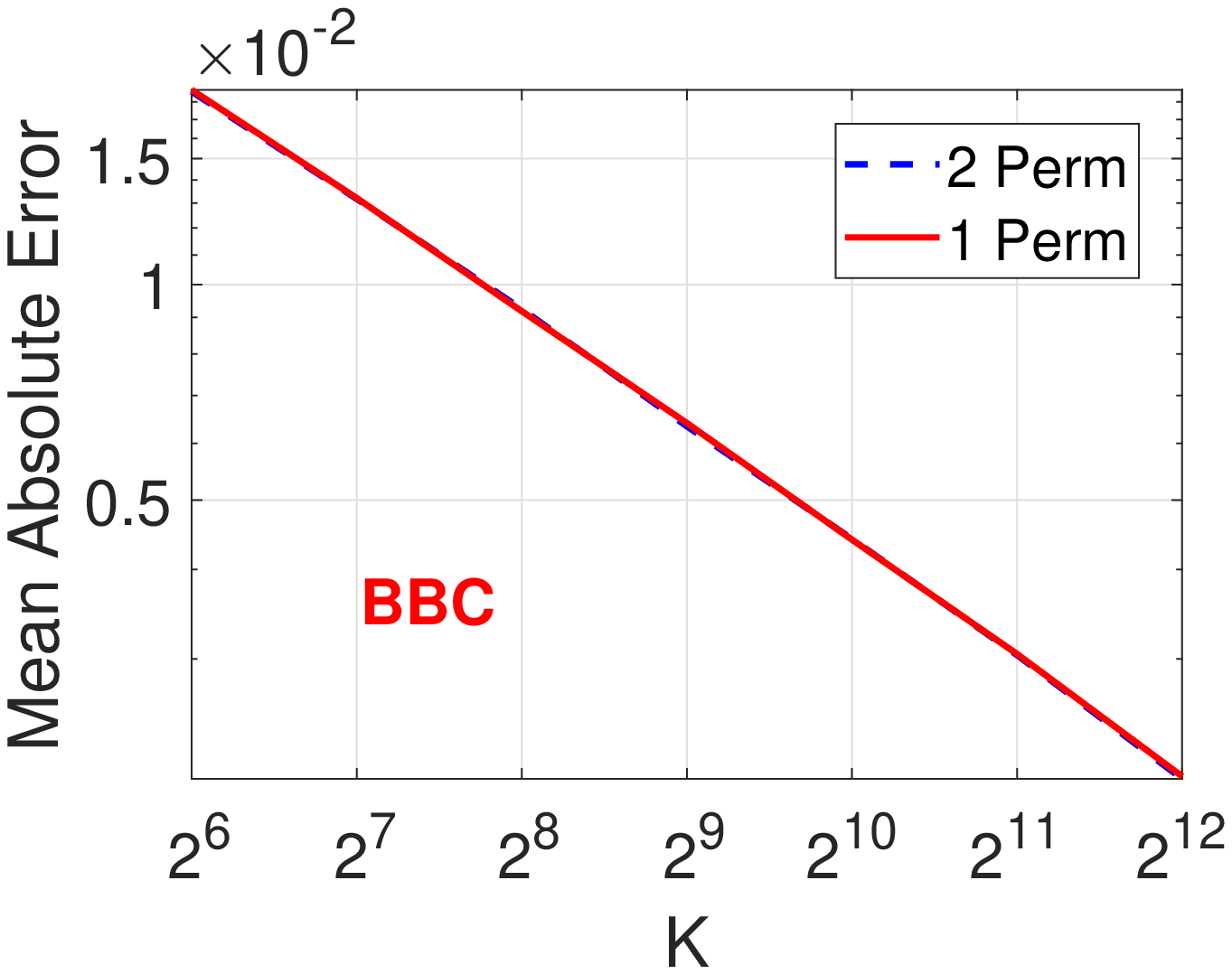}\hspace{-0.1in}
    \includegraphics[width=2.1in]{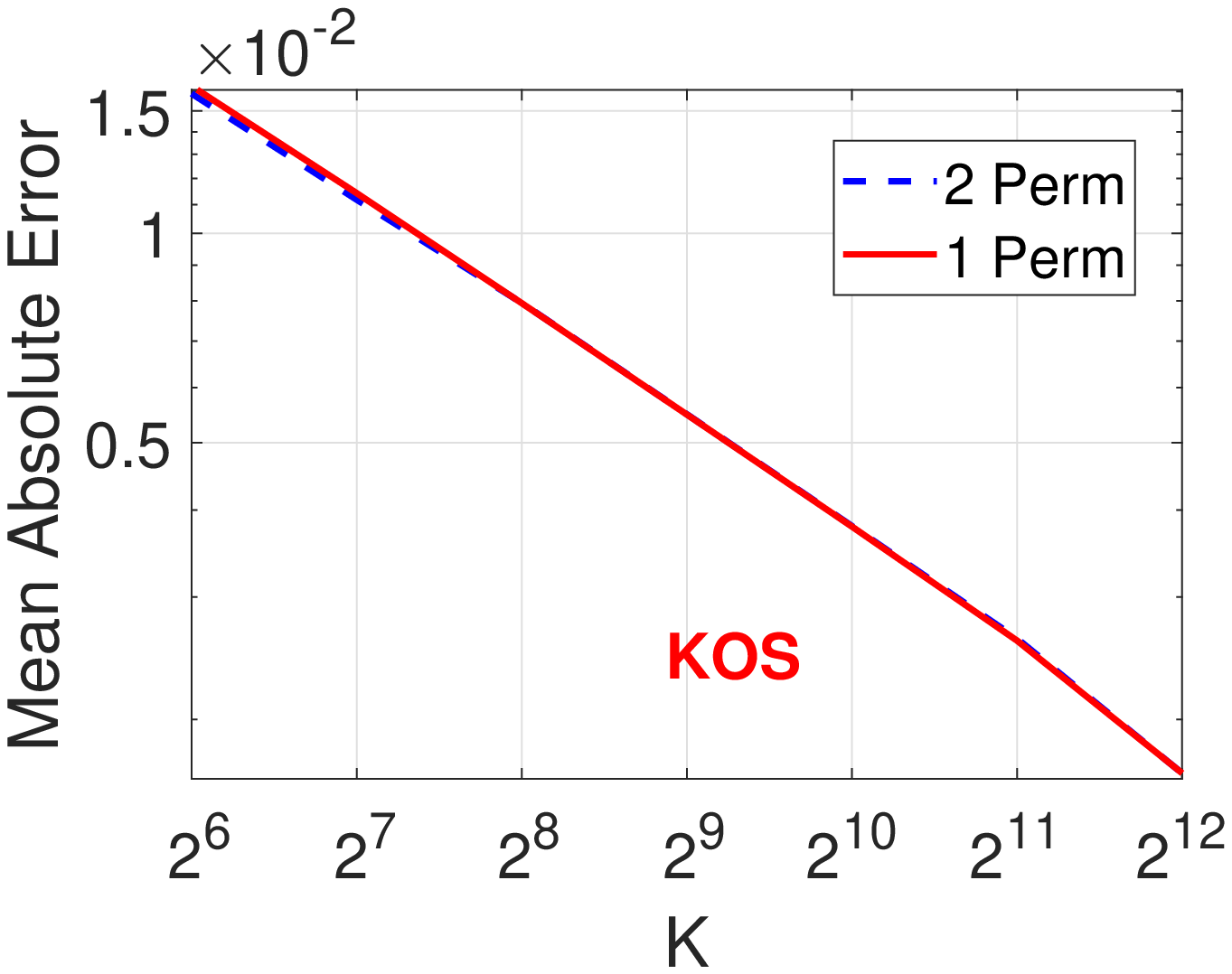}\hspace{-0.1in}
    \includegraphics[width=2.1in]{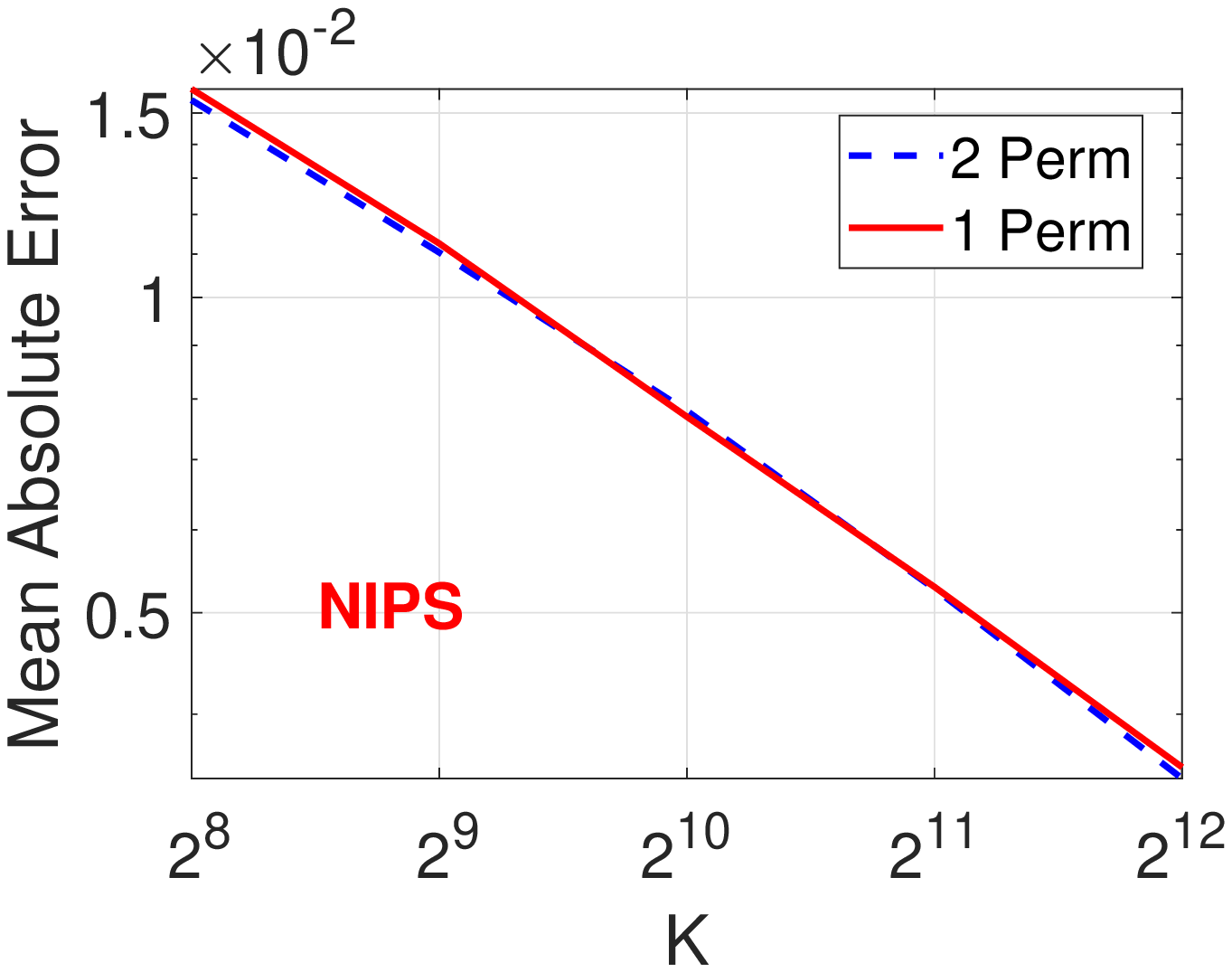}
    }
    \mbox{\hspace{-0.1in}
    \includegraphics[width=2.1in]{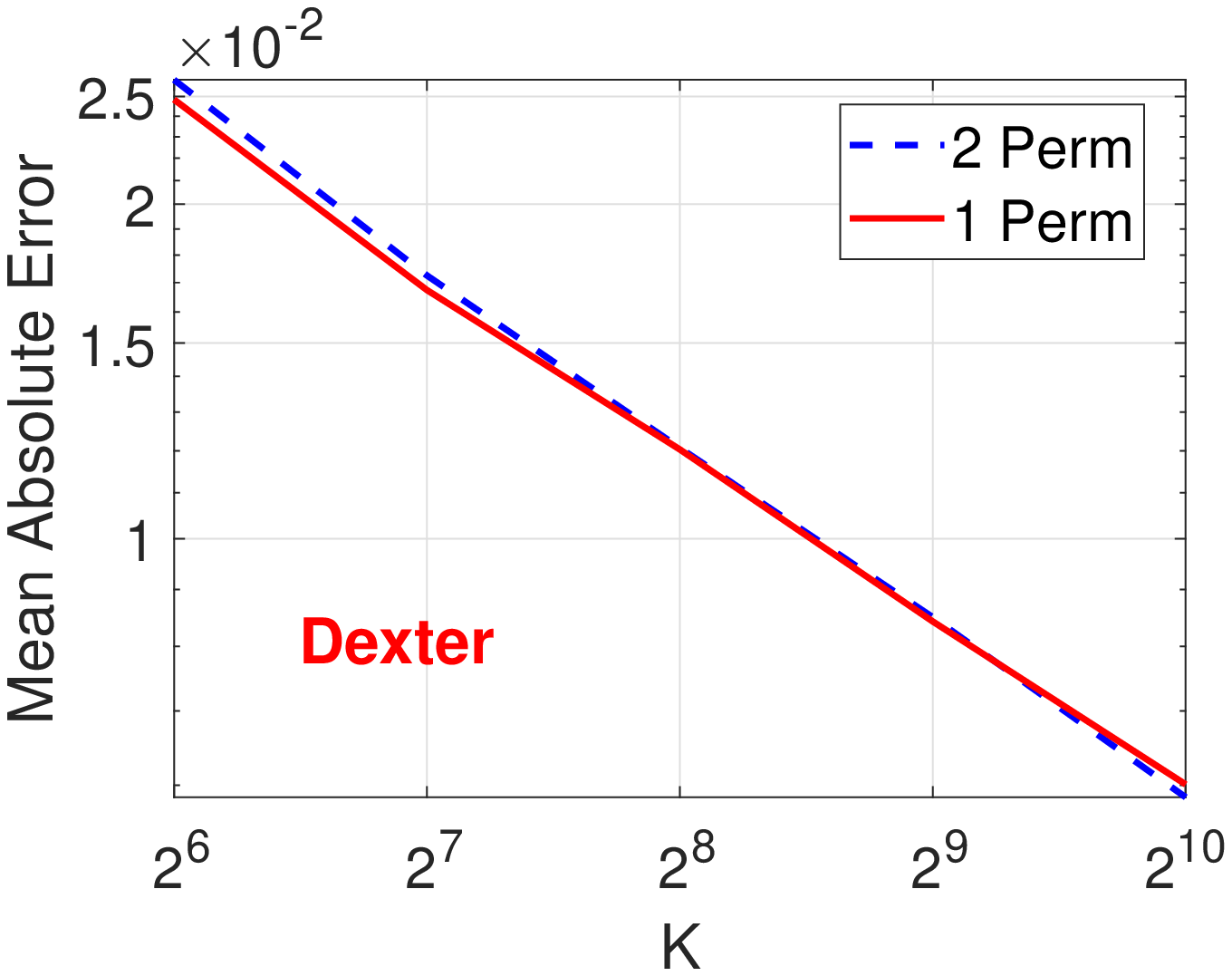}\hspace{-0.1in}
    \includegraphics[width=2.1in]{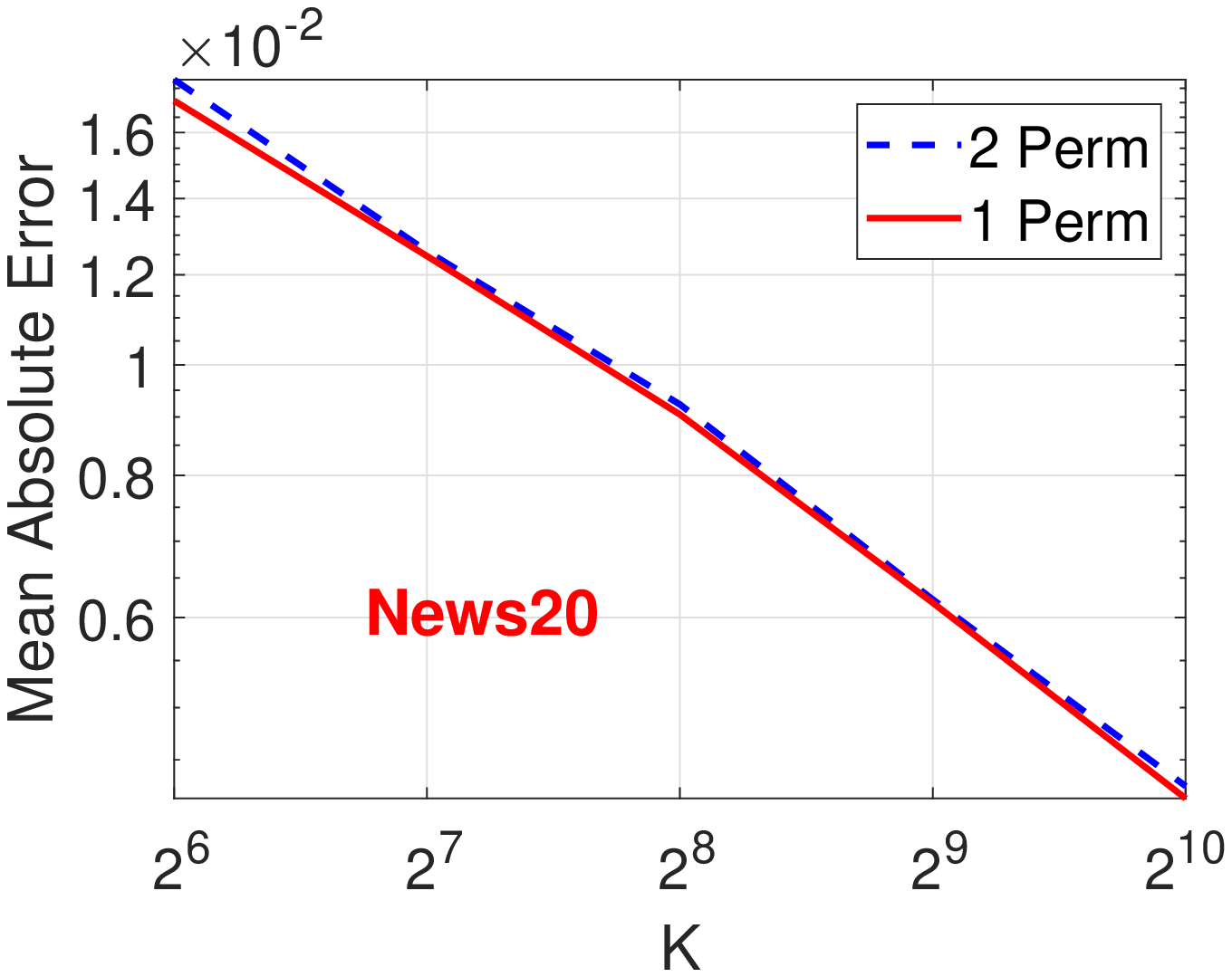}\hspace{-0.1in}
    \includegraphics[width=2.1in]{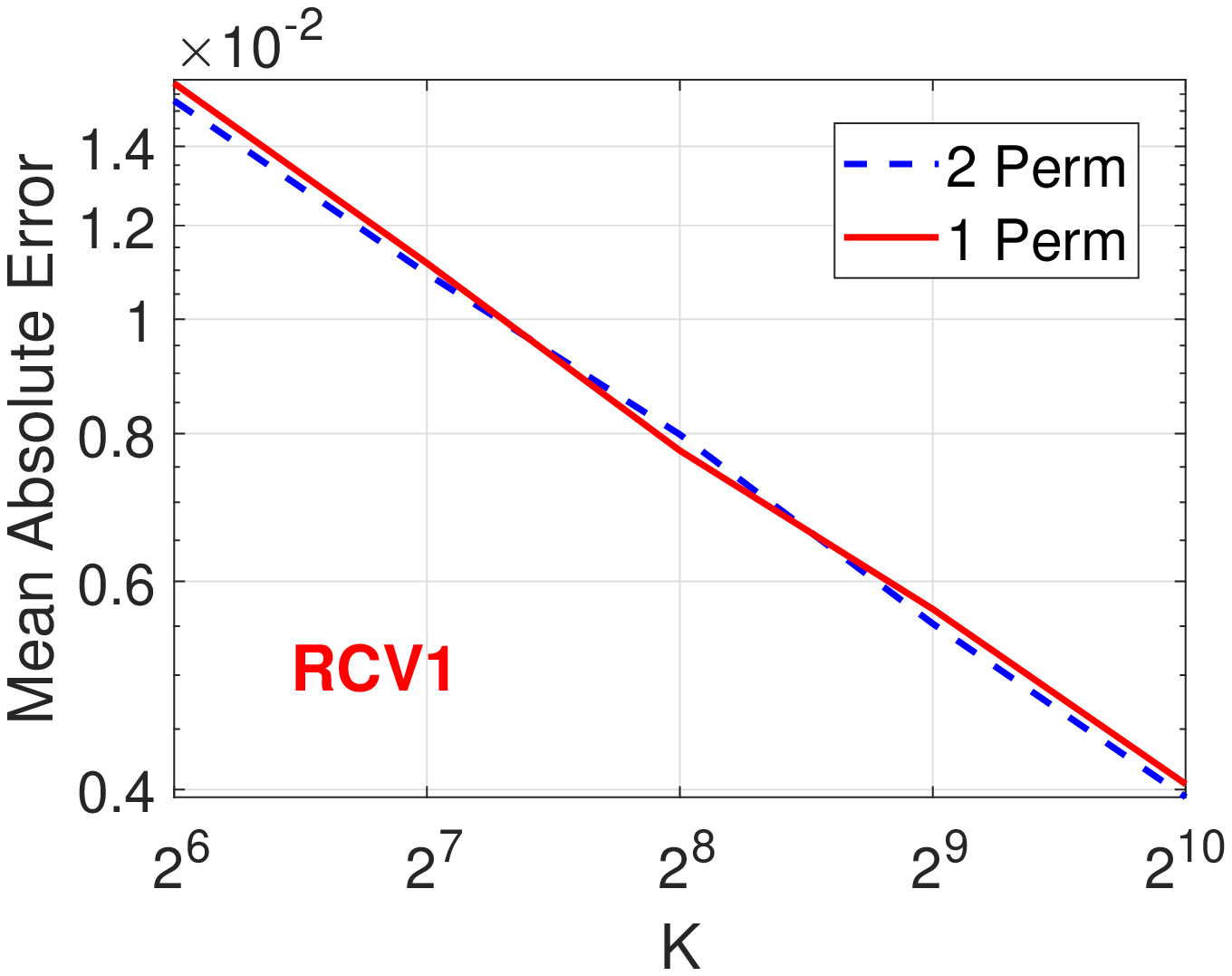}
    }
    \mbox{\hspace{-0.1in}
    \includegraphics[width=2.1in]{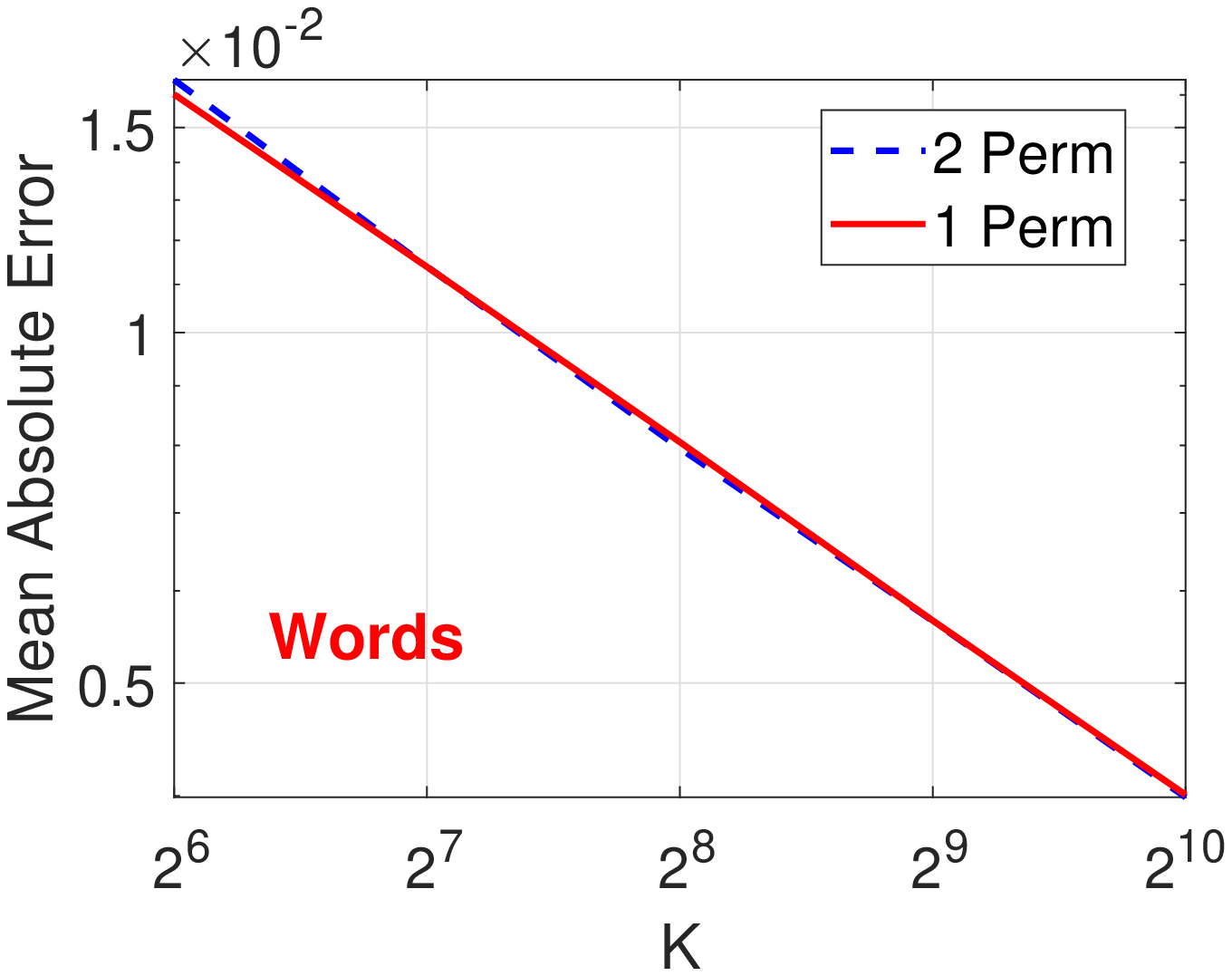}\hspace{-0.1in}
    \includegraphics[width=2.1in]{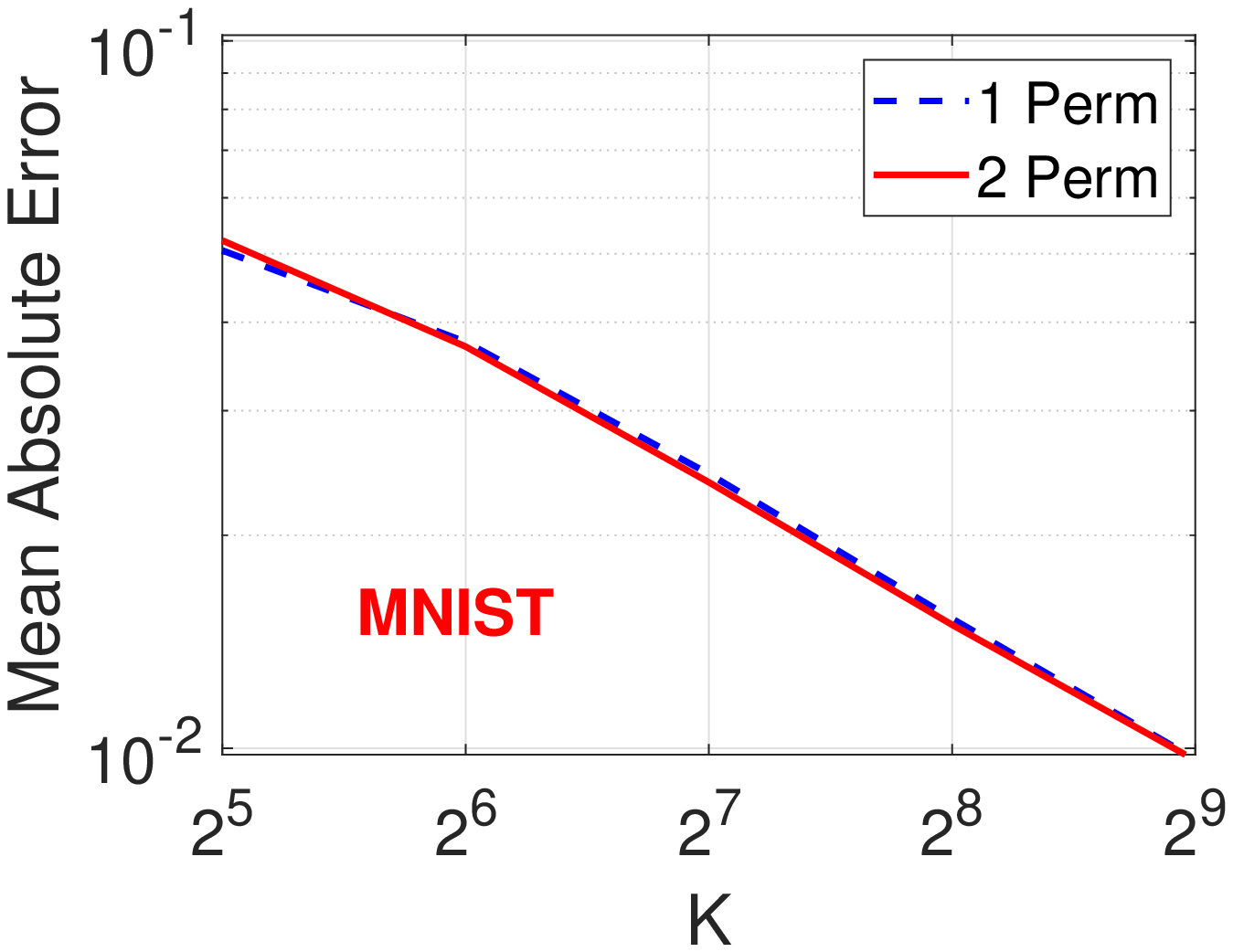}\hspace{-0.1in}
    \includegraphics[width=2.1in]{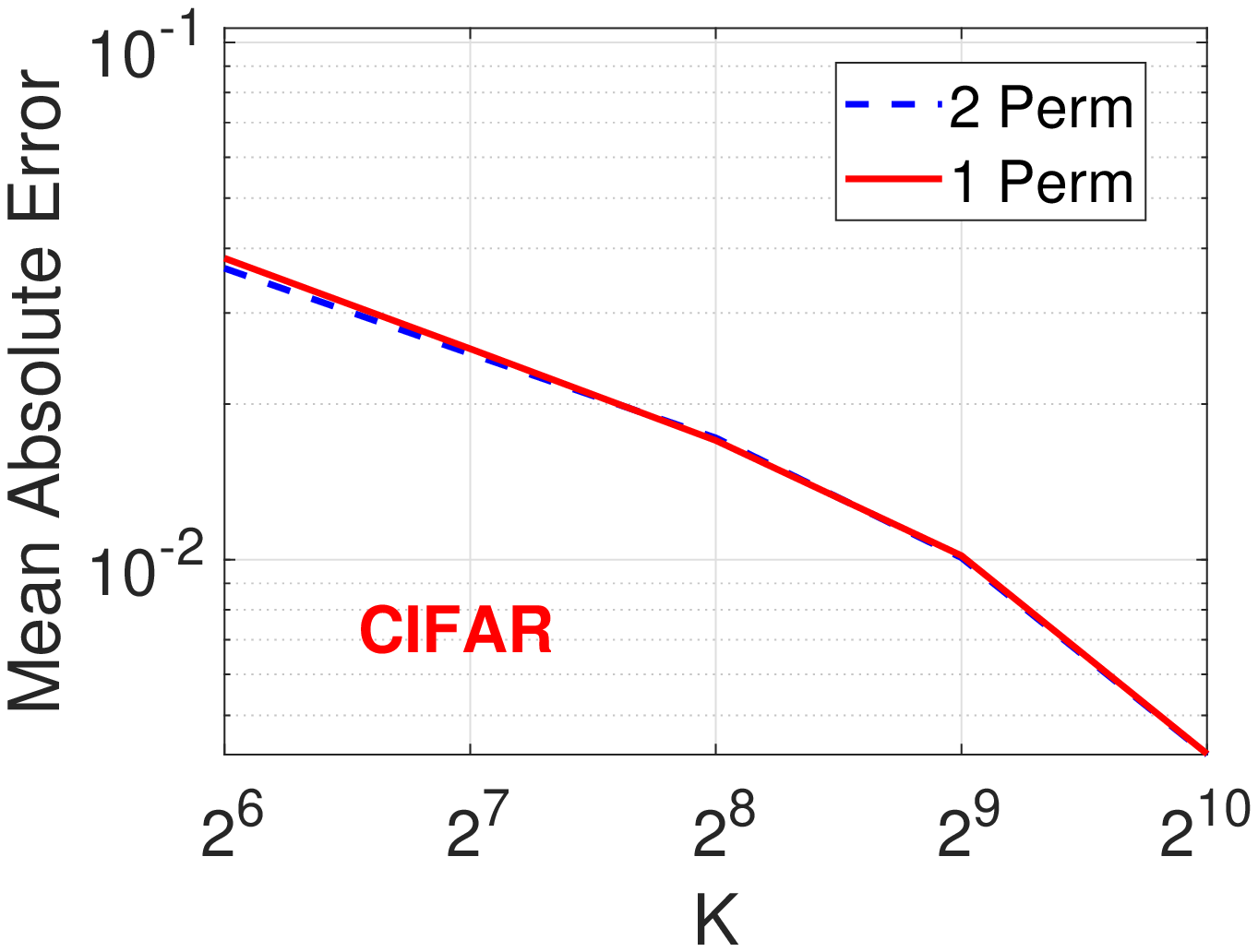}
    }

  \end{center}
  \vspace{-0.1in}
  \caption{Empirical mean absolute errors (MAEs) of C-MinHash-$(\pi,\pi)$ (``1 Perm'', red, solid) vs. C-MinHash-$(\sigma,\pi)$ (``2 Perm'', blue, dashed) on  9 datasets.  }
  \label{fig:dataset}
\end{figure}


\subsection{Comparing Estimation Errors in All Pairs for Several Datasets}  \label{sec:experiment dataset}

A total of 9 datasets are used in this set of experiments, including the  ``Words''  dataset ($D=65,536$)~\citep{Proc:Li_Church_EMNLP05}. The ``KOS'' blog entries dataset ($D=6,906$), the ``NIPS'' full paper dataset ($D=12,419$), and the ``Dexter'' dataset ($D=20,000$) are publicly available from the UCI machine learning repository~\citep{UCI}. We also use the ``BBC'' news dataset~\citep{greene06icml}  ($D=9,635$). In addition, we also test the ``News20'' dataset ($D=1,355,191$) and ``RCV1'' dataset ($D=47,236$) from the LIBSVM website~\citep{Article:LIBSVM}. Lastly, we also use the popular ``MNIST'' dataset~\citep{mnist} ($D=784$) and the ``CIFAR'' dataset~\citep{cifar} ($D=1,024$), by binarizing every entry.

\vspace{0.1in}

For this set of experiments, to report the accuracy, we choose the measure of ``mean absolute error (MAE)'', which is different from MSE. Given a dataset with $n$ data vectors, there are in total $n(n-1)/2$ pairs. Unless $n$ is  small, we cannot repeat the experiments many (say $10^4$) times in order to reliably estimate the MSE for every data vector pair. Thus, for each data vector pair, we compute the absolute error $|\hat J - J|$ and average the errors over all $n(n-1)/2$ pairs to obtain the MAE for this dataset. Finally, we report the averaged MAE from 10 repetitions for each dataset, as presented in Figure~\ref{fig:dataset}, for both C-MinHash-$(\sigma,\pi)$ and C-MinHash-$(\pi,\pi)$. The plots show that the curves for these two estimators ($\hat J_{\sigma,\pi}$, and $\hat J_{\pi,\pi}$) match well. Note that, with only 10 repetitions, it is expected that the two curves on each plot should have some (very small) discrepancies.

\section{Conclusion}

The classical MinHash requires applying $K$ independent permutations on the data where $K$, depending on applications, can be several hundreds or even several thousands. The recently proposed hashing algorithm  C-MinHash-$(\sigma,\pi)$~\citep{CMH2Perm2021} needs just two permutations: an initial permutation $\sigma$ is applied as a pre-processing step to break whatever structures which might exist in the original data, and a second permutation $\pi$ is re-used $K$ times to generate $K$ hash values, in a circulant shifting manner. It was shown in~\cite{CMH2Perm2021} that the prepocessing step is crucial and should not be skipped.

\vspace{0.1in}

\noindent In this paper, we develop another variant named C-MinHash-$(\pi,\pi)$. That is, we use the same permutation $\pi$ for both the initial pre-processing step and the subsequent hashing step. While the idea is intuitive, the theoretical analysis of C-MinHash-$(\pi,\pi)$ becomes  sophisticated. Nevertheless, we are able to derive the expectation of the estimator for C-MinHash-$(\pi,\pi)$. Although the estimator is slightly biased, the bias (and bias$^2$) is so small that it can be safely ignored. An extensive experimental study has confirmed that, in terms of the estimation accuracy, C-MinHash-$(\pi,\pi)$ behaves essentially the same as C-MinHash-$(\sigma,\pi)$.

\newpage
\appendix

\section{Proof of Theorem \ref{theo:mean-M2}}  \label{sec:append proof}

\begin{proof}

Denote $\mathbb E[\mathbbm 1_k]\vcentcolon = \mathbb E[\mathbbm 1\{h_k(\bm v)=h_k(\bm w)\}]$ for any $1\leq k\leq K$. We first recall some notations. We have $\bm v,\bm w\in \{0,1\}^D$, and $a$ and $f$ are defined in (\ref{def:a,f}). Denote $\mathcal B_1=\{i:\bm x_i=O\}$, $\mathcal B_2=\{i:\bm x_i=\times\}$ and $\mathcal B_3=\{i:\bm x_i=-\}$ as the sets of three types of points, respectively. For $a\leq j\leq D$ and $1\leq k\leq K$, define
\begin{align*}
    &\mathcal A_-(j)=\{\bm x_i:(i+k-1\  mod\ D)+1\leq j\},\\
    &\mathcal A_+(j)=\{\bm x_i:(i+k-1\  mod\ D)+1> j\}.
\end{align*}
Let $n_{-,1}(j)=|\{\bm x_i=O:i\in\mathcal A_-(j)\}|$ be the number of ``$O$'' points in $\mathcal A_-(j)$. Analogously let $n_{-,2}(j),n_{-,3}(j)$ be the number of ``$\times$'' and ``$-$'' points in $\mathcal A_-(j)$, and $n_{+,1}(j),n_{+,2}(j),n_{+,3}(j)$ be the number of ``$O$'', ``$\times$'' and ``$-$'' points in $\mathcal A_+(j)$. For any $i$, denote $i^*=(i+k-1\ mod\ D)+1$,  $i^\#=(i-k-1\ mod\ D)+1$.

Our analysis starts with the decomposition of hash collision probability,
\begin{align}\label{thm4-eqn1}
    \mathbb E[\mathbbm 1_k]=P\Big[h_k(\bm v)=h_k(\bm w)\Big]=\sum_{j=1}^D P\Big[h_k(\bm v)=h_k(\bm w)=j\Big],
\end{align}
where recall $h(\cdot)$ is the hash sample. Consider the process for generating the hash. As before, we look at the location vector $\bm x$. In Method 2, we first permute $\bm x$ by $\pi$ to get $\pi(\bm x)$. Then the $k$-th hash samples collide if the minimum of $\pi_{\rightarrow k}(\pi(\bm x))$ is ``$O$''. One key observation is that, when applying $\pi_{\rightarrow k}$, the random index for the $i$-th element in $\pi(\bm x)$ is exactly the one used for $\bm x_{i^\#}$ (shifted backwards) in the initial permutation. A toy example in provided in Figure \ref{fig:collision} to help understand the reasoning.

\begin{figure}[h!]
\centering
\includegraphics[width=2.5in]{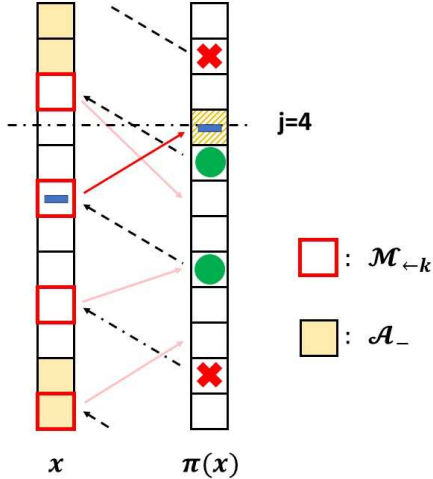}

\caption{Illustration of C-MinHash-$(\pi,\pi)$ hash collision, with $k=2$. Here, ``circulant right'' means ``circulant down''. Small indices correspond to upper elements.}
\label{fig:collision}
\end{figure}

Further denote set $\mathcal M=\{\pi(i):\pi(i)\notin \mathcal B_3\}$ be the collection of indices of initially permuted vector $\pi(\bm x)$ that are not ``$-$'' points, and $\mathcal M_{\leftarrow k}=(\mathcal M-k-1\ mod\ D)+1$ be the corresponding indices shifted backwards. In Figure \ref{fig:collision}, $\mathcal M=\{2,5,8,11\}$. Also, $\pi(6)=4$, and the permutation maps are described by the red arrows. Consequently, in $\pi_{\rightarrow k}(\pi(\bm x))$, the $8$-th element (``$O$'') in $\pi(x)$ will be permuted to the same index of the $8-2=6$-th element in $\bm x$, which equals to $4$. It is important to notice that, when considering the $k$-th collision, only points in $\mathcal M_{\leftarrow k}$ matters. Hence, we deduct:
\begin{itemize}
    \item \textbf{(Collision condition)} Denote $i=\argmin_{t\in\mathcal M_{\leftarrow k}} \pi(\bm x_t)$ be the location of minimal permutation index in $\mathcal M_{\leftarrow k}$. The $k$-th collision occurs at $j$ i.f.f. $\pi(i)=j$, and the $i^*$-th element in $\pi(\bm x)$ must be a ``O'' point. Recall the definition $i^*=(i+k-1\ mod\ D)+1$.
\end{itemize}

In Figure \ref{fig:collision}, consider $i=6$ and $j=4$ for example. Above condition means that the $i=6$-th element in $\bm x$ (``$-$'' in red bold border) is permuted to the $j=4$-th position, and it is above all other permuted elements with red bold borders. Meanwhile, the $i+k=6+2=8$-th element in $\pi(\bm x)$ must be a ``$O$''. Figure \ref{fig:collision} exactly satisfies the condition, so it depicts a collision. Mathematically, we have
\begin{align}
    \mathbb E[\mathbbm 1_k]&=\sum_{j=1}^D P\Big[h_k(\bm v)=h_k(\bm w)=j\Big] \nonumber\\
    &=\sum_{j=1}^D \sum_{i=1}^D P\Big[\pi(i)=j,\pi^{-1}(i^*)\in\mathcal B_1\Big], \label{thm4-eqn2}
\end{align}
where $i=\argmin_{t\in\mathcal M_{\leftarrow k}} \pi(\bm x_t)$. In this expression, everything is random of $\pi$, except for the set $\mathcal B_1$ which is fixed given the data.

Now we will focus on deriving the probability for a fixed $i$ and $j$ in (\ref{thm4-eqn2}). Our analysis will be conditional on the collection of variables $Z$ which is defined as follows. Let $z_{-,1}$, $z_{-,2}$ and $z_{-,3}$ be the number of ``$O$'', ``$\times$'' and ``$-$'' points in $\mathcal A_-(j)\cap \mathcal M_{\leftarrow k}^c$, and $z_{+,1}$, $z_{+,2}$ and $z_{+,3}$ be the number of ``$O$'', ``$\times$'' and ``$-$'' points in $\mathcal A_+(j)\cap \mathcal M_{\leftarrow k}^c$, respectively. Here $\mathcal M_{\leftarrow k}^c$ represents the complement of $\mathcal M_{\leftarrow k}$. Notice that $Z$ (and its density function) depends on different $j$ since $\mathcal A_-(j)$ and $\mathcal A_+(j)$ depends on $j$. For the ease of notation we suppress the information of $j$ in $Z$ (and $z$'s). It is easy to see that $Z=(z_{-,k}|_1^3,z_{+,k}|_1^3)$ follows hyper($D,D-f,n_{-,k}(j)|_1^3,n_{+,k}(j)|_1^3$). Denote the domain of $Z$ and $\Theta_j$. Conditional on $Z$, we obtain
\begin{align}
    \mathbb E[\mathbbm 1_k]&=\sum_{j=1}^D \sum_{Z\in\Theta_j} \sum_{i=1}^D P\Big[\pi(i)=j,\pi^{-1}(i^*)\in\mathcal B_1|Z\Big]P_j\Big[Z\Big] \nonumber\\
    &=\sum_{j=1}^D \sum_{Z\in\Theta_j} \sum_{i=1}^D P\Big[\pi^{-1}(i^*)\in\mathcal B_1|\pi(i)=j,Z\Big]P\Big[\pi(i)=j|Z\Big]P_j\Big[Z\Big] \nonumber\\
    &\triangleq \sum_{j=1}^D \sum_{Z\in\Theta_j} \sum_{i=1}^D \Gamma(i,j) P_j\Big[Z\Big], \label{thm4-E1s}
\end{align}
with $i=\argmin_{t\in\mathcal M_{\leftarrow k}} \pi(\bm x_t)$. We will carefully compute the probabilities in the summation. Basically, the key is that elements in $\mathcal M$ need to be controlled, i.e. smaller than $j$, and other positions can be arbitrary. Given $Z$, this means that we need to put $r_1=a-z_{-,1}-z_{+,1}$ type ``$O$'' points, $r_2=f-a-z_{-,2}-z_{+,2}$ type ``$\times$'' points and $r_3=D-f-z_{-,3}-z_{+,3}$ type ``$-$'' points no smaller than $j$, with $\pi(i)=j$ exactly. Also note that there are fixed $b_0=\sum_{k=1}^3z_{+,k}$ type ``$-$'' points no smaller than $j$.

With all these definitions and reasoning, we are ready to proceed with the proof. Based on $\bm x_i$, we have three general cases.

\textbf{1) $\bm x_i\in\mathcal B_1$.}\hspace{0.1in}The first case is that $\bm x_i=O$.

\textbf{Case 1a) $j<i^*$.} Firstly, we consider the case where $j<i^*$. By combinatorial theory we have
\begin{align}
    P\Big[\pi(i)=j|Z\Big]&=P\Big[\pi(i)=j|\pi^{-1}(j)\notin \mathcal B_3,Z\Big]P\Big[\pi^{-1}(j)\notin \mathcal B_3|Z\Big] \nonumber\\
    &=\frac{1}{r_1+r_2}\frac{\binom{b_0}{r_3}\binom{D-j-b_0}{r_1+r_2-1}}{\binom{D-f}{r_3}\binom{f}{r_1+r_2}}P\Big[\pi^{-1}(j)\notin \mathcal B_3|Z\Big] \nonumber\\
    &\triangleq \tilde P_1\cdot P\Big[\pi^{-1}(j)\notin \mathcal B_3|Z\Big], \label{thm4-eqn3}
\end{align}
where the second probability is that the $j$-th element in $\pi(\bm x)$ is not ``$-$''. Conditional on $Z$, the probability is dependent on $j^\#$:
\begin{align*}
    P\Big[\pi^{-1}(j)\notin B_3|Z\Big]=\sum_{p=1}^3 \mathbbm 1\{j^\#\in \mathcal B_p\} (1-\frac{z_{-,p}}{n_{-,p}(j)}).
\end{align*}
Combining with (\ref{thm4-eqn3}) we obtain
\begin{align}
    P\Big[\pi(i)=j|Z\Big]=\tilde P_1\sum_{p=1}^3 \mathbbm 1\{j^\#\in \mathcal B_p\} (1-\frac{z_{-,p}}{n_{-,p}(j)}). \label{thm4-eqn4}
\end{align}
Next we compute $P[\pi^{-1}(i^*)\in\mathcal B_1|\pi(i)=j,Z]$. Note that given the conditions, $\pi^{-1}(i^*)$ has two cases: 1) it comes from $\mathcal M_{\leftarrow k}$ (i.e. it is one of the elements with red bold border); 2) Otherwise. We then can write
\begin{align}
    &P\Big[\pi^{-1}(i^*)\in\mathcal B_1|\pi(i)=j,Z\Big] \nonumber\\
    &=P\Big[\pi^{-1}(i^*)\in\mathcal B_1|\pi^{-1}(i^*)\in\mathcal M_{\leftarrow k},\pi(i)=j,Z\Big]P\Big[\pi^{-1}(i^*)\in\mathcal M_{\leftarrow k}|\pi(i)=j,Z\Big] \nonumber\\
    &\hspace{0.5in}+P\Big[\pi^{-1}(i^*)\in\mathcal B_1|\pi^{-1}(i^*)\notin\mathcal M_{\leftarrow k},\pi(i)=j,Z\Big]P\Big[\pi^{-1}(i^*)\notin\mathcal M_{\leftarrow k}|\pi(i)=j,Z\Big] \nonumber\\
    &=(1-\frac{z_{+,1}}{n_{+,1}(j)})\left[\frac{r_1+r_2-1}{D-j-b_0}\frac{r_1-1}{r_1+r_2-1}+(1-\frac{r_1+r_2-1}{D-j-b_0})\frac{a-r_1}{f-r1-r2}\right] \nonumber\\
    &\triangleq (1-\frac{z_{+,1}}{n_{+,1}(j)})\left[\frac{r_1-1}{D-j-b_0}+(1-\frac{r_1+r_2-1}{D-j-b_0})J^*\right] \nonumber\\
    &\triangleq (1-\frac{z_{+,1}}{n_{+,1}(j)}) \bar J_1. \label{thm4-eqn5}
\end{align}
Combining (\ref{thm4-eqn4}) and (\ref{thm4-eqn5}), we obtain when $i\in\mathcal B_1$ and $j<i^*$,
\begin{align}
    \Gamma(i,j)=\sum_{p=1}^3 \mathbbm 1\{j^\#\in \mathcal B_p\} (1-\frac{z_{-,p}}{n_{-,p}(j)})(1-\frac{z_{+,1}}{n_{+,1}(j)}) \tilde P_1 \bar J_1. \label{thm4-1<}
\end{align}

\textbf{Case 1b) $j=i^*$.} Similarly approach also applies to the situation with $j=i^*$. In this case,
\begin{align}
    P\Big[\pi^{-1}(j)\notin \mathcal B_3|Z\Big]=(1-\frac{z_{-,1}}{n_{-,1}(j)})\tilde P_1, \hspace{0.2in} P[\pi^{-1}(i^*)\in\mathcal B_1|\pi(i)=j,Z]=1. \label{thm4-1=}
\end{align}
The equations are because $(i^*)^\#=i\in \mathcal B_1$, and equivalently, $\pi^{-1}(i^*)=\pi^{-1}(j)=i\in \mathcal B_1$.

\textbf{Case 1c) $j>i^*$.} I this case, we still have
\begin{align*}
    P\Big[\pi(i)=j|Z\Big]=\tilde P_1\sum_{p=1}^3 \mathbbm 1\{j^\#\in \mathcal B_p\} (1-\frac{z_{-,p}}{n_{-,p}(j)}),
\end{align*}
but the probability of $\pi_{-1}(i^*)$ being ``$O$'' is different. Since $j>i^*$, this event now depends on $z_{-,p}$, $p=1,2,3$. More specifically,
\begin{align*}
    &P[\pi^{-1}(i^*)\in\mathcal B_1|\pi(i)=j,Z] \\
    &=\left[\mathbbm 1\{j^\#\in \mathcal B_1\}(1-\frac{z_{-,1}}{n_{-,1}(j)-1})+\sum_{p=2,3} \mathbbm 1\{j^\#\in \mathcal B_p\}(1-\frac{z_{-,p}}{n_{-,p}(j)})\right] J^*.
\end{align*}
Therefore, when $i\in\mathcal B_1$ and $j>j^*$, it holds that
\begin{align}
    \Gamma(i,j)=(1-\frac{z_{-,1}}{n_{-,1}(j)})\left[\mathbbm 1\{j^\#\in \mathcal B_1\}(1-\frac{z_{-,1}}{n_{-,1}(j)-1})+\sum_{p=2,3} \mathbbm 1\{j^\#\in \mathcal B_p\}(1-\frac{z_{-,p}}{n_{-,p}(j)})\right] J^*.
    \label{thm4-1>}
\end{align}

\textbf{2) $\bm x_i\in\mathcal B_2$.} The case where $\bm x_i\in\mathcal B_2$ can be analyzed using similar arguments. For conciseness, we mainly present the final results.

\textbf{Case 2a) $j<i^*$.} The calculation ends up in the same form. We have
\begin{align*}
    P\Big[\pi(i)=j|Z\Big]=\tilde P_2\sum_{p=1}^3 \mathbbm 1\{j^\#\in \mathcal B_p\} (1-\frac{z_{-,p}}{n_{-,p}(j)}),
\end{align*}
with $\tilde P_2=\tilde P_1$. In addition,
\begin{align*}
    P\Big[\pi^{-1}(i^*)\in\mathcal B_1|\pi(i)=j,Z\Big]= (1-\frac{z_{+,2}}{n_{+,2}(j)}) \bar J_2,
\end{align*}
where $\bar J_2=\frac{r_1}{D-j-b_0}+(1-\frac{r_1+r_2-1}{D-j-b_0})J^*$. Hence, when $\bm x_i\in\mathcal B_2$ and $j<i^*$, we have
\begin{align}
    \Gamma(i,j)=\sum_{p=1}^3 \mathbbm 1\{j^\#\in \mathcal B_p\} (1-\frac{z_{-,p}}{n_{-,p}(j)})(1-\frac{z_{+,2}}{n_{+,2}(j)}) \tilde P_2 \bar J_2. \label{thm4-2<}
\end{align}

\textbf{Case 2b) $j=i^*$.} In this case, $\Gamma(i,j)$ simply equals to $0$, since $\pi^{-1}(i^*)=\pi^{-1}(j)\in \mathcal B_2$. The probability of $\pi^{-1}(i^*)$ being ``$O$'' is $0$.

\textbf{Case 2c) $j>i^*$.} Omitting the details, we have
\begin{align}
    \Gamma(i,j)=(1-\frac{z_{-,2}}{n_{-,2}(j)})\left[\mathbbm 1\{j^\#\in \mathcal B_2\}(1-\frac{z_{-,2}}{n_{-,2}(j)-1})+\sum_{p=1,3} \mathbbm 1\{j^\#\in \mathcal B_p\}(1-\frac{z_{-,p}}{n_{-,p}(j)})\right] J^*.
    \label{thm4-2>}
\end{align}

\textbf{2) $\bm x_i\in\mathcal B_3$.}

\textbf{Case 3a) $j<i^*$.} The expression is different from previous two, in that we need $\pi^{-1}(j)\in B_3$.
\begin{align*}
    P\Big[\pi(i)=j|Z\Big]&=P\Big[\pi(i)=j|\pi^{-1}(j)\in B_3,Z\Big]P\Big[\pi^{-1}(j)\in B_3|Z\Big] \\
    &=\tilde P_3\sum_{p=1}^3 \mathbbm 1\{j^\#\in \mathcal B_p\} \frac{z_{-,p}}{n_{-,p}(j)},
\end{align*}
where
\begin{align*}
    \tilde P_3=\frac{1}{r_3}\frac{\binom{b_0}{r_3-1}\binom{D-j-b_0}{r_1+r_2}}{\binom{D-f}{r_3}\binom{f}{r_1+r_2}}.
\end{align*}
Moreover, we have
\begin{align*}
    P\Big[\pi^{-1}(i^*)\in\mathcal B_1|\pi(i)=j,Z\Big]= (1-\frac{z_{+,3}}{n_{+,3}(j)}) \bar J_3,
\end{align*}
with $\bar J_3=\frac{r_1}{D-j-b_0}+(1-\frac{r_1+r_2}{D-j-b_0})J^*$. Combining parts together we obtain
\begin{align}
    \Gamma(i,j)=\sum_{p=1}^3 \mathbbm 1\{j^\#\in \mathcal B_p\} \frac{z_{-,p}}{n_{-,p}(j)}(1-\frac{z_{+,3}}{n_{+,3}(j)}) \tilde P_3 \bar J_3. \label{thm4-3<}
\end{align}

\textbf{Case 3b) $j=i^*$.} By same reasoning as Case 2a), $\Gamma(i,j)=0$.

\textbf{Case 3c) $j>i^*$.} We have in this case
\begin{align}
    \Gamma(i,j)&=\left[\mathbbm 1\{j^\#\in \mathcal B_3\}\frac{z_{-,3}}{n_{-,3}(j)}\frac{n_{-,3}(j)-z_{-,3}}{n_{-,3}(j)-1}+(1-\frac{z_{-,3}}{n_{-,3}(j)})\sum_{p=1,2} \mathbbm 1\{j^\#\in \mathcal B_p\}\frac{z_{-,p}}{n_{-,p}(j)}\right] J^*  \nonumber\\
    &=\sum_{p=1}^3 \mathbbm 1\{j^\#\in \mathcal B_p\}\frac{z_{-,p}}{n_{-,p}(j)}(1-\frac{z_{-,3}-\mathbbm 1\{p=3\}}{n_{-,3}(j)}-\mathbbm 1\{p=3\}) J^*.
    \label{thm4-3>}
\end{align}
Finally, combining (\ref{thm4-E1s}), (\ref{thm4-1<}), (\ref{thm4-1=}), (\ref{thm4-1>}), (\ref{thm4-2<}), (\ref{thm4-2>}), (\ref{thm4-3<}) and (\ref{thm4-3>}), and re-organizing terms, the proof is complete.

\end{proof}

\vspace{0.5in}

\bibliography{standard}
\bibliographystyle{plainnat}

\end{document}